\documentclass[a4paper,10pt]{article}
\pdfoutput=1 %

\usepackage[T1]{fontenc}
\usepackage[english]{babel}
\usepackage{amssymb,amsmath}
\usepackage[top=1in, bottom=1in, left=1in, right=1in]{geometry}\usepackage{float}
\usepackage{graphicx}
\usepackage[font=scriptsize]{subcaption}
\usepackage[font=small]{caption}
\usepackage{float}
\usepackage{pgf}
\usepackage{tikz}
\usepackage{physics}
\usetikzlibrary{arrows.meta}
\usepackage{mathtools}
\usepackage{sidecap}
\usepackage{xfrac}

\usepackage{relsize}

\usetikzlibrary{positioning}
\tikzset{main node/.style={circle,fill=blue!20,draw,pattern=crosshatch dots,pattern color=grey,minimum size=1.3cm,inner sep=0pt},
            }
\tikzset{main node2/.style={circle,fill=blue!40,draw,pattern=crosshatch dots,pattern color=grey,minimum size=3cm,inner sep=0pt},
            }
            
\usepackage{tikz-feynman}
            
\tikzfeynmanset{
  fermion1/.style={
    /tikz/postaction={
      /tikz/decoration={
        markings,
        mark=at position 0.5 with {
          \node[
            transform shape,
            xshift=-0.5mm,
            fill,
            inner sep=0.5pt,
            draw=none,
            dart
          ] { };
        },
      },
      /tikz/decorate=true,
    },
  },
  fermion2/.style={
    /tikz/postaction={
      /tikz/decoration={
        markings,
        mark=at position 0.5 with {
          \arrow{[length=4pt, width=3pt]};
        },
      },
      /tikz/decorate=true,
    },
  },
}

\usetikzlibrary{backgrounds,calc}
\usepackage{multicol}
\usetikzlibrary{positioning}
\tikzfeynmanset{compat=1.1.0}
\definecolor{grey}{rgb}{0.7,0.7,0.7}
\usetikzlibrary{patterns}
\usepackage{lscape}
\usepackage{xcolor}
\usepackage{multirow}

\usepackage{jheppub} %
\gdef\@fpheader{Prepared for submission to JHEP}
\gdef\@journal{jhep}

\usetikzlibrary{positioning}
\tikzset{main node/.style={circle,fill=blue!20,draw,pattern=crosshatch dots,pattern color=grey,minimum size=1.3cm,inner sep=0pt},
            }
\tikzset{main node2/.style={circle,fill=blue!40,draw,pattern=crosshatch dots,pattern color=grey,minimum size=3cm,inner sep=0pt},
            }
\usepackage{tikz-feynman}
\usetikzlibrary{backgrounds,calc}
\usepackage{multicol}
\usetikzlibrary{positioning}
\tikzfeynmanset{compat=1.1.0}
\definecolor{grey}{rgb}{0.7,0.7,0.7}
\usetikzlibrary{patterns}
\usepackage{xcolor}
\usepackage{slashed}

\def\LTD2{Local Unitarity}
\def\alphaLoop{$\alpha${\sc\small Loop}}
\def\SE{self-energy}
\def\DT{double-triangle}

\chardef\MyArticleWithColor=\pdfcolorstackinit page direct{0 g}
\def\cmtVH#1{}
\def\cmtZC#1{}
\def\cmtBR#1{}
\def\cmtAP#1{}
\def\cmtZC#1{\emph{\pdfcolorstack\MyArticleWithColor push {0 1 0 rgb} Z.C. : #1 \pdfcolorstack\MyArticleWithColor pop}}

\long\def\vSWITCH#1#2{\ifdefined\vPUBLISHED #1 \else #2 \fi} %

\title{Local Unitarity: \\
A representation of differential cross-sections that is locally free of infrared singularities at any order}
\date{Oct 2020}

\author[]{Zeno Capatti,}
\author[]{Valentin Hirschi,}
\author[]{Andrea Pelloni}
\author[]{and Ben Ruijl}

\affiliation[]{ETH Z\"urich,\\
R\"amistrasse 101, %
8092 Z\"urich, Switzerland}

\emailAdd{zeno.ca@gmail.com}
\emailAdd{valentin.hirschi@gmail.com}
\emailAdd{apelloni90@gmail.com}
\emailAdd{benruyl@gmail.com}

\abstract{
We propose a novel representation of differential scattering cross-sections that locally realises the direct cancellation of infrared singularities exhibited by its so-called real-emission and virtual degrees of freedom. We take advantage of the Loop-Tree Duality representation of each individual forward-scattering diagram and we prove that the ensuing expression is locally free of infrared divergences, applies at any perturbative order and for any process without initial-state collinear singularities. Divergences for loop momenta with large magnitudes are regulated using local ultraviolet counterterms that reproduce the usual Lagrangian renormalisation procedure of quantum field theories.
Our representation is especially suited for a numerical implementation and we demonstrate its practical potential by computing fully numerically and without any IR counterterm the next-to-leading order accurate differential cross-section for the process $e^+ e^- \rightarrow d \bar{d}$.
We also show first results beyond next-to-leading order by computing interference terms part of the N$^4$LO-accurate inclusive cross-section of a $1\rightarrow 2+X$ scalar scattering process.
}

\begin{document}

\maketitle
\section{Introduction}

The ever-increasing need for generic and accurate Monte-Carlo simulations for collider experiments spurred the emergence of an entire subfield of the high energy physics community whose research activities are to a large extent motivated by fulfilling this demand. Over the last three decades, physicists pursued this goal by developing new computational techniques and studying the mathematical structure of perturbative computations in Quantum Field Theories (QFTs).

The 1990s saw the development of the first tools\footnote{e.g. {\sc\small form}~\cite{vanOldenborgh:1989wn,Vermaseren:2000nd}, {\sc\small grace}~\cite{Tanaka:1990wn,Fujimoto:2002sj}, {\sc\small FeynArts}~\cite{Kublbeck:1990xc}, {\sc\small MadGraph}~\cite{Stelzer:1994ta}, {\sc\small CompHEP}~\cite{Pukhov:1999gg} and {\sc\small amegic}~\cite{Krauss:2001iv}} 
for the automatic evaluation of tree-level scattering amplitudes. These efforts were naturally followed up by independent groups working on the corresponding phase-space integration programs, also called event generators, for automatically computing the associated differential cross-sections\footnote{e.g. {\sc\small Whizhard}~\cite{Ohl:2000pr}, {\sc\small MadEvent}~\cite{Maltoni:2002qb}, {\sc\small Sherpa}~\cite{Gleisberg:2003xi} and {\sc\small CalcHEP}~\cite{Pukhov:2004ca}}.
The following decade then witnessed the push for the automation of the computation of Next-to-Leading-Order (NLO) corrections by many members of the aforecited groups. This resulted in the so-called ``NLO revolution'' that cemented the arguably artificial divide between the task of computing loop amplitudes\footnote{e.g. {\sc\small MadLoop}~\cite{Hirschi:2011pa,Alwall:2014hca}, {\sc\small OpenLoops}~\cite{Cascioli:2011va}, {\sc\small BlackHat}~\cite{Bern:2013pya}, {\sc\small GoSam}~\cite{Cullen:2011ac} and {\sc\small recola}~\cite{Actis:2016mpe}} 
and that of regulating the infrared (IR) divergences of the phase-space integral by means of dedicated counterterms. Over the years, various strategies have been elaborated to this end, first at NLO~\cite{Frixione:1995ms,Catani:1996vz,Somogyi:2009ri} and then at Next-to-Next-to-Leading Order
(NNLO)~\cite{Currie:2016bfm,Czakon:2010td,Boughezal:2015dra,Somogyi:2005xz,DelDuca:2016ily,Grazzini:2017mhc,Cieri:2018oms,Boughezal:2016wmq,Cacciari:2015jma,Caola:2017dug,Herzog:2018ily,Magnea:2018hab}. These efforts have been complemented by advances in the analytic computation of (multi-)loop amplitudes that mostly follow a pipeline of distinct processing steps. Amplitudes are first projected onto scalar form factors that are reduced using integration-by-parts identities~\cite{Tkachov:1981wb,Chetyrkin:1981qh,Baikov:1996iu,Smirnov:1999wz,Larin:1991fz,Anastasiou:2000mf,Laporta:2001dd,Anastasiou:2004vj,Smirnov:2005ky,Lee:2008tj,Kant:2013vta,vonManteuffel:2014ixa,Mastrolia:2016dhn,Smirnov:2014hma,Ruijl:2017cxj,Maierhoefer:2017hyi,Smirnov:2019qkx,Peraro:2019svx,Frellesvig:2019uqt,Peraro:2016wsq,Boehm:2017wjc,Kosower:2018obg} 
so as to be expressed in terms of a smaller set of \emph{master} integrals. These irreducible loop integrals are then computed by means of differential equations~\cite{Gehrmann:1999as,Kotikov:1990kg,Papadopoulos:2014hla} which under certain conditions can be solved numerically~\cite{,Bonciani:2019jyb,Francesco:2019yqt,Czakon:2020vql} or by leveraging a detailed understanding of the mathematical structure of iterated integrals~\cite{Remiddi:1999ew,Goncharov:1998kja,Brown:2004ugm,Duhr:2012fh,Broedel:2017siw,Passarino:2016zcd,Ablinger:2017bjx,Broedel:2019hyg,Remiddi:2017har} yielding special functions with efficient numerical representations.

The large body of work cited in this concise summary of the state-of-the-art for the computation of higher-order corrections to differential cross-sections is a testament to its many successes and importance for collider phenomenology. 
However, its more recent progression also signals that the traditional pipeline is arguably facing a complexity barrier that is unlikely to be overcome by incremental progress. Instead, the situation calls for a radical change in methodology for addressing the root cause of this complexity increase, that is IR divergences, more efficiently than the canonical approach.
One such alternative is to part ways with the IR subtraction paradigm and aim at accommodating a more direct cancellation of real and virtual degrees of freedom. 
A possible avenue in that regard is that of the Reverse Unitarity~\cite{Anastasiou:2002yz,Anastasiou:2002wq,Anastasiou:2015ema,Anastasiou:2016cez,Duhr:2019kwi,Chen:2019lzz} approach. This technique turns the phase-space integrals of real-emission contributions into loop integrals so as to reduce the complete set of higher order contributions down to a small set of master integrals and eventually combine their divergences at the integrated level. This method produced milestone predictions for the N$^3$LO corrections to the Higgs hadro-production cross-section~\cite{Dulat:2018bfe} as well as N$^4$LO corrections to inclusive two-body decays~\cite{Herzog:2017dtz}.
However, because of its reliance on a completely analytical treatment of the loop integrals, reverse-unitarity cannot accommodate arbitrary observable functions or complicated multi-scale processes.

Another possible direction is to turn all loop integrals into phase-space integrals that can be performed numerically if the phase-space measure of resolved and unresolved degrees of freedom can be aligned so as to guarantee a local cancellation of all infrared divergences. This is the main goal of our paper and, perhaps contrary to the expectations of many, we show that it is possible to write the differential cross-section for an arbitrary process (without initial-state singularities) and at any perturbative order as an expression that is locally free of any IR singularities.
We refer to this rewriting of the differential cross-section as its \emph{Local Unitarity} (LU) representation. Its $\emph{Local}$ aspect stresses the applicability to arbitrary (IR-safe) observables, whereas \emph{Unitarity} highlights the direct combination of real and virtual degrees of freedom, i.e. emission and no-emission evolutions, into finite transition probabilities.

A first attempt at this programme at NLO accuracy goes back to 1999 with the avant-gardist work of D. Soper~\cite{Soper:1999xk,Soper:2001hu,Kramer:2002cd,Kramer:2003jk,Soper:2003ya,BeowulfProgram} who demonstrated its practical viability by applying it to a particular differential observable of the lepto-production of up to three partonic jets. However, the success of the competing approaches discussed earlier in this introduction diverted attention away from this limited form of Local Unitarity.
More importantly, its generalisation to arbitrary perturbative orders and processes as well as the computational techniques and resources necessary for its practical implementation were not available at the time.
Our work aims at leveraging recent progresses in the generalisation of the multi-Loop Tree Duality (LTD) relation~\cite{Catani:2008xa,Bierenbaum:2010cy,Runkel:2019yrs,Capatti:2019ypt,Capatti:2019edf,Driencourt-Mangin:2019aix,Verdugo:2020kzh,Aguilera-Verdugo:2020kzc, Capatti:2020ytd}, in order to provide solid theoretical foundations to Local Unitarity and demonstrate its potential for numerical applications. 
In particular, the alternative Manifestly Causal LTD (cLTD) representation recently presented in ref.~\cite{Capatti:2020ytd} plays a key role in our proof of local IR cancellations in Local Unitarity.
In what follows, we will discuss how the radically different perspective adopted by Local Unitarity allows to not \emph{solve} but rather \emph{avoid} altogether traditional core issues of perturbative computations, of which the regularisation of infrared singularities is a prime example.

Since R. Feynman drew his first eponymous diagram in 1948, his graphical representations have been the cornerstone of perturbative computations. Even seventy years later, the theoretical physics community still leans on Feynman diagrams to guide its intuition, although modern techniques use them mostly as a starting point for building the amplitude integrand which is then subject to heavy symbolic manipulations.
Instead, our Local Unitarity representation applies individually to each forward scattering diagram, called a \emph{supergraph}, which encompasses a particular subset of interferences of Feynman diagrams.
The LU representation enjoys a straightforward graphical interpretation as it is constructed entirely from the identification and characterisation of the singular surfaces of each supergraph.
The type of post-processing and analytic transformations of the integrand of supergraphs in LU implies that it retains a direct correspondence to the diagrammatic origin of each contribution. For example, this helps identify contributions that are non-abelian, $n_f$-dependent, planar or of electroweak origin.
Moreover, LU is formulated in momentum space which facilitates the study of any particular kinematic regime.
we apply LTD to analytically integrate over all loop energies of the supergraph.
We show that the resulting expression for each supergraph can be made locally finite, even though it is not gauge invariant nor does it satisfy any of the usual universal factorisation properties~\cite{Collins:1984kg,Collins:1988ig,Collins:1989gx,Akhoury:1978vq,Catani:1998bh,
Sterman:2002qn,Sterman:1995fz,Dixon:2008gr,Gardi:2009zv,Becher:2009cu,Feige:2014wja,Anastasiou:2018rib,Becher:2019avh,Anastasiou:2020sdt} relating processes of different multiplicities close to infrared limits. 
By relying solely on energy-momentum conservation, Local Unitarity exposes that the mechanism underlying the cancellation of infrared divergences~\cite{Bloch:1937pw,Kinoshita:1962ur,Lee:1964is} in QFT is of purely kinematic origin.

The LU representation realises local IR cancellation \emph{by construction} so that there is no need for explicitly listing and regularising all singular limits and their overlaps, thereby rendering it de facto valid for arbitrary perturbative orders, both conceptually and in the practical context of numerical computations.
We view this characteristic as unique to Local Unitarity and it is directly responsible for LU's universal applicability as the formalism does not depend on the particular theory, scattering process or observables considered.
Another defining property of Local Unitarity is that it does not require dimensional regularisation except for its unavoidable introduction when computing the integrated countertpart of the local UV counterterms and imposing that they reproduce results in the commonly used $\overline{\textrm{MS}}$ renormalisation scheme.
Dimensional regularisation~\cite{tHooft:1972tcz,Bollini:1972ui,Ashmore:1972uj} is a fundamental development from the 1970s which played an essential role in providing a convenient regulator preserving most symmetries. While it allows one to consistently characterise the divergent pieces entering perturbative computations, considering infinitesimal dimensions obfuscates some core properties of amplitudes such as chiral symmetry and the definition of asymptotic states. It also significantly complicates intermediate results and it is not amenable to numerical computations.

The unorthodox approach of Local Unitarity not only provides new theoretical insights on many aspects of the perturbative expansion of scattering cross-sections but also offers practical prospects for computing deeper perturbative corrections.
For each relevant section, we first provide a detailed account of the general concepts introduced for the specific illustrative example of the NLO correction to the scattering process $e^+ e^- \rightarrow d \bar{d}$.
In sect.~\ref{sec:supergraph}, we set our notation and introduce the core concept of organising the computation in terms of supergraphs.
We then present our main result in eq.~\eqref{lfmo} of sect.~\ref{sec:general_IR_cancellation_proof} together with the proof of the cancellation of IR singularities that do not involve initial-state splittings. In sect.~\ref{sec:UV_treatment_in_LU} we discuss the regularisation of the remaining UV singularities in the Local Unitarity representation.
Sect.~\ref{sec:generalisation} provides details about the future generalisation and automated implementation of our method. In sect.~\ref{sec:numerical_results}, we give quantitative results about the performance of our numerical implementation of LU for the computation of a next-to-leading accurate differential cross-section. We also explicitly verify local IR cancellations beyond next-to-leading order by computing several scalar supergraphs featuring up to five loops.
Finally, we give our conclusion in sect.~\ref{sec:conclusion}.

\clearpage

\section{Foundations of the Local Unitarity representation}
\label{sec:supergraph}

The fundamental object underlying any QFT observable calculated perturbatively is the supergraph~\cite{Soper:1999xk}, an entity that generalizes interference diagrams used in scattering amplitudes. More specifically, the supergraph can be seen as the representative of a class of interference diagrams which we will prove to be IR-finite when summed together.

In the following section we will show
\begin{itemize}
    \item how to rewrite any differential cross-section as a sum over cuts of supergraphs, and give a local representation of each individual supergraph that provides some first insight on how an equivalent local and IR-finite object can be defined. In particular, we will show how to accommodate a global routing common to all interference diagrams referring to the same supergraph class.
    
    \item that there is a direct correspondence between connected subgraphs of a supergraph and its threshold (pinched or non-pinched) singularities. This implies that the singularities of an amplitude can be fully characterised by couplets of connected subgraphs of a supergraph (loosely speaking, one subgraph identifies the Cutkosky cut and the other one the actual singular surface).
    
  \item that it is possible to fully characterise the notion of pinching in the LTD formalism by identifying singular surfaces for which no valid causal deformation of the loop kinematics is possible because it would either violate the continuity constraints or the singular surface has no well-defined oriented normal.
    
    \item that is it possible to accurately study the pattern of IR cancellations, described in more detail in sect.~\ref{sec:general_IR_cancellation_proof}, by considering a simple functional analogue of interference diagrams. We will show that the sum of this analogue of each interference diagram corresponding to the same supergraph is locally finite.
    
\end{itemize}

In each section of our paper, including this one, we choose to first address its content by applying it to the explicit example of the NLO computation of the differential cross-section of the scattering process $e^+ e^- \rightarrow d \bar{d}$. We find that this approach facilitates the introduction of our notation and the more abstract concepts presented in later parts of each section. This also provides the reader with some first intuition about the inner workings of the method and of its graphical interpretation.

\subsection{Illustrative example: NLO correction to $e^+ e^- \rightarrow d \bar{d}$}
\label{sec:foundation_example}

Our example process only involves two supergraphs from which arise all interferences of Feynman diagrams contributing to the amplitudes. We then analyse the threshold structure of these two supergraphs in their LTD representation, and write a formula for cross-sections using the LTD formalism.
A locally regularised expression for the NLO correction to the $e^+ e^- \rightarrow d \bar{d}$ cross-section has been studied before in ref.~\cite{Soper:1999xk} and in refs.~\cite{Sborlini:2016gbr, Sborlini:2016hat, Hernandez-Pinto:2015ysa}.

\begin{figure}[ht!]
\centering

\begin{subfigure}[b]{.48\linewidth}
\includegraphics[width=\linewidth]{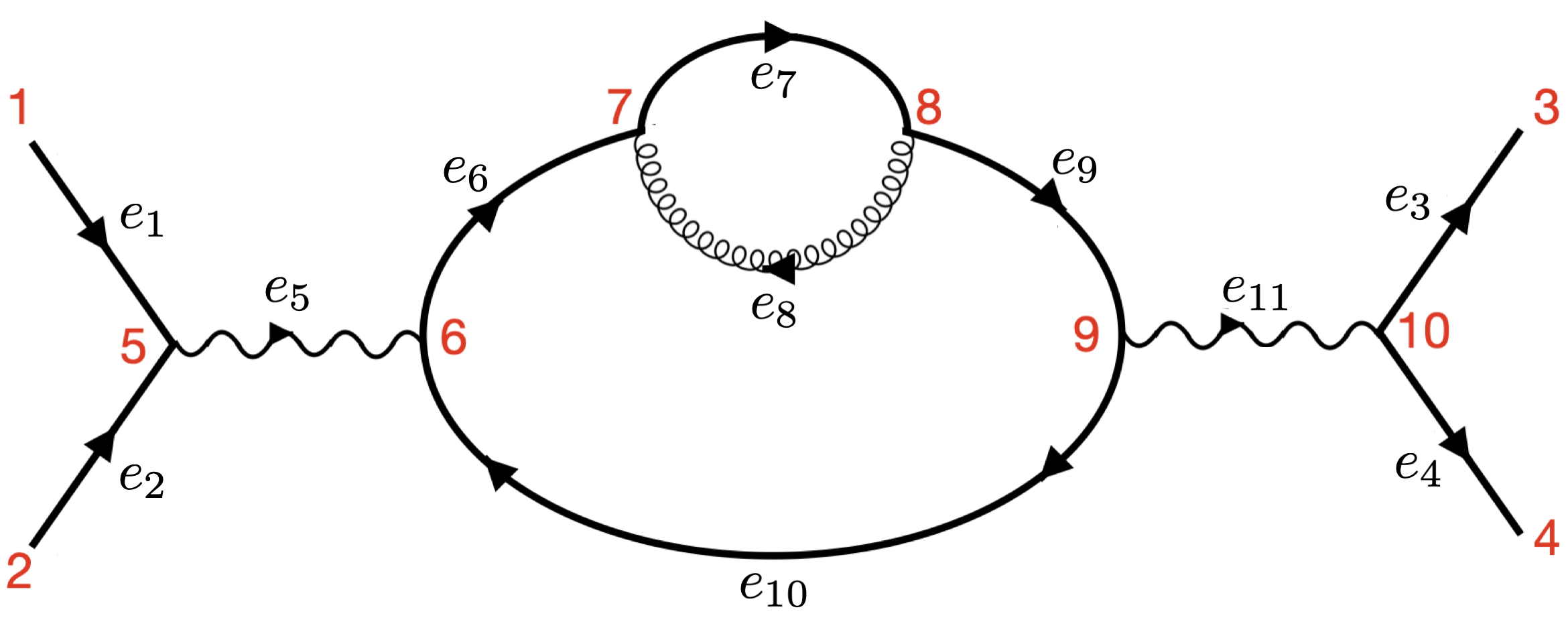}
\caption{}\label{fig:se_edge_labelling}
\end{subfigure}
\begin{subfigure}[b]{.48\linewidth}
\includegraphics[width=\linewidth]{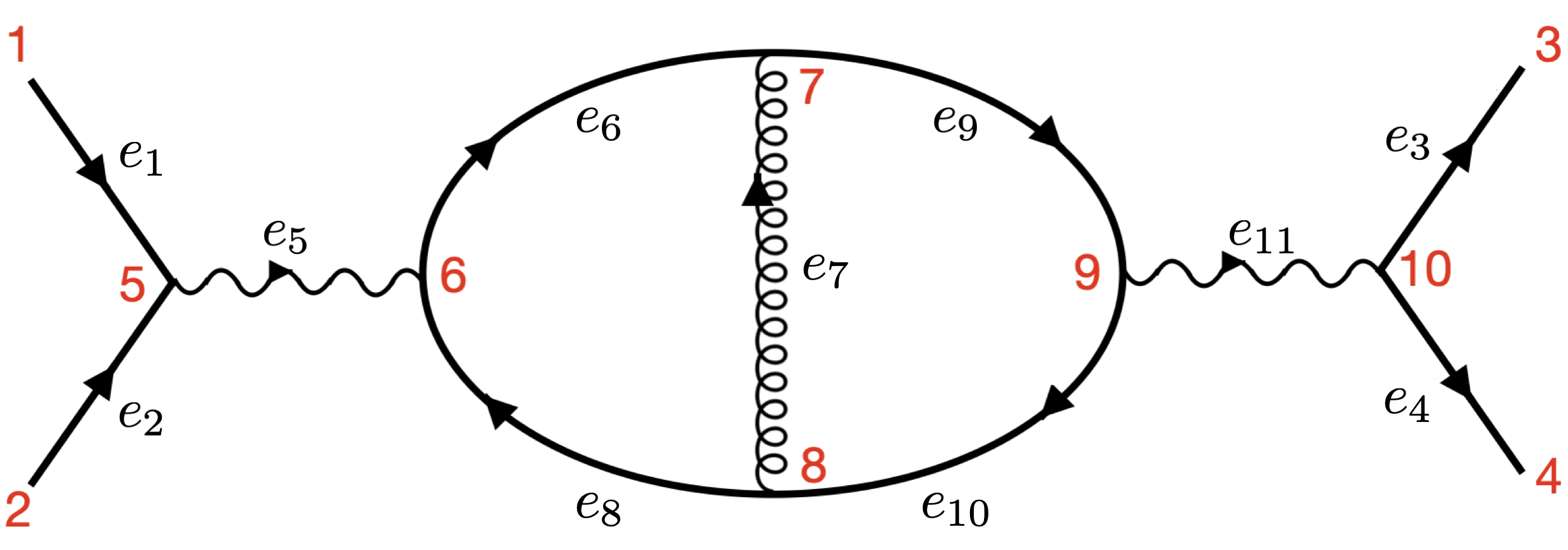}
\caption{}\label{fig:dt_edge_labelling}
\end{subfigure}
\caption{\label{fig:edge_labelling_NLO} The two supergraphs Self-Energy (SE,~\ref{fig:se_edge_labelling}) and Double-Triangle (DT,~\ref{fig:dt_edge_labelling}) contributing to the NLO QCD correction to the cross-section of the process $e^+ e^- \rightarrow \gamma^\star \rightarrow d \bar{d}$.}
\end{figure}

\subsubsection{The double-triangle and self-energy supergraphs}

Our first task is to enumerate all supergraphs contributing to our process of interest. Each supergraph encompasses a number of interference diagrams, and the collection of all interference diagrams stemming from all supergraphs reproduces the complete set of interference diagrams whose sum yields the scattering probability. In the specific case of the NLO correction to the $e^+ e^- \rightarrow \gamma^\star \rightarrow d \bar{d}$ cross-section, we identify only two distinct supergraphs, which we refer to as the nested Self-Energy (SE) supergraph and the Double-Triangle (DT) supergraph
, shown in fig.~\ref{fig:edge_labelling_NLO}.
We note that the \SE{} supergraph has two isomorphic occurences which we combine into a single representative weighted by a symmetry factor of two (see details in sect.~\ref{sec:technical_implementation}).

The two supergraphs $\Gamma^\textsc{se}$ and $\Gamma^\textsc{dt}$ can be described formally by a couplet of a graph and a set of incoming edges. More precisely, $\Gamma\in\{\Gamma^\textsc{se},\Gamma^\textsc{dt}\}$ can be rewritten as $\Gamma=(G^\Gamma,\mathbf{a}^\Gamma)$, where $G^\Gamma=(\mathbf{v}^\Gamma,\mathbf{e}^\Gamma)$ is a graph and $\mathbf{a}^\Gamma$ is a set of initial states. We encode the graph as a couplet of a set of vertices and oriented edges connecting them. 
We then write $e
^\Gamma=e^\Gamma_{\text{int}}\cup e^\Gamma_{\text{ext}}$ so as to distinguish between exterior edges that are connected to a degree-1 exterior vertex and the other interior edges. Since degree-1 vertices are in one-to-one correspondence with external edges, we will exclude them from $\mathbf{v}^\Gamma$. 
In summary, both \SE{} and \DT{} supergraphs can be characterised as follows: 
\begin{equation}
\mathbf{a}^\Gamma=\{ (1,5),(2,5)\},\hspace{0.2cm}
\mathbf{v}^\Gamma=\{5,6,7,8,9,10\},\hspace{0.2cm}
\mathbf{e}^\Gamma_{\textrm{ext}}=\{(1,5), (2,5), (10,3), (10,4)\}\\,, \end{equation}
\begin{equation}
\mathbf{e}^{\Gamma}_{\textrm{in}}=\begin{cases}
\{(5,6), (6,7), (7,8), (8,7), (8,9), (9,6), (9,10)\} \text{ if } \Gamma=\Gamma^{\textsc{se}} 
\\
\{(5,6), (6,7), (8,7), (8,6), (7,9), (9,8), (9,10)\} \text{ if } \Gamma=\Gamma^{\textsc{dt}}
\end{cases}.
\end{equation}
Whereas edges in $\mathbf{e}$ can in principle be identified from the two vertex labels they connect, we instead choose the more concise single label specified in fig.~\ref{fig:edge_labelling_NLO}.

\begin{figure}[t]
\centering
\begin{subfigure}[b]{.48\linewidth}
\includegraphics[width=\linewidth]{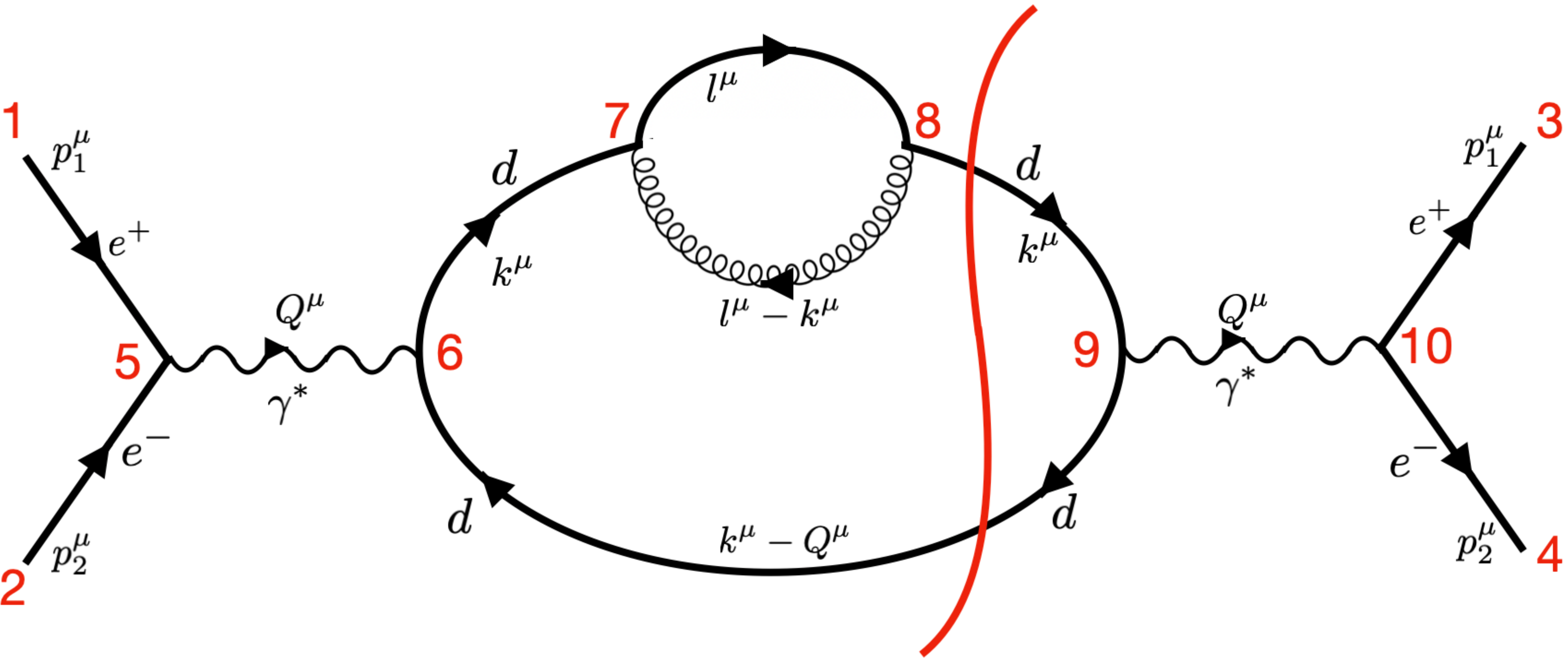}
\caption{$\mathbf{s}_2^{\text{v}}=\{5,6,7,8\}$ }\label{fig:Two_SG_epemddx_NLO:SE_LxBstar}
\end{subfigure}
\begin{subfigure}[b]{.48\linewidth}
\includegraphics[width=\linewidth]{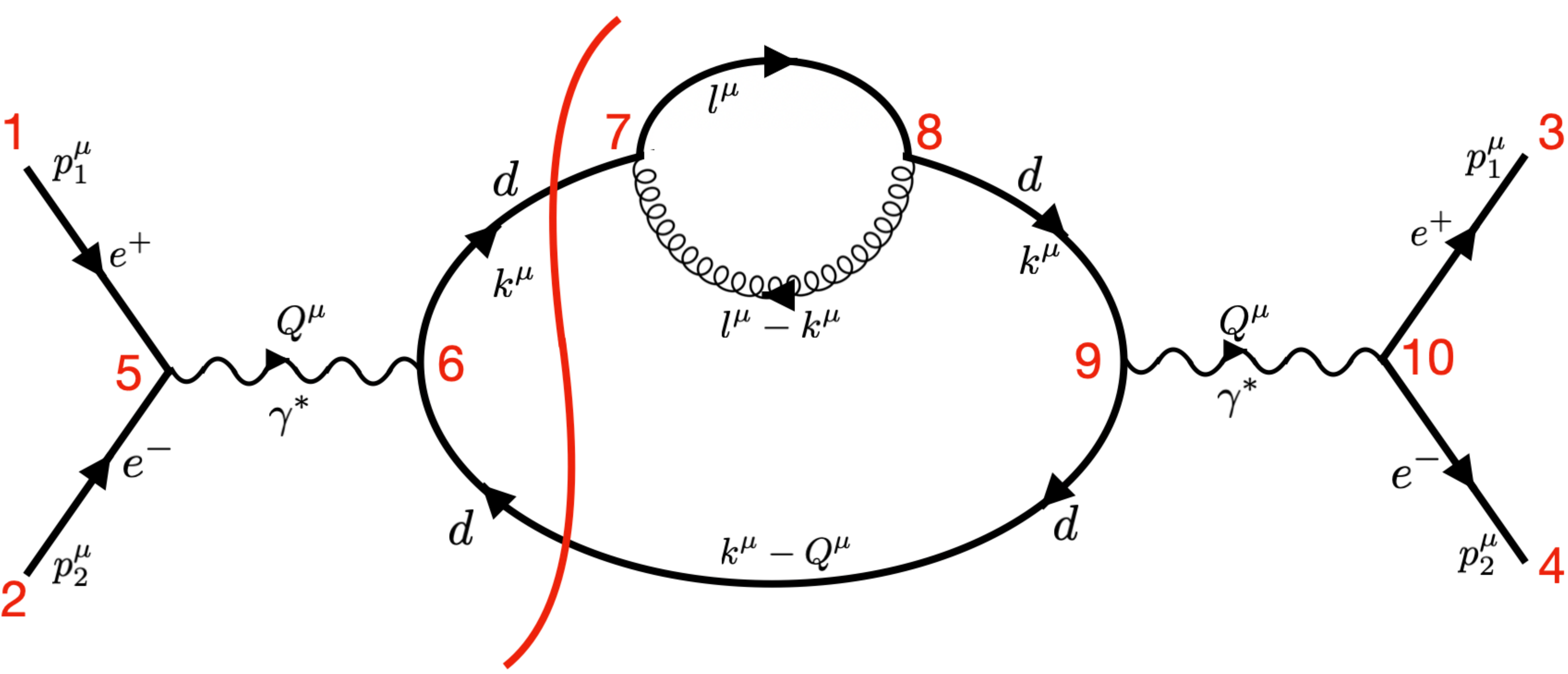}
\caption{$\mathbf{s}_1^{\text{v}}=\{5,6\}$ }\label{fig:Two_SG_epemddx_NLO:SE_BxLstar}
\end{subfigure}\\

\begin{subfigure}[b]{.48\linewidth}
\includegraphics[width=\linewidth]{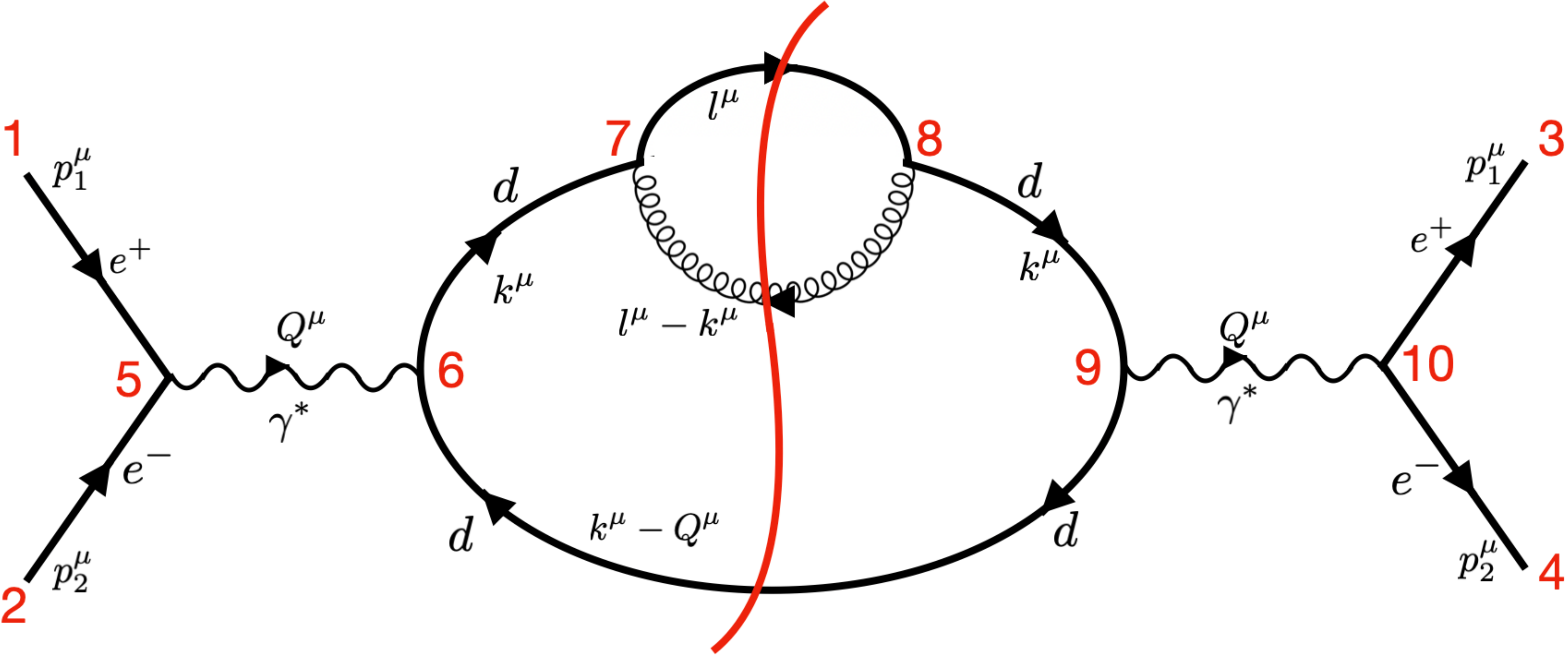}
\caption{$\mathbf{s}^{\text{r}}=\{5,6,7\}$ }\label{fig:Two_SG_epemddx_NLO:SE_R}
\end{subfigure}\\
\hspace{0.5cm}

\begin{subfigure}[b]{.48\linewidth}
\includegraphics[width=\linewidth]{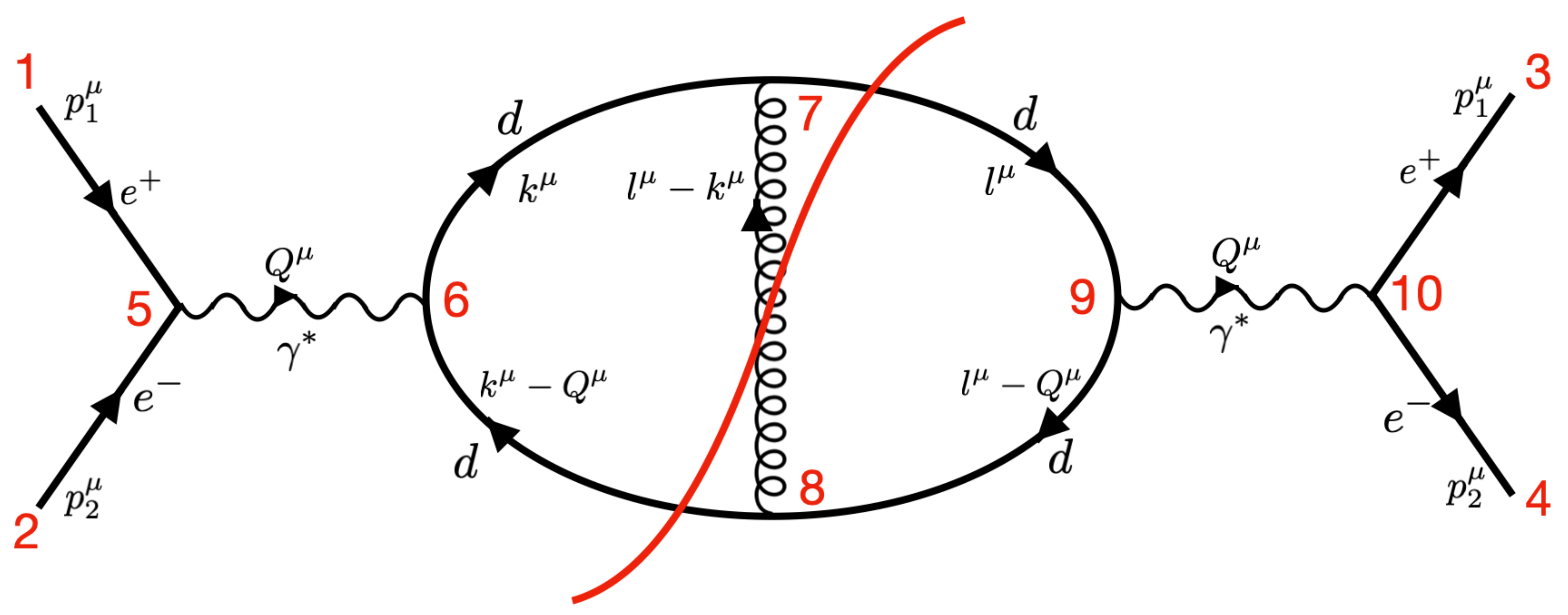}
\caption{$\mathbf{s}_2^{\text{r}}=\{5,6,7\}$}\label{fig:Two_SG_epemddx_NLO:DT_RxRA}
\end{subfigure}
\begin{subfigure}[b]{.48\linewidth}
\includegraphics[width=\linewidth]{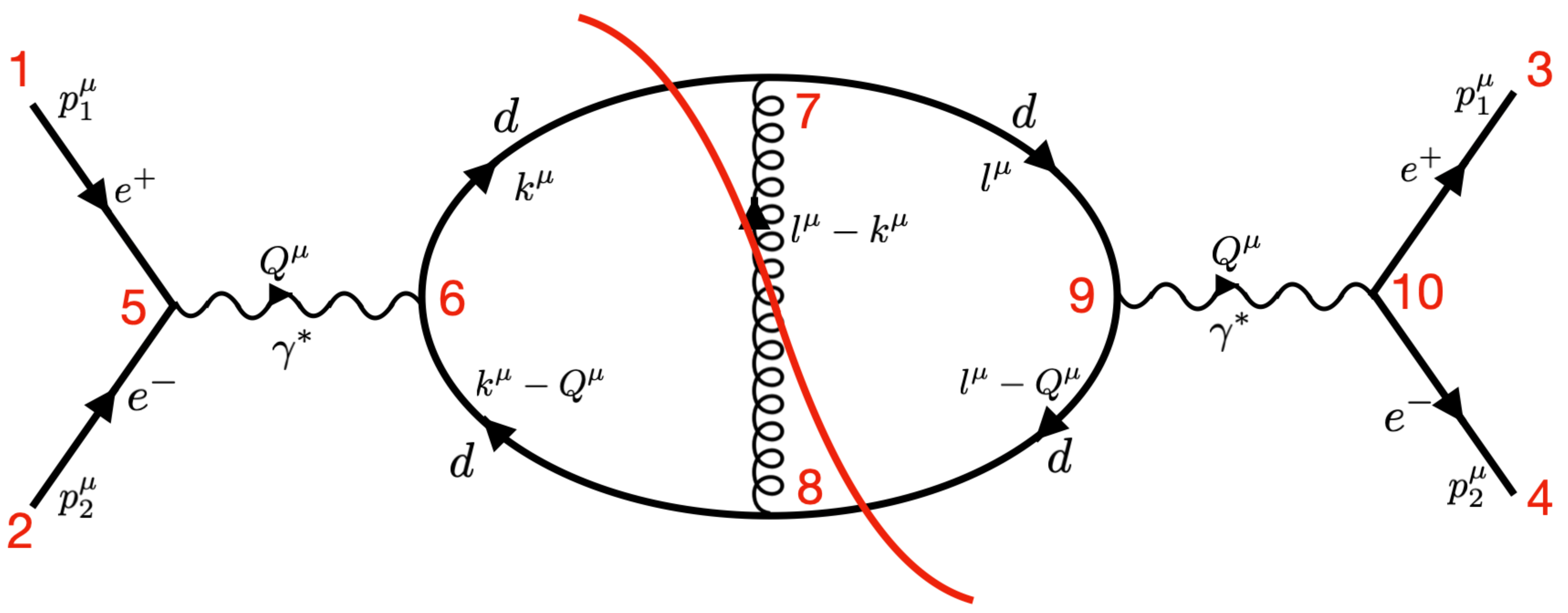}
\caption{$\mathbf{s}_1^{\text{r}}=\{5,6,8\}$ }\label{fig:Two_SG_epemddx_NLO:DT_RxRB}
\end{subfigure}\\

\begin{subfigure}[b]{.48\linewidth}
\includegraphics[width=\linewidth]{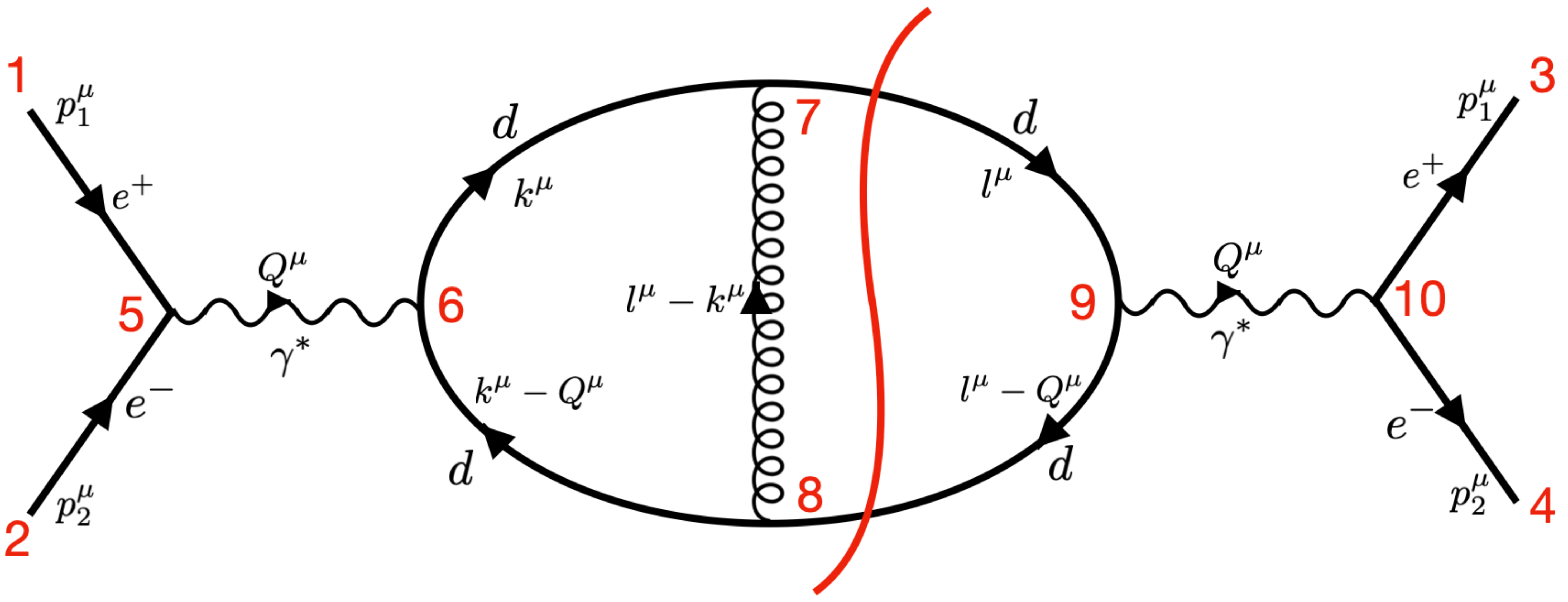}
\caption{$\mathbf{s}_2^{\text{v}}=\{5,6,7,8\}$ }\label{fig:Two_SG_epemddx_NLO:DT_LxBstar}
\end{subfigure}
\begin{subfigure}[b]{.48\linewidth}
\includegraphics[width=\linewidth]{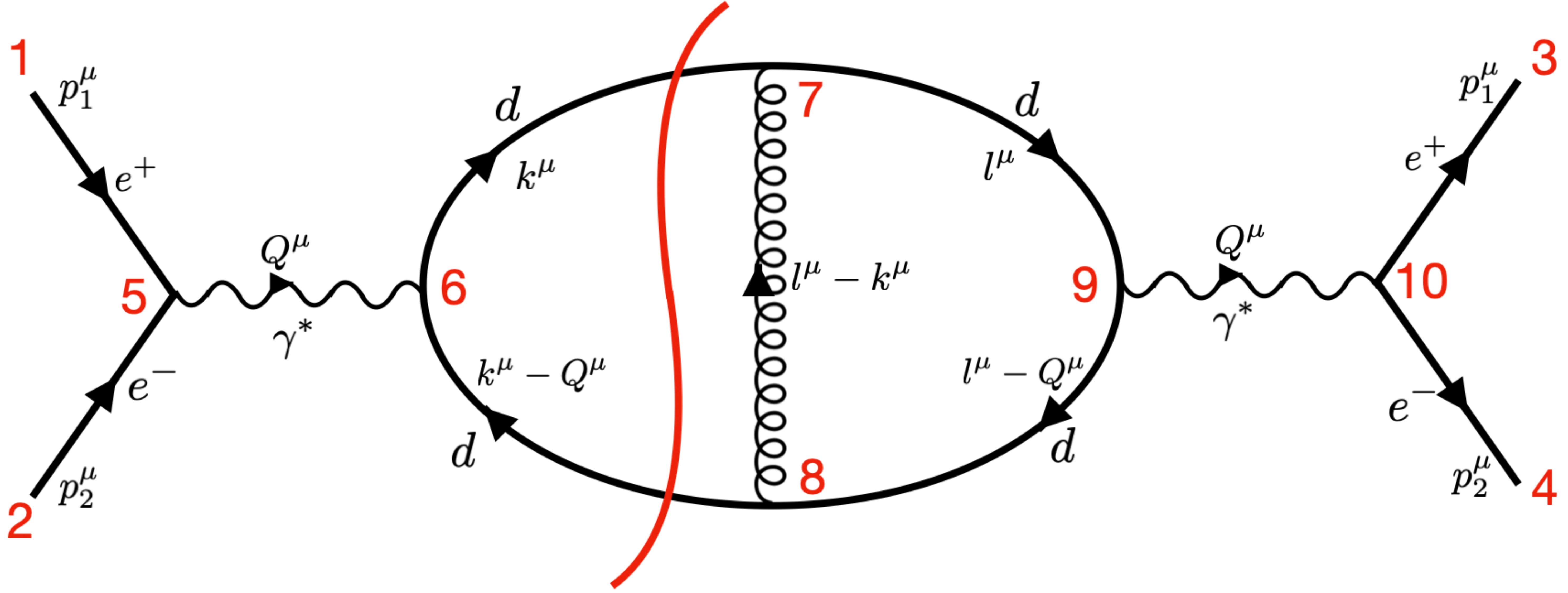}
\caption{$\mathbf{s}_1^{\text{v}}=\{5,6\}$ }\label{fig:Two_SG_epemddx_NLO:DT_BxLstar}
\end{subfigure}

\caption{\label{fig:Two_SG_epemddx_NLO} All three Cutkosky cuts of the \SE{} supergraph (\ref{fig:Two_SG_epemddx_NLO:SE_LxBstar},~\ref{fig:Two_SG_epemddx_NLO:SE_BxLstar} and~\ref{fig:Two_SG_epemddx_NLO:SE_R}) and all four cuts of the \DT{} supergraph (\ref{fig:Two_SG_epemddx_NLO:DT_RxRA},~\ref{fig:Two_SG_epemddx_NLO:DT_RxRB},~\ref{fig:Two_SG_epemddx_NLO:DT_LxBstar} and~\ref{fig:Two_SG_epemddx_NLO:DT_BxLstar}).}
\end{figure}

Each oriented edge $e_i$ is assigned a four-vector $q_{e_i}$ specifying the momentum it carries. We also define the characteristic vector of a set of edges $\mathbf{x}$ as follows:
\begin{equation}
    \chi^{\mathbf{x}}\in\{ 0,1\}^{|\mathbf{e}^\Gamma|}\hspace{0.5cm}\chi^{\mathbf{x}}_e=\begin{cases}
    1 &\text{ if } e\in\mathbf{x}\\
    0 &\text{ otherwise}
    \end{cases},
\end{equation}
as well as the following sets of edges for a given set $\mathbf{s}$ of vertices:

\begin{subequations}
\begin{align}
    \delta^{+}(\mathbf{s})&=\left\{(v,v')\in\mathbf{e}^\Gamma\middle|v\in\mathbf{s}, \ v'\in\mathbf{v}^\Gamma\setminus\mathbf{s} \right\} \\
    \delta^{-}(\mathbf{s})&=\left\{(v,v')\in\mathbf{e}^\Gamma\middle|v'\in\mathbf{s}, \ v\in\mathbf{v}^\Gamma\setminus\mathbf{s} \right\}, \\
    \delta(\mathbf{s})&=\delta^{+}(\mathbf{s})\cup\delta^{-}(\mathbf{s}),\\
    \delta^{\circ}(\mathbf{s})&=\left\{(v,v')\in\mathbf{e}^\Gamma\middle|v, v'\in\mathbf{s}\right\}.
\end{align}
\end{subequations}
The set $\delta^{\pm}(\mathbf{s})$ consists of all edges of the supergraph with only one of the two vertices it connects being part of the set $\mathbf{s}$.
$\delta^{\pm}(\mathbf{s})$ can loosely be defined as the list of edges ``entering'' (resp. ``exiting'') the set of edges $\mathbf{s}$. 
$\delta^{\circ}(\mathbf{s})$ consists of all edges connecting two vertices in $\mathbf{s}$, or loosely speaking the ``interior'' of $\mathbf{s}$. The momentum conservation condition  reads as follows for the explicit example of the subset of vertices $\mathbf{s}=\{7,8,9\}$ of the \DT{} supergraph:
\begin{equation}
\sum_{e\in \mathbf{e}} X_e^\mathbf{s} q_e := \sum_{e\in \mathbf{e}} (\chi_e^{\delta^+(\mathbf{s})}-\chi_e^{\delta^-(\mathbf{s})}) q_e  = q_{e_{6}}-q_{e_{8}}-q_{e_{11}}=0,
\end{equation}
which simply expresses the constraint that the momenta of the edges in $\delta(\{7,8,9\})=\{e_6,e_8,e_{11}\}$ have to sum up to zero. More specifically, the momenta of the incoming edges $\delta^{+}(\{7,8,9\})=\{e_6\}$ must sum up to the momenta of the outgoing edges $\delta^{-}(\{7,8,9\})=\{e_8,e_{11}\}$. 

As mentioned earlier, each supergraph is effectively the representative of a class of interference diagrams, each associated to a particular Cutkosky cut of their reference supergraph. Each Cutkosky cut of the supergraph $\Gamma$ is then defined using a subset of vertices $\mathbf{s}\subset\mathbf{v}^\Gamma$ that identifies two connected subgraphs $\delta^\circ(\mathbf{s})$ and $\delta^\circ(\mathbf{v}^\Gamma\setminus\mathbf{s})$, with the extra constraint that the initial states are contained in $\mathbf{s}$, that is $\mathbf{a}^\Gamma \cap \delta^{\circ}(\mathbf{s})=\mathbf{a}^\Gamma$.
Perhaps more intuitively, the Cutkosky cut can equivalently be identified using the set of internal ``cut'' edges $\mathbf{c}_\mathbf{s}^\Gamma=\delta(\mathbf{s})\setminus\mathbf{a}^\Gamma$ whose removal divides the supergraph into two connected amplitude graphs.

Let $\mathcal{E}_{\text{s-ch}}^\Gamma$ be the set of all possible Cutkosky cuts of a given supergraph $\Gamma$. This set reads as follows for the two distinct supergraphs of our example process:
\begin{equation}
    \mathcal{E}_{\text{s-ch}}^\Gamma=\begin{cases}
    \{\mathbf{s}_1^{\text{v}}=\{5,6\},\mathbf{s}_2^{\text{v}}=\{5,6,7,8\},\mathbf{s}_1^{\text{r}}=\{5,6,8\},\mathbf{s}_2^{\text{r}}=\{5,6,7\}\} &\text{ if }\quad \Gamma=\Gamma^{\textsc{dt}} \\
     \{\mathbf{s}_1^{\text{v}}=\{5,6\},\mathbf{s}_2^{\text{v}}=\{5,6,7,8\},\mathbf{s}^{\text{r}}=\{5,6,7\}\} &\text{ if }\quad \Gamma=\Gamma^{\textsc{se}}.
    \end{cases}
\end{equation}
Observe that we have intentionally left out the Cutkosky cuts $\{5\}$ and $\mathbf{v}\setminus\{10\}$ as these two contributions are vanishing because, on top of having no phase-space support, they can be thought of as being subject to an observable function that is  in this case identically zero on cuts containing $\gamma$. Using the definition of $\mathbf{c}_\mathbf{s}^\Gamma$, graphically represented as a line crossing all the edges contained in it, we list in fig.~\ref{fig:Two_SG_epemddx_NLO} the three (resp. four) Cutkosky cuts of the \SE{} (resp. \DT{}) supergraph.

In most sections featuring this illustrative example process, we focus on the \DT{} supergraph only for simplicity in which case we suppress the upper index in $\Gamma^{\textsc{dt}}$ when not ambiguous.

\subsubsection{LTD representation and thresholds of the double-triangle supergraph}

\begin{figure}[ht!]
\centering
\begin{subfigure}[b]{.24\linewidth}
\includegraphics[width=\linewidth]{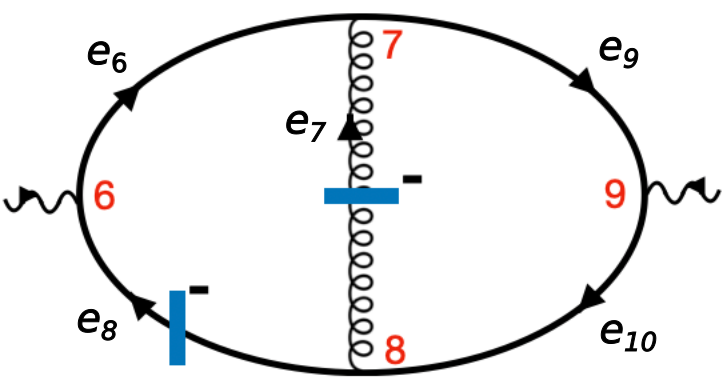}
\caption{$\mathbf{b}_1=\{e_{7}^-,e_{8}^-\}$ }\label{fig:DT_LTD:b1}
\end{subfigure}
\begin{subfigure}[b]{.24\linewidth}
\includegraphics[width=\linewidth]{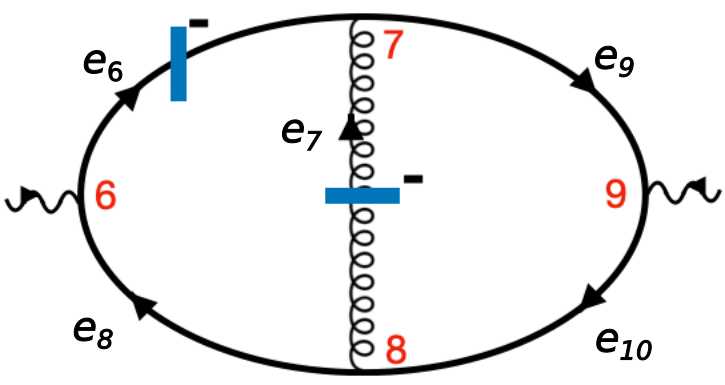}
\caption{$\mathbf{b}_2=\{e_{6}^-,e_{7}^-\}$ }\label{fig:DT_LTD:b2}
\end{subfigure}
\begin{subfigure}[b]{.24\linewidth}
\includegraphics[width=\linewidth]{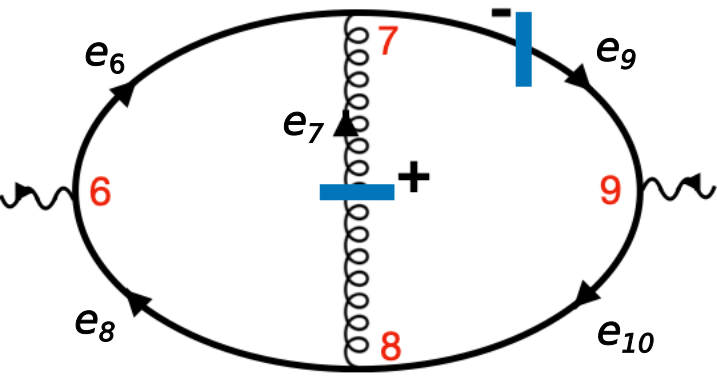}
\caption{$\mathbf{b}_3=\{e_{7}^+,e_{9}^-\}$ }\label{fig:DT_LTD:b3}
\end{subfigure}
\begin{subfigure}[b]{.24\linewidth}
\includegraphics[width=\linewidth]{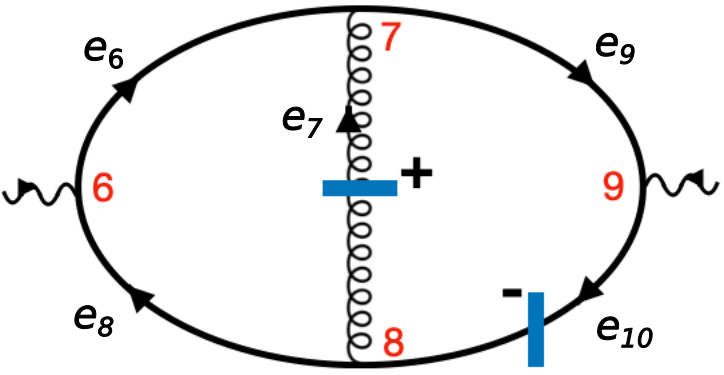}
\caption{$\mathbf{b}_4=\{e_{7}^+,e_{10}^-\}$ }\label{fig:DT_LTD:b4}
\end{subfigure} \\

\hspace{0.5cm}

\begin{subfigure}[b]{.24\linewidth}
\includegraphics[width=\linewidth]{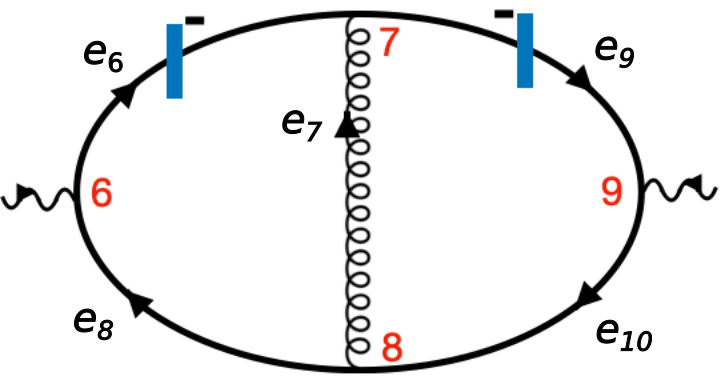}
\caption{$\mathbf{b}_5=\{e_{6}^-,e_{9}^-\}$ }\label{fig:DT_LTD:b5}
\end{subfigure}
\begin{subfigure}[b]{.24\linewidth}
\includegraphics[width=\linewidth]{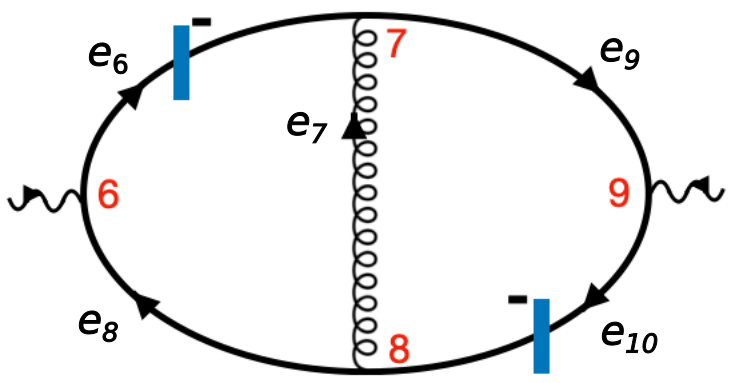}
\caption{$\mathbf{b}_6=\{e_{6}^-,e_{10}^-\}$ }\label{fig:DT_LTD:b6}
\end{subfigure}
\begin{subfigure}[b]{.24\linewidth}
\includegraphics[width=\linewidth]{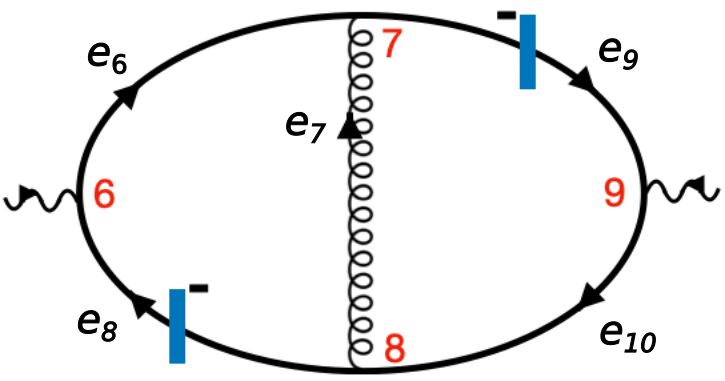}
\caption{$\mathbf{b}_7=\{e_{8}^-,e_{9}^-\}$ }\label{fig:DT_LTD:b7}
\end{subfigure}
\begin{subfigure}[b]{.24\linewidth}
\includegraphics[width=\linewidth]{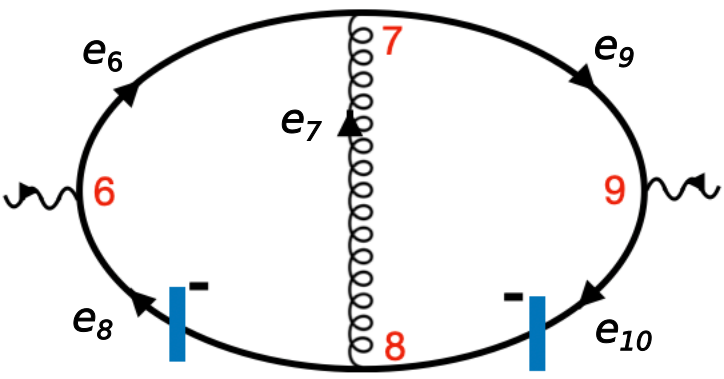}
\caption{$\mathbf{b}_8=\{e_{8}^-,e_{10}^-\}$ }\label{fig:DT_LTD:b8}
\end{subfigure}

\caption{\label{fig:DT_LTD} The eight loop momentum bases (the inverse of spanning trees) resulting from applying the LTD expression of ref.~\cite{Capatti:2019ypt} to the \DT{} topology. The cut structure $\sigma^{\mathbf{b}_i}$ is reported both in the drawings as signs placed next to the LTD cuts as well in the captions as a superscript of the edge labels.
}
\end{figure}

Our goal is to demonstrate the local cancellation of the IR soft and collinear divergences of the \DT{} supergraph. To this end, we first identify its IR limits by re-expressing the \DT{} \emph{supergraph} integral using its Loop-Tree Duality (LTD) representation~\cite{Capatti:2020ytd, Bierenbaum:2010cy}, where the energy components of the loop momenta are integrated out analytically using residue theorem:
\begin{equation}
M(p_1^\mu, p_2^\mu)=\sum_{\mathbf{b}\in\mathcal{B}}\int \frac{d^4\vec{k}'}{(2\pi)^3}\frac{d^4\vec{l}'}{(2\pi)^3}  N\frac{\prod_{e\in\mathbf{b}} \delta^{(\sigma^\mathbf{b}_e)}(q_e^2-m_e^2)}{\prod_{e\in \mathbf{e}\setminus \mathbf{b}}(q_e^2-m_e^2)},
\label{eq:DT_SG_expression}
\end{equation}
where the set $\mathcal{B}$ enumerates all possible momentum bases (or equivalently spanning trees) of the \DT{} supergraph:
\begin{align}
\begin{split}\label{eq:DT_LTD}
\mathcal{B}=\{&\mathbf{b}_1=\{e_8,e_7\}, \mathbf{b}_2=\{e_6,e_7\}, \mathbf{b}_3=\{e_7,e_9\}, \mathbf{b}_4=\{e_7,e_{10}\},\\
&\mathbf{b}_5=\{e_6,e_{9}\}, \mathbf{b}_6=\{e_6,e_{10}\}, \mathbf{b}_7=\{e_8,e_{9}\}, \mathbf{b}_8=\{e_8,e_{10}\}\}
\end{split}
\end{align}
and $\sigma^{\mathbf{b}_i}$ are the \emph{cut-structure} signs (see ref.~\cite{Capatti:2019ypt}) assigned to each of the edges in $\mathbf{b}_i$.
We show in fig.~\ref{fig:DT_LTD} all eight momenta bases $\mathbf{b}\in\mathcal{B}$ of the \DT{} LTD representation together with their corresponding cut structure $\sigma^{\mathbf{b}}$\footnote{The specific cut structure reported in fig.~\ref{fig:DT_LTD} is obtained by analytically integrating over ${l'}^0$ and ${k'}^0$ using the loop momentum routing of the \DT{} supergraph depicted in fig.~\ref{fig:Two_SG_epemddx_NLO:DT_RxRA} and choosing to close all energy integral contours in the upper half of complex plane}. Each element in the basis corresponds to a particle becoming on-shell, i.e. it is \emph{cut}. The cut structure $\sigma^{\mathbf{b}}_e$ sign that appears as superscript in the Dirac delta $\delta^{(\sigma^{\mathbf{b}}_e)}$ determines on which sheet the on-shell particle resides (the positive or negative energy solution), and is represented in fig.~\ref{fig:DT_LTD} as a plus or minus sign associated to each cut.

We note that we opted here to use the original LTD representation of ref.~\cite{Capatti:2019ypt}, and not the Manifestly Causal (cLTD) variant of ref.~\cite{Capatti:2020ytd}. While we will see that the latter plays an important role both in the proof of local IR cancellations and for the numerical stability of our implementation of LU, the former is better suited to highlight the connection of LTD with Cutkosky cuts.

The LTD representation consists of a sum of tree-like graphs that is only singular on bounded, convex (ellipsoid-like) surfaces called E-surfaces in ref.~\cite{Capatti:2019ypt}. Even though each individual summand (referred to as \emph{dual integrand}) building this representation is also singular on (hyperboloid-like) H-surfaces, their sum is regular on these surfaces in virtue of a mechanism known as \emph{dual cancellations}~\cite{LTDRodrigo2019,Catani:2008xa,Capatti:2020ytd}.

We can now relate the E-surfaces of the LTD representation of the \DT{} supergraph with its Cutkosky cuts. To this end, we first recall what the elements of $\mathcal{E}_\textrm{s-ch}^{\Gamma^\textsc{DT}}$, i.e. the set of all Cutkosky cuts of the \DT{} topology (represented in figs.~\ref{fig:Two_SG_epemddx_NLO:DT_RxRA}-\ref{fig:Two_SG_epemddx_NLO:DT_BxLstar}) are:
\begin{equation}\label{DT:sv_sr}
    \mathcal{E}_\textrm{s-ch}^{\Gamma^\textsc{DT}} = \{\mathbf{s}_1^{\text{v}}, \mathbf{s}_2^{\text{v}}, \mathbf{s}_1^{\text{r}}, \mathbf{s}_2^{\text{r}}\}.
\end{equation}
We also introduce the following notation for the on-shell energy of edge $e$: 
\begin{equation}
E_{e}= \sqrt{\lVert\vec{q}_{e}\rVert^2+m_{e}^2},
\end{equation}
where for this particular \DT{} supergraph the propagator masses $m_e$ are 0. An element $\mathbf{s}\in\mathcal{E}_{\text{s-ch}}$ can be associated to a threshold which, in Minkowski space, corresponds to a singularity of the integrand of the supergraph when energies of the corresponding cut edges take the following on-shell values:
\begin{align}
    q_e^0=X_e^{\mathbf{s}}E_e, \ \  \forall e\in \mathbf{c}_\mathbf{s}, 
\end{align}
whereas in the LTD representation, the same singularity is characterised by the following E-surface:
\begin{equation}
    \sum_{i\in \mathbf{c}_\mathbf{s}}E_i-\sum_{j\in \mathbf{a}}E_j=\sum_{i\in \mathbf{c}_\mathbf{s}}E_i-Q^0=0.
\end{equation}
We can thus list the LTD representations of all thresholds of the \DT{} supergraph:
\begin{eqnarray}\label{dt_esurf}
\begin{split}
    \eta_{\mathbf{s}^{\text{r}}_2}(\vec{k'},\vec{l'},Q^0) =& E_{e_9}+E_{e_7}+E_{e_8}-Q^0,& \qquad
    \eta_{\mathbf{s}^{\text{r}}_1}(\vec{k'},\vec{l'},Q^0) =& E_{e_6}+E_{e_7}+E_{e_{10}}-Q^0 \\ 
    \eta_{\mathbf{s}^{\text{v}}_1}(\vec{k'},\vec{l'},Q^0) =& E_{e_6}+E_{e_8}-Q^0,& \qquad
    \eta_{\mathbf{s}^{\text{v}}_2}(\vec{k'},\vec{l'},Q^0) =& E_{e_9}+E_{e_{10}}-Q^0 .
\end{split}
\end{eqnarray}
As we shall see, the particular signs selected for the on-shell energies of the edges being cut stems from the fact that these correspond to the only singular surfaces of the multi-loop LTD representation of the \DT{} supergraph (other than soft configurations).  

The LTD expression of the \DT{} also involves two additional E-surfaces that we refer to as \emph{internal} as they do not involve $Q^0$. Their implicit equation is:
\begin{align}
\begin{split}
    E_{e_6}+E_{e_9}+E_{e_7}=0 \\
    E_{e_8}+E_{e_{10}}+E_{e_7}=0 \,,\label{eq:DT_internal_E_surfaces}
\end{split}
\end{align}
which corresponds to the two (non-Cutkosky) cuts identified from the subgraph with the set of vertices $\{7\}$ and $\{8\}$ respectively. The defining equations~\eqref{eq:DT_internal_E_surfaces} can only be satisfied at soft points so that internal E-surfaces do not correspond to Cutkosky cuts since their phase-space support has no volume.
We define
\begin{equation}\label{DT:s0}
    \mathcal{E}_{\text{int}}=\{\mathbf{s}^{\circ}_1=\{7\},\mathbf{s}^{\circ}_2=\{8\}\}.
\end{equation}

In order to show more precisely how this correspondence between Cutkosky cuts and thresholds naturally arises within the LTD formalism, we now consider the eight terms of fig.~\ref{fig:DT_LTD} whose sum $M$ corresponds to the LTD expression of the \DT{} given in eq.~\eqref{eq:DT_LTD}.

\begin{figure}[ht!]
\centering
\begin{subfigure}[b]{.47\linewidth}
\includegraphics[width=\linewidth]{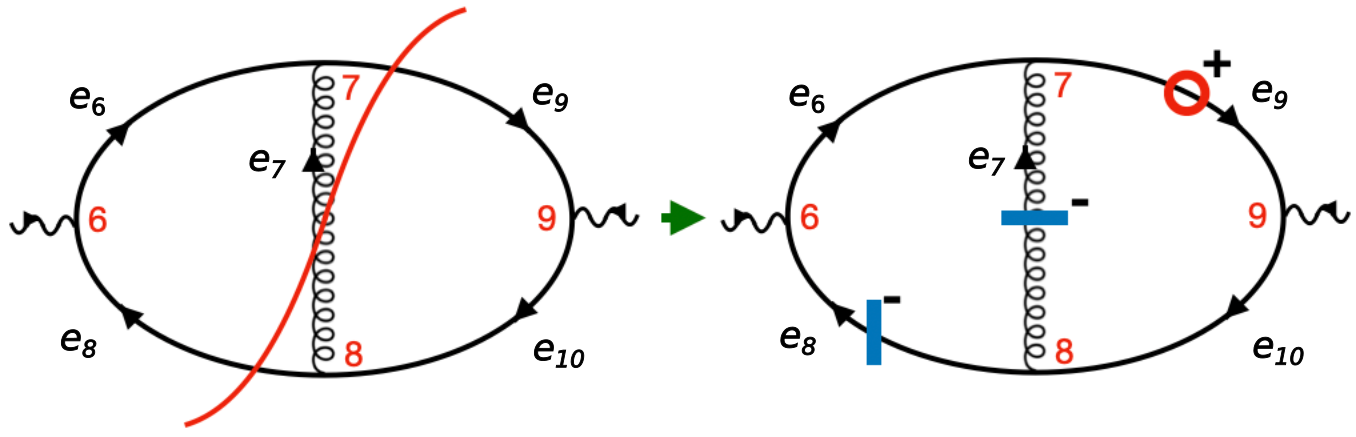}
\caption{$\mathbf{c}_{\mathbf{s}_{2}^{\text{r}}}= \mathbf{b}_1\cup\{e_{9}^+\}$ }\label{fig:CC_to_LTD:RA}
\end{subfigure},
\begin{subfigure}[b]{.47\linewidth}
\includegraphics[width=\linewidth]{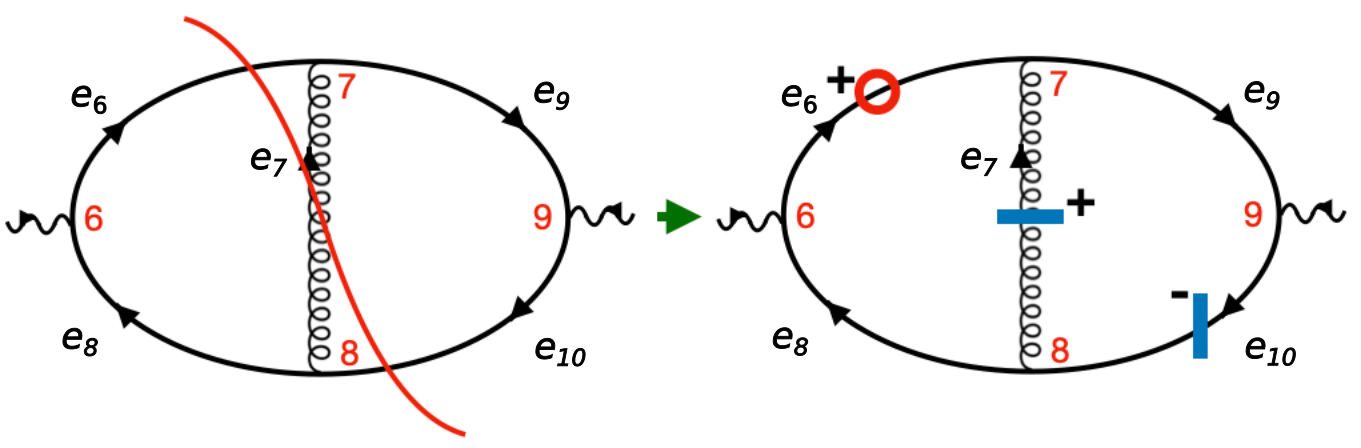}
\caption{$\mathbf{c}_{\mathbf{s}_{1}^{\text{r}}} = \mathbf{b}_4\cup\{e_{6}^+\}$ }\label{fig:CC_to_LTD:RB}
\end{subfigure} \\
\begin{subfigure}[b]{.95\linewidth}
\includegraphics[width=\linewidth]{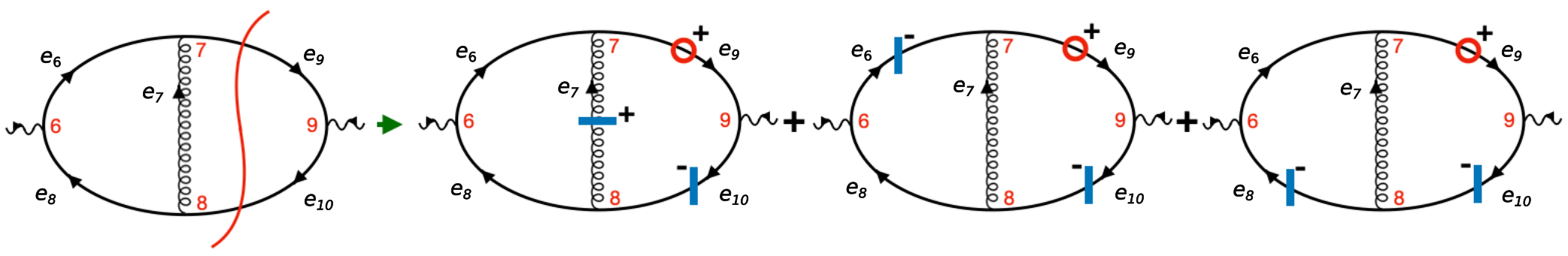}
\caption{$\mathbf{c}_{\mathbf{s}_{2}^{\text{v}}}= \{e_{9}^+,e_{10}^-\} $ }\label{fig:CC_to_LTD:LxBstar}
\end{subfigure} \\
\begin{subfigure}[b]{.95\linewidth}
\includegraphics[width=\linewidth]{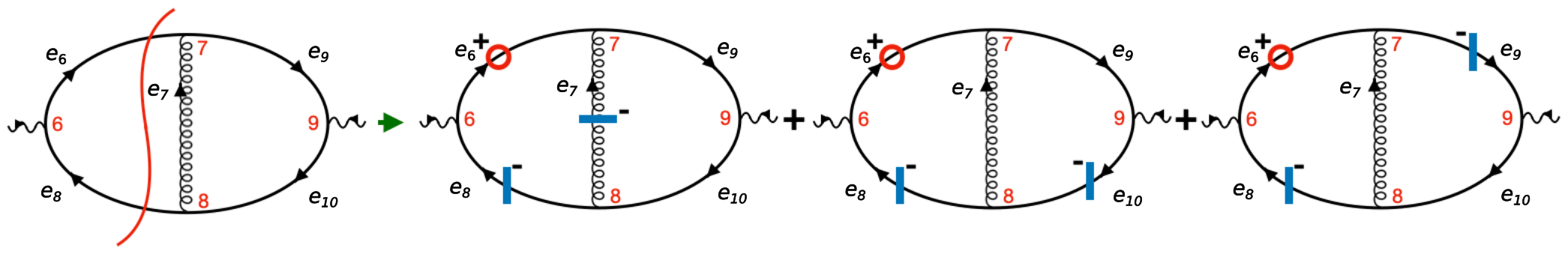}
\caption{$\mathbf{c}_{\mathbf{s}_{1}^\textrm{v}} = \{e_{6}^+,e_{8}^-\} $ }\label{fig:CC_to_LTD:BxLstar}
\end{subfigure}
\caption{\label{fig:DT_CC_to_LTD} Explicit correspondence between the four Cutkosky cuts of the \DT{} supergraph and the causal threshold surfaces of its 2-loop LTD representation. The loop momenta bases $\mathbf{b}_i$ that are absent from this figure correspond to summands of the LTD expression that contain no E-surface. The propagators circled in red are the ones inducing the causal threshold singularity when becoming on-shell (with a positive or negative on-shell energy depending on the sign above the red circle).
}
\end{figure}

We remind the reader that in virtue of dual-cancellations, H-surfaces are not singularities of $M$, which thus consist of only the E-surfaces of the form given in eq.~\eqref{dt_esurf}. Graphically, we identify in fig.~\ref{fig:DT_CC_to_LTD} which LTD summands involve each of these four thresholds by separately highlighting the on-shell cuts due to the LTD treatment and the cut indicating a vanishing propagator. This clearly illustrates the correspondence between the Cutkosky cuts and the terms of the LTD representation of the \DT{} supergraph.
We observe that the LTD terms containing the thresholds singularities corresponding to the Cutkosky cuts $\mathbf{c}_{\mathbf{s}_1^{\text{v}}}$ and $\mathbf{c}_{\mathbf{s}_2^{\text{v}}}$ already express the remaining triangle loop as a one-loop LTD expression, also including the correct energy sign for the cut $e_{8,7}^+$ in the first term of fig.~\ref{fig:CC_to_LTD:LxBstar}.

We stress that the key fact underlying the supergraph expression of eq.~\eqref{eq:DT_SG_expression} is that the parametric equations of the four Cutkosky cuts, $\eta_{\mathbf{s}}=0$ with $\mathbf{s}\in\mathcal{E}_{\text{s-ch}}$, given in eq.~\eqref{dt_esurf} are the \emph{only} (non-spurious) singular threshold surfaces of the 2-loop LTD representation of the \DT{}, due to the aformentioned dual cancellations. The only other divergences arise from internal propagators having a zero on-shell energy, typically leading to an integrable singularity.

\subsubsection{Construction of the cross-section}

In this section, we will define the cross section of the process in relation to the singular structure of the two supergraphs. Starting from the golden rule for cross-sections, we collect all possible supergraphs $\{\Gamma^{\textsc{se}},\Gamma^{\textsc{dt}}\}$ and their s-channel thresholds $\mathcal{E}_{\text{s-ch}}^{\Gamma^{\textsc{se}}}$, $\mathcal{E}_{\text{s-ch}}^{\Gamma^{\textsc{dt}}}$. We recall that s-channel thresholds of the supergraphs are in one-to-one correspondence with Cutkosky cuts and thus with interference diagrams contributing to the NLO correction to the cross-section of the process. 

For every interference diagram, we express each of the two amplitudes (to the left and right of the Cutkosky cut) in their LTD representation. Furthermore, we choose a consistent loop momentum routing for all the interference diagrams corresponding to the same supergraph. As a consequence, all interference diagrams can be expressed using a common integration measure. This procedure is carefully carried out in sect.~\ref{sec:supergraphs}. For the moment, we state that after having performed these operations, the NLO correction to the inclusive cross-section of the process $e^+ e^- \rightarrow d \bar d$ can then be written as
\begin{equation}
\label{eq:dt_and_se_xsection}
    \sigma^{(1)}_{e^+e^-\rightarrow d \bar d}=\sum_{a\in\{\textsc{se},\textsc{dt}\}} \sum_{\mathbf{s}\in\mathcal{E}_{\text{s-ch}}^{\Gamma^a}} \sigma_{\Gamma^a,\mathbf{s}}
\end{equation}
with
\begin{equation}
\label{eq:dt_and_se_xsection_cut}
\sigma_{\Gamma^a,\mathbf{s}}=\sum_{\mathbf{b}\in\mathcal{B}_\mathbf{s}}\sum_{\mathbf{b}'\in\mathcal{B}_{\mathbf{v}\setminus\mathbf{s}}}\int \frac{d^4k '}{(2\pi)^4}\frac{d^4l'}{(2\pi)^4} \frac{N_{\Gamma^a} \prod_{i\in \mathbf{b}\cup\mathbf{b}' \cup \mathbf{c}_\mathbf{s}}\delta^{\sigma_i^{\mathbf{s}\mathbf{b}\mathbf{b}'}}(q_i^2-m_i^2)}{\prod_{i\in \delta^\circ(\mathbf{s})\setminus \mathbf{b}}(q_i^2-m_i^2+\mathrm{i}\epsilon)\prod_{i\in \delta^\circ(\mathbf{v}\setminus\mathbf{s})\setminus \mathbf{b}'}(q_i^2-m_i^2-\mathrm{i}\epsilon) }.
\end{equation}
where $\mathcal{B}_\mathbf{s}$ is the collection of all the loop momentum basis of the subgraph $(\mathbf{s},\delta^\circ(\mathbf{s}))$ and $N_{\Gamma^a}$ is the appropriately defined numerator, which may include non-trivial symmetry factors. The vector $\sigma^{\mathbf{s}\mathbf{b}\mathbf{b}'}$ fixes the energy flow of the on-shell particles consistently with fig.~\ref{fig:DT_CC_to_LTD}, by identifying $\delta(\mathbf{s})\cup\mathbf{b}\cup\mathbf{b}'$ with the respective set of edges which are crossed by a cut or a circle in one of the diagrams of fig.~\ref{fig:DT_CC_to_LTD}, and the components $\sigma_i^{\mathbf{s}\mathbf{b}\mathbf{b}'}\in\{\pm 1\}$ with the signs associated to those circles or cuts.

The identification between the terms summed in $\sigma_{\Gamma^a,\mathbf{s}}$ and the cut diagrams of fig.~\ref{fig:DT_CC_to_LTD} is now clear. More specifically, $\sigma_{\Gamma^{\textsc{dt}},\mathbf{s}_1^{\text{r}}}$ corresponds to the sum of diagrams in fig.~\ref{fig:CC_to_LTD:RA}, which involves one term only, since in this case the Cutkosky cut does not leave loops on either side. Thus $\mathcal{B}_\mathbf{s}=\mathcal{B}_{\mathbf{v}\setminus\mathbf{s}}=\emptyset$. $\sigma_{\Gamma^{\textsc{dt}},\mathbf{s}_1^{\text{v}}}$ is the sum of three terms, corresponding to the three loop momentum basis of the remaining triangle loop, as depicted in fig.~\ref{fig:CC_to_LTD:BxLstar}.
Note that in eq.~\eqref{eq:dt_and_se_xsection_cut}, the propagator denominators in $\delta^\circ(\mathbf{v}\setminus\mathbf{s})\setminus \mathbf{b}'$ take the causal prescription $-\mathrm{i}\epsilon$ because of the complex conjugation applied to the amplitude to the right of the Cutkosky cut.

We now generalise the concepts of supergraph and E-surface identification beyond this example process.

\newpage
\subsection{Supergraphs}
\label{sec:supergraphs}

A Final-State Radiation (FSR) supergraph is a couplet $\Gamma=(G,\mathbf{a})$, where $G=(\mathbf{v}=\mathbf{v}_{\text{int}}\cup\mathbf{v}_{\text{ext}},\mathbf{e}=\bf{e}_{\text{int}}\cup\bf{e}_{\text{ext}})$ is a directed and connected graph with a set $\bf{e}_{\text{ext}}$ as external legs, with $a\subset e_{\text{ext}}$,  $|\mathbf{a}|=|\bf{e}_{\text{ext}}| / \rm 2$ and $\delta(\mathbf{v}_{\text{int}})=\mathbf{e}_{\text{ext}}$. Roughly speaking, the edges in $\mathbf{a}$ correspond to the incoming particles of the process, and $\mathbf{e}_{\text{ext}}$ contains two copies of the incoming particles of the process at hand. In the following, we will suppress the set $\mathbf{v}_{\text{ext}}$ and just refer to $\mathbf{v}_{\text{int}}$ as $\mathbf{v}$. 

It is convenient to describe features of the supergraph in terms of cuts, that is subsets of the set of all vertices of the graph, their boundary, that is the collections of edges connecting vertices in the cuts with vertices not in the cut and their interior, that is the subgraphs identified by the edges whose vertices are contained in the cut. Given any subset of vertices $\mathbf{s}\subseteq \mathbf{v}$, we define the following operators:
\begin{align} 
\begin{split}
    \delta^{+}(\mathbf{s})&=\left\{(v,v')\in\mathbf{e}^\Gamma\middle|v\in\mathbf{s}, \ v'\in\mathbf{v}^\Gamma\setminus\mathbf{s} \right\}, \\
    \delta^{-}(\mathbf{s})&=\left\{(v,v')\in\mathbf{e}^\Gamma\middle|v'\in\mathbf{s}, \ v\in\mathbf{v}^\Gamma\setminus\mathbf{s} \right\}, \\
    \delta(\mathbf{s})&=\delta^{+}(\mathbf{s})\cup\delta^{-}(\mathbf{s}),\\
    \delta^{\circ}(\mathbf{s})&=\left\{(v,v')\in\mathbf{e}^\Gamma\middle|v, v'\in\mathbf{s}\right\}.
\end{split}
\end{align}
As for most of the notation introduced in this section, $\delta$ always implicitly carries a dependency on the super graph $\Gamma$, which our notation will often omit for brevity. Furthermore, each subset of the edges can be fully characterised by a binary vector whose entries are $1$ if the corresponding edges are in the subset and $0$ otherwise. Given any subset of edges $\bf{e}'\subseteq \bf{e}$, the \emph{characteristic vector} of $\bf{e}'$, is defined as $\chi^{\bf{e}'}\in\{0,1\}^{|\mathbf{e}|}$ with 
\begin{equation}
\chi^{\bf{e}'}_e=\begin{cases}
1 \text{ if } e\in \bf{e}' \\
0 \text{ otherwise}
\end{cases} \,.
\end{equation}
Characteristic vectors allow to compactly write momentum-conservation conditions, which are interpreted as a conserving network flow. Each edge of the supergraph can be assigned with a weight $q_e\in\mathbb{R}^{4}, \ e\in\mathbf{e}$ that corresponds to the momentum carried by the edge, and momentum conservation constraints can then be written as follows:
\begin{equation}\label{momentum-conservation}
\sum_{e\in \mathbf{e}} X_e^\mathbf{s} q_e :=\sum_{e\in \mathbf{e}} (\chi^{\delta^{\mathsmaller{+}}(\mathbf{s})}_e-\chi^{\delta^{\mathsmaller{-}}(\mathbf{s})}_e) q_e =0 \hspace{1cm} \forall \mathbf{s}\subseteq \mathbf{v}.
\end{equation}

Most of the constraints of eq.~\eqref{momentum-conservation} are linearly dependent. In order to eliminate redundancies and obtain a minimal set of momentum-conservation constraints, it is possible to reduce the system and obtain the set of minimal constraints that is in one-to-one correspondence with the edges of a spanning tree. This fact alone is sufficient to show that the kinematic space of the virtual momenta (the linear space where momentum conservation constraints hold) is spanned by the momentum weights associated to the edges not contained in a given spanning tree. The basis corresponding to a given spanning tree can be mapped via a totally unimodular matrix to a basis corresponding to a different spanning tree.

In this framework, a Cutkosky cut admits an especially simple representation as a connected subset of the vertices, which thus allows to divide the supergraph in two connected subgraphs of it (i.e. two interfering amplitudes). The energy flow across the Cutkosky cut is enforced to be such that every edge in the boundary of the Cutkosky cut and not in $\mathbf{e}_\mathbf{\text{ext}}$ has an energy flow that is opposite to that of the edges in $\mathbf{e}_\mathbf{\text{ext}}$ themselves. More precisely, a Cutkosky cut on the FSR supergraph is a subset $\mathbf{s}\subseteq \mathbf{v}$ with the following properties:
\begin{itemize}
\item the graphs $G'=(\mathbf{s}, \mathbf{e}_\mathbf{s})$ and $G'=(\mathbf{v}\setminus\mathbf{s}, \mathbf{e}_{\mathbf{v}\setminus \mathbf{s}})$, with $\mathbf{e}_\mathbf{s}=\sum_{v\in \mathbf{s}} \delta(v)$ are connected, 
\item $\delta(\mathbf{s})\cap\mathbf{e}_{\text{ext}}=\mathbf{a}$.
\end{itemize}
The Cutkosky cut can be equivalently identified with the subset of edges $\mathbf{c}_\mathbf{s}=\delta(\mathbf{s})\setminus\delta(\mathbf{v}_{\text{int}})$, and graphically represented as a line crossing all edges in $\mathbf{c}_\mathbf{s}$. Removal of the edges in $\mathbf{c}_\mathbf{s}$ from the graph $G$ yields two connected components $\delta^\circ(\mathbf{s})\cup \mathbf{a}$ and $\delta^\circ(\mathbf{s}^c)\cup(\mathbf{e}_{\text{ext}}\setminus\mathbf{a})$ with $\mathbf{s}^c=\mathbf{v}\setminus\mathbf{s}$ containing the external edges $\mathbf{a}$ and $\mathbf{e}_{\text{ext}}\setminus \mathbf{a}$ respectively. As already mentioned, these correspond to the two interfering amplitudes that form the supergraph when stitched back together. We observe that there is an apparent two-fold degeneracy since $\mathbf{c}_\mathbf{s}$ and $\mathbf{c}_{\mathbf{v}\setminus\mathbf{s}}$ identify the same Cutkosky cut.
Let $\mathcal{E}_{\text{s-ch}}$ be the set of all Cutkosky cuts modulo this two-fold symmetry (the name of this set will become clear later when we relate it to a subset of the threshold singularities of the supergraph in its LTD representation).

The couplet formed by an FSR supergraph and one of its Cutkosky cut $\mathbf{c}_\mathbf{s}$ is an interference diagram. In the perturbative formulation of relativistic quantum mechanics, the all-order cross-section is obtained by summing all supergraphs for which one sums over all possible interference diagrams arising from its Cutkosky cuts, each weighted by a density of states (i.e. observable):
\begin{equation}
\sigma_\mathcal{O}(\{q_e\}_{e\in\mathbf{a}})=\sum_{\Gamma\in \mathcal{G}}\sum_{\mathbf{s}\in\mathcal{E}_{\text{s-ch}}}(-i)^{|\mathbf{c}_\mathbf{s}|}\sigma^\mathcal{O}_{\Gamma,\mathbf{s}}(\{q_e\}_{e\in\mathbf{a}}),
\label{eq:cross_section_master_formula}
\end{equation}
with
\begin{equation}
\sigma^\mathcal{O}_{\Gamma,\mathbf{s}}(\{q_e\}_{e\in\mathbf{a}})=\int\Bigg(\prod_{e\in \mathbf{c}_\mathbf{s}}\frac{d^4 q_e}{(2\pi)^3}\delta^{(\sigma_e^{\mathbf{c}_\mathbf{s}})}(q_e^2-m_e^2)\Bigg)\delta\Bigg(\sum_{e\in \delta(\mathbf{s})}X^\mathbf{s}_e q_e\Bigg) [\mathcal{A}_{\mathbf{s}}\mathcal{A}_{\mathbf{s}^c}^\star]\mathcal{O}_\mathbf{s},
\end{equation}
where $\mathcal{O}_\mathbf{s}:\mathbb{R}^{4|\mathbf{c}_\mathbf{s}|}\rightarrow \mathbb{R}$ is the observable function, and
\begin{equation}
\sigma^{\mathbf{c}_\mathbf{s}}_i=-X^{\mathbf{s}}_i, \ \ i\in\mathbf{c}_\mathbf{s}.
\end{equation}
The formula for $\sigma_\mathcal{O}$ matches the usual formula for semi-differential cross-sections, rewritten in terms of supergraphs. The phase-space integration over the final state particles is constrained by the on-shell conditions associated to them. The amplitudes, however, in the Minkowski representation, feature unconstrained four-dimensional integrations for each of the loops of the graph. This apparent asymmetry in the treatment of virtual and real particles is partially lifted within the LTD representation, in which the amplitude is re-expressed as follows:
\begin{equation}
\mathcal{A}_{\mathbf{s}}^{\vec{\mu}}(\{q_e\}_{e\in\mathbf{a}\cup C})=\sum_{\mathbf{b}\in\mathcal{B}}(-i)^{|\mathbf{b}|}\int \prod_{i\in \mathbf{b}_\emptyset}\frac{d^4 k_i}{(2\pi)^3}\prod_{j\in \mathbf{b}}\delta^{(\sigma^\mathbf{b}_j)}(q_j^2-m_j^2)\prod_{e\in \delta^\circ(\mathbf{s})\setminus\mathbf{b}}\frac{N_\mathbf{s}^{\vec{\mu}}}{q_e^2-m_e^2+i\epsilon}.
\end{equation}
where $\sigma^\mathbf{b}\in\{-1,1\}^{|\mathbf{b}|}$ is the cut structure for the basis $\mathbf{b}\in\mathcal{B}$ ($\mathcal{B}$ is the set of all possible loop momenta basis of the subgraph $\delta^\circ(\mathbf{s})$) with respect to the reference basis $\mathbf{b}_\emptyset$, and $N_\mathbf{s}^{\vec{\mu}}$ is a tensor polynomial in the loop variables whose (colour, spinor and Lorentz) indices are collected in the symbol $\vec{\mu}$. 

When rewriting each amplitude using their LTD representations and factoring out the common loop integration measures, we find that the virtual and real degrees of freedom now appear undifferentiated. 
In particular, the procedure defines an extended Lorentz-invariant measure that encompasses both loop and phase-space integration:
\begin{equation}\label{Mformula}
\sigma^\mathcal{O}_{\Gamma,\mathbf{s}}(\{q_e\}_{e\in\mathbf{a}})=\sum_{\mathbf{b}\in\mathcal{B}_\mathbf{s}}\sum_{\mathbf{b}'\in\mathcal{B}_{\mathbf{s}^c}}\int [d\Pi^\mathbf{s}_{\mathbf{b},\mathbf{b}'}]\Bigg( \prod_{e\in \delta^\circ(\mathbf{s})\setminus\mathbf{b}}\frac{N_{\mathbf{s}\vec{\mu}}}{q_e^2-m_e^2+i\epsilon}\prod_{e\in\delta^\circ(\mathbf{s}^c)\setminus\mathbf{b}'}\frac{N_{\mathbf{v}\setminus\mathbf{s}
}^{\vec{\mu}}}{q_e^2-m_e^2-i\epsilon} \Bigg)\mathcal{O}_\mathbf{s} \,,
\end{equation}
where the integration measure is the extended Lorentz invariant phase-space measure
\begin{multline}
[d\Pi^\mathbf{s}_{\mathbf{b},\mathbf{b}'}]=\delta^{(4)}\Bigg(\sum_{e\in \delta(\mathbf{s})}X^\mathbf{s}_e q_e\Bigg)\prod_{i\in \mathbf{c}_\mathbf{s}} \frac{d^4 k_i}{(2\pi)^3}\delta^{(\sigma^{\mathbf{c}_\mathbf{s}}_i)}(k_i^2-m_i^2) \\ \prod_{j \in \mathbf{b}} \frac{d^4 k_j}{(2\pi)^3}\delta^{(\sigma^{\mathbf{b}}_j)}(k_j^2-m_j^2)\prod_{n \in \mathbf{b}'} \frac{d^4 k_n}{(2\pi)^3}\delta^{(\sigma^{\mathbf{b}'}_n)}(k_n^2-m_i^2).
\end{multline}

This shows that quantum corrections are equivalent to performing phase-space integrals of tree processes in which virtual particles are substituted by on-shell particles whose momentum is allowed to range over all possible kinematic values (contrary to external particles, whose momenta are naturally constrained by the collision energy).
Energy-momentum conservation conditions can be solved jointly for both graphs on the left and right of the Cutkosky cut by directly considering one loop momentum basis $\mathbf{b}_\emptyset$ of the complete supergraph.
This allows one to write each of the extended phase-space integration measures, for varying $\mathbf{s}, \ \mathbf{b}, \ \mathbf{b}'$, as arising naturally from the loop integration measure of the supergraph, plus an extra energy conservation delta on the momenta crossing the Cutkosky cuts. It is then possible to solve all but one of the Dirac deltas and adopt a common basis for all spatial degrees of freedom:
\begin{multline}\label{Sigmaformula}
\sigma^\mathcal{O}_{\Gamma,\mathbf{s}}(\{q_e\}_{e\in\mathbf{a}})=\int [d\Pi^\mathbf{s}]\sum_{\mathbf{b}\in\mathcal{B}_\mathbf{s}}\sum_{\mathbf{b}'\in\mathcal{B}_{\mathbf{s}^c}} \frac{1}{\prod_{i\in\mathbf{c}_\mathbf{s}\cup \mathbf{b} \cup \mathbf{b}'} 2E_i} \\ 
\times\Bigg( \prod_{e\in \delta^\circ(\mathbf{s})\setminus\mathbf{b}}\frac{N_{\mathbf{s}\vec{\mu}}}{q_e^2-m_e^2+i\epsilon}\prod_{e\in\delta^\circ(\mathbf{s}^c)\setminus\mathbf{b}'}\frac{N_{\mathbf{v}\setminus\mathbf{s}
}^{\vec{\mu}}}{q_e^2-m_e^2-i\epsilon} \Bigg)\mathcal{O}_\mathbf{s}\Bigg|_{C_{\mathbf{b},\mathbf{b}'}^\mathbf{s}} \,,
\end{multline}
with
\begin{equation}
C_{\mathbf{b},\mathbf{b}'}^\mathbf{s}=\{q_i^0=\sigma^{\mathbf{b}}_iE_i\}_{i\in\mathbf{b}} \cup\{q_i^0=\sigma^{\mathbf{b}'}_iE_i\}_{i\in\mathbf{b}'}\cup \{q_i^0=\sigma^{\mathbf{c}_\mathbf{s}}_iE_i\}_{i\in\mathbf{c}_\mathbf{s}},
\end{equation}
and
\begin{equation}\label{measure-interference}
[d\Pi^\mathbf{s}]=\delta\Bigg(\sum_{i\in\mathbf{c}_\mathbf{s}} E_i - \sum_{j\in \mathbf{a}}E_j \Bigg)\prod_{i\in \mathbf{b}^\Gamma_\emptyset} \frac{d^3 \vec{k}_i}{(2\pi)^3},\end{equation}
given a reference momentum basis $\mathbf{b}^\Gamma_\emptyset$ of the supergraph $\Gamma$. Recall that $\mathbf{a}$ is the set of incoming particles, so that the delta in eq.~\eqref{measure-interference} establishes the conservation of on-shell energies of incoming and outgoing particles.

We have partially aligned the integration measures of all interference diagrams arising from the same supergraph by rerouting each of them according to a fixed loop momentum basis $\mathbf{b}^\Gamma_\emptyset$ of the corresponding supergraph $\Gamma$, which is a first important step towards proving local cancellation of IR singularities.
We observe that, if a reference basis of the supergraph is fixed, the only dependence on $\mathbf{s}$ left in the measure element is due to the Dirac delta enforcing the conservation of on-shell energies across each Cutkosky cut.
Aligning this last element of the measures requires a more advanced mathematical construction, presented in detail in sect.~\ref{sect:IR_cancellation_proof}.
In the rest of this section, we first discuss the singular structure of amplitudes and derive a heuristic cancellation argument, as both of these aspects do not strictly rely on the complete alignment of the measures $[d\Pi^\mathbf{s}]$.

\newpage
\subsection{Identification of E-surfaces with cuts}
\label{sec:threshold_to_cc_cuts}

The singular surfaces of interference diagrams are also singular surfaces of the supergraph. More specifically, E-surfaces of amplitudes correspond to intersections of E-surfaces of the supergraph, as one can think of E-surfaces of the amplitudes as E-surfaces of the supergraph evaluated on the E-surface corresponding to the energy conservation delta.
Indeed, we observe that after solving the energy conservation Dirac delta in eq.~\eqref{measure-interference}, the integrand is evaluated on the zeros of the following E-surface:
\begin{equation}\label{E-surface}
    \eta_\mathbf{s}=\sum_{i\in\mathbf{c}_\mathbf{s}} E_i - \sum_{j\in \mathbf{a}}E_j, 
\end{equation}
which is itself an E-surface of the supergraph in its LTD representation. 

Eq.~\eqref{E-surface} also suggests a useful identification of E-surfaces with connected subsets of the vertices. This graphical representation of thresholds, and the notion of connectivity, encodes the causal ordering of the scattering events (the vertices). The particles connecting two vertices on different sides of the cut generate a singularity by simultaneously lying on their respective mass shell. Cutkosky cuts, in particular, correspond to thresholds in which the incoming momenta enter the boundary of a subset of vertices and produce $|\mathbf{c}_\mathbf{s}|$ outgoing (that is, with opposite energy flow) on-shell particles. 
More specifically, let
\begin{equation}
\mathcal{E}=\left\{\boldsymbol{\tau}\subset \mathbf{v}_{\text{int}}\middle| \ \delta^\circ(\mathbf{\boldsymbol{\tau}}), \ \delta^\circ(\mathbf{v}_{\text{int}}\setminus \boldsymbol{\tau}) \ \text{are connected}\right\}.
\end{equation}
Then an E-surface of the supergraph can be represented as a tuple $(\boldsymbol{\tau},\alpha)$, with $\mathbf{\boldsymbol{\tau}}\in\mathcal{E}$ and $\alpha\in\{\pm 1\}$. The parametric equation is
\begin{equation}\label{e-surface-equation-cut} 
\eta_{(\mathbf{\boldsymbol{\tau}},\alpha)}=\sum_{j\in \mathbf{a}\cap \delta(\boldsymbol{\tau})} E_j - \sum_{j\in (\mathbf{e}_{\text{ext}}\setminus\mathbf{a})\cap \delta(\boldsymbol{\tau})} E_j  +\sum_{i\in \delta(\boldsymbol{\tau})\setminus \mathbf{e}_{\text{ext}}}\alpha\;E_i=0,
\end{equation}
which is a general equation establishing the generic form of a threshold singularity on the supergraph, for an on-shell energy flow assigned to the external edges $\mathbf{e}_{\text{ext}}$ according to its bipartition in $\mathbf{a}$ and $\mathbf{e}_{\text{ext}}\setminus \mathbf{a}$. The set $\mathcal{E}$ can be mapped surjectively to the set of singularities of the supergraph. This is one of the results of the Manifestly Causal LTD representation~\cite{Capatti:2020ytd, Aguilera-Verdugo:2020kzc, Verdugo:2020kzh, Ramirez-Uribe:2020hes}.

According to our earlier definition of Cutkosky cuts, the E-surfaces of the supergraph which correspond to interference diagrams are only those corresponding to cuts whose boundary contains all incoming edges. 
It is useful to decompose the set $\mathcal{E}$ as the disjoint union of the following four sets:
\begin{itemize}
     \item $\mathcal{E}_{\text{s-ch}}=\left\{\boldsymbol{\tau}\in\mathcal{E}\ \middle|\ \delta(\mathbf{\boldsymbol{\tau}})\cap\mathbf{e}_{\text{ext}}=\mathbf{a} \ \text{or}  \ \delta(\mathbf{\boldsymbol{\tau}})\cap\mathbf{e}_{\text{ext}}=\mathbf{e}_{\text{ext}}\setminus\mathbf{a} \right\}$; an element $\boldsymbol{\tau}\in\mathcal{E}_{\text{s-ch}}$, is said to be \it s-channel-like\rm.
    \item $\mathcal{E}_{\text{t-ch}}=\left\{\boldsymbol{\tau}\in\mathcal{E}\ \middle|\ |\mathbf{a}|>|\delta(\mathbf{\boldsymbol{\tau}})\cap\mathbf{a}|>0, \ |\mathbf{e}_{\text{ext}}\setminus \mathbf{a}|>|\delta(\mathbf{\boldsymbol{\tau}})\cap(\mathbf{e}_{\text{ext}}\setminus \mathbf{a})|>0\right\}$; an element $\boldsymbol{\tau}\in\mathcal{E}_{\text{t-ch}}$ is said to be \it t-channel-like\rm.
    \item $\mathcal{E}_{\text{int}}=\left\{\boldsymbol{\tau}\in\mathcal{E}\ \middle|\ |\delta(\mathbf{\boldsymbol{\tau}})\cap\mathbf{e}_{\text{ext}}|=0 \text{ or }|\delta(\mathbf{\boldsymbol{\tau}})\cap\mathbf{e}_{\text{ext}}|=|\mathbf{e}_{\text{ext}}|\right\}$; an element $\boldsymbol{\tau}\in\mathcal{E}_{\text{int}}$, is said to be \it internal-like\rm,
    \item $\mathcal{E}_{\text{isr}}=\mathcal{E}\setminus (\mathcal{E}_{\text{s-ch}}\cup\mathcal{E}_{\text{t-ch}}\cup\mathcal{E}_{\text{int}})$; an element $\boldsymbol{\tau}\in\mathcal{E}_{\text{isr}}$, is said to be \it Initial-State-Radiation(ISR)-like\rm.
 \end{itemize}
so that $\mathcal{E}=\mathcal{E}_{\text{s-ch}}\cup\mathcal{E}_{\text{t-ch}}\cup \mathcal{E}_{\text{isr}}\cup\mathcal{E}_{\text{int}}$. We will consider all these sets to be modulo conjugation, that is a cut $\mathbf{s}$ and a cut $\mathbf{v}\setminus\mathbf{s}$ are equivalent. The E-surfaces in $\mathcal{E}_{\text{s-ch}}$ are the only ones that divide the supergraph into two graphs that have all incoming particles or none of them. 
\vSWITCH{
}{ %
The treatment of E-surfaces in $\mathcal{E}_{\text{isr}}$ leads to interesting theoretical observations that we discuss further in sect.~\ref{sec:ISR_outlook}.
} %

The singularities of the interference diagrams with Cutkosky cut $\mathbf{c}_{\boldsymbol{\tau}}$, can be described as the intersection of two E-surfaces, one describing the on-shell energy conservation condition associated with the Cutkosky cut, and the other being a singularity of an amplitude. More specifically, the location of the singularities of the interference diagram is characterised by the set of points which, when embedded in $\mathbb{R}^{3L}$, satisfies the following equations:
\begin{equation}
\label{singularities-interferences}
    \eta_{(\boldsymbol{\tau},1)}=0, \ \eta_{(\mathbf{s},\alpha)}=0, \ \ \boldsymbol{\tau}\in \mathcal{E}_{\text{s-ch}}, \ \mathbf{s}\in\mathcal{E}, \text{ with } \mathbf{s}\subset \boldsymbol{\tau} \text{ or } \boldsymbol{\tau} \subset \mathbf{s}.
\end{equation}
This follows from a repeated application of the principle identifying connected cuts and E-surfaces to the amplitudes participating in the interference diagram.
One interesting consequence of this claim is that no t-channel can be a singularity of the interference diagram.
\eqref{singularities-interferences} plays a crucial role in the determination of the cancellation pattern of IR singularities and precisely relates the singular structure of interference diagrams to those of the corresponding supergraph.
For two s-channel E-surfaces corresponding to the cuts $\boldsymbol{\tau}$ and $\mathbf{s}$, one can also define
\begin{equation}
    \eta_{(\boldsymbol{\tau},1)}|_\mathbf{s}=\sum_{i\in \mathbf{c}_{\boldsymbol{\tau}}} E_i-\sum_{j\in \mathbf{c}_\mathbf{s}} E_j.
\end{equation}
which is an alternative representation for the E-surface $\eta_{\boldsymbol{\tau}}$ when evaluated at the zeros of $\eta_\mathbf{s}$. In the following, if the index $\alpha$ is suppressed, it is assumed to take the value $\alpha=1$. 
As we shall see in sect.~\ref{sec:XOR_and_heuristics}, the property $\eta_{(\boldsymbol{\tau},1)}|_\mathbf{s}=-\eta_{(\mathbf{s},1)}|_{\boldsymbol{\tau}}$ plays a crucial role in the local cancellation of pinched singularities in LU since both $\mathbf{c}_\mathbf{s}$ and $\mathbf{c}_{\boldsymbol{\tau}}$ are valid Cutkosky cuts of the supergraph.

\subsection{Pinched E-surfaces and their properties}
\label{sec:pinched_E_surface_discussion}

The notion of pinching has a direct physical interpretation as it is associated with the degeneracy of massless particles in collinear or soft configurations. 
A general definition of pinched E-surfaces characterises them as poles that cannot be deformed around via a complex contour.

In order to study the conditions for which an E-surface becomes pinched, we provide each integral $\sigma^\mathcal{O}_\mathbf{s}$ with its own deformation, constructed by analytically continuing the spatial degrees of freedom of the loop variables forming the momentum basis of the left and right subgraphs $\delta^\circ (\mathbf{s})$ and $\delta^\circ (\mathbf{v}\setminus\mathbf{s})$. 
Given a basis $\mathbf{b}$ of the supergraph, we consider the contour
\begin{equation}
    \vec{q}_e=p_0+\sum_{j=1}^L \alpha_{ej} \vec{k}_j\rightarrow \vec{q}_e - \mathrm{i} \sum_{j=1}^L \alpha_{ej} \vec{\kappa}_j, \ e\in \mathbf{b},
\end{equation}    
where 
\begin{equation}
    \sum_{j=1}^L \alpha_{e'j} \vec{\kappa}_j=0, \ \forall e'\in \delta(\mathbf{s}),
\end{equation}
which establishes that the momenta of the particles in the Cutkosky cut is kept real as they enter the observable function.
It follows that understanding the pinching condition for interference diagrams is equivalent to understanding it in the context of amplitudes. 
More explicitly, we can analyze pinching within the object $\mathcal{A}_\mathbf{s}^{\vec{\mu}}(\vec{k}-i\vec{\kappa})$. Since the deformation only affects the amplitudes, it must satisfy the four constraints laid out in ref.~\cite{Capatti:2019edf}, where a general deformation for amplitudes in the LTD representation is constructed: the continuity constraint, the causal constraint, the expansion validity constraint, and the complex pole constraint.

 The continuity constraint imposes that any valid deformation must go to zero faster than $E_i$ on the zeros of $E_i$. We thus conclude that any surface in
\begin{equation}\label{soft-region}
    \text{S}_{\mathbf{x}}=\left\{\vec{k}\in\mathbb{R}^{3L} \ \middle|\  \lVert\vec{q}_i(\vec{k})\rVert^2+m_i^2=0,\  \forall i \in \mathbf{x}\right\}, \ \mathbf{x}\subset \mathbf{e}
\end{equation}
is a pinched surface due to a soft configuration (when $m_i^2=0$). We now turn to the causal constraint. An E-surface is pinched if it is impossible to satisfy the causal constraint for points on it, that is
\begin{equation}\label{pinch-condition}
   \text{Im}[\eta(\vec{k}-i\vec{\kappa})]=\mathcal{O}(\lVert \vec{\kappa}\rVert^2), \hspace{0.3cm}\forall \vec{k} \text{ with } \eta(\vec{k})=0, \ \forall \vec{\kappa} \,,
\end{equation}
with the understanding that there exists no contour deformation winding around the thresholds and that is compatible with the causal constraints. 

An intuitive understanding of this condition (e.g. see fig.~6 of ref.~\cite{Capatti:2019edf}) is obtained by analysing the complex zeros of the E-surface.
The real space is entirely sandwiched between two complex surfaces denoting the zeros of $\eta$. 
The two surfaces become purely real simultaneously at the location of the pinches, as established by the complex pole constraint~\cite{Gong:2008ww,Collins:2020euz,Sterman:1994ce}. Any valid contour deformation is thus constrained to be identically zero at the location of the pinches. 

\begin{SCfigure}[1][ht!]
\centering
 \caption{\label{fig:amplitude_E_surf} Example of a supergraph E-surface $\eta_{\boldsymbol{\tau}}|_\mathbf{s}$ ($=-\eta_{\mathbf{s}}|_{\boldsymbol{\tau}}$) identified by the two sets of vertices $\mathbf{s}=\{2,3,4,5,6,7\}$ and $\boldsymbol{\tau}=\{4,5,6,7\}$ defining the two Cutkosky cuts $\mathbf{c}_\mathbf{s}=\{e_1,e_6\}$ and $\mathbf{c}_{\boldsymbol{\tau}}=\{e_2,e_3,e_4,e_6\}$. The two-loop amplitude $\mathcal{A}^{\vec{\mu}}_\mathbf{s}$ on the left of the Cutkosky cut $\mathbf{c}_\mathbf{s}$ has an amplitude E-surface $\eta_{(\boldsymbol{\tau}',1)}^{\mathbf{c}_\mathbf{s}}$, with $\boldsymbol{\tau}'=\mathbf{s}\setminus\boldsymbol{\tau}=\{2,3\}$, which is pinched for
 $-\vec{q}_i=x_i \vec{p}_{\boldsymbol{\tau}'}, \ \vec{q}_m=-x_m \vec{p}_{\boldsymbol{\tau}'}, \ \vec{q}_l=-x_l \vec{p}_{\boldsymbol{\tau}'},\ \forall i\in \{e_1\}, \ \forall m\in \emptyset, \ \forall l \in \{e_2,e_3,e_4\}$, with $\vec{p}_{\boldsymbol{\tau}'}=-q_1$. More concisely, for $x_i>0$, the pinch happens for $\vec{q}_1//\vec{q}_2//\vec{q}_3//\vec{q}_4$. 
 The complex-conjugate of the tree-level amplitude $\mathcal{A}^{\vec{\mu}}_{\mathbf{v}\setminus\boldsymbol{\tau}}$ on the right of the Cutkosky cut $\mathbf{c}_{\boldsymbol{\tau}}$ has no amplitude E-surfaces, but instead features the familiar real-emission triple-collinear phase-space singularity which, within the LU framework, cancels $\mathcal{A}^{\vec{\mu}}_\mathbf{s}$ on the pinch of its amplitude E-surface.}
\includegraphics[width=0.5\textwidth]{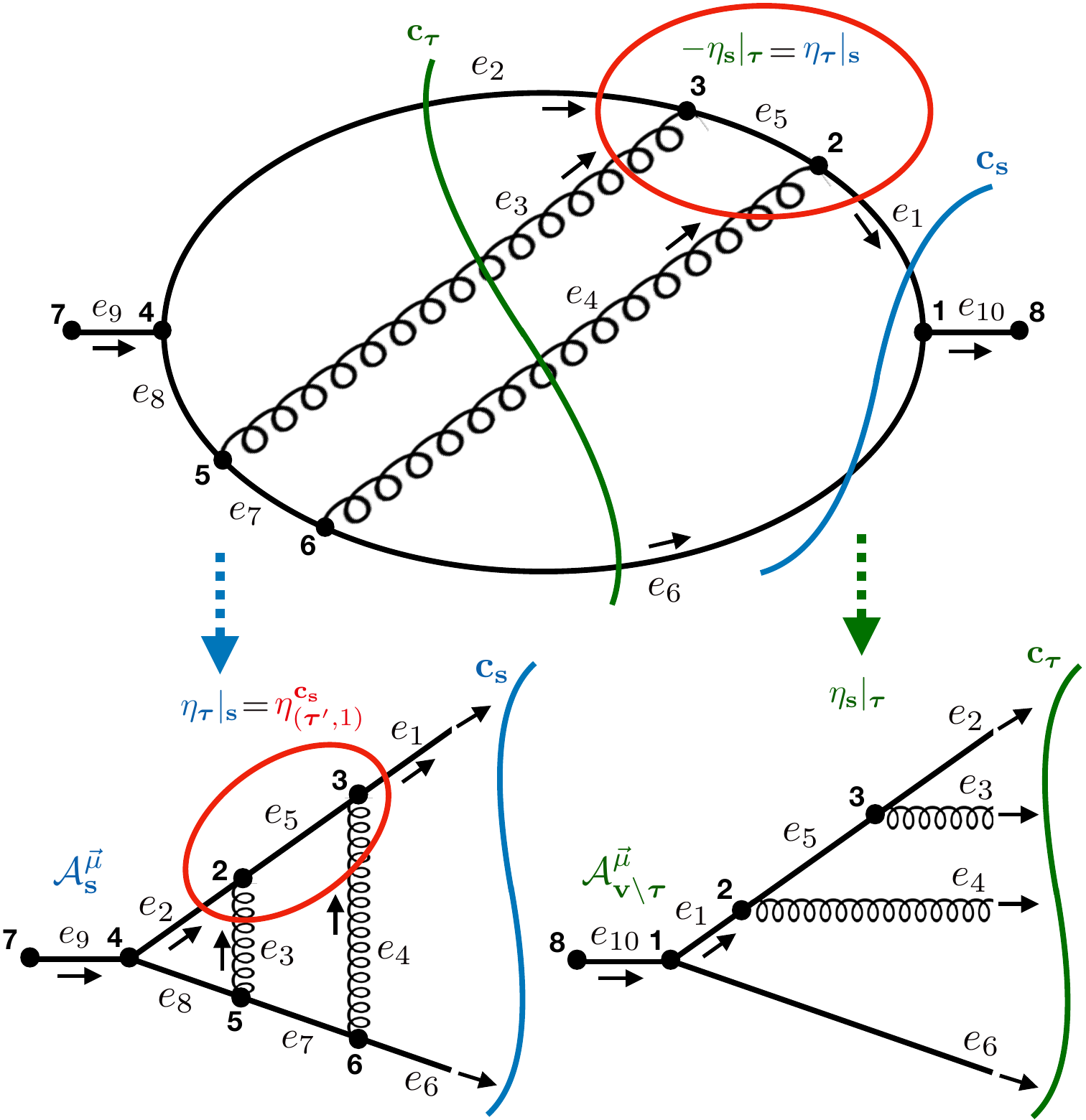}
\end{SCfigure}

We now turn to the identification of the E-surfaces within the (loop) amplitude $\mathcal{A}^{\vec{\mu}}_\mathbf{s}$ on the left of a particular Cutkosky cut $\mathbf{c}_\mathbf{s}$. In order to be able to characterise all possible E-surfaces of such amplitudes, we identify them with $\eta_{(\boldsymbol{\tau}',\alpha)}^{\mathbf{c}_\mathbf{s}}$, and $\boldsymbol{\tau}'\subset \mathbf{s}$ , which is part of an interference diagram corresponding to the supergraph $(G,\mathbf{a})$, that is 
\begin{equation}\label{e-surface-equation-cut2} 
\eta_{(\boldsymbol{\tau}',\alpha)}^{\mathbf{c}_\mathbf{s}}=\sum_{j\in \mathbf{a}\cap \delta(\boldsymbol{\tau}')} E_j - \sum_{j\in \mathbf{c}_\mathbf{s}\cap \delta(\boldsymbol{\tau}')} E_j  +\sum_{i\in \delta(\boldsymbol{\tau}')\setminus \delta(\mathbf{s})}\alpha\;E_i=0,
\end{equation}
with $\alpha=\pm 1$. We choose, without loss of generality, that all the edges in $\delta(\boldsymbol{\tau}')$ have the same orientation. The imaginary part of such E-surface can be written conveniently by expanding to first order in the Taylor expansion. For any $j\in \delta(\boldsymbol{\tau}')\setminus \delta(\mathbf{s})$ identifying the dependent edge (for example, for the amplitude E-surface give in fig.~\ref{fig:amplitude_E_surf}, one can choose $j=e_2$ and write $q_2=q_1-q_3-q_4$), we can define $\mathbf{b}_{\boldsymbol{\tau}'}=\delta(\boldsymbol{\tau}')\setminus \delta(\mathbf{s})\setminus\{j\}$ and:
\begin{equation}\label{im-part-e-surf}
\text{Im}[\eta_{(\boldsymbol{\tau}',\alpha)}^{\mathbf{c}_\mathbf{s}}]=\sum_{i\in \mathbf{b}_{\boldsymbol{\tau}'}} \vec{Q}_i(\vec{\kappa})\cdot\Bigg(\frac{\vec{q}_i}{E_i}+\frac{\vec{p}_{\boldsymbol{\tau}'}+\sum_{m\in \mathbf{b}_{\boldsymbol{\tau}'}}\vec{q}_m}{E_{j}}\Bigg)+\mathcal{O}(\lVert \kappa \rVert^3).
\end{equation}
where 
\begin{equation}
    \vec{p}_{\boldsymbol{\tau}'}=\sum_{m\in \delta(\boldsymbol{\tau}')\cap\delta(\mathbf{s})} \vec{q}_m.
\end{equation}
The condition of vanishing imaginary part of eq.~\eqref{pinch-condition} is trivially satisfied for E-surfaces arising from real emissions only (e.g. $\eta_{\mathbf{s}}|_{\boldsymbol{\tau}}$ of fig.~\ref{fig:amplitude_E_surf}), as there are no loop variables that can be contour deformed in this case.
Applying the pinched condition in eq.~\eqref{pinch-condition} to eq.~\eqref{im-part-e-surf}, and using that the $\vec{Q}_i$'s are linearly independent by construction,
we see that the requirement that the imaginary part must be identically zero for every value of the deformation field, implies that each of the vector that $\vec{Q}_i$ multiplies must themselves be identically zero. That is:   
\begin{equation}\label{cond_pinch1}
\frac{\vec{q}_i}{E_i}+\frac{\vec{q}_j}{E_{j}}=0, \hspace{0.3cm} \forall i\in\mathbf{b}_{\boldsymbol{\tau}'}
\end{equation}
which can only be satisfied when $\vec{q}_i$ and $\vec{q}_j$ are anticollinear. Observe that for massless on-shell momenta, the two vectors are normalized to unity. On the other hand, in the massive case the two vectors have varying norms that range between zero and one. 
This prohibits pinching of E-surfaces with non-zero masses.
We recall that soft emission of massless particles from massive ones still lead to pinching in virtue of eq.~\eqref{soft-region}.

Using eq.~\eqref{cond_pinch1} and eq.~\eqref{e-surface-equation-cut2} we provide a necessary and sufficient condition for $\eta_{(\boldsymbol{\tau}',\alpha)}^{\mathbf{c}_\mathbf{s}}$ to be pinched:
\begin{itemize}
    \item $m_j=0, \ \forall j\in\delta(\boldsymbol{\tau}')$, that is, all particles participating in the threshold are massless,
    \item $ \vec{q}_i=x_i \vec{p}_{\boldsymbol{\tau}'}, \ \vec{q}_m=-x_m \vec{p}_{\boldsymbol{\tau}'}, \ \vec{q}_l=-\alpha x_l \vec{p}_{\boldsymbol{\tau}'},\ \forall i\in \delta(\boldsymbol{\tau}')\cap \mathbf{c}_\mathbf{s}, \ \forall m\in \delta(\boldsymbol{\tau}')\cap \mathbf{a}, \ \forall l \in \delta(\boldsymbol{\tau}')\setminus \delta(\mathbf{s})$, with all the $x_i\ge 0, \forall i \in \delta(\boldsymbol{\tau}')$, that is, the external particles participating in the threshold are all simultaneously collinear to a given direction and all the virtual particles participating in the threshold are all simultaneously anti-collinear to that direction (when assuming that all edges in $\delta(\boldsymbol{\tau})$ are either flowing inward or outward of the set $\boldsymbol{\tau}$).
\end{itemize}
These conditions allow to completely characterise pinched surfaces of amplitudes in their LTD representation, on top of providing the precise locations of the singularities. 
We recall that thresholds of interference diagrams can be seen as intersections of E-surfaces of the supergraph. 
Analogously, pinched points of interference diagrams, are contained in (but do not coincide with) such intersections. 
We show in fig.~\ref{fig:amplitude_E_surf} and explicit example of the correspondence between the E-surface $\eta_{\boldsymbol{\tau}}|_\mathbf{s}$ of a supergraph and its counterpart $\eta_{(\boldsymbol{\tau}',1)}^{\mathbf{c}_\mathbf{s}}$ in the amplitude obtained when imposing the Cutkosky cut $\mathbf{c}_\mathbf{s}$.
We stress that the set of vertices $\boldsymbol{\tau}'$ defining the amplitude E-surface $\eta_{(\boldsymbol{\tau}',1)}^{\mathbf{c}_\mathbf{s}}$ relates to the two sets $\boldsymbol{\tau}$ and $\mathbf{s}$ identifying its corresponding supergraph E-surface $\eta_{\boldsymbol{\tau}}|_\mathbf{s}$ with $\boldsymbol{\tau}'=\mathbf{s}\setminus\boldsymbol{\tau}$ (when $\boldsymbol{\tau}\subset\mathbf{s}$).

The t-channel type of amplitude E-surfaces never induce divergences in LU as their pinched configuration is either excluded by the observable~\cite{Melnikov_1997} and/or regulated by the propagator width assigned to unstable particles.
We are therefore interested in s-channel type of supergraph E-surfaces and assume each particle to be massless, then the set of points at which any \cmtVH{We just limited ourselves to s-channel here, so we must say "any such" no?}E-surface of any interference diagram corresponding to a fixed supergraph $\Gamma$ is pinched can be written as
\begin{equation}
    \mathcal{P}=\text{H}\cap\Bigg(\bigcup_{\boldsymbol{\tau}\in\mathcal{E}_{\text{s-ch}}}\delta\eta_{\boldsymbol{\tau}}\Bigg) 
\text{
 with 
}
\text{H}=\bigcup_{\substack{\mathbf{s},\boldsymbol{\tau}\in\mathcal{E}_{\text{s-ch}}\\ \boldsymbol{\tau}\subset \mathbf{s}}}\text{H}_{\mathbf{s}\boldsymbol{\tau}},
\end{equation}
and
\begin{equation}
\text{H}_{\mathbf{s}\boldsymbol{\tau}}=\left\{\vec{k}\in\mathbb{R}^{3L}\middle| \vec{q}_i(\vec{k})= x_i\vec{p}, \ \vec{q}_j(\vec{k})=-x_j\vec{p}, \ \forall i\in \mathbf{c}_\mathbf{\boldsymbol{\tau}}\setminus\mathbf{c}_\mathbf{s}, \ \forall j\in\mathbf{c}_\mathbf{s}\setminus\mathbf{c}_{\boldsymbol{\tau}}, \ \forall \vec{p}\in\mathbb{R}^3\right\}.
\end{equation}
Adding masses to particles of the interference diagrams decreases the size of $\text{H}$ and, consequently, that of $\mathcal{P}$.

This concludes the study of pinched E-surfaces in the LTD framework. It is worth mentioning that a proper classification of the pinched E-surfaces requires studying their intersections. However, this study is unnecessary to prove FSR cancellations and is outside the scope of this work.

\subsection{Local cancellations for final-state radiation within a toy model}
\label{sec:XOR_and_heuristics}

\begin{figure}
\centering
\input{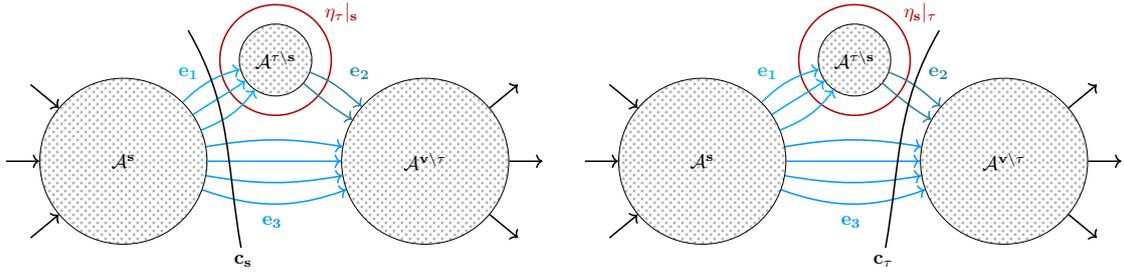}
\caption{Two different interference diagrams, corresponding to Cutkosky cuts $\mathbf{c}_\mathbf{s}$ and $\mathbf{c}_{\boldsymbol{\tau}}$. The two Cutkosky cuts identify a singularity of the amplitudes, $\eta_{\boldsymbol{\tau}}|_\mathbf{s}=-\eta_\mathbf{s}|_{\boldsymbol{\tau}}$. The direction of the edges identify the on-shell energy flow necessary for the particles in $\delta(\boldsymbol{\tau}\setminus\mathbf{s})$ to participate in a physical threshold.}\label{fig:IR_cancellation_pattern}
\end{figure}

In this section, we render the cancellation shown fig.~\ref{fig:amplitude_E_surf} more systematic by investigating it within a toy model.
Such local cancellation requires a specific alignment of the integration measures in order to solve the Dirac deltas enforcing on-shell energy conservation. This treatment is carried out in detail in sect.~\ref{sec:causal_flows} and we first discuss here the cancellation mechanism for a simplified model.
In order to carry out the argument, we consider an analogue for the integrand of interference diagrams which is constructed in the following way: each interference diagram, identified by $\mathbf{s}\in\mathcal{E}_{\text{s-ch}}$, shall be associated to a function $\omega^\mathbf{s}:\mathbb{R}^{3L}\rightarrow \mathbb{R}$ which is the product of the LTD representations of the graphs $\delta
^\circ(\mathbf{s})$ and $\delta^\circ(\mathbf{v}\setminus\mathbf{s})$ times the product of the inverse energies of all the particles in the Cutkosky cut $\mathbf{c}_\mathbf{s}$,
\begin{equation}
    \omega^\mathbf{s}(\{E_i\}_{i\in\mathbf{e}}, \vec{k})=\frac{1}{\prod_{i\in\mathbf{c}_\mathbf{s}}E_i}A_\mathbf{s}A_{\mathbf{v}\setminus \mathbf{s}}\Bigg|_{\{q_i^0=X^{\mathbf{s}}_iE_i\}_{i\in\mathbf{c}_\mathbf{s}}}
\end{equation}
with
\begin{equation}
    A_\mathbf{s}=\int \prod_{i=1}^L \frac{dk_i^0}{(2\pi)}\frac{N}{\prod_{j\in\mathbf{e}}(q_j^2-m_j^2)} \,,
\end{equation}
where we take the amplitude to have a scalar numerator, that is $A^{\vec{\mu}}_\mathbf{s}=A_\mathbf{s}$, for simplicity. The singularities of $\omega^\mathbf{s}$ are the same as those of the two amplitudes integrands $A_\mathbf{s}$, $A_{\mathbf{v}\setminus\mathbf{s}}$, plus the integrable singularities due to the inverse energies of the particles in the Cutkosky cut. More specifically, the E-surfaces of $\omega^\mathbf{s}$ correspond to the zeros of E-surfaces satisfying eq.~\eqref{singularities-interferences}; thus $\omega^\mathbf{s}$ is a valid analogue of the integrand of an interference diagram. Furthermore, $\omega^\mathbf{s}$ exhibits an interesting local factorisation property in the neighbourhood of such singularities: following the convention set in figure~\ref{fig:IR_cancellation_pattern}, in the limit $\eta_{\boldsymbol{\tau}}|_\mathbf{s}\rightarrow 0$, each amplitude integrand can be factorised as a product of inverse energies for each element of $\mathbf{c}_{\boldsymbol{\tau}}$ multiplied by the LTD representation of the two subgraphs $\delta
^\circ(\mathbf{s})$ and $\delta
^\circ(\boldsymbol{\tau}\setminus\mathbf{s})$.  We observe that such a local factorisation property relates to analogous ones holding at the integrated level and also playing an important role in traditional computational methods.

Let $A^\mathbf{s}$ be divergent at a location identified by the E-surface $\eta_{\boldsymbol{\tau}}$.
Then this local factorisation property simply reads:
\begin{equation}
     \omega^\mathbf{s}(\{E_i\}_{i\in\mathbf{e}}, \vec{k})=\frac{1}{\prod_{i\in c_\mathbf{s}\cup c_\mathbf{\boldsymbol{\tau}}} 2E_i}A_\mathbf{s}A_{\boldsymbol{\tau}\setminus\mathbf{s}}A_{\mathbf{v}\setminus \boldsymbol{\tau}}\Bigg|_{C_{\mathbf{s}\boldsymbol{\tau}}} \frac{1}{\eta_{\boldsymbol{\tau}}|_{\mathbf{s}}}+\mathcal{O}(1), \ \eta_{\boldsymbol{\tau}}|_{\mathbf{s}}\rightarrow 0 \,,
\end{equation}
with
\begin{equation}
    C_{\mathbf{s}\boldsymbol{\tau}}=\{q_i^0=X_i^\mathbf{s}E_i\}_{i \in c_\mathbf{s}}\cup \{q_i^0=X_i^{\boldsymbol{\tau}}E_i\}_{i \in\cup c_{\boldsymbol{\tau}}}\,.
\end{equation}

We can now show the explicit cancellation pattern. Let us consider the sum of the functions $\omega^\mathbf{s}$ for all $\mathbf{s}\in\mathcal{E}_{\text{s-ch}}$, which is the analogue of the sum of all interference diagrams of a single supergraph. Such sum, if $\boldsymbol{\tau}\in\mathcal{E}_{\text{s-ch}}$, also contains the term $\omega^{\boldsymbol{\tau}}$. We observe that
\begin{equation}
     \omega^\mathbf{s}
     +\omega^{\boldsymbol{\tau}}=\frac{1}{\prod_{i\in c_\mathbf{s}\cup c_\mathbf{\boldsymbol{\tau}}} 2E_i}A_{\mathbf{s}}A_{\boldsymbol{\tau}\setminus\mathbf{s}}A_{\mathbf{v}\setminus \boldsymbol{\tau}}\Bigg|_{C_{\mathbf{s}\boldsymbol{\tau}}}\Bigg(\frac{1}{\eta_{\boldsymbol{\tau}}|_\mathbf{s}}+\frac{1}{\eta_\mathbf{s}|_{\boldsymbol{\tau}}}\Bigg)+\mathcal{O}(1), \ \eta_{\boldsymbol{\tau}}|_{\mathbf{s}}, \eta_\mathbf{s}|_{\boldsymbol{\tau}}\rightarrow 0 \,.
\end{equation}
Since $\eta_\mathbf{s}|_{\boldsymbol{\tau}}=-\eta_{\boldsymbol{\tau}}|_\mathbf{s}$, it follows that $\omega^\mathbf{s}+\omega^{\boldsymbol{\tau}}=\mathcal{O}(1)$ when $\eta_{\boldsymbol{\tau}}, \ \eta_\mathbf{s}\rightarrow 0$. Thus, the singularity at $\eta_{\boldsymbol{\tau}}|_\mathbf{s}$ of an interference diagram whose Cutkosky is identified by $\mathbf{s}$ cancels pairwise with the singularity at  $\eta_\mathbf{s}|_{\boldsymbol{\tau}}$ of an interference diagram whose Cutkosky is identified by $\boldsymbol{\tau}$. This heuristic argument can easily be generalised to an arbitrary number of E-surfaces which vanish simultaneously, by iterating the factorisation argument and using the fact that each s-channel E-surface $\eta_\mathbf{s}|_{\boldsymbol{\tau}}$ can be written as the difference of two variables, each being the sum of energies in the Cutkosky cut $\mathbf{c}_{\boldsymbol{\tau}}$ or $\mathbf{c}_\mathbf{s}$. This mechanism will be studied in more detail in sect.~\ref{sect:IR_cancellation_proof}.

In the next section, we will show how the cancellations unfold when including observables and construct a proof. As already mentioned, this will require re-expressing the integrand of interference diagrams in a different way by solving the energy conservation delta explicitly (or equivalently, by considering a contour integration of the LTD representation of the supergraph). After the integrands are rewritten in this fashion, the cancellations will be realised algebraically, similarly as for dual cancellations. 

\clearpage

\section{Local cancellations of threshold (IR) singularities}
\label{sec:general_IR_cancellation_proof}

In this section we present the major steps in defining a local representation of differential cross-sections that is manifestly free of IR singularities. As the paper is focused on final-state radiation, we will show that such a representation is integrable on the whole $\mathbb{R}^{3L}$ excluding the initial state radiation (ISR) contributions 
\vSWITCH{
.
}{ %
(see sect.~\ref{sec:ISR_outlook} for treatment of ISR).
} %

The conceptual unfolding of the proof is summarised by the following prescriptions
\begin{itemize}
    \item Given a supergraph, construct a flow $\vec{\phi}$ satisfying an ODE involving a reference vector field $\kappa$ satisfying the causal constraints laid out in ref.~\cite{Capatti:2019edf}. $\kappa$ is then called the \it causal vector field \rm and $\vec{\phi}$ is called the \it causal flow\rm. Then change variables so as to make it possible to perform a contour deformation in the flow parameter $t$, that is along the flow, thus allowing the use of the ordinary one-dimensional residue theorem. 
    
    \item Construct a local representation of differential cross-sections, that is a function $\sigma_{\text{d}}:\mathbb{R}^{3L}\rightarrow \mathbb{R}$. Such a function is locally equivalent to summing over the discontinuities of s-channel E-surfaces of the supergraph along a flow line.
    
    \item Show that $\sigma_{\text{d}}$ allows for a cancellation mechanism analogous to the type described in sect.~\ref{sec:XOR_and_heuristics}, and that is mathematically summarised by the partial fractioning identity
    \begin{equation}\label{p-fractioning-rel}
        \sum_{j=1}^N\prod_{\substack{i=1 \\ i\neq j}}^N\frac{1}{t_i-t_j}=0.
    \end{equation}
    
    \item Use this cancellation pattern to derive analytic constraints on observables by requiring that $\sigma_\text{d}$ is finite on $\mathbb{R}^{3L}$ excluding all the regions at which an initial-state E-surface vanishes. Next, we show that they are satisfied by observables that cluster particles with energy or relative direction under a mathematically well-defined scale $\delta$. In other words, these constraints match the usual requirement of IR-safety for collider observables. 
    
    \item Derive the scaling of $\sigma_{\text{d}}$ near soft points, and show that it depends on the scaling of the deformation field around soft points, thus relating the request of integrability of $\sigma_{\text{d}}$ with a constraint on the scaling of the causal flow.
    
    \item Perform power-counting and argue that soft points are always integrable in physical theories.
    
\end{itemize}
Before detailing the proof in sect.~\ref{sec:proof}, we construct the LU representation of the $e^+ e^- \rightarrow d \bar{d}$ example, which we use to unfold explicitly the steps presented above.

\subsection{Illustrative example: NLO correction to $e^+ e^- \rightarrow d \bar{d}$}
\label{sec:cancellation_example}

We start by recalling the LTD representation of the \DT{} supergraph, and rewrite eq.~\eqref{eq:dt_and_se_xsection} and eq.~\eqref{eq:dt_and_se_xsection_cut} directly in terms of the LTD representation of that supergraph.

    Observe that in the rest frame of $\gamma^\star$, the two threshold E-surfaces $\eta_{\mathbf{s}_1^{\text{r}}}$ and $\eta_{\mathbf{s}_2^{\text{r}}}$ have the same exact functional dependence in the loop variables. Accounting for this accidental degeneracy, we can write eq.~\eqref{eq:DT_SG_expression} as
\begin{equation}
M=\int \frac{d^3\vec{k'}}{(2\pi)^3}\frac{d^3\vec{l'}}{(2\pi)^3} \frac{f(\vec{k'},\vec{l'})}{\prod_{j\in\mathbf{e}_{\textrm{in}}}E_j \prod_{{\boldsymbol{\tau}}\in\{\mathbf{s}_1^{\text{v}},\mathbf{s}_2^{\text{v}},\mathbf{s}_1^{\text{r}}\}}\eta_{\boldsymbol{\tau}} \prod_{{\boldsymbol{\tau}}\in\{\mathbf{s}_1^{\circ},\mathbf{s}_2^{\circ}\}}\eta_{\boldsymbol{\tau}} }
\end{equation}
with
\begin{equation}
f(\vec{k'},\vec{l'})
=\left[\prod_{j\in\mathbf{e}_{\textrm{in}}}E_j\right]
\left[\prod_{{\boldsymbol{\tau}}\in\{\mathbf{s}_1^{\text{v}},\mathbf{s}_2^{\text{v}},\mathbf{s}_1^{\text{r}},\mathbf{s}_1^{\circ},\mathbf{s}_2^{\circ}\}}\eta_{\boldsymbol{\tau}} \right]
\sum_{\mathbf{b}\in\mathcal{B}}\int d {k'}^0 d {l'}^0  N\frac{\prod_{e\in\mathbf{b}} \delta^{(\sigma^\mathbf{b}_e)}(q_e^2-m_e^2)}{\prod_{e\in \mathbf{e}\setminus \mathbf{b}}(q_e^2-m_e^2)} \,.
\end{equation}
where the Cutkosky cuts $\{\mathbf{s}_1^{\text{v}},\mathbf{s}_2^{\text{v}},\mathbf{s}_1^{\text{r}}\}$ of the double-triangle are defined in eq.~\eqref{DT:sv_sr} and the additional $E$-surfaces $\{\mathbf{s}_1^{\circ},\mathbf{s}_2^{\circ}\}$ in eq.~\eqref{DT:s0}. A study of the divergences of $f$ reveals that $f$ is infinitely differentiable on $\mathbb{R}^{6}$. 

We now extract the residue of the threshold singularities of the LTD representation of the DT listed as $\mathbf{s}\in\mathcal{E}_{\text{s-ch}}$ by substituting $\eta_{\mathbf{s}}^{-1}$ with $\delta(\eta_\mathbf{s})\mathcal{O}_{\mathbf{s}}$, which is what is commonly referred to as the "application of the Cutkosky cuts.
This yields a representation that is trivially equivalent to that of eq.~\eqref{eq:dt_and_se_xsection} and eq.~\eqref{eq:dt_and_se_xsection_cut}, after all the Dirac deltas except for the one imposing the conservation of on-shell energy flowing across the Cutkosky cut $\mathbf{s}$ have been solved. We have
\begin{equation}\label{LTD_sigma}
\sum_{\mathbf{s}\in\mathcal{E}_{\text{s-ch}}^{\Gamma^{\textsc{DT}}}} \sigma_{\Gamma^{\textsc{DT}},\mathbf{s}}^{\mathcal{O}}=\sum_{\mathbf{s}\in\{\mathbf{s}_1^{\text{v}},\mathbf{s}_2^{\text{v}},\mathbf{s}_1^{\text{r}}\}}\int \frac{d^3 \vec{k}'d^3 \vec{l}' }{(2\pi)^6}\frac{ \mathcal{O}_\mathbf{s}\ \eta_\mathbf{s}\ \delta(\eta_\mathbf{s}) \ f(\vec{k'},\vec{l'})}{\prod_{j\in\mathbf{e}_{\textrm{in}}}E_j \prod_{{\boldsymbol{\tau}}\in\{\mathbf{s}_1^{\text{v}},\mathbf{s}_2^{\text{v}},\mathbf{s}_1^{\text{r}}\}}\eta_{\boldsymbol{\tau}} \prod_{{\boldsymbol{\tau}}\in\{\mathbf{s}_1^{\circ},\mathbf{s}_2^{\circ}\}}\eta_{\boldsymbol{\tau}}},
\end{equation}
where the sum runs over all possible Cutkosky cuts, and $\mathcal{O}_\mathbf{s}$ is, for now, a non-specified function whose functional form depends on the cut $\mathbf{s}$. It is clear that eq.~\eqref{LTD_sigma} can be obtained from the usual form of $\frac{d\sigma}{d\mathcal{O}}$ by applying LTD to the energy integrals left after applying the phase-space cuts. 
At the same time, eq.~\eqref{LTD_sigma} also expresses the well-known fact that Cutkosky cuts can be seen as the residues of the supergraph acquired by contour-deforming around its thresholds, which is also the core of the original derivation by Cutkosky~\cite{Cutkosky:1960}. 

We would like to solve the Dirac deltas on the right-hand side of eq.~\eqref{LTD_sigma} simultaneously for all Cutkosky cuts of the \DT{} topology, in a way that allows to write 
\begin{equation}
    \frac{d\sigma_{\Gamma^{\textsc{dt}}}}{d\mathcal{O}}=\sum_{\mathbf{s}\in\mathcal{E}_{\text{s-ch}}^{\Gamma^{\textsc{DT}}}}\sigma_{\Gamma^{\textsc{DT}},\mathbf{s}}^{\mathcal{O}}=\int \frac{d^3 \vec{k}'d^3 \vec{l}' }{(2\pi)^6} \sigma_{\text{d}}^{\Gamma^{\textsc{DT}}}.
\end{equation}
where $\sigma_{\text{d}}$ now contains no Dirac delta. In particular, the contribution from all interference diagrams stemming from the \DT{} topology should be written as a single integral over $\mathbb{R}^{3}\times\mathbb{R}^{3}$ of a particular integrand.
In following three sections we will discuss a general method to solve the remaining delta function encoding energy conservation.
We denote with $d\sigma_\Gamma / d\mathcal{O}$ the sum of all the interference diagrams arising from Cutkosky cuts of the supergraph $\Gamma$ and we suppress the $\Gamma$ superscript henceforth.

\subsubsection{Soper's rescaling for solving conservation of on-shell energies}
\label{sec:sopertrick}

For $1\rightarrow N$ processes, such as (effectively) $e^+ e^- \rightarrow d \bar{d}$, final-state singularities can be aligned at any perturbative order in QCD using Soper's $\delta$ solving strategy~\cite{Soper:1999xk} which offers an easy way to rewrite the phase-space measure in a form where there is no Dirac delta anymore.
It was presented for the first time for integrands with conformal symmetries. In the following we will generalise it to multi-scale integrands, for arbitrary masses, loop momentum routings and Lorentz frames. 

Consider the integral~\eqref{LTD_sigma}, and multiply it by the integral in $t$ of a normalized function $h(t)$:
\begin{equation}
\frac{d\sigma_{\Gamma^{\textsc{DT}}}}{d\mathcal{O}}=\sum_{\mathbf{s}\in\{\mathbf{s}_1^{\text{v}},\mathbf{s}_2^{\text{v}},\mathbf{s}_1^{\text{r}}\}} \int \frac{d^3 \vec{k}'d^3 \vec{l}' }{(2\pi)^6} \int_0^{\infty} dt\ h(t)\frac{ \mathcal{O}_\mathbf{s} \ \eta_\mathbf{s}\ \delta(\eta_\mathbf{s}) \ f(\vec{k'},\vec{l'})}{\prod_{j\in\mathbf{e}_{\textrm{in}}}E_j \prod_{{\boldsymbol{\tau}}\in\{\mathbf{s}_1^{\text{v}},\mathbf{s}_2^{\text{v}},\mathbf{s}_1^{\text{r}}\}}\eta_{\boldsymbol{\tau}} \prod_{{\boldsymbol{\tau}}\in\{\mathbf{s}_1^{\circ},\mathbf{s}_2^{\circ}\}}\eta_{\boldsymbol{\tau}}}
\label{eq:normalised_integral}
\end{equation}
We can now change variables from $(\vec{k}',\vec{l}')$ to $\vec{\phi}(t; \vec{k}',\vec{l}')$.
We call $\vec{\phi}$ the causal flow, because for any fixed $\vec{k}'$ and $\vec{l}'$, $\vec{\phi}$ denotes a curve which always flows outwards with respect to the Cutkosky cut E-surfaces. This new object will be described in more detail in sect.~\ref{sec:causal_flows}.
In the case of the example process considered in this section, we can choose\footnote{In this section, contrary to sect.~\ref{sec:general_IR_cancellation_proof}, we choose the argument of the causal flow to be $\log(t)$ and not $t$ for simplicity .} $\vec{\phi}(t; \vec{k}',\vec{l}')=t(\vec{k}',\vec{l}')$.
An illustration of this causal flow is given in fig.~\ref{fig:DT_causal_flow}.
This change of variables then yields:
\begin{equation}
\frac{d\sigma_{\Gamma^{\textsc{DT}}}}{d\mathcal{O}}=\sum_{\mathbf{s}\in\{\mathbf{s}_1^{\text{v}},\mathbf{s}_2^{\text{v}},\mathbf{s}_1^{\text{r}}\}}\int \frac{d^3 \vec{k}'d^3 \vec{l}' }{(2\pi)^6}ds\frac{ \mathcal{O}_\mathbf{s}\ \eta_\mathbf{s}\ f(t(\vec{k'},\vec{l'}))\  h(t)\ t^{6}}{\prod_{j\in\mathbf{e}_{\textrm{in}}}E_j \prod_{{\boldsymbol{\tau}}\in\{\mathbf{s}_1^{\text{v}},\mathbf{s}_2^{\text{v}},\mathbf{s}_1^{\text{r}},\mathbf{s}_1^{\circ},\mathbf{s}_2^{\circ}\}}\eta_{\boldsymbol{\tau}} }\delta\Bigg(\sum_{i\in\mathbf{c}_\mathbf{s}} E_i( t(\vec{k'},\vec{l'})) -Q^0\Bigg)
\label{eq:Soper_change_of_variables}
\end{equation}
and by solving Dirac's delta explicitly, we find
\begin{equation}\label{eq:xsection_ddbar_scaling}
\frac{d\sigma_{\Gamma^{\textsc{DT}}}}{d\mathcal{O}}=\sum_{\mathbf{s}\in\{\mathbf{s}_1^{\text{v}},\mathbf{s}_2^{\text{v}},\mathbf{s}_1^{\text{r}}\}} \int \frac{d^3 \vec{k}'d^3 \vec{l}' }{(2\pi)^6}\frac{ \mathcal{O}_\mathbf{s}\ \eta_\mathbf{s}\ f(t(\vec{k'},\vec{l'}))\  h(t)\ t^6}{|\partial_t \eta_\mathbf{s}|\prod_{j\in\mathbf{e}_{\textrm{in}}}E_j \prod_{{\boldsymbol{\tau}}\in\{\mathbf{s}_1^{\text{v}},\mathbf{s}_2^{\text{v}},\mathbf{s}_1^{\text{r}},\mathbf{s}_1^{\circ},\mathbf{s}_2^{\circ}\}}\eta_{\boldsymbol{\tau}} }\Bigg|_{t=t^\star_\mathbf{s}(\vec{k})},
\end{equation}
where $t^\star_\mathbf{s}$, is the unique (as we shall argue in sect.~\ref{sec:causal_flows}) value of $t$ such that the E-surface identified by the energy conservation delta vanishes. 

From that point onward, we will abbreviate our notation by defining $\vec{k}:=(\vec{k}',\vec{l}')$.
For a given $\vec{k}$, $t^\star_\mathbf{s}$ is a function satisfying
\begin{equation}
\eta_\mathbf{s}(t^\star_\mathbf{s}\vec{k})=\sum_{i\in \mathbf{c}_\mathbf{s}} E_i(t^\star_\mathbf{s}\vec{k}) - Q^0=0 .
\end{equation}
Observe that for every $\vec{k}$, $t_\mathbf{s}^\star$ is the factor with which to rescale the loop momenta in order for $t_\mathbf{s}^\star\vec{k}$ to lie on an E-surface.
Alternatively, one can think of fixing a point in loop momentum space and dilate or contract the E-surface by a quantity $1/t_\mathbf{s}^\star$ so that the point $\vec{k}$ lies on it.
If the E-surface contains the origin, then for every $\vec{k}$ there is only one positive value of $t^\star_\mathbf{s}$ such that $t_\mathbf{s}^\star\vec{k}$ lies on the E-surface.
This is a consequence of the convexity of $\eta$ as a function of $t$. 
Knowing that there is at most two solutions if $t$ is allowed to take positive and negative values, and since a solution $t_\mathbf{s}^\star$ satisfies
\begin{equation}\label{bounds_newton}
|t_\mathbf{s}^\star|<\frac{\sum_{j\in \mathbf{a}}E_j}{\sum_{i\in\mathbf{c}_\mathbf{s}}\lVert \vec{Q}_i(\vec{k}) \rVert}+\sum_{i\in\mathbf{c}_\mathbf{s}}\Bigg|\frac{\vec{Q}_i(\vec{k}) }{\lVert \vec{Q}_i(\vec{k}) \rVert^2}\cdot \vec{p}_i \Bigg|,
\end{equation}
\cmtVH{Did we specify th $Q_i$ notation anywhere?}
we conclude that the equation in $t$ can be solved numerically by using Newton's method with seeds provided by the bounds of the inequality in~\eqref{bounds_newton}. Since there are at most two solutions, Newton's method is guaranteed to converge. Thus, it follows that Soper's  $\delta$ solving strategy is numerically straightforward to implement.

\subsubsection{LU representation of \DT{} interferences}
\label{sec:IR_cancellation_example}

The next step is to relate the expression of the \DT{} supergraph given in eq.~\eqref{eq:DT_SG_expression} to that of the traditional expression $\frac{d\sigma}{d\mathcal{O}}$ of the NLO QCD accurate differential cross-section of the scattering process ${e^+ e^- \rightarrow d \bar{d}}$. The correspondence between the contributing threshold singularities of the LTD expression of the \DT{} supergraph and its Cutkosky cuts (fig.~\ref{fig:DT_CC_to_LTD}) shows that $\frac{d\sigma}{d\mathcal{O}}$ can be obtained by computing the residues associated with each of the causal surfaces for which we must however make sure to assign the observable function with the appropriate dependence. The fact that each Cutkosky cut involves the observable function with a \emph{different} functional dependence on the kinematics is the very reason why the residues from each of the singular surfaces of the supergraph must be computed separately. Indeed, when only interested in the \emph{fully inclusive} cross-section of $1\rightarrow2$ processes, one can instead consider directly computing the imaginary part of the supergraphs and extract from it the inclusive cross-section via the optical theorem (see, e.g. ref.~\cite{Herzog:2017dtz}).

Eq.\eqref{eq:xsection_ddbar_scaling} can then be written as
\begin{equation}
    \frac{d\sigma_\Gamma}{d\mathcal{O}}=\int \frac{d^3 \vec{k}' d^3\vec{l}'}{(2\pi)^6} \sigma_{\text{d}}(\vec{k})
\end{equation}
where we recall that we define $\vec{k}:=(\vec{k}',\vec{l'})$ and
\begin{equation} \label{eq:DT_diff_xsec_unevaluated}
    \sigma_{\text{d}}(\vec{k}) =
    \sum_{\mathbf{s}\in\mathcal{E}_{\vec{k},\vec{\phi}}} \text{Ind}_{\mathbf{s}} \lim_{t\rightarrow t^\star_\mathbf{s}} (t-t^\star_\mathbf{s}) \Bigg[\frac{f(t\vec{k})\ h(t)\ |\text{det}\mathbb{J}\vec{\phi}|\ \mathcal{O}_{\mathbf{s}}}{ \prod_{j\in\mathbf{e}_{\textrm{in}}}E_j \prod_{{\boldsymbol{\tau}}\in\{\mathbf{s}_1^{\text{v}},\mathbf{s}_2^{\text{v}},\mathbf{s}_1^{\text{r}},\mathbf{s}_1^{\circ},\mathbf{s}_2^{\circ}\}}\eta_{\boldsymbol{\tau}} }\Bigg],
\end{equation}
\cmtVH{It is a bit weird and not ideal to have a generic causal field $\vec{\phi}$ above when we're already restricting ourselves in this section to the rescaling change of variable.}
where we used the formal definition of the residue of a single pole located at $t_\mathbf{s}^\star$, and $\text{Ind}_{\mathbf{s}}=+\text{sign}[\partial_t \eta_\mathbf{s}(t^\star_\mathbf{s})]$. The symbol $\mathcal{E}_{\vec{k},\vec{\phi}}$ describes the ensemble of all threshold surfaces, i.e. Cutkosky cuts, which have a solution in $t$ given the change of variables induced by the causal field $\vec{\phi}(t;\vec{k})$ and for the particular sampling point $\vec{k}=(\vec{k}',\vec{l}')$ considered. Due to convexity, the number of solutions in $t$ of the E-surface equation is limited to one or potentially zero. However, in our simple $1\rightarrow N$ case, this change of variable amounts to a trivial rescaling which always offers exactly one solution for each threshold, so that we can effectively consider a summation over the complete set $\{\mathbf{s}_1^{\text{v}},\mathbf{s}_2^{\text{v}},\mathbf{s}_1^{\text{r}}\}$.

We choose the normalised function $h(t)$ to be
\begin{equation}
h(t)=\frac{1}{2 K_1(2 \sigma)} e^{-\sigma\frac{t^2 + 1}{t}} ,
\label{eq:ht_definition}
\end{equation}
where $K_1$ is the Bessel function of the second kind. 
This particular choice of a normalised function $h(t)$ is motivated by the fact that it vanishes exponentially fast at zero and infinity while being maximal at $t=1$. We fix the tunable parameter $\sigma$ to 1. These features naturally drive the integrator to probe points in space that are close to at least one Cutkosky surface (see sect.~\ref{sec:technical_multi_channeling} for more details regarding integration efficiency) and also avoids spurious integrable singularities at $t=0$. This also implies that the norms of the loop momenta before and after the rescaling treatment are of similar order of magnitude, thereby maintaining the direct interpretation of the location of the IR and UV domain.
We are now equipped to delve into the details of the cancellation of IR singularities and the numerical implementation of the \DT{} and \SE{} supergraphs within the formalism of Local Unitarity.

\subsubsection{E-surface cancellations for the double-triangle supergraph}

 In general, the direct evaluation of eq.~\eqref{eq:DT_diff_xsec_unevaluated} reads:
\begin{equation}\label{eq:DT_diff_xsec}
\sigma_{\text{d}} = \sum_{\mathbf{s}\in\{\mathbf{s}_1^{\text{v}},\mathbf{s}_2^{\text{v}},\mathbf{s}_1^{\text{r}}\}}
\frac{
    (t^\star_\mathbf{s})^6\ 
    f(t^\star_\mathbf{s}\vec{k})\ 
    h(t^\star_\mathbf{s})\ 
    \mathcal{O}_\mathbf{s}(\{\vec{q}_e(t^\star_\mathbf{s}\vec{k})\}_{e\in \mathbf{c}_\mathbf{s}})
}{\displaystyle{
    \abs{\partial_t \eta_\mathbf{s}(t^\star_\mathbf{s}\vec{k})}
    \prod_{i\in\mathbf{e}_{\textrm{in}}}E_i
    \prod_{{\boldsymbol{\tau}}\in\{\mathbf{s}_1^{\text{v}},\mathbf{s}_2^{\text{v}},\mathbf{s}_1^{\text{r}}\}\setminus\{\mathbf{s}\}} \left[\sum_{j>0}
    \frac{(t^\star_\mathbf{s}-t^\star_{{\boldsymbol{\tau}}})^j}{j!}
    \partial_t^j \eta_{\boldsymbol{\tau}} (t^\star_{\boldsymbol{\tau}}\vec{k})
    \right]
    \prod_{{\boldsymbol{\tau}}\in\{\mathbf{s}_1^{\circ},\mathbf{s}_2^{\circ}\}}\eta_{\boldsymbol{\tau}}
}} , 
\end{equation}
where we substituted each Cutkosky cut threshold surface $\eta_{\boldsymbol{\tau}}, \ {\boldsymbol{\tau}}\in\mathcal{E}_{\text{s-ch}}$ in the denominator by its Taylor expansion around its on-shell solution $t^\star_{\boldsymbol{\tau}}$ (so that the zeroth-order term of this expansion is necessarily zero).
We use the short-hand notation $\partial_t \eta_\mathbf{s}(t^\star_\mathbf{s}\vec{k}):=\partial_t \eta_\mathbf{s}(t)|_{t=t^\star_\mathbf{s}\vec{k}}$.
Thanks to the simple functional form of the rescaling change of variable as well as the fact that all propagators of the \DT{} supergraph are massless and that we are considering an incoming momentum configuration at rest, i.e. $Q^\mu=(Q^0,\vec{0})$, we can give a simple expression for the rescaling solution $t^\star_\mathbf{s}$ as well as the derivative function $ \partial_t^j \eta_{\boldsymbol{\tau}} (t\vec{k})$ of the Cutkosky surfaces:
\begin{eqnarray}
t^\star_\mathbf{s} &=& \frac{Q^0}{\eta_\mathbf{s}(\vec{k},0)} \\
\partial_t^j \eta_{\boldsymbol{\tau}} ( t\vec{k}, Q^0) &=& \begin{cases} 
\eta_{\boldsymbol{\tau}} ( \vec{k}, 0) & \mbox{if } j = 1 \\
0 & \mbox{if } j > 1
\end{cases},
\end{eqnarray}
so that:
\begin{equation}\label{eq:DT_simple_taylor}
\sum_{j>0}
\frac{(t^\star_\mathbf{s}-t^\star_{{\boldsymbol{\tau}}})^j}{j!}
\partial_t^j \eta_{\boldsymbol{\tau}} (t^\star_\mathbf{s})=
\left( \frac{Q^0}{\eta_\mathbf{s}(\vec{k},0)}-\frac{Q^0}{\eta_{\boldsymbol{\tau}}(\vec{k},0)} \right)\eta_{\boldsymbol{\tau}}(\vec{k},0)
,
\end{equation}
In our case, this expression of course reproduces the \emph{exact} expression of $\eta_{\boldsymbol{\tau}} ( t^\star_\mathbf{s} \vec{k}, Q^0)$ since the Taylor expansion terminates, but this is in general not the case for more complicated causal flows or when in presence of massive propagators. 

Proving the \emph{local} cancellation of IR singularities amounts to demonstrating that the expression of the differential cross section $\sigma_\text{d}$ in eq.~\eqref{eq:DT_diff_xsec} is integrable except for remaining UV divergences. From our earlier discussion, we can already argue that $f$ is free of any singularity. The causal nature of the field $\vec{\phi}$ inducing our change of variable also insures that $|\partial_t \eta_\mathbf{s}(t^\star_\mathbf{s})| \neq 0$ (see sect.~\ref{sec:causal_flows}). In our case, we have:
\begin{equation}
    |\partial_t \eta_\mathbf{s}(t^\star_\mathbf{s})| = \eta_\mathbf{s}(\vec{k},0)
\end{equation}
Finally the product of on-shell energies $\prod_{i\in\mathbf{e}_{\textrm{in}}}E_i$ and the product of inverse internal E-surfaces $\prod_{{\boldsymbol{\tau}}\in\{\mathbf{s}^\circ_1,\mathbf{s}^\circ_2\}}\eta_{\boldsymbol{\tau}}$ can be rewritten using the relation
\begin{align}
    E_i(t^\star_\mathbf{s}\vec{k})=\frac{Q^0 E_i(\vec{k})}{\eta_\mathbf{s}(\vec{k},0)}, \\
    \eta_{{\boldsymbol{\tau}}}(t^\star_\mathbf{s}\vec{k})=\frac{Q^0 \eta_{{\boldsymbol{\tau}}}(\vec{k})}{\eta_{\mathbf{s}}(\vec{k},0)}.
\end{align}
In conclusion, we can rewrite $\sigma_{\text{d}}$ by underlining that the denominator can be written as a polynomial in the energies and in the $\eta_\mathbf{s}$, $\mathbf{s}\in\mathcal{E}_{\text{s-ch}}$:
\begin{equation}\label{sigma-d-dt}
    \sigma_{\text{d}}=\frac{1}{(Q^0)^3\prod_{e\in\mathbf{e}_{\text{int}} }E_e(\vec{k})\prod_{{\boldsymbol{\tau}}\in\{\mathbf{s}^\circ_1,\mathbf{s}^\circ_2\}} \eta_{\boldsymbol{\tau}}(\vec{k})}\sum_{\mathbf{s}\in\{\mathbf{s}_1^{\text{v}},\mathbf{s}_2^{\text{v}},\mathbf{s}_1^{\text{r}}\}}\frac{ \eta_\mathbf{s}(\vec{k},0)^2 f (t^\star_\mathbf{s}\vec{k}) h (t^\star_\mathbf{s}) \mathcal{O}_\mathbf{s}(t^\star_\mathbf{s}\vec{k})}{\prod_{{\boldsymbol{\tau}}\in\{\mathbf{s}_1^{\text{v}},\mathbf{s}_2^{\text{v}},\mathbf{s}_1^{\text{r}}\}\setminus\mathbf{s}}(\eta_{\boldsymbol{\tau}}(\vec{k},0)-\eta_\mathbf{s}(\vec{k},0)) } \,.
\end{equation}

We can now rewrite $\sigma_{\text{d}}$ and show that the cancellations can be made explicit \emph{algebraically} in the denominators $(\eta_{\boldsymbol{\tau}}(\vec{k},0)-\eta_\mathbf{s}(\vec{k},0))$, thereby sidestepping their complicated kinematic dependence. One peculiar implication of this proof is then that it can be made for any parametric kinematic point $\vec{k}$, that is without taking any limit. This is particularly convenient given that enumerating all possible IR divergent kinematic limits becomes cumbersome at high perturbative orders.

The summand of $\sigma_{\text{d}}$ corresponding to $\mathbf{s}$, as given in eq.~\eqref{sigma-d-dt}, is clearly singular at the locations $\eta_{\boldsymbol{\tau}}|_{\mathbf{s}}=\eta_{\boldsymbol{\tau}}(\vec{k},0)-\eta_\mathbf{s}(\vec{k},0)=0$ and at soft points. That is, regions of kinematics space at which any energy vanishes. In order to make the notation less heavy, let us call $x_\mathbf{s}=\eta_{\mathbf{s}}(\vec{k},0)$ and suppress the dependence of functions unless their dependence is itself dependent on $\mathbf{s}$, which is an index that is summed over. 

In the absence of numerators, we could immediately show that the sum in eq.~\eqref{sigma-d-dt} is identically zero thanks to the general partial fractioning identity given in eq.~\eqref{p-fractioning-rel}. If observables are non-trivial, instead, we have to expand the numerator in the variable $t$. Let us assume, in the following, that
\begin{equation}\label{obs}
    \mathcal{O}_\mathbf{s}(\vec{k})=\mathcal{O}_{\boldsymbol{\tau}}(\vec{k}), \ \ \forall (\vec{k}) \text{ s.t. } x_{\boldsymbol{\tau}}=x_\mathbf{s}+\epsilon,
\end{equation}
for fixed $\epsilon>0$. This condition, which will be discussed in detail later, allows to state that in a neighbourhood of problematic points $[f\ h\  \mathcal{O}_\mathbf{s}](Q_0/x_\mathbf{s})=g(x_\mathbf{s})$, $\forall \mathbf{s}\in \mathcal{E}_{\text{s-ch}}$, where $g$ is a continuous function on $\mathbb{R}\setminus\{0\}$ in virtue of the continuity of the observable, of the numerator and of the normalising function. Its singularity at the origin is not problematic if we use that the integrand must initially be UV convergent, either on itself or with the aid of counterterms. With this in mind, we can write
\begin{equation}
    \sigma_{\text{d}}=\frac{1}{Q_0^3\prod_{e\in\mathbf{e}_{\text{int}} }E_e\prod_{{\boldsymbol{\tau}}\in\{\mathbf{s}^\circ_1,\mathbf{s}^\circ_2\}} \eta_{\boldsymbol{\tau}}}\sum_{\mathbf{s}\in\{\mathbf{s}_1^{\text{v}},\mathbf{s}_2^{\text{v}},\mathbf{s}_1^{\text{r}}\}}\frac{ x_\mathbf{s}^2 g (x_\mathbf{s})}{\prod_{{\boldsymbol{\tau}}\in\{\mathbf{s}_1^{\text{v}},\mathbf{s}_2^{\text{v}},\mathbf{s}_1^{\text{r}}\}\setminus\mathbf{s}}(x_{\boldsymbol{\tau}}-x_\mathbf{s}) }, \ \forall \vec{k} \text{ s.t. } x_{\boldsymbol{\tau}}=x_\mathbf{s}+\epsilon
\end{equation}
We will now rearrange the sum in a way that makes cancellations manifest. 
\begin{equation}\label{sigma-d-canc}
  \sigma_{\text{d}}=\frac{1}{Q_0^3\prod_{e\in\mathbf{e}_{\text{int}} }E_e\prod_{{\boldsymbol{\tau}}\in\{\mathbf{s}^\circ_1,\mathbf{s}^\circ_2\}} \eta_{\boldsymbol{\tau}}}\frac{\frac{x_{\mathbf{s}^{\text{v}}_1}^2 g(x_{\mathbf{s}^{\text{v}}_1})-x_{\mathbf{s}^{\text{r}}_1}^2 g(x_{\mathbf{s}^{\text{r}}_1})}{x_{\mathbf{s}^{\text{v}}_1}-x_{\mathbf{s}^{\text{r}}_1}} - \frac{x_{\mathbf{s}^{\text{v}}_2}^2 g(x_{\mathbf{s}^{\text{v}}_2})-x_{\mathbf{s}^{\text{r}}_1}^2 g(x_{\mathbf{s}^{\text{r}}_1})}{x_{\mathbf{s}^{\text{v}}_2}-x_{\mathbf{s}^{\text{r}}_1}} }{x_{\mathbf{s}^{\text{v}}_1}-x_{\mathbf{s}^{\text{v}}_2}}
\end{equation}
Written in this form, it is clear that away from soft points, when $x_{\mathbf{s}}-x_{\boldsymbol{\tau}}\rightarrow 0$, for ${\boldsymbol{\tau}},\mathbf{s}\in\{\mathbf{s}_1^{\text{v}},\mathbf{s}_2^{\text{v}},\mathbf{s}_1^{\text{r}}\}$, $\sigma_\text{d}$ is finite. A power-counting procedure can be set up to show that this integrand has at most integrable singularities at soft points. Such an analysis, however, requires studying the structure of $g$ and is performed rigorously in sect.~\ref{sec:power_counting_soft_integrable_singularities}. The straightforward cancellation structure that manifests itself in eq.~\eqref{sigma-d-canc} already alludes to its generalisation.
Also, except for considerations regarding the observable dependence, this cancellation pattern does not discriminate between types of singularities (pinched or non-pinched) and holds on intersection of singular surfaces.

Finally, we go back to the condition shown in eq.~\eqref{obs}, which enforces the IR safety of the observable. We will assume that the observables only depend on the size of the cut and the masses of the particles belonging to the Cutkosky cut, other than their momentum, that is:
\begin{equation}
\mathcal{O}_\mathbf{s}(\{\vec{q}_e(t^\star_\mathbf{s}\vec{k})\}_{e\in \mathbf{c}_\mathbf{s}})=\mathcal{O}_{|\mathbf{c}_\mathbf{s}|}(\{\vec{q}_e(t^\star_\mathbf{s}\vec{k}), m_e\}_{e\in \mathbf{c}_\mathbf{s}}) \,.
\end{equation}
When rewriting~\eqref{obs} explicitly for a pinched singularity and the momentum routing shown in fig.~\ref{fig:Two_SG_epemddx_NLO:DT_RxRA}, with $Q^\mu=(Q^0,\vec{0})$, we find:
\begin{eqnarray}
    &&\lim_{ (\vec{l}'-\vec{k}')\cdot \vec{k}' \;\rightarrow\; 0 }
 \mathcal{O}_3\left(\vec{p}_g=\vec{k}'-\vec{l}'\;;\;\vec{p}_{\bar{d}}=-\vec{k}'\;;\;\vec{p}_d=\vec{l}'\;\right) \nonumber\\
 &=&  \lim_{ (\vec{l}'-\vec{k}')\cdot \vec{k}' \;\rightarrow\; 0 } \mathcal{O}_2\left(\vec{p}_{\bar{d}}=\vec{k}'-\vec{l}'+(-\vec{k}')=-\vec{l}'\;;\;\vec{p}_d=\vec{l}'\;\right),\label{DT:IR_safety_condition}
\end{eqnarray}
which is the familiar IR-safety condition that relates observable functions $\mathcal{O}_{n}$ acting on kinematic configurations of different multiplicities $n$ in the soft and/or collinear limits of its massless constituents. Thus the constraint on observables implies by the request of finiteness of $\sigma_{\text{d}}$ on pinched singularities can be satisfied in the usual way of constructing observables.

\begin{figure}[ht!]
\centering
\begin{subfigure}[b]{.45\linewidth}
\centering
\includegraphics[width=.9\linewidth]{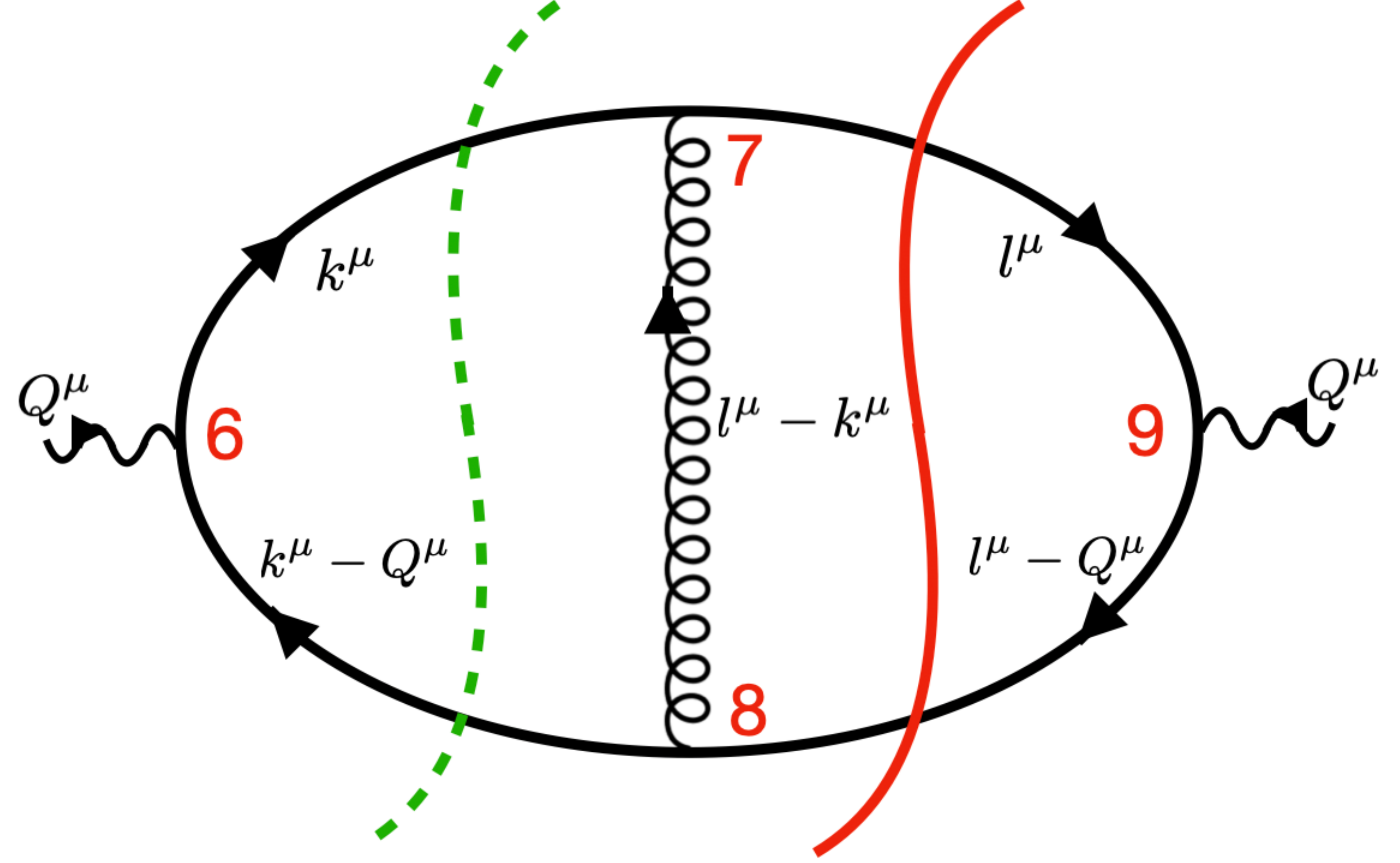}
\caption{Threshold of the $\mathbf{c}_{\mathbf{s}_{2}^{\text{v}}}$ cut at $|\vec{k}'|=\frac{Q^0}{2}$  }\label{DT:fig:threshold_LxBstar}
\end{subfigure}
\begin{subfigure}[b]{.45\linewidth}
\centering
\includegraphics[width=.9\linewidth]{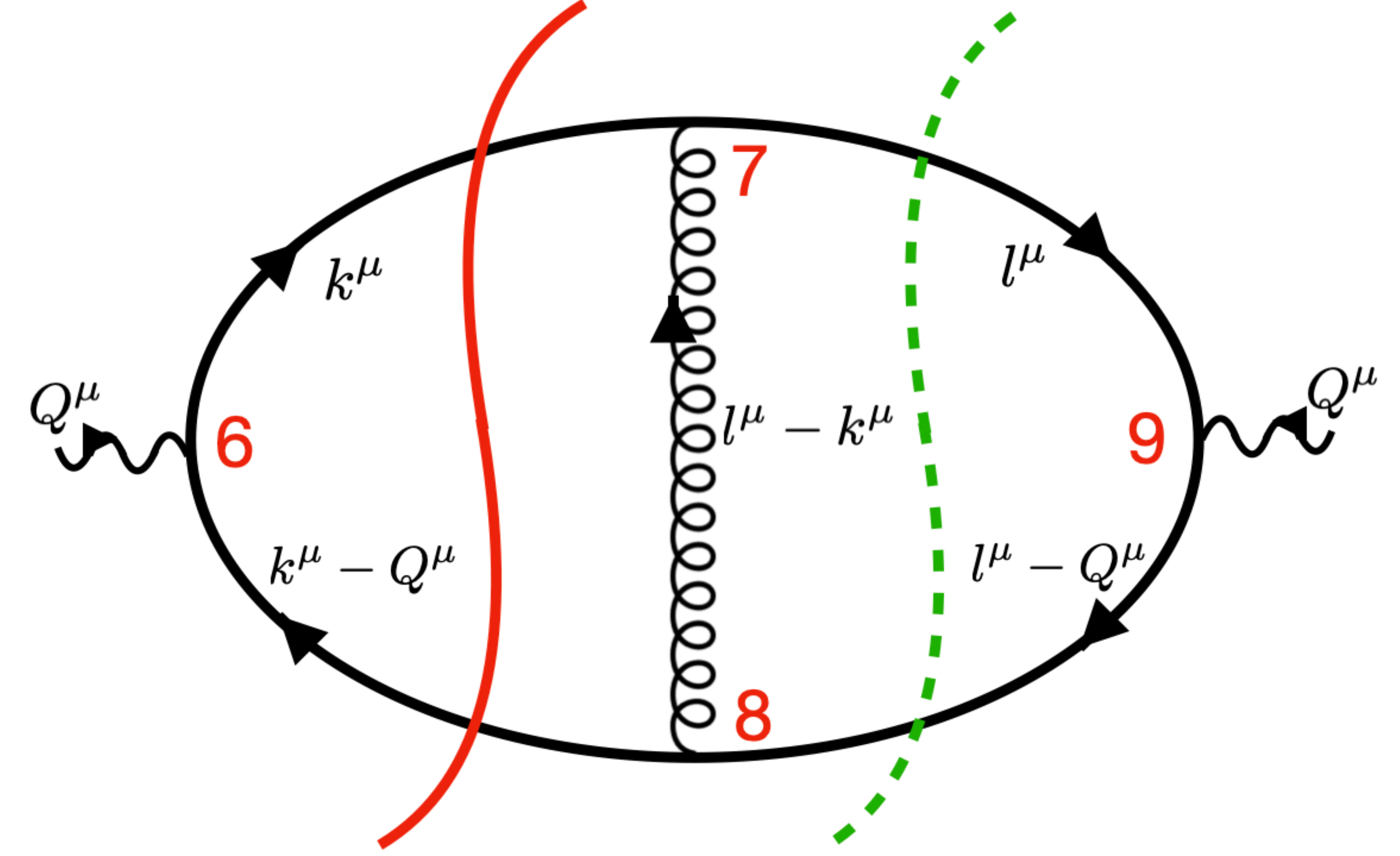}
\caption{Threshold of the $\mathbf{c}_{\mathbf{s}_{1}^{\text{v}}}$ cut at $|\vec{l}'|=\frac{Q^0}{2}$  }\label{DT:fig:threshold_BxLstar}
\end{subfigure}
\caption{\label{DT:fig:threshold_singularities} Pair of cancelling contributions from the E-surfaces (dashed green line) $|\vec{k}'|+|\vec{k}'-\vec{Q}|-Q^0=0$ and $|\vec{l}'|+|\vec{l}'-\vec{Q}|-Q^0=0$ of the Cutkosky cuts (solid red line) $\mathbf{c}_{\mathbf{s}_{2}^{\text{v}}}$ and $\mathbf{c}_{\mathbf{s}_{1}^{\text{v}}}$ respectively. 
When considering $\vec{Q}=\vec{0}$, these two singularities are reached when $x_{\mathbf{s}_{2}^{\text{v}}}-x_{\mathbf{s}_{1}^{\text{v}}}=0$, that is whenever $|\vec{k}'|=|\vec{l}'|$ because of the rescaling change of variables with $t=t^\star_{\mathbf{s}_{1}^{\text{v}}}$ (or $t=t^\star_{\mathbf{s}_{2}^{\text{v}}}$) which will map any such point to $t |\vec{k}'| = t |\vec{l}'|= \frac{Q^0}{2}$. 
}
\end{figure}

We can also consider the potential singularity at $x_{\mathbf{s}_1^{\text{v}}}-x_{\mathbf{s}_2^{\text{v}}}=0$, identifying the phase-space points satisfying  $|\vec{k}'|=|\vec{l}'|$, (see eq.~\eqref{dt_esurf}). This corresponds to a configuration on the \emph{non-pinched} E-surface of the one-loop triangle on the left (right) of the Cutkosky cut $\mathbf{c}_{\mathbf{s}_{1}^{\text{v}}}$ ($\mathbf{c}_{\mathbf{s}_{2}^{\text{v}}}$).
The study of the regularity of the differential cross-section at these threshold singularities seems at first glance to follow completely analogously to that of the IR pinched singularities. Perhaps surprisingly, this implies that we also expect cancellation between the two threshold singularities of the Cutkosky virtual contributions illustrated in fig.~\ref{DT:fig:threshold_singularities}.
We can follow the exact same study of cancellation between the IR pinched surfaces in the sum of terms in the r.h.s of eq.~\eqref{sigma-d-dt}, with the \emph{only} qualitative difference being the resulting condition on the observable function:
\begin{equation}
    \lim_{ |\vec{l}'|\;\rightarrow\;|\vec{k}'| }
 \mathcal{O}_{2}\left(\vec{p}_d=\vec{l}'\;;\;\vec{p}_{\bar{d}}=-\vec{l}'\;\right) = \lim_{ |\vec{l}'|\;\rightarrow\;|\vec{k}'| } \mathcal{O}_{2}\left(\vec{p}_d=\vec{k}'\;;\;\vec{p}_{\bar{d}}=-\vec{k}'\;\right).\label{DT:E_surface_cancelation_condition}
\end{equation}
It is clear that this condition is of a completely different nature than the IR safety condition obtained in~\eqref{DT:IR_safety_condition}.
Indeed, the limit $\lim_{ |\vec{l}'|\;\rightarrow\;|\vec{k}'|}$ does not imply any degeneracy in the experimental signature since one can obviously resolve the directional information of the quark and anti-quark in the final state.
Consequently, one should in general expect observable functions \emph{not} to satisfy eq.~\eqref{DT:E_surface_cancelation_condition}, implying that for non-trivial observable functions, considering the contour deformation discussed in sect.~\ref{sec:contour_deformation} is necessary.
A couple of key additional points are in order:
\begin{itemize}
    \item When considering \emph{fully} inclusive cross-sections, observable conditions similar to eq.~\eqref{DT:E_surface_cancelation_condition} are \emph{always} satisfied, so that the computation can be performed \emph{without} considering a deformation.
    Note however that omitting altogether certain Cutkosky cut contributions effectively amounts to setting the observable to exactly zero for them. This implies that even for the computation of the inclusive NLO cross-section of the scattering process $e^+ e^- \rightarrow \gamma^\star \rightarrow H t \bar{t}$, a deformation would be warranted since all Cutkosky cuts involving \emph{only} the two top quarks are not considered given that the observable demands a final-state Higgs.
    
    \item Because the two observable functions on each side of the condition eq.~\eqref{DT:E_surface_cancelation_condition} do \emph{not} share any loop momentum dependency, it is clear that in the case of the \DT{} supergraph, only a \emph{constant} observable function can satisfy it (i.e. inclusive measurement).
    
    \item Observable functions often consists of only products of Heaviside functions (e.g. implementing phase-space cuts and/or binning into histograms), in which case this opens the possibility of investigating dynamically at run time and for each integration sampling point what are the E-surfaces whose pair of canceling Cutkosky cuts are \emph{not} either both selected or both removed by the observable definition. Then, the contour deformation for this point only needs to consider those surviving E-surfaces.
\end{itemize}

\subsection{Proof of local cancellations of threshold (IR) singularities}
\label{sec:proof}

In the following sections we will construct the LU representation for a generic differential cross-section and require it to be free of non-integrable singularities. We show that the resulting constraint on observable functions is satisfied by IR-safe observables. 
This results in a systematic proof of local IR cancellations within the LU representation of differential cross-sections.  

\subsubsection{Causal flows}
\label{sec:causal_flows}
In section~\ref{sec:supergraph} we introduced cross-sections by referring to them as weighted sums of interference diagrams, each of which is associated to a well-defined Cutkosky cut, which in turn corresponds to a Dirac delta imposing the conservation of on-shell energies. For a fixed and positive center of mass energy, the equation imposing conservation of the on-shell energies is the equation of an E-surface. These deltas, in Cutkosky's original derivation~\cite{Cutkosky:1960}, arise from contour deforming the energy around thresholds of the diagrams. However, such a derivation obscures the important subtlety that the energy variables of the particles in the Cutkosky cut are linearly dependent and thus cannot be used as independent integration variables. In the following, we will provide an alternative derivation of Cutkosky cuts, by reducing the integration along thresholds to a one-dimensional problem.

This derivation of Cutkosky cuts will expose the local structure of the integrands and the location of their singularities. It is formulated such that the cancellations of all divergences related to E-surfaces are local, thus allowing one to construct an integrand which is locally free of divergences by aligning the integration measures supported by the different Cutkosky cuts. Furthermore, it shows how considering transition probabilities, rather than amplitudes, is required for the infrared structure of observables to be completely understood.

Let
\begin{equation}
M_\Gamma=\int \Bigg(\prod_{i=1}^L\frac{d^3k_i}{(2\pi)^3}\Bigg)\frac{f(\vec{k})}{\prod_{j\in\mathbf{e}}E_j \prod_{{\boldsymbol{\tau}}\in\mathcal{E}}\eta_{\boldsymbol{\tau}} } \,,
\label{eq:LTD_SG_expression}
\end{equation}
be the LTD representation of a supergraph $\Gamma$, with
\begin{equation}
    f(\vec{k})=\prod_{{\boldsymbol{\tau}}\in\mathcal{E}}\eta_{\boldsymbol{\tau}} \sum_{\mathbf{b}\in\mathcal{B}}\frac{N(\vec{k})\prod_{j\in\mathbf{e}\setminus \mathbf{b}}E_j}{\prod_{i\in\mathbf{e}\setminus\mathbf{b}} (q_i^2-m_i^2)}\Bigg|_{\{q_j^0=\sigma^\mathbf{b}_j E_j\}_{j\in\mathbf{b}}} \,.
\end{equation}
The discontinuities of $M_\Gamma$ across its s-channel thresholds represent distinct summands that contribute to define the total probability of the process whose initial states are fixed to be the particles in $\mathbf{a}$ and whose final states are, for now, unspecified.  The function $f$, which is the LTD representation of the supergraph multiplied by the product of all energies and all E-surfaces appearing in the representation itself, is finite for any value of the loop momenta.

We introduce a dummy integration variable $t$ by introducing unity as the integral of a normalized function
\begin{equation}
M_\Gamma=\int \Bigg(\prod_{i=1}^L\frac{d^3\vec{k}_i}{(2\pi)^3}\Bigg) dt \frac{f(\vec{k})h(t)}{\prod_{j\in\mathbf{e}}2E_j \prod_{{\boldsymbol{\tau}}\in\mathcal{E}}\eta_{\boldsymbol{\tau}} }.
\end{equation}
and then introduce the change of variables $\vec{k}=\vec{\phi}(t,\vec{k}')$, with $\vec{\phi}$ being the solution to the following first-order system of differential equations:
\begin{equation}
\begin{cases}
    \partial_t \vec{\phi}(t,\vec{k})=\kappa(\vec{\phi}(t,\vec{k})) \\
    \vec{\phi}(0,\vec{k})=\vec{k}
    \end{cases},
    \label{eq:causal_ode}
\end{equation}
where we introduce the vector field $\kappa:\mathbb{R}^{3L}\rightarrow \mathbb{R}^{3L}$ which we require to be Lipschitz-continuous. We will use the map $\vec{\phi}_t:\vec{k}\rightarrow \vec{\phi}(t,\vec{k})$ to change the parametrisation of the phase space integral. If the zeros of $\kappa$ form a zero-measured subset of $\mathbb{R}^{3L}$ with respect to the Lebesgue measure in $\mathbb{R}^{3L}$, then the change of variables is well-defined. Thus, we can exclude all the sinks, sources and ridges that the flow $\vec{\phi}_t$ may have  from the integration.
Therefore, we can write
\begin{equation}
M_\Gamma=\int \Bigg(\prod_{i=1}^L\frac{d^3{k'}_i}{(2\pi)^3}\Bigg) dt |\text{det}[\mathbb{J}_{\vec{k}'}\vec{\phi}]| \frac{f(\vec{\phi}(t,\vec{k}'))h(t)}{\prod_{j\in\mathbf{e}}E_j \prod_{{\boldsymbol{\tau}}\in\mathcal{E}}\eta_{\boldsymbol{\tau}} } \,.
\end{equation}
 We are now interested in performing contour integration in the variable $t$. In order to do this, observe that for each $\vec{k}$, the curve $\vec{\phi}(t,\vec{k})$ crosses a number of E-surfaces.
However this alone does not allow to determine how many times a specific E-surface is intersected by a curve in the flow, and if approaching an E-surface along a curve yields a simple pole in the integrand.
If the pole is simple, then there is a well-defined principal value procedure associated with it, and the sign of the imaginary part acquired in the contour integration is fixed by the sign of the Feynman prescription.
 
It is possible, however, to construct a flow whose properties make the answer to these two questions manifest. Specifically, consider solutions to flow ODEs where the vector field $\kappa$ is chosen to be causal, that is
 \begin{equation}\label{causal_prescription}
     \kappa\cdot \nabla_{\vec{k}} \eta_\mathbf{s}>0,\  \ \forall \mathbf{s} \in \mathcal{E}_{\text{s-ch}}, \ \ \forall \vec{k} \text{ s.t. } \eta_\mathbf{s}=0 .
 \end{equation}
If that is the case, then the flow will consequently have three properties:
 \begin{itemize}
    \item $\forall \vec{k}, \ \text{there exists at most one value } t^\star_\mathbf{s}  \text{ s.t. }\eta_\mathbf{s}(\vec{\phi}(t^\star_\mathbf{s},\vec{k}))=0$, which follows from the fact that the curve $\vec{\phi}(t,\vec{k})$ cannot flow outward \emph{and} inward of the E-surface $\eta_{\mathbf{s}}$ without violating the causal prescription of eq.~\eqref{causal_prescription}.
    \item$\partial_t\eta_\mathbf{s}(\vec{\phi}(t^\star_\mathbf{s},\vec{k}))\neq 0$, since $\partial_t\eta_\mathbf{s}(\vec{\phi}(t^\star_\mathbf{s},\vec{k}))=\nabla_{\vec{k}}\eta_\mathbf{s} \cdot \partial_t \vec{\phi} |_{t=t^\star_\mathbf{s}}=\nabla_{\vec{k}}\eta_\mathbf{s} \cdot \kappa|_{t=t^\star_\mathbf{s}}$, which is guaranteed to be non zero by eq.~\eqref{causal_prescription} for any $\eta_\mathbf{s}$ with $\mathbf{s}\in\mathcal{E}_{\text{s-ch}}$.
   \item $ \text{sign}[\partial_t \eta_\mathbf{s}]=\text{sign}[\partial_t \eta_\tau], \ \ \forall  \vec{\phi}(t,\vec{k})\in\delta\eta_\mathbf{s}\cap \delta\eta_\tau$, also trivially guaranteed by eq.~\eqref{causal_prescription},
\end{itemize}
where we define $\delta\eta_\mathbf{s}:=\left\{\vec{k}\in\mathbb{R}^{3L}\middle|\eta_\mathbf{s}(\vec{k})=0\right\}$. With a slight abuse of notation, we write $\partial_t\eta_\mathbf{s}(\vec{\phi}(t^\star_\mathbf{s},\vec{k}))=\partial_t\eta_\mathbf{s}(\vec{\phi}(t,\vec{k}))|_{t=t^\star_\mathbf{s}}$.

The first property determines the number of intersections a curve has with a determined E-surface. The second determines that all poles appearing on the real $t$ axis for fixed $\vec{k}$ are simple. The last one, although momentarily obscure, will be fundamental in order to realize local cancellations of pinch singularities.
We stress that the first two conditions are not strictly necessary in order to build a valid contour integration in $t$. Indeed, regarding the first condition, there is nothing in the principal value procedure that forbids us to contour integrate around two distinct poles. Regarding the second condition, one could think of excluding from integration the regions of space at which $\partial_t \eta(\vec{\phi}(t,\vec{k}))=0$, which lie on a zero-measured set, and then establish if the integration is finite.

Given the first property, for every $\vec{k}$ one can define the set
\begin{equation}
    \mathcal{E}_{\vec{k},\vec{\phi}}:=\left\{\mathbf{s}\in\mathcal{E}_{\text{s-ch}}\middle| \exists t\in\mathbb{R} \text{ with } \eta_\mathbf{s}(\vec{\phi}(t,\vec{k}))=0\right\} \,,
\label{eq:intersected_e_surfaces}
\end{equation}
which contains all the E-surfaces which are intersected once by the curve $\vec{\phi}(t,\vec{k})$. Thus, for each $\mathbf{s}\in\mathcal{E}_{\vec{k},\vec{\phi}}$, we can write the expansion of $\eta_\mathbf{s}$ around its unique zero, $t^\star_{\mathbf{s}}$:
\begin{equation}
    \eta_\mathbf{s}(\vec{\phi}(t,\vec{k}))=(t-t^\star_\mathbf{s})\partial_t\eta_\mathbf{s}(\vec{\phi}(t,\vec{k}))|_{t=t^\star_\mathbf{s}}+
     \order{(t-t^\star_\mathbf{s})^2} \,.
\label{eq:linearized_eta}
\end{equation}
and the first order in the expansion is ensured to be non-zero by the second property. 
In conclusion, we observe that the existence of a causal vector field $\kappa$ being causal on all the E-surfaces of the supergraph $\Gamma$ is guaranteed by the work carried out in ref.~\cite{Capatti:2019edf}. Since the causal vector field, as constructed therein, is infinitely differentiable, Sard's theorem also ensures that its zeros lie on a zero-measured surface. For future convenience, given a set of points $V\subset \mathbb{R}^{3L}$, we define a set containing all the points that can be mapped into $V$ by the causal flow
\begin{equation}
    \vec{\phi}^{-1}[V]=\left\{\vec{k}\in\mathbb{R}^{3L}\middle| \exists t\in\mathbb{R} \text{ with } \vec{\phi}(t,\vec{k})\in V\right\} \,.
\end{equation}
The inverse image of the causal flow is fundamental in determining the analytic properties of the supergraph and how they relate to those of residues in $t$ of the supergraph as parametrised along the flow. 

\subsubsection{Visualisation of the causal flow}
\label{sec:DT_causal_flow}

\begin{figure}
\centering
\begin{subfigure}[b]{.8\linewidth}
\includegraphics[width=\linewidth]{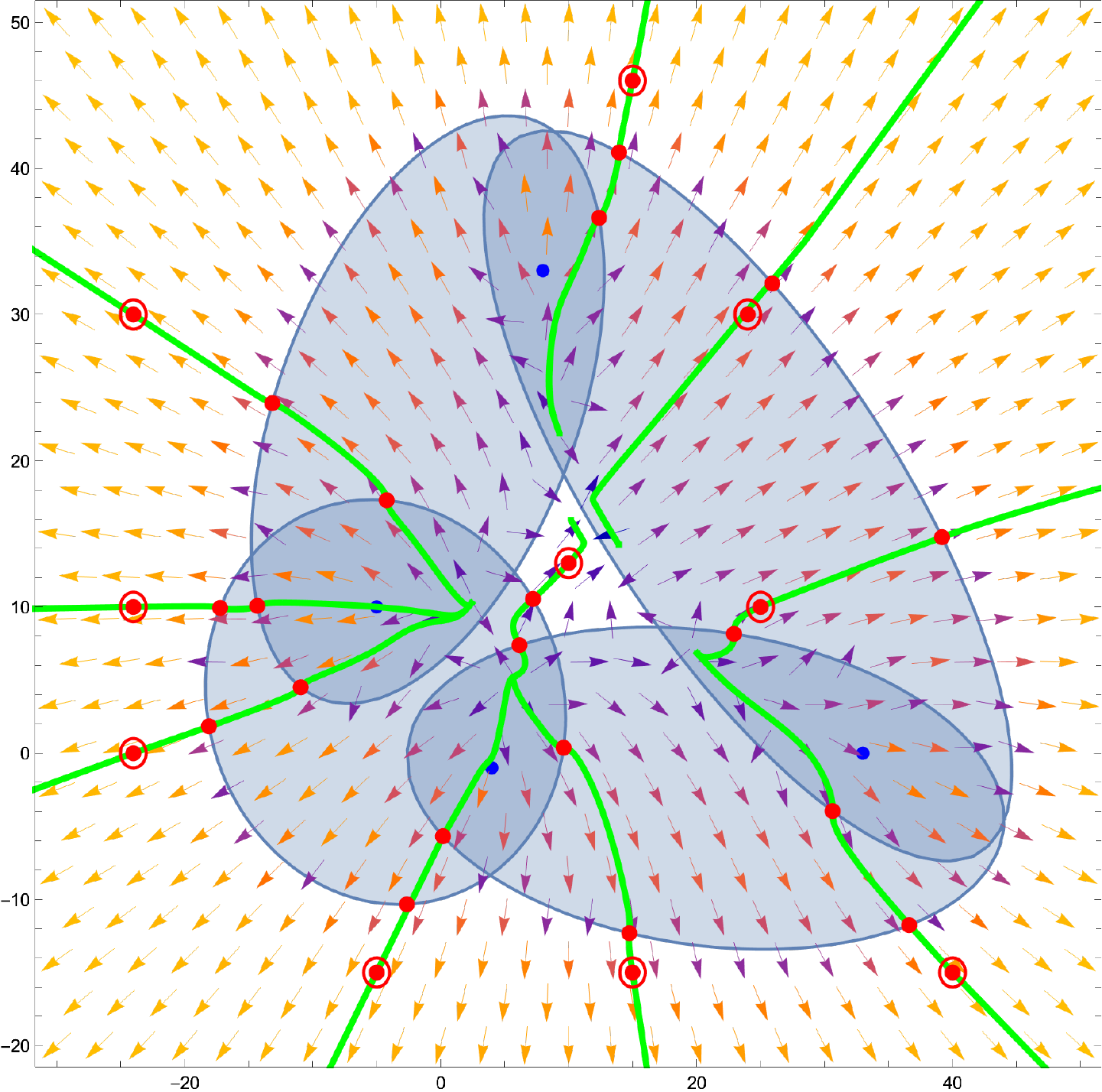}
\caption{Causal flow for the {\texttt Box\_4E} example of sect. 3.1 of ref.~\cite{Capatti:2019edf}.  }\label{fig:causal_flow_Box_4E}
\end{subfigure} \\

\begin{subfigure}[b]{.45\linewidth}
\includegraphics[width=\linewidth]{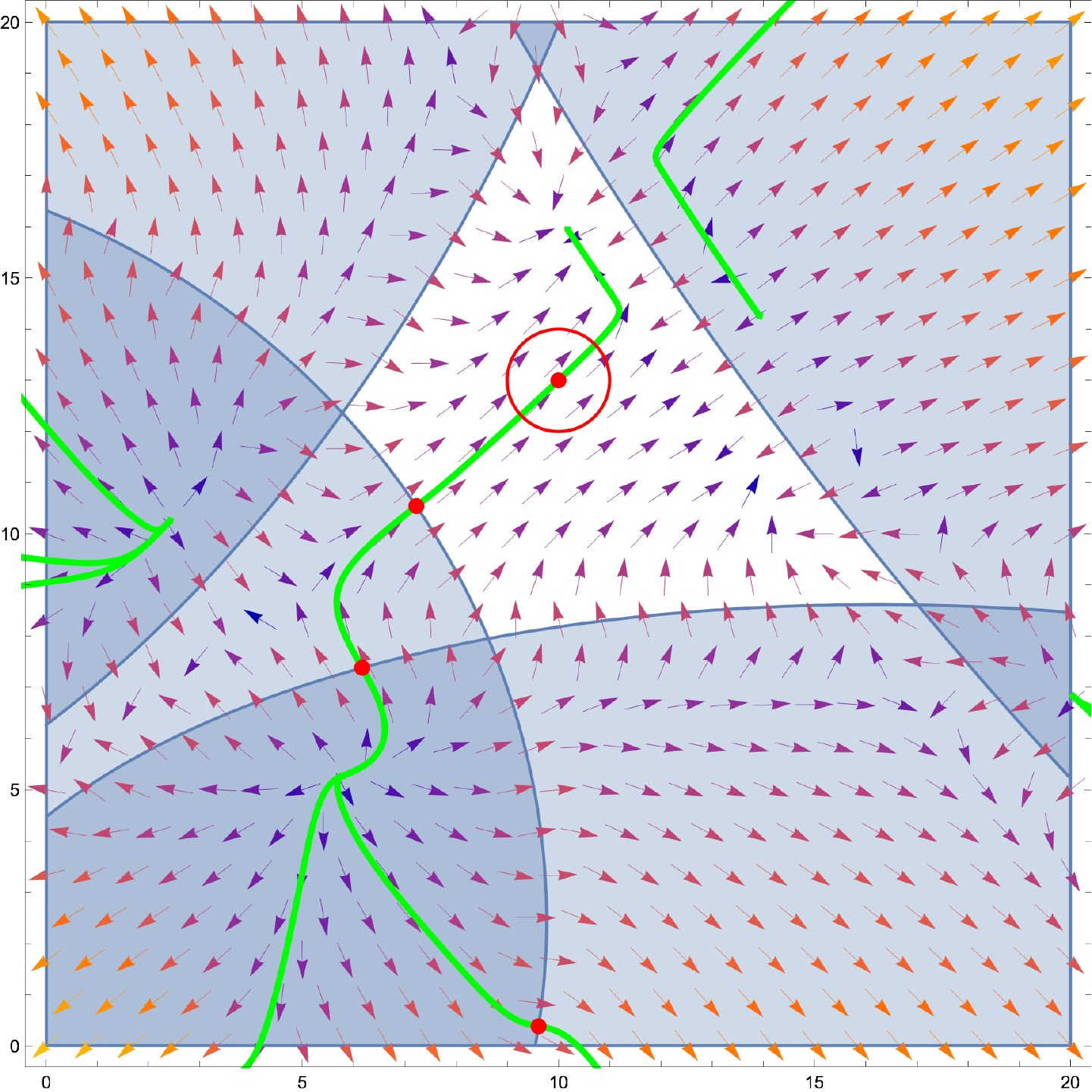}
\caption{Close-up on the central region of fig.~\ref{fig:causal_flow_Box_4E}. }\label{fig:causal_flow_Box_4E_zoom}
\end{subfigure}
\begin{subfigure}[b]{.45\linewidth}
\includegraphics[width=\linewidth]{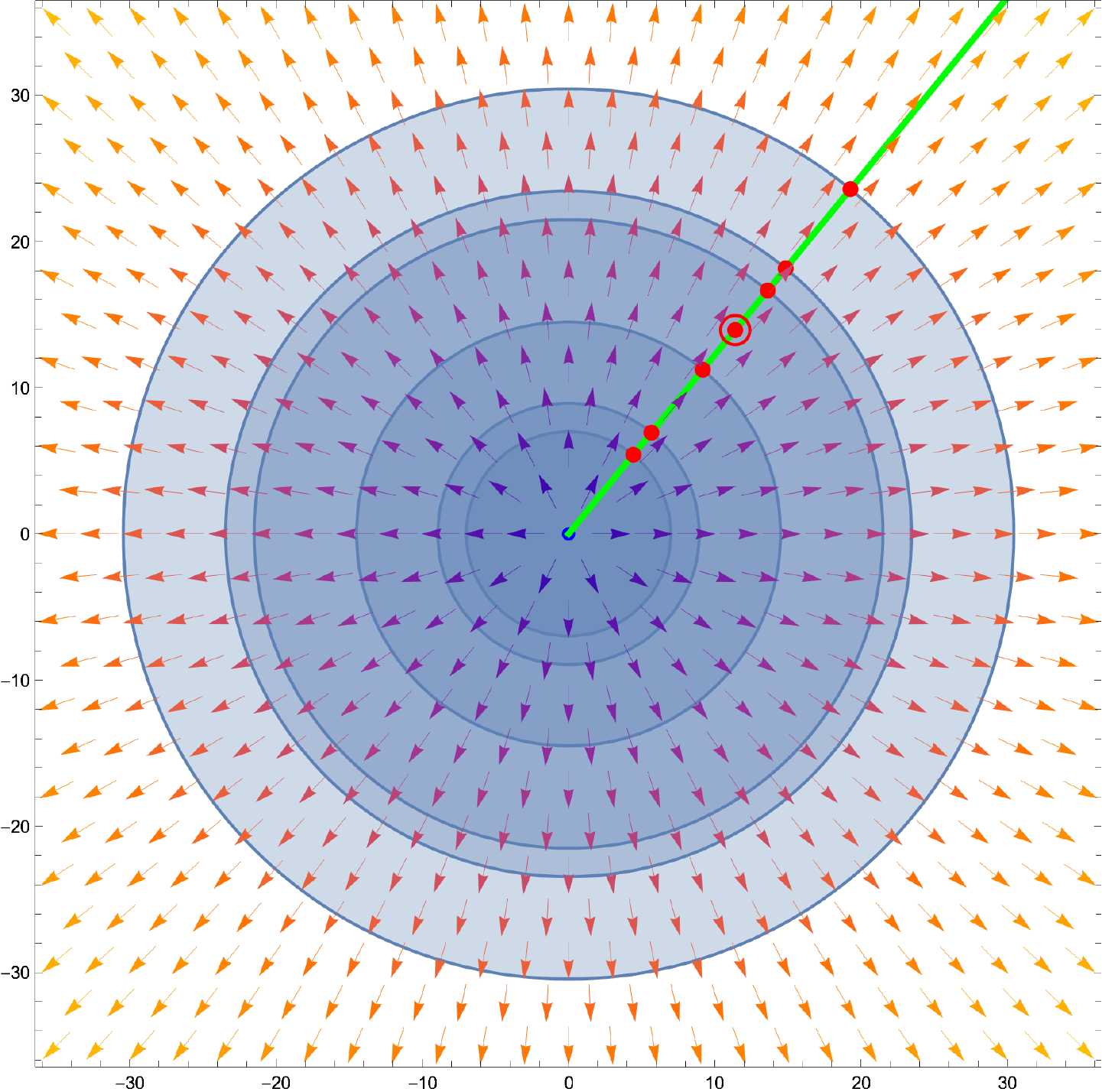}
\caption{Rescaling flow for {\texttt Box\_4E} with $\vec{p}_i=\vec{0}$. }\label{fig:causal_flow_Box_4E_decay}
\end{subfigure}

\caption{
\label{fig:general_causal_flow_Box_4E}
Causal flow for the complicated maximal overlap structure of the {\texttt Box\_4E} one-loop example of sect. 3.1 of ref.~\cite{Capatti:2019edf} (figs.~\ref{fig:causal_flow_Box_4E} and~\ref{fig:causal_flow_Box_4E_zoom}) and its much simpler pure rescaling counterpart when setting all spatial parts of the external momenta of {\texttt Box\_4E} to zero (fig.~\ref{fig:causal_flow_Box_4E_decay}). The x and y axes indicate the $k_x$ and $k_y$ component of the {\texttt Box\_4E} loop momentum. The circled red point indicates possible input sampling points that the LU representation depends on and the non-circled red points depict the various projections induced by the causal flow on the contributing E-surfaces.
The green lines represent $\vec{\phi}^{-1}[V]$ with $V$ being the circled dots.
The vector field depicts the orientation of the contour deformation vector constructed according to ref.~\cite{Capatti:2019edf} (using the four deformation sources indicated with blue dots), and its colour pertains to its relative magnitude.}
\end{figure}

In figs.~\ref{fig:causal_flow_Box_4E},~\ref{fig:causal_flow_Box_4E_zoom} and ~\ref{fig:causal_flow_Box_4E_decay} the causal flow of a particular one-loop example, called {\texttt Box\_4E} in ref.~\cite{Capatti:2019edf}, is shown.
Since the origin is not in the interior of all E-surfaces, the rescaling strategy defined in section~\ref{sec:sopertrick} cannot be applied.
For all $1\rightarrow N$ processes the simple rescaling flow is applicable, and we show the causal flow of {\texttt Box\_4E} only for the purpose of illustrating the complications arising for a non-trivial overlap structure of E-surfaces, such as the one that can appear for challenging supergraph topologies of $2\rightarrow N$ processes. In that case, it is likely that the system of ODE defining the causal flow requires a numerical solution, but all properties of the LU representation discussed in this paper would still hold.
In particular, we observe that each sampling point (circled in red) of the LU integrand still yields exactly one or zero projection onto the contributing E-surfaces. This is thanks to the presence of sinks and sources in the causal flow that its parameterisation can only asymptotically approach, but never reach. Moreover, it is clear that since the \emph{same} causal flow is used to project sampling points onto all reachable E-surfaces, then E-surface intersections are reached simultaneously by the corresponding projections, which is key for the local cancellation of the corresponding singularities. Finally, the orientation of the deformation vector field on these intersections makes it clear why the conditions discussed below eq.~\eqref{causal_prescription} on the derivative $\partial_t$ of the flow induced are fulfilled.

\begin{SCfigure}[1][ht!]
\centering
 \caption{\label{fig:DT_causal_flow} Causal flow of the DT supergraph with $Q^\mu=(2,0,0,0)$ in the phase-space plane $(\vec{k}',\vec{l}')=\frac{1}{2}((0,0,\lambda_k),(0,\lambda_l,\lambda_l))$. Drawing conventions follow those of fig.~\ref{fig:general_causal_flow_Box_4E} and the points labeled "i)" are integrable singularities while the ones analoguous to "e)" correspond to the intersection of the two non-pinched E-surfaces of the \DT{} supergraph.}
\includegraphics[width=0.6\textwidth]{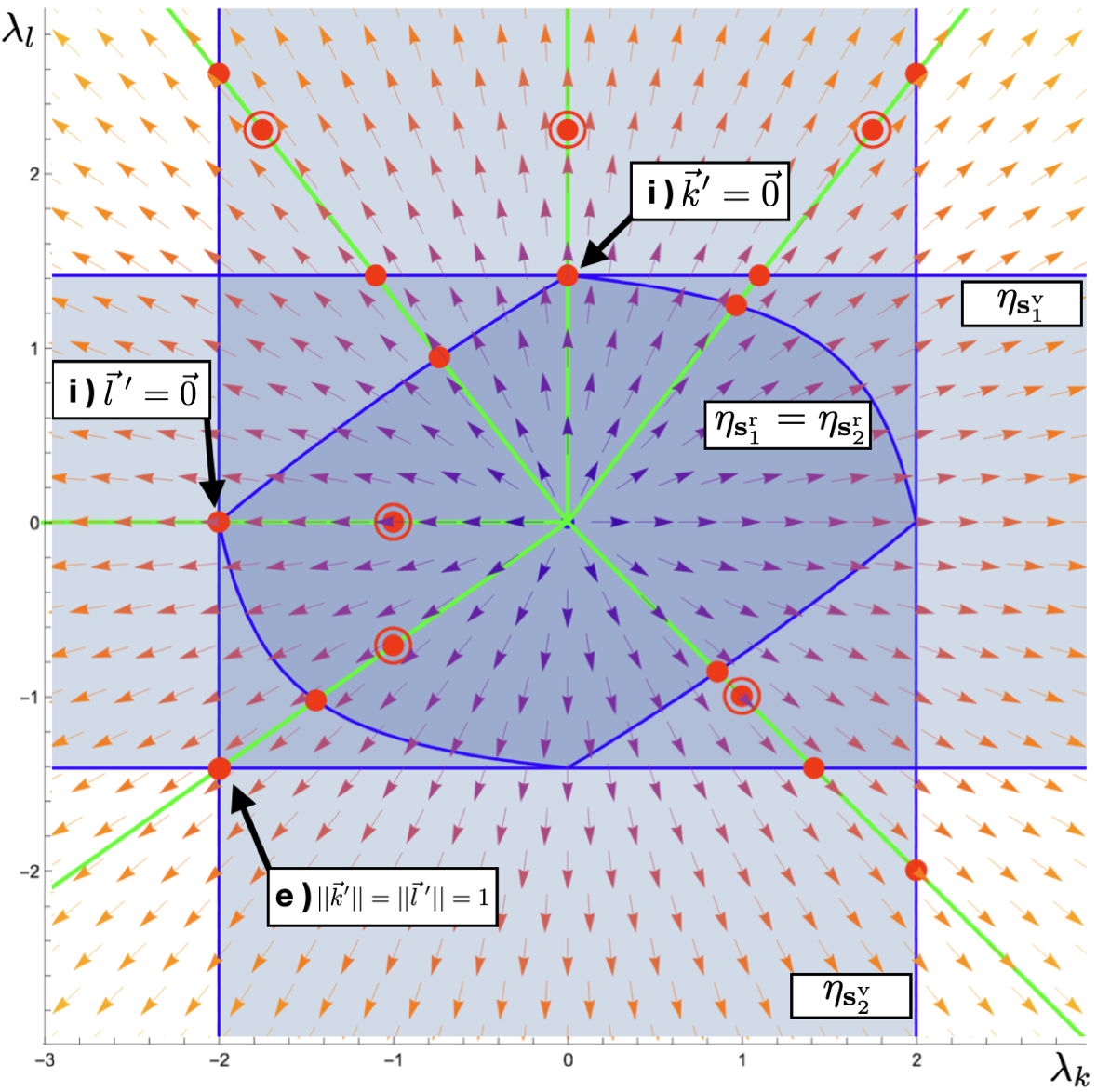}
\end{SCfigure}

The causal flow of the double-triangle supergraph is more delicate to represent than that of {\texttt Box\_4E}, since there are two loop momenta in that case.
We choose to project the six-dimensional input space $(\vec{k}',\vec{l}')$ of the DT topology with $Q^\mu=(2,0,0,0)$ onto the parametric plane $\lambda_k \hat{e}_k + \lambda_l \hat{e}_l$ with $\hat{e}_k=((0,0,\frac{1}{2}),(0,0,0))$ and $\hat{e}_l=((0,0,0),(0,\frac{1}{2},\frac{1}{2}))$. This section involving a non-constant momentum component is convenient since it necessarily contains the image of a rescaling flow, thus allowing to render the flow within this same plane, like it was the case in fig.~\ref{fig:DT_causal_flow}.
An important remark is that the E-surfaces $\eta_{\mathbf{s}_1^{\text{v}}}$ and $\eta_{\mathbf{s}_2^{\text{v}}}$ are unbounded. This is a result of their independence of $\lambda_k$ (resp. $\lambda_l$) which only controls the one-loop integration volume of the triangle loop remaining on the left (resp. right) of the Cutkosky cuts $\eta_{\mathbf{s}_1^{\text{v}}}$ and $\eta_{\mathbf{s}_2^{\text{v}}}$. This is what allows the LU representation of the DT supergraph to probe the UV regime with a rescaling parameter of $\mathcal{O}(1)$.
In contrast, the volume described by the E-surface $\eta_{\mathbf{s}_1^{\text{r}}}$ results from an equation involving the sum of three square roots and is therefore quite complicated but still bounded since it corresponds to a particular hyper-plane of the three-body decay phase-space volume $Q^\mu \rightarrow (k'^\mu, k'^\mu-l'^\mu, l'^\mu-Q^\mu)$, which we know to be necessarily contained within a sphere of radius $\sqrt{Q^2}$.
Notice however, that even for a sampling point $(\vec{k}',\vec{l}')$ with arbitrary large moduli $|\vec{k}'|$ and $|\vec{l}'|$, the global rescaling flow will always yield a contribution for the real-emission $\eta_{\mathbf{s}_1^{\text{r}}}$ and $\eta_{\mathbf{s}_2^{\text{r}}}$ Cutkosky cuts. 
However these will be associated with very small values of the parameter $t^\star_{\mathbf{s}_1^{\text{r}}}$ and $t^\star_{\mathbf{s}_2^{\text{r}}}$ yielding a contribution exponentially suppressed for our choice of normalised function $h(t)$ of eq.~\eqref{eq:normalised_integral}. For the particular projection chosen for fig.~\ref{fig:DT_causal_flow}, the integrable singularities at $\vec{k}'=\vec{0}$ and $\vec{l}'=\vec{0}$ also correspond to vanishing values of the corresponding rescaling solution and will thus also be suppressed.
These observations underline the appealing feature of LU that the change of variable $\vec{k}\rightarrow \vec{\phi}(t,\vec{k})$ does not affect the interpretation of what kinematic region of the cross-section is probed since contributions from $t^\star_i$ solutions far from one are exponentially suppressed.
Moreover, in the particular case of a global rescaling causal flow, we even observe that the change of variable retains the collinear and soft properties of the momenta of the (pairs of) partons in the input configuration. 
This allows one to easily probe potentially singular regions, which we investigate in sect.~\ref{sec:DT:integrand_visualisation} by choosing a different projection of the \DT{} sampling space. 

\subsubsection{The Local Unitarity representation of differential cross-sections}

In his original 1960 paper~\cite{Cutkosky:1960}, Cutkosky recognised that interference diagrams can be seen as the imaginary part acquired by contour deforming around specific thresholds of supergraphs. The equation of the threshold appears in the energy conserving delta as a result of the principal value identity that establishes the functional form of the imaginary part acquired by contour deforming around a pole on the real axis. Moreover, every supergraph $\Gamma$ gives rise to a number of squared amplitudes obtained by summing over the imaginary parts acquired by contour integrating along all of its thresholds corresponding to cuts $\mathbf{s}\in\mathcal{E}_{\text{s-ch}}$. Such a derivation does not detail the subtleties of setting up a contour deformation programme for the supergraph. For example, it does not specify what constraints the chosen contour needs to satisfy and what integration variables can easily be chosen to transform the real integration into a contour integration.

We will now show that is possible to see Cutkosky cuts as the discontinuities of the LTD representation of the supergraph along the one-dimensional flow line of a causal flow. This allows us to explicitly construct a local representation of differential cross section from which the second golden rule arises naturally as a consequence of identifying picking up the residues of the LTD representation of the supergraph along the flow with explicitly solving the Dirac delta associated to Cutkosky cuts. Thus the work in this section should be considered to be equivalent to that of aligning the integration measures $d\Pi^\mathbf{s}$ and, specifically, resolving Dirac deltas expressing energy conservation across on-shell physical particles.

In the following, we will assume that we have a causal flow $\vec{\phi}(t,\vec{k})$. For every fixed $\vec{k}$, the unique curve $\vec{\phi}(t,\vec{k})$ passing through it will intersect a subset of all E-surfaces. Recall that we defined $\mathcal{E}_{\vec{k}, \vec{\phi}}$ to be the set of such E-surfaces in eq.~\eqref{eq:intersected_e_surfaces}, which can be expanded as shown in eq.~\eqref{eq:linearized_eta}. For all s-channel E-surfaces not in $\mathcal{E}_{\vec{k}, \vec{\phi}}$, that is for every ${\boldsymbol{\tau}}\in\mathcal{E}_{\text{s-ch}}\setminus\mathcal{E}_{\vec{k}, \vec{\phi}}$, we have that  $\eta_{\boldsymbol{\tau}}(\vec{\phi}(t,\vec{k}))\neq 0$ for every $t$. Thus, every E-surface not contained in $\mathcal{E}_{\vec{k}, \vec{\phi}}$ never has a vanishing zeroth order in the expansion on the flow line. Thus, given a supergraph
\begin{equation}
    M_\Gamma=\int \Bigg(\prod_{i=1}^L\frac{d^3{k}_i}{(2\pi)^3}\Bigg)\frac{f(\vec{k})}{\prod_{j\in\mathbf{e}}2E_j \prod_{{\boldsymbol{\tau}}\in\mathcal{E}}\eta_{\boldsymbol{\tau}} } \,,
\end{equation}
one can define a locally finite volume form as
\begin{equation}\label{lfmo}
    \sigma_{\text{d}}^\Gamma[\vec{\phi}]=\sum_{\mathbf{s}\in\mathcal{E}_{\vec{k}, \vec{\phi}}} \text{Ind}_{\mathbf{s}} \lim_{t\rightarrow t^\star_\mathbf{s}} (t-t^\star_\mathbf{s})\frac{f(\vec{\phi}(t,\vec{k}))|\text{det}[\mathbb{J}_{\vec{k}'}\vec{\phi}]|}{\eta_{\mathbf{s}} \prod_{i\in\mathbf{e}}2E_i \prod_{{\boldsymbol{\tau}}\in\mathcal{E}_{\mathbf{s}}^{L}}\eta_{\boldsymbol{\tau}} \prod_{{\boldsymbol{\tau}}\in\mathcal{E}_{\mathbf{s}}^{R}}\bar{\eta}_{\boldsymbol{\tau}}}\frac{h(t) \mathcal{O}_{\mathbf{s}}}{\big[\prod_{{\boldsymbol{\tau}}\in\mathcal{E}_\mathbf{s}
   ^\emptyset} \eta_{\boldsymbol{\tau}}\big]_{\epsilon=0}} \,.
\end{equation}
where
\begin{align}
\begin{split}
\mathcal{E}_\mathbf{s}^L=\{{\boldsymbol{\tau}}\in\mathcal{E}| {\boldsymbol{\tau}}\subset\mathbf{s}\}\\
\mathcal{E}_\mathbf{s}^R=\{{\boldsymbol{\tau}}\in\mathcal{E}| \mathbf{s}\subset{\boldsymbol{\tau}}\} \\
\mathcal{E}_\mathbf{s}^\emptyset=\mathcal{E}\setminus (\mathcal{E}_\mathbf{s}^L \cup \mathcal{E}_\mathbf{s}^R)\setminus\{\mathbf{s}\}
\end{split}
\end{align}
and $\mathcal{O}_\mathbf{s}$ defines the observable and is a function of $\vec{k}$ through the dependence on cut edge momenta, that is
\begin{equation}\label{observables-form}
    \mathcal{O}_\mathbf{s}(\vec{k})=\mathcal{O}_{|\mathbf{c}_\mathbf{s}|}\big(\{(\lambda_e,m_e,\vec{q}_e)\}_{e\in \mathbf{c}_\mathbf{s}}\big).
\end{equation}
where $\lambda_e$ denotes the spin or helicity of the particle.
At the integrand level, $\sigma_\text{d}^\Gamma$ reproduces what we obtained in sect.~\ref{sec:supergraph} through LTD and the golden rule for differential cross-sections in eq.~\eqref{Mformula}.
The index associated to the residue is fixed by the convention $\text{Ind}_{\mathbf{s}}=+\text{sign}[\partial_t \eta_\mathbf{s}(\vec{\phi}(t^\star_\mathbf{s},\vec{k}))]$. Fixing this choice is equivalent to choosing a Feynman prescription for the s-channel thresholds which are included in the definition of $\sigma_\text{d}^\Gamma$; this arbitrary choice was already discussed in ref.~\cite{Cutkosky:1960}. The relation
\begin{equation}
    \frac{d\sigma}{d\mathcal{O}}=\sum_{\Gamma\in\mathcal{G}}\int \prod_{i=1}^L\frac{d^3\vec{k}_i}{(2\pi)^3}\sigma_\text{d}^\Gamma \,,
\end{equation}
defines the Local Unitarity (LU) representation of the differential cross-section $d\sigma /d\mathcal{O}$. In the following, we will sometimes refer to $\sigma_\text{d}^\Gamma$ as $\sigma_\text{d}$, without the explicit reference to the supergraph $\Gamma$.

Yet another feature of eq.~\eqref{lfmo} is the presence of unregulated E-surfaces, that is E-surfaces for which the prescription can be set to zero. This is a consequence of $\sigma_{\text{d}}$ not being divergent at the location of these singularities, that is:
\begin{equation}\label{eq:double_limit}
    \lim_{t_\mathbf{s}^\star\rightarrow t^\star_{\boldsymbol{\tau}}}(t^\star_\mathbf{s}-t^\star_{\boldsymbol{\tau}})\lim_{t\rightarrow t^\star_\mathbf{s}}(t-t^\star_\mathbf{s})\frac{f(\vec{\phi}(t,\vec{k}))}{\eta_{\mathbf{s}} \prod_{i\in\mathbf{e}}E_i \prod_{{\boldsymbol{\tau}}\in\mathcal{E}_{\mathbf{s}}^{L}}\eta_{\boldsymbol{\tau}} \prod_{{\boldsymbol{\tau}}\in\mathcal{E}_{\mathbf{s}}^{R}}\bar{\eta}_{\boldsymbol{\tau}}}\frac{h(t) \mathcal{O}_{\mathbf{s}}}{\big[\prod_{{\boldsymbol{\tau}}\in\mathcal{E}_\mathbf{s}
   ^\emptyset} \eta_{\boldsymbol{\tau}}\big]_{\epsilon=0}}=0, \ \ \forall {\boldsymbol{\tau}} \in \mathcal{E}^\emptyset_\mathbf{s} \,.
\end{equation}
This is related to the property of $f$ that establishes a necessary condition for the multiple limits of the LTD expression when approaching the intersection of many E-surfaces to be zero. More specifically, given a set of E-surfaces $\mathcal{E}'\subseteq \mathcal{E}$ intersecting on a surface, one can ask how $f(\vec{k})$ behaves in a neighbourhood of any point on the intersection surface. We will show that approaching the intersection $\delta\eta_\mathbf{s}\cap\delta\eta_\mathbf{{\boldsymbol{\tau}}}$, $f$ does not vanish if and only if $\mathbf{s}\cap{\boldsymbol{\tau}}\in\{\mathbf{s},{\boldsymbol{\tau}},\emptyset\}$, that is if one of the corresponding set of vertices defining the cuts is contained within the other, or their intersection is the empty set. This property was assumed in the usual construction of Cutkosky cuts, and was used extensively in sect.~\ref{sec:supergraph}. 

In order to generalise and properly describe this property we define \emph{cross-free} families of subsets. A cross-free family is a \emph{laminar} family of subsets of the vertices such that no subset in it can be written as the union of two or more sets in the family itself. A laminar family is a family of sets such that for any two sets, the intersection is either one of the two sets or the empty set. These two definitions imply that a cross-free family is a set of $|\mathbf{v}|-1$ subsets of the vertices. %
\begin{figure}
    \centering
    \input{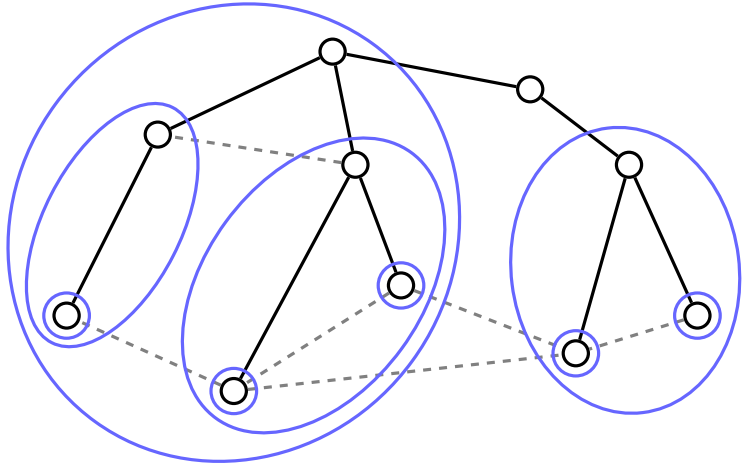}
    \caption{A cross-free family of connected subsets of the vertices, shown as blue circles, which contain the corresponding vertices. The cross-free family of subsets is associated to a unique (if existing) spanning tree (solid line). Observe that by construction, the family allows at least one choice of orientations of the edges (interpreted as energy flow) such that each cut has either only an incoming energy flow or only an outgoing energy flow.}
    \label{fig:cross-free-families}
\end{figure}
Let us define the set of all cross-free families of connected cuts of the graph:
\begin{equation}
    \mathcal{F}=\{F\subseteq \mathcal{E}| F \text{ is a cross-free family}\}.
\end{equation}
An example of a cross-free family is illustrated in fig.~\ref{fig:cross-free-families}. We then write
\begin{equation}\label{asymptotic-behaviour}
    M_\Gamma=\int \Bigg(\prod_{i=1}^L\frac{d^3{k}_i}{(2\pi)^3}\Bigg)\frac{1}{\prod_{i\in\mathbf{e}}2E_i}\sum_{F\in\mathcal{F}}\sum_{\vec{\alpha}\in\{\pm 1\}^{|F|}}\frac{c_{F\vec{\alpha}}(\vec{k})}{\prod_{\mathbf{s}\in F}\eta_{(\mathbf{s},\alpha_{\mathbf{s}})} } \,,
\end{equation}
where $c_F$ is a polynomial in the energies and the spatial loop momenta. Observe that we do not claim $c_F$ to be non-zero; instead, eq.~\eqref{asymptotic-behaviour} should be considered as a constraint on the functional form of $M_\Gamma$ which guarantees that eq.~\eqref{eq:double_limit} holds. Determining the coefficients $c_F$ is a non-trivial task which can be carried out with the treatment of ref.~\cite{Capatti:2020ytd}, and is not strictly necessary for the proof. 

Finally, we observe that it is also arbitrary that only a subset of thresholds of the supergraph should be selected to contribute with their discontinuities to $\sigma_{\text{d}}$. In the current way of defining differential cross-sections, all the residues associated to thresholds with $\delta(\mathbf{s})\cap\mathbf{a}\neq\mathbf{a}$ are completely ignored: the sum in eq.~\eqref{lfmo} runs over elements which also are in $\mathcal{E}_{\text{s-ch}}$, and thus all the thresholds in $\mathcal{E}_{\text{isr}}$ are left out of the definition.

\vSWITCH{
The price to pay for this choice is the appearance of Initial-State Radiation (ISR) singularities in observables, which are commonly treated within the collinear mass factorisation paradigm. 
}{ %
The price to pay for this choice is the appearance of Initial-State Radiation (ISR) singularities in observables, which are commonly treated within the collinear mass factorisation paradigm. In sect.~\ref{sec:ISR_outlook} we will consider how the construction of $\sigma_{\text{d}}$ changes and what could happen if observables which allow for all thresholds of the supergraph are included. 
} %

For now, as anticipated, we show that $\sigma_{\text{d}}$ is bounded on $\mathbb{R}^{3L}\setminus B_{\text{isr},\epsilon}$ where
\begin{equation}
B_{\text{isr},\epsilon}=B_\epsilon(\vec{\phi}^{-1}[\text{S}_{\text{isr}}]), \hspace{0.3cm} \text{S}_{\text{isr}}=\Bigg(\bigcup_{{\boldsymbol{\tau}}\in\mathcal{E}_{\text{s-ch}}} \delta\eta_{\boldsymbol{\tau}} \Bigg)\cap \Bigg(\bigcup_{\mathbf{s}\in\mathcal{E}_{\text{isr}}} \delta\eta_\mathbf{s}\Bigg),
\end{equation}
with
\begin{equation}
    B_\epsilon(V)=\left\{\vec{k}\in\mathbb{R}^{3L}\middle|\lVert\vec{k}-\vec{v}\rVert <\epsilon, \ \forall \vec{v}\in V\right\}.
\end{equation}
An important subtlety in the definition of $\sigma_{\text{d}}$, in eq.~\eqref{lfmo}, is the possibility of taking the limit for $(t-t^\star_\mathbf{s})\rightarrow 0$ of a quantity that has raised propagators, that is higher powers of the propagators: this makes $\sigma_{\text{d}}$ ill-defined as the limit would yield infinity. Raised propagators and their treatment varies according to their origin, following a general principle: that the degenerate case is obtained as the limit of the non-degenerate case. More specifically:
\begin{itemize}
    \item Raised propagators that are due to external self-energy insertions can be dealt with by showing that after mass renormalisation in the on-shell scheme, the contribution of a self-energy diagram equates that of terms of order one or higher in a Taylor expansion around the on-shell condition. This realisation allows us to effectively lower the raised propagator by one power, thus solving the problem. An extensive analysis of this method is performed in sect.~\ref{sec:UV_treatment_in_LU}. Raised propagators that are due to internal energy insertions can be dealt with by adding a different fictitious shift in each of the raised propagators, calculating $\sigma_{\text{d}}$ with this configuration, and then taking the limit in which all the fictitious shits vanish.
    \item Two or more distinct E-surfaces intersecting in a non-empty region can give rise to a locally raised propagator (the first-order expansion of the E-surfaces in the flow variables $t$ is proportional to $(t-t^\star)$ for all the E-surfaces which vanish at the intersection, with $t^\star$ being the flow location of the intersection point). In this case we impose the behaviour of $\sigma_{\text{d}}$ to be the continuation of the non-intersecting case. More specifically, since the deformation field $\kappa$, related to the causal flow by eq.~\eqref{eq:causal_ode}, is non-zero on any point lying on a non-pinched E-surface, we are ensured that the set of curves passing through an intersection point is zero-measured. If we define
    \begin{equation}
        \mathcal{I}=\vec{\phi}^{-1}[ \text{S}_{\text{s-ch}} ],\hspace{0.3cm} \text{S}_{\text{s-ch}}=\bigcup_{\substack{\mathbf{s},{\boldsymbol{\tau}}\in\mathcal{E}_{\text{s-ch}} \; \mathbf{s}\neq {\boldsymbol{\tau}}}}\big(\delta\eta_{\mathbf{s}} \cap \delta\eta_{{\boldsymbol{\tau}}}\big)
    \end{equation}
    then $\mathcal{I}$ is zero-measured with respect to $\mathbb{R}^{3L}$, as argued before, and thus proving that $\sigma_{\text{d}}$ is integrable on $\mathbb{R}^{3L}\setminus B_{\text{isr},\epsilon}\setminus \mathcal{I}$ itself proves that the integral of $\sigma_{\text{d}}$ on $\mathbb{R}^{3L}\setminus B_{\text{isr},\epsilon}$ is finite.
    \item 
    For a particular choice of the collision reference frame there can be E-surfaces corresponding to different Cutkosky cuts but which take the same functional form (e.g. see $\eta_{\mathbf{s}^{\text{r}}_1}$ and $\eta_{\mathbf{s}^{\text{r}}_2}$ in the example process discussed in sect.~\ref{sec:cancellation_example}). These raised poles are spurious and can be eliminated by taking advantage of Lorentz invariance to change the frame.

\end{itemize}
Finally, we observe that $\sigma_{\text{d}}$ is continuous on $\mathbb{R}^{3L}\setminus B_{\text{isr},\epsilon}$ only if soft configurations are also excluded, since at these locations the inverse energies (other than, possibly, elements of $\mathcal{E}_{\text{int}}$) of massless particles diverge.
We define
\begin{equation}
    \mathcal{W}=\vec{\phi}^{-1}[\text{S}_{\text{soft}}], \hspace{0.3cm} \text{S}_{\text{soft}}=\Bigg(\bigcup_{{\boldsymbol{\tau}}\in\mathcal{E}_{\text{s-ch}}} \delta\eta_{\boldsymbol{\tau}} \Bigg)\cap \bigcup_{\mathbf{x}\subseteq\mathbf{e}}\text{S}_{\mathbf{x}} \,,
\end{equation}
where $\text{S}_{\mathbf{x}}$ is defined as in eq.~\eqref{soft-region}. The definition of $\mathcal{W}$ is manifestly a zero-measured set with respect to $\mathbb{R}^{3L}$. We then consider $\sigma_{\text{d}}$ to be defined on $\mathbb{R}^{3L}\setminus B_{\text{isr},\epsilon}\setminus (\mathcal{I}\cup \mathcal{W} )$. We will always make a distinction between the sets $\mathcal{I}$ and $\mathcal{W}$, as it makes the unfolding of the proof easier to understand. 

We shall now show that $\sigma_{\text{d}}$ is integrable on the $\mathbb{R}^{3L}\setminus B_{\text{isr},\epsilon}\setminus (\mathcal{I}\cup \mathcal{W} )$ space, which implies that $\sigma_{\text{d}}$ has a finite integral on the whole $\mathbb{R}^{3L}\setminus B_{\text{isr},\epsilon}$ space.

\subsubsection{Cancellation of pinched surfaces}
\label{sect:IR_cancellation_proof}
We now investigate the mechanism that leads to the cancellation of pinched E-surfaces (and E-surfaces in general, in the fully inclusive case). As anticipated previously, the mechanism relies on the generalisation of the straightforward algebraic relation in eq.~\eqref{p-fractioning-rel}:
\begin{equation}\label{ext-p-relation}
    \lim_{t_1\rightarrow t_N}\ldots\lim_{t_{N-1}\rightarrow t_N}\sum_{i=1}^N \frac{f(t_i)}{\prod_{j\neq i} (t_j-t_i)}=\frac{1}{(N-1)!}\partial_t^{N-1} f \,.
\end{equation}
Cancellations are explicitly shown by expanding all the summands in $\sigma_{\text{d}}$ in the proximity parameters $t^\star_\mathbf{s}-t
^\star_{\boldsymbol{\tau}}$. This leads to large expressions, with multiple sums and apparently complicated coefficients; however, these expressions allow to factorise sums in the form of eq.~\eqref{ext-p-relation}, and thus to immediately prove cancellations. This proof thus has the remarkable advantage of applying to any type of singularity. Differentiating pinched and non-pinched thresholds is the only important distinction and is only instrumental for characterising what are the IR-safety constraints imposed on the observable definition.

We stress that cancellation of pinched E-surfaces, by itself, does not guarantee that $\sigma_{\text{d}}$ is integrable on $\mathbb{R}^{3L}\setminus (\mathcal{I}\cup \mathcal{W} )\setminus B_{\text{isr},\epsilon}$, and specifically at the locations of soft singularities. However, it shows that purely collinear singularities (that is, excluding the boundaries of a pinched singularity, the soft points) of the summands of $\sigma_{\text{d}}$ are subject to cancellations, and that at soft locations, the resulting power-counting is less severe. The proof is thus concluded only after a power-counting analysis of soft singularities (which is carried in subsect.~\ref{sec:power_counting_soft_integrable_singularities}).

Another important feature of $\sigma_{\text{d}}$ is that the E-surfaces are associated to Feynman prescriptions which could complicate the proof for complex kinematics, e.g., in the presence of a deformation. In the following we will drop the prescription and discuss the case of a non-zero deformation in sect.~\ref{sec:contour_deformation}.

The limits in the definition of $\sigma_\text{d}$, for $\vec{k}\in\mathbb{R}^{3L}\setminus (\mathcal{I}\cup \mathcal{W} )\setminus B_{\text{isr},\epsilon}$, can be performed explicitly and yield
\begin{align}\label{eq:master_formulat_without_w}
    \sigma_{\text{d}}=\sum_{\mathbf{s}\in\mathcal{E}_{\vec{k}, \vec{\phi}}} \Bigg[\frac{g(\vec{\phi}(t^\star_\mathbf{s}, \vec{k}))|\text{det}[\mathbb{J}_{\vec{k}'}\vec{\phi}]|}{\prod_{{\boldsymbol{\tau}}\in\mathcal{E}_{\vec{k}, \vec{\phi}}\setminus\{\mathbf{s}\}}\big[\sum_{j>0}(t^\star_\mathbf{s}-t^\star_{{\boldsymbol{\tau}}})^j \partial_t^j \eta_{\boldsymbol{\tau}}(\vec{\phi}(t^\star_{\boldsymbol{\tau}}, \vec{k}))/j! \big]}\Bigg]\frac{h(t^\star_\mathbf{s})\mathcal{O}_\mathbf{s}}{|\partial_t \eta_\mathbf{s}(\vec{\phi}(t^\star_\mathbf{s}, \vec{k}))|} \,,
\end{align}
where
\begin{equation}
    g=\frac{f}{\prod_{i\in\mathbf{e}}2E_i \prod_{{\boldsymbol{\tau}}\in \mathcal{E}\setminus \mathcal{E}_{\vec{k},\vec{\phi}}}\eta_{\boldsymbol{\tau}}} \,.
\end{equation}
We remind the reader that the explicit unfolding of the dense eq.~\eqref{eq:master_formulat_without_w} for the \DT{} supergraph of our example process is given in eq.~\eqref{eq:DT_diff_xsec}. We will now rewrite $\sigma_{\text{d}}$ in a more convenient form, in which the divided differences~\eqref{ext-p-relation} are manifest:
\begin{align}\label{my-best}
    \sigma_{\text{d}}=\frac{1}{\prod_{\mathbf{s}\in\mathcal{E}_{\vec{k},\vec{\phi}}}|\partial_t \eta_\mathbf{s}(\vec{\phi}(t^\star_\mathbf{s}, \vec{k}))|}\sum_{\mathbf{s}\in\mathcal{E}_{\vec{k}, \vec{\phi}}} \Bigg[\frac{g(t^\star_\mathbf{s}, \vec{k}) w(t^\star_\mathbf{s},\vec{k})\mathcal{O}_\mathbf{s}}{\prod_{{\boldsymbol{\tau}}\in\mathcal{E}_{\vec{k}, \vec{\phi}}\setminus\{\mathbf{s}\}}(t^\star_\mathbf{s}-t^\star_{{\boldsymbol{\tau}}})}\Bigg] \,,
\end{align}
where
\begin{equation}
    w(t',\vec{k})=\lim_{t\rightarrow t'}\frac{h(t) |\det[\mathbb{J}_{\vec{k}'}\vec{\phi}]|}{\displaystyle{\prod_{\tau\in\mathcal{E}_{\vec{k},\vec{\phi}}} \frac{\eta_\tau(\vec{\phi}(t, \vec{k}))}{(t-t^\star_\tau)\partial_t \eta_\tau(\vec{\phi}(t^\star_\tau, \vec{k}))}}} \,.
\end{equation}
We observe that, under appropriate conditions on $h$, $w(t^\star_\mathbf{s},\vec{k})$ is bounded for every $\vec{k}$. Furthermore, observe that we have factored out the product of the absolute value of the first order derivative of the E-surfaces in $\mathcal{E}_{\vec{k}\vec{\phi}}$, which can only be done when the following condition is satisfied:
\begin{equation}
    \text{sign}\big[\partial_t \eta_{{\boldsymbol{\tau}}_1}(\vec{\phi}(t^\star_{{\boldsymbol{\tau}}_1},\vec{k}))\big]=\text{sign}\big[\partial_t \eta_{{\boldsymbol{\tau}}_2}(\vec{\phi}(t^\star_{{\boldsymbol{\tau}}_2},\vec{k}))\big], \ \ \forall {\boldsymbol{\tau}}_1,  {\boldsymbol{\tau}}_2\in\mathcal{E}_{\vec{k},\vec{\phi}} \,,
\end{equation}
which is guaranteed by the choice of a causal flow. Specifically, the causal flow is constructed from a deformation which is explicitly constructed to satisfy the causal prescription on all E-surfaces \it and on their intersections\rm. In this approach, we see how the realisation of IR cancellations is interlocked with causality, enforced through the definition of the Feynman prescription.

Eq.~\eqref{my-best} is exactly in the form needed to apply divided differences. We will now set the stage to derive the conditions under which the observable function $\mathcal{O}_\mathbf{s}$ preserves the IR cancellations established by the divided differences, and thus can be considered IR-safe. At first, let $\mathcal{O}_\mathbf{s}=1, \ \forall \mathbf{s}\in\mathcal{E}_{\text{s-ch}}$, so that cancellations are trivially realised, since we can immediately apply the divided difference relation of eq.~\eqref{ext-p-relation} to
\begin{equation}
    \sigma_{\text{d}}=\frac{1}{\prod_{\mathbf{s}\in\mathcal{E}_{\vec{k},\vec{\phi}}}|\partial_t \eta_\mathbf{s}(\vec{\phi}(t^\star_\mathbf{s}, \vec{k}))|}\sum_{\mathbf{s}\in\mathcal{E}_{\vec{k}, \vec{\phi}}} \Bigg[\frac{g(t^\star_\mathbf{s}, \vec{k}) w(t^\star_\mathbf{s},\vec{k})}{\prod_{{\boldsymbol{\tau}}\in\mathcal{E}_{\vec{k}, \vec{\phi}}\setminus\{\mathbf{s}\}}(t^\star_\mathbf{s}-t^\star_{{\boldsymbol{\tau}}})}\Bigg] \,
\end{equation}
and argue that all the multiple singular limits other than soft limits yield a finite result.
Thus, in the fully inclusive case, $\sigma_{\text{d}}$ is integrable on $\mathbb{R}^{3L}\setminus B_{\text{isr},\epsilon}\setminus(\mathcal{I}\cup B_\epsilon(\mathcal{W}))$, that is on $\mathbb{R}^{3L}\setminus B_{\text{isr},\epsilon}\setminus B_\epsilon(\mathcal{W})$. The simple pattern of cancellations in eq.~\eqref{ext-p-relation} immediately applies for an arbitrary perturbative order and in the case of fully inclusive observables. For non-trivial observables, the situation is more complicated. In particular, we observe that observables carry an explicit dependence on $\mathbf{s}$, and not just indirectly through their dependence on $t^\star_\mathbf{s}$: the functional form of the observables is dependent on the Cutkosky cut. Thus for arbitrary observables, the pattern of cancellations in eq.~\eqref{ext-p-relation} does not necessarily hold. Instead, the request of IR-finiteness implies constraints on observables.
\subsubsection{IR-safe observables and infrared scales}

In this section we investigate the role that non-trivial observables play in the mechanism of cancelling divergences discussed in the previous section. Since observables discriminate Cutkosky cuts and are intrinsically local objects, proving that the sum is finite for non-trivial observables is not a purely algebraic matter like in the fully inclusive case, but can only be done in a neighbourhood of the problematic points.
In principle, for what concerns the proof, an observable can depend on the supergraph $\Gamma$, the Cutkosky cut $\mathbf{s}$ it corresponds to, the whole set of loop variables of the supergraph $\vec{k}$, and the kinematic configuration of the initial states. In practice however, we defined observables as having a dependence on the momentum, mass and quantum numbers of only the particles in the Cutkosky cut. Thus, we will first work out a general sufficient condition for the cancellation pattern to be realised, and then enquire if and how observables satisfying
eq.~\eqref{observables-form} satisfy these general constraints.
We will impose these constraints to be flow-independent, since the choice of a causal vector field should not affect the physical properties of observables.

We start by observing that $\sigma_\text{d}$ is bounded on $\mathbb{R}^{3L}\setminus B_{\text{isr},\epsilon}\setminus(\mathcal{I}\cup B_\epsilon(\mathcal{W}))$ if 
\begin{equation}\label{mmm}
    \partial_t^n \mathcal{O}_{{\boldsymbol{\tau}}_1}(\vec{\phi}(t_{{\boldsymbol{\tau}}_1}, \vec{k}))=\partial_t^n \mathcal{O}_{{\boldsymbol{\tau}}_2}(\vec{\phi}(t_{{\boldsymbol{\tau}}_2}, \vec{k})), \ \ \forall \vec{k} \in \vec{\phi}^{-1}[\delta\eta_{{\boldsymbol{\tau}}_1}\cap \delta\eta_{{\boldsymbol{\tau}}_2}], \ \ \forall {\boldsymbol{\tau}}_1, {\boldsymbol{\tau}}_2\in \mathcal{E}_{\vec{k},\vec{\phi}}\setminus I, \ \forall n\in\mathbb{N} \,.
\end{equation}
This is required for the divided differences in eq.~\eqref{ext-p-relation} to be applicable. 
This condition is manifestly dependent on the flow. We will use the following relation: if a function $f$ satisfies a property on an open set $V$, then the function $f\circ \vec{\phi}$ satisfies it on $\vec{\phi}^{-1}[V]$. Indeed the flow projects a set $B_\epsilon(\mathcal{I})$ on a neighbourhood of an intersection $\delta\eta_{{\boldsymbol{\tau}}_1}\cap \delta\eta_{{\boldsymbol{\tau}}_2}$, and we conclude that a sufficient condition for eq.~\eqref{mmm} to hold is that
\begin{equation}\label{observable-condition}
    O_{{\boldsymbol{\tau}}_1}(\vec{k})=O_{{\boldsymbol{\tau}}_2}(\vec{k}), \ \ \forall \vec{k} \in B_{\epsilon}(\delta\eta_{{\boldsymbol{\tau}}_1}\cap \delta\eta_{{\boldsymbol{\tau}}_2}) \,,
\end{equation}
where $B_{\epsilon}(\delta\eta_{{\boldsymbol{\tau}}_1}\cap \delta\eta_{{\boldsymbol{\tau}}_2})$ is a neighbourhood of the intersection between the two E-surfaces, and $\epsilon$ is a quantity dependent on the experimental setup. The interpretation of $\epsilon$ is that of a natural resolution below which the experimental setup is unable to distinguish degenerate kinematic configurations of final states, identified by ${\boldsymbol{\tau}}_1$ and ${\boldsymbol{\tau}}_2$.
In conclusion, the notion of degeneracy and its relation with experimental resolution is derived here as a result of the analytic properties of the local representation of differential cross-sections. 

We explicitly verified that \emph{constant} observables of the form of eq.~\eqref{observables-form} can retain local cancellation of non-pinched E-surface in the entirety of the integration volume.
Instead of performing the same analysis for pinched E-surfaces and showing that IR-safe observables can always be defined, we will limit ourselves to stating how the condition relates to the common definitions of IR-safe collider observables.

It may now seem that for non-trivial observables it is not possible to make $\sigma_{\text{d}}$ integrable on the whole $\mathbb{R}^{3L}\setminus B_{\text{isr},\epsilon}$, as non-pinched threshold are not subject to cancellations. 
However, non-pinched E-surfaces can be integrated using an ad hoc contour deformation, which cannot be done for pinched E-surfaces (for which the cancellation pattern must still hold, because of IR-safety). The deformation of amplitudes was discussed in ref.~\cite{Capatti:2019edf} and we discuss how to extend it to cross sections in sect.~\ref{sec:contour_deformation}. In the following, we will assume that such a contour deformation can be constructed and that proving $\sigma_{\text{d}}$ to be integrable on $B_\epsilon(\mathcal{P})$ implies that $\sigma_{\text{d}}$ can be analytically continued and its non-pinched thresholds contour integrated so that it is regular on the entirety of $\mathbb{R}^{3L}\setminus B_{\text{isr},\epsilon} \setminus B_\epsilon(\mathcal{W})$.

We observe that in order to preserve the cancellations at the pinched points, it is sufficient to require that
\begin{equation}
\exists \epsilon>0 \text{ s.t. }\mathcal{O}_{\boldsymbol{\tau}}(\vec{k})=\mathcal{O}_\mathbf{s}(\vec{k}), \ \forall \vec{k}\in B_\epsilon(\text{H}_{{\boldsymbol{\tau}}\mathbf{s}}) \,,
\end{equation}
where we require $\mathbf{c}_\mathbf{s}\setminus\mathbf{c}_{\boldsymbol{\tau}}$ and $\mathbf{c}_{\boldsymbol{\tau}}\setminus\mathbf{c}_{\mathbf{s}}$ to only contain massless particles, for otherwise there is no pinch singularity associated with the points in $\text{H}_{{\boldsymbol{\tau}}\mathbf{s}}$.
If we use the constraint in eq.~\eqref{observables-form}, this equation takes a more illuminating form:
\begin{equation}\label{IR-safe-cond}
    \mathcal{O}_{{|\mathbf{c}_{\boldsymbol{\tau}}|}}(\{\alpha_i \vec{p}+\epsilon_i\}_{i\in\mathbf{c}_{\boldsymbol{\tau}}\setminus \mathbf{c}_{\mathbf{s}}};\{\vec{q}_j\}_{j\in\mathbf{c}_{\boldsymbol{\tau}}\cap \mathbf{c}_{\mathbf{s}}})=\mathcal{O}_{|\mathbf{c}_{\mathbf{s}}|}(\{-\beta_i \vec{p}+\epsilon_i\}_{i\in\mathbf{c}_\mathbf{s}\setminus \mathbf{c}_{{\boldsymbol{\tau}}}};\{\vec{q}_j\}_{j\in\mathbf{c}_{\boldsymbol{\tau}}\cap \mathbf{c}_{\mathbf{s}}}), \ \ \forall \epsilon_i, \ \lVert \epsilon_i \rVert <\epsilon \,,
\end{equation}
with $\alpha_i,\ \beta_j, \ \epsilon_i$ subject to momentum conservation, and we have suppressed the explicit dependence of the observable on the mass and spin of the particles. Eq.~\eqref{IR-safe-cond} makes manifest that the ability of observables to satisfy IR-safety constraints is necessarily associated with momentum-conservation laws on the graph; it can only be satisfied if the observables themselves cluster together all the particles of their defining Cutkosky cut that are collinear or soft, identifying them with one object which has the sum of the momenta of the constituent particles. The dependence of the observable on each of the momenta of particles in the Cutkosky cut individually morphs into a unique dependence on the sum of the momenta of collinear or soft particles in the neighbourhood of a pinched point and the usual independence on individual momenta of the particles that are not collinear or soft:
\begin{equation}\label{Ir-safety}
     \mathcal{O}_{{|\mathbf{c}_{\boldsymbol{\tau}}|}}(\{\alpha_i \vec{p}+\epsilon_i\}_{i\in\mathbf{c}_{\boldsymbol{\tau}}\setminus \mathbf{c}_{\mathbf{s}}};\{\vec{q}_j\}_{j\in\mathbf{c}_{\boldsymbol{\tau}}\cap \mathbf{c}_{\mathbf{s}}})= \mathcal{O}_{{|\mathbf{c}_{\boldsymbol{\tau}}\cap \mathbf{c}_{\mathbf{s}}|+1}}\Bigg(\sum_{i\in\mathbf{c}_{\boldsymbol{\tau}}\setminus \mathbf{c}_{\mathbf{s}}}\alpha_i \vec{p}+\epsilon_i;\{\vec{q}_j\}_{j\in\mathbf{c}_{\boldsymbol{\tau}}\cap \mathbf{c}_{\mathbf{s}}}\Bigg), \ \forall \lVert \epsilon_i \rVert<\epsilon \,.
\end{equation}
Ultimately, eqs.~\eqref{Ir-safety} and~\eqref{IR-safe-cond} establishes whether an observable satisfies sufficient conditions for the cancellation of the pinched singularities associated to the points in $\text{H}_{{\boldsymbol{\tau}}\mathbf{s}}$ and can be used in defining a locally finite differential cross-section. The volume form $\sigma_{\text{d}}$, with observables satisfying eqs.~\eqref{Ir-safety} and~\eqref{IR-safe-cond} at any pinched location is thus locally finite on $\mathbb{R}^{3L}\setminus B_{\text{isr},\epsilon}\setminus B_\epsilon(\mathcal{W})$. In order to conclude the proof, we study the soft scaling of $\sigma_{\text{d}}$ and determine under which conditions it has at most integrable singularities.

\subsubsection{Soft scaling from the causal flow}
\label{sec:soft_scaling_from_the_causal_flow}

We now study the integrability of $\sigma_{\text{d}}$ in a neighbourhood $B_\epsilon(\mathcal{W})$. Since soft points are at the extrema of pinched singularities, the IR-safety of observables allows to write $\sigma_\text{d}$ as eq.~\eqref{ext-p-relation}
 in a neighbourhood of the soft points. Furthermore, we see that in the limit of eq.~\eqref{ext-p-relation} the remaining function has a scaling which is at worst the scaling of $g$ or any of its derivatives in $t$, assuming $\mathcal{O}_\mathbf{s}$ is engineered in such a way that, approaching the soft point from any direction, pinched singularities are all cancelled, and thus do not contribute to any worsening of the scaling of the integrand.

Thus, in order to analyse the scaling of the integrand in the neighbourhood of soft singularities, we have to understand the scaling of $g$ and its derivatives in $t$. We now identify extra constraints on the causal flow such that derivatives of $g$ have the same scaling as $g$, assuming the observables and the $h$ function are bounded on such neighbourhood. 

We recall that $g$ can be defined in terms of the manifestly causal (cLTD) representation, and particularly in terms of the simplified constrained form of eq.~\eqref{asymptotic-behaviour}, such that
\begin{equation}\label{g-expr}
    g=\frac{\prod_{{\boldsymbol{\tau}}\in\mathcal{E}_{\vec{k},\vec{\phi}}} \eta_{\boldsymbol{\tau}}}{\prod_{i\in\mathbf{e}}2E_i}\sum_{F\in\mathcal{F}}\sum_{\vec{\alpha}\in\{\pm 1\}^{|F|}}\frac{c_{F\vec{\alpha}}(\vec{k})}{\prod_{\mathbf{s}\in F}\eta_{(\mathbf{s},\alpha_{\mathbf{s}})} } \,.
\end{equation}
Let us start by observing that, by definition, $g$ cannot be singular on any s-channel E-surface, as  $g$ is bounded on $\mathbb{R}^{3L}\setminus B_{\text{isr},\epsilon}\setminus\mathcal{I}\setminus B_\epsilon(\mathcal{W})$ (and excluding the UV region too). Near soft points the scaling of $g$ is entirely determined by the inverse E-surfaces in $\mathcal{E}_{\text{int}}$ and by the inverse energies (since $g$ is not singular on any s-channel E-surface and since ISR singularities are excluded, thus leaving internal E-surfaces only).

We can now state a sufficient condition on the causal flow $\vec{\phi}$ and the causal vector field $\vec{\kappa}$ for derivatives of $g$ with respect to $t$ to have the same (or less singular) scaling as $g$:
\begin{equation}\label{soft-scaling}
    \lim_{\vec{k}\rightarrow \vec{k}^\star}\Bigg|\frac{\partial_t^n g (t^\star_\mathbf{s},\vec{k})}{g(t^\star_\mathbf{s},\vec{k})}\Bigg|\in\mathbb{R}^+, \ \forall\vec{k}^\star \in B_\epsilon(\mathcal{W})
\end{equation}
if the deformation field satisfies the continuity constraint of eq.~(3.37) of ref.~\cite{Capatti:2020ytd}:
\begin{equation}\label{eq:branchcut_scaling}
\vec{q}_j(\vec{k})^2+m_j^2-\vec{Q}_j(\kappa)^2\ge 0 \hspace{0.3cm} \forall j\in \mathbf{e} \, ,
\end{equation}
This condition is obtained by iteratively applying Leibniz's rule on the $n$-th derivative of $g$, thus isolating products of multiple derivatives of inverse energies, of inverse internal E-surfaces and the numerator. Then, we require derivatives of inverse energies and of inverse internal E-surfaces to have the same or better soft scaling than inverse energies and inverse E-surfaces themselves. Observe that 
\begin{align}
\begin{split}
    \pdv{}{t} \frac{1}{E_i(\vec{\phi}(t,\vec{k}))}&=-\frac{\vec{Q}_i(\kappa)\cdot \vec{q}_i}{E_i^3}, \\
    \pdv{}{t} \frac{1}{\eta_{\boldsymbol{\tau}}(\vec{\phi}(t,\vec{k}))}&=-\frac{1}{\eta_{\boldsymbol{\tau}}^2}\sum_{i\in\delta({\boldsymbol{\tau}})}\frac{\vec{Q}_i(\kappa)\cdot \vec{q}_i}{E_i} \,.
\end{split}
\end{align}
If the derivative of the inverse energy and the inverse energy itself are required to have the same soft scaling, $\vec{Q}_i(\vec{\kappa})$ is required to vanish as fast or faster than the on-shell energy $E_i$. The same holds for the inverse E-surface. In the construction of the deformation field of ref.~\cite{Capatti:2019edf} this constraint already appears and is called the continuity constraint. The role of this constraint in that work was to make sure that the deformation is continuous and does not cross a branch cut.

Since it is shown in ref.~\cite{Capatti:2019edf} that such causal vector field can be constructed, we conclude that thanks to eq.~\eqref{soft-scaling} the scaling of $\sigma_\text{d}$ is the same as the worst scaling between $g$ and its derivatives, assuming all pinched singularities cancel and do not contribute to the soft scaling when approaching the soft point from any direction. We are now able to show that $\sigma_{\text{d}}$ is integrable on a neighbourhood of soft points $B_\epsilon(\mathcal{W})$. Because of the previous work, this is equivalent to showing that the scaling of $g(\vec{k})$ at soft points only leads to integrable singularities.
These findings can be summarised in the following claim:
\begin{equation}
\label{eq:scaling_LU}
    \lim_{\vec{k}'\rightarrow \vec{k}^\star}\frac{\sigma_{\text{d}}(\vec{k}')}{\max_{\mathbf{s}\in\mathcal{E}_{\vec{k},\vec{\phi}}}\{g(t_\mathbf{s}^\star,\vec{k}')\}}\in\mathbb{R}\setminus\{0\}, \ \ \forall \vec{k}^\star\in B_\epsilon(\mathcal{W}).
\end{equation}
The scaling of $g$, on the other hand, follows from its explicit expression in eq.~\eqref{g-expr} and from the property in eq.~\eqref{asymptotic-behaviour}. In order to establish that the LU representation is integrable on soft singularities, one must still show that the soft scaling of $g$, and therefore of $\sigma_{\text{d}}$ as well in virtue of eq.~\eqref{eq:scaling_LU}, is sufficiently tame for physical theories. This is the object of the next section.

\subsubsection{Power-counting of soft singularities}
\label{sec:power_counting_soft_integrable_singularities}

In the following, we will analyse the scaling of $\sigma_\text{d}$ close to points at which connected clusters of particles simultaneously become soft (any configuration that is not the union of disconnected clusters can easily be shown to have a better scaling). Consider a connected subgraph of $\Gamma$, $(\mathbf{s},\mathbf{e}_\mathbf{s}=\delta^\circ(\mathbf{s})\cup\delta(\mathbf{s}))$ closed under momentum-conservation conditions (that is, if the momenta of edges in the subgraph uniquely determine any other edge's momentum, then that edge is also in the subgraph).

Using eq.~\eqref{g-expr} and eq.~\eqref{eq:scaling_LU}, we concluded that the worst possible scaling of $\sigma_{\text{d}}$ near a soft singularity (assuming that the observable functions allow for the cancellation of all pinched singularities in the neighbourhood of the soft region), can be at most $\sigma_{\text{d}}\sim g \sim \delta^{-(|\mathbf{s}|+|\mathbf{e}_\mathbf{s}|+N)}$, since each inverse energy corresponding to an edge in $\mathbf{e}_\mathbf{s}$ contributes one power, the numerator contributes with $N$ powers and, in the worst-case scenario, there exists a term in the sum of eq.~\eqref{asymptotic-behaviour} which has $|\mathbf{s}|$ E-surfaces of $\mathcal{E}_{\text{int}}$ vanishing (such a case would correspond to a term of eq.~\eqref{asymptotic-behaviour} whose respective cross-free family contains a cross-free family of subsets of $(\mathbf{s},\mathbf{e}_\mathbf{s})$). 

Since the actual soft scaling of the numerator is theory-dependent, we expect integrability at soft points to be conditional on the specific theory one is considering. We will now simplify the discussion by proving integrability for the physical theory of Yang-Mills coupled with massless fermions theories. Since the Higgs boson is massive, including it does not change the reasoning, along with any type of massive particles. We can distinguish between fermion edges and vector boson edges by writing $\mathbf{e}_{\mathbf{s}}=\mathbf{e}_f\cup\mathbf{e}_b$. Furthermore, we write  $\mathbf{s}_{fb}
^{(3)}$ for 3-vertices at which two fermions and a vector boson meet, $\mathbf{s}^{(3)}_b$ for vertices at which three vector bosons meet, $\mathbf{s}^{(4)}_b$ for vertices at which four vector bosons meet, so that $\mathbf{s}=\mathbf{s}^{(3)}_b\cup\mathbf{s}_{fb}
^{(3)}\cup\mathbf{s}^{(4)}_b$. Finally, if all momenta in the subgraph are set to zero, the degree of divergence associated to this subgraph is
\begin{equation}
    d=3L_\mathbf{s}-|\mathbf{e}_{\mathbf{s}}|-|\mathbf{s}|+|\mathbf{s}^{(3)}_b|+|\mathbf{e}_f|=2|\mathbf{e}_{\mathbf{s}}|-3|\mathbf{s}|+|\mathbf{s}^{(3)}_b|+|\mathbf{e}_f|-|\mathbf{s}| \,.
\end{equation}
We substitute the following relations between the number of vertices, their degree and the edges of a graph, $|\mathbf{e}|=3/2|\mathbf{s}^{(3)}_b|+2|\mathbf{s}^{(4)}_b|+3/2|\mathbf{s}^{(3)}_{fb}|+|\delta(\mathbf{s})|/2$ and $|\mathbf{e}_f|\ge|\mathbf{s}^{(3)}_{fb}|$, such that we obtain:
\begin{equation}\label{degree}
    d\ge|\mathbf{s}^{(4)}_b|+|\mathbf{s}^{(3)}_b|+|\mathbf{s}^{(3)}_{fb}|+|\delta(\mathbf{s})|-|\mathbf{s}|=|\delta(\mathbf{s})| ,
\end{equation}
which confirms the inexistence of soft singularities, and confirms that the power-counting at soft points, in the absence of pinched E-surfaces, which are here assumed to all cancel when approaching the soft points from any direction, yields integrable singularities at most. Eq.~\eqref{degree} also establishes that the soft power-counting could only yield a non-integrable divergence for an amplitude only if there exists a set of pinched singularities with size equal to the number of particles external to the cluster $\mathbf{s}$ which add an extra $-|\delta(\mathbf{s})|$ to the power-like behaviour of the integrand when approaching the soft point from a certain direction; this fact can be used to define infrared-safe observables that are less restrictive, at least in neighbourhoods of the soft points. 
For a cubic scalar theory, the power-counting formula for soft clusters reads
\begin{equation}
    d=|\delta(\mathbf{s})|-|\mathbf{s}| \,
\end{equation}
which can become negative. This reproduces the known fact~\cite{Sterman:1978bi} that the IR region of super-renormalisable theories such as cubic scalar theory exhibits soft divergences that do not cancel. 
As an example of a graph that is still divergent within LU, we give an example of a supergraph in cubic scalar theory that features a quadratic soft divergence ($d=-2$) on the grey lines:
\begin{center}
\begin{tabular}{c}
    \begin{tikzpicture}
    \begin{feynman}

    \tikzfeynmanset{every vertex={dot,minimum size=0.9mm}}    
    \vertex (lc);

        \vertex[right=1cm of lc] (rc);   
     \tikzfeynmanset{every vertex={empty dot,minimum size=0mm}}    
    \vertex[right=0.5cm of lc] (cc);      
    \vertex[below=0.5cm of cc] (bc);
    \vertex[above=0.5cm of cc] (ac);

    \vertex[right=0.15cm of rc] (extrc);
    \vertex[left=0.15cm of lc] (extlc);

    \vertex[above=0.15cm of bc] (mc);
    \vertex[left=0.17cm of cc] (llc);
    \vertex[right=0.17cm of cc] (rrc);

        \vertex[left=0.147cm of rc] (a);
        \vertex[right=0.147cm of lc] (b);

\tikzfeynmanset{every vertex={dot,minimum size=0.mm}}

    \vertex[below=0.35355339cm of a] (newrc);    
    
\tikzfeynmanset{every vertex={empty dot,minimum size=0mm}}

    \vertex[below=0.1cm of newrc] (aaa);
    \vertex[right=0.1cm of aaa] (bbb);   
    
        \vertex[right=0.353553cm of cc] (ccaa);
    \vertex[above=0.353553cm of ccaa] (aa);
    
            \vertex[left=0.353553cm of cc] (ccll);
    \vertex[above=0.353553cm of ccll] (d);
    
        \diagram*[large]{	
        (lc)--[quarter left,black!35!white](ac) -- [quarter left,black!35!white](rc) -- [quarter left](bc)-- [quarter left](lc),
        (mc)--[black!35!white](aa),
        (llc)--[black!35!white](rrc),
        (mc)--[black!35!white](d),
        (mc)--[black!35!white](bc),
	(lc)--(extlc),
	(rc)--(extrc),
        }; 
    \end{feynman}
    \end{tikzpicture} \\ $d=|\delta(\mathbf{s})|-|\mathbf{s}|=-2$ \\
\end{tabular} \,.
\end{center}

This concludes the proof, as we have now shown that $\sigma_{\text{d}}$ is integrable on $\mathbb{R}^{3L}\setminus B_{\text{isr},\epsilon}$ for a physical theory.

\newpage

\section{Self-energies and IR cancellations in non-abelian gauge theories}
\label{sec:UV_treatment_in_LU}

External self-energy corrections, and specifically one particle irreducible insertions on edges that belong to a Cutkosky cut lead to the presence of spurious divergences that can be eliminated through careful on-shell renormalisation of the propagators. Furthermore, the nature of the IR cancellation pattern requires that the computation be done with non-truncated amplitudes.

The problem of self-energy corrections also makes very clear another issue in preserving IR cancellations: external physical boson propagators correspond to a different Feynman rule than internal boson propagators. The difference in tensor structure also leads to miscancellations. In order to address this issue, we allow for ghosts and unphysical bosons to be external particles.

\subsection{Propagator renormalisation and IR cancellations}
\label{sec:selfenergy_treatment_in_LU}

When a supergraph features raised propagators, which in physical theories appear as a result of a self-energy insertion, 
the naive substitution of propagators with Dirac deltas leads to manifestly ill-defined interference diagrams, as performing the substitution for one of the raised propagators exactly evaluates the remaining repeated propagators on their mass shell. 

The first possible solution to this problem relies on recognising the origin of the Cutkosky rule as an analogue for performing contour integration. This suggests that one should use the residue formula for higher-order poles, and modify the Cutkosky rule accordingly so that each propagator, raised to the power $n$, is substituted by an appropriately normalised $(n-1)$-th derivative of Dirac's delta function. However, considering this higher-order residue formula would require taking derivatives of the observable function, which is not always possible.
As we discuss in this section, it is instead possible to apply the derivative to the self-energy subgraph only.

In ref.~\cite{Soper:1999xk}, D. Soper circumvents the problem of taking derivatives for addressing raised propagators by engineering an alternative representation of self-energy correction. Through the use of a dispersion relation and algebraic manipulation of the numerator, he was able to show that the raised propagators can be re-absorbed into a propagator with power one. However, the extension of this procedure to more complicated topologies is challenging, especially when the self-energy diagram features UV divergences, since the alternate representation does not allow to construct UV counterterms in four dimensions, but requires to derive them directly from the three-dimensional representation. 

In this section we derive a novel treatment of self-energy diagrams that operates in Minkowsi space and does not require taking derivatives of observables.
As the raised propagator issue arises when one of the propagators adjacent to a self-energy corrections is cut (that is, it is set to be an external particle), it is clear that any solution should bear some relation to the LSZ formalism, Dyson resummation and propagator renormalization. We will show that raised propagators disappear after mass renormalization in the on-shell scheme.

Let us start by defining the renormalized 1PI graph as
\begin{equation}
    \Sigma_R(p^2)=\Sigma(p^2)-\delta Z_m - (p^2-m^2)\delta Z_{\psi} \,.
\end{equation}
In the following, we consider $\Sigma$ to be derived from a scalar theory. 
\vSWITCH{
The extension to physical theories requires extra attention and is left for future work.
}{%
The extension to physical theories requires extra attention and is discussed in sect.~\ref{sec:ghosts}. 
}%
The extra issues due to considering physical theories are however related to the role of numerator algebra and gauge invariance in the IR cancellation mechanism and do not directly affect the general treatment of raised propagators hereby presented. 

Consider now the Taylor expansion of the renormalized one-particle irreducible function
\begin{equation}
    \frac{\Sigma_R(p^2)}{(p^2-m^2)^2}=\frac{\Sigma_R(m^2)}{(p^2-m^2)^2}+\frac{1}{2p_0(p^2-m^2)}\frac{d\Sigma_R}{dp_0}\Bigg|_{p^2=m^2}+\order{(p^2-m^2)^0}.
\end{equation}
The zero-th order of the expansion contains the raised propagator. By imposing the first renormalization condition of the on-shell scheme, $\Sigma_R(m^2)=0$, we obtain that  
\begin{equation}
    \frac{\Sigma_R(p^2)}{(p^2-m^2)^2}=\frac{1}{2p_0(p^2-m^2)}\frac{d\Sigma_R}{dp_0}\Bigg|_{p^2=m^2}+\order{(p^2-m^2)^0}.
    \label{eq:expanded_1PI}
\end{equation}
Thus, each self-energy correction of the graph only brings one power of the external particle propagator. In practice, this allows us to perform Cutkosky cuts by using the rule associated to simple poles. We observe that all terms $\order{(p^2-m^2)^0}$ are not singular at $p^2=m^2$, and therefore their associated residue is zero. 

The modified Cutkosky rule, when applied to a propagator that has a self-energy insertion then reads:
\begin{equation}
\label{modified-cutkosky-rule}
    \frac{\Sigma(p^2)}{(p^2-m^2)^2}\rightarrow \frac{1}{2 p_0}\frac{d\Sigma_R}{dp_0} \delta(p^2-m^2)\,.
\end{equation}
It is important to observe that performing renormalization selectively for the external self-energy corrections does not break the pattern of IR cancellations, since
\begin{equation}
\label{ir-free-two-point}
    \Delta\Sigma=\frac{1}{4(\lVert \vec{p} \rVert^2+m^2)}\frac{d\Sigma}{dp_0}\Bigg|_{p_0=\sqrt{\lVert \vec{p} \rVert^2+m^2}}-\sum_{\mathbf{s}\in\mathbf{v}_\Sigma}\frac{\eta_\mathbf{s}\Sigma}{(p^2-m^2)^2}\Bigg|_{p_0=\eta_\mathbf{s}-p_0}
\end{equation}
is free of IR singularities. 
This statement implies that the IR singularities of two-point functions are fully contained in its derivative when evaluated on the on-shell hyperboloid, and that its singular structure is equal to that of the sum over all possible Cutkosky cuts of the two-point function. One can also think of eq.~\eqref{ir-free-two-point} as establishing a counterterm for the self-energy correction. However, in doing so, the relation with on-shell renormalization is lost.

Finally, we can discuss how field renormalization participates in the clear separation of UV and IR divergences. The poles obtained through the dimensional regularization of the self-energy insertions are usually renormalized, indifferently of their origin (UV or IR), and reabsorbed in the coupling constant. 
Considering truncated amplitudes and renormalizing away IR singularities of external self-energies obscures the presence of a deeper cancellation mechanism for IR divergences.
In the on-shell scheme, required for eq.~\eqref{eq:expanded_1PI} to hold, the two-point function is substituted with a local representation of its pure IR pole that eventually cancels with its real-emission counterpart, while the finite part is set to zero, as prescribed by the renormalization condition.
We start with the on-shell scheme constraint on the finite part of the energy derivative:
\begin{equation}\label{field-ren}
    \text{finite}\Bigg[\frac{d\Sigma_R}{dp_0}\Bigg]=0\,.
\end{equation}
 Eq.~\eqref{field-ren} implies that $\frac{d\Sigma_R}{dp_0}$ must be a local representation of the pure IR pole (in dimensional regularisation). This leads to the following formula:
\begin{equation}
    \frac{d\Sigma_R}{dp^0}=\Bigg[\frac{d\Sigma}{dp^0}-\text{CT}_\text{uv}\Bigg[\frac{d\Sigma}{dp^0}\Bigg]\Bigg]_{d=4}-\Bigg[\text{finite}\Bigg(\frac{d\Sigma}{dp^0}\Bigg)-\text{finite}\Bigg(\text{CT}_\text{uv}\Bigg[\frac{d\Sigma}{dp^0}\Bigg]\Bigg)\Bigg]_d \,. \label{eq:general_UV_regularised_SE}
\end{equation}
$d\Sigma_R/dp^0$ is finite in the UV region. It is divergent in the IR region, so that its expression in dimensional regularization features IR poles. Finally, it has zero finite part. Because of eq.~\eqref{ir-free-two-point}, its IR poles locally cancel with the real counterparts.

\begin{figure}
    \centering
    \input{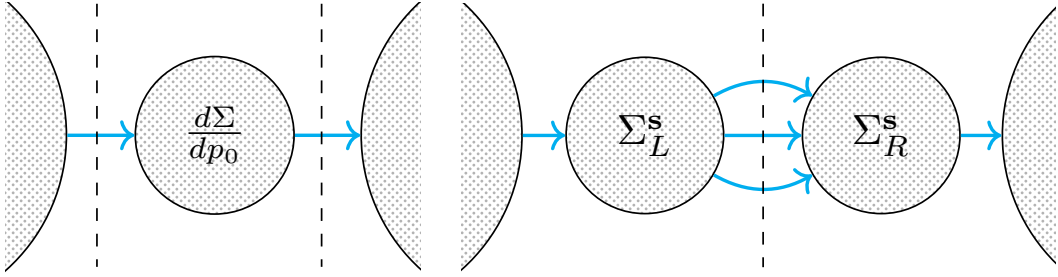}
    \caption{On the left, the derivative of the two point function with cuts on both of the propagators adjacent to it, obtained from the modified Cutkosky rule~\eqref{modified-cutkosky-rule}. It has the same IR structure as the sum of contributions which can be written as the diagram on the right. }
    \label{fig:two-point-functions}
\end{figure}

\vSWITCH{
}{ %
\subsection{Preserving IR cancellations in non-abelian gauge theories}
\label{sec:ghosts}
In the derivations of sect.~\ref{sec:supergraph} and sect.~\ref{sec:general_IR_cancellation_proof} we have assumed that the procedure of performing Cutkosky cuts on the supergraph does not affect the numerator, i.e., that the numerator of an interference diagram is the same as that of the supergraph with some energies set to their on-shell value. This property is fundamental for showing local IR cancellations, and was used in the heuristic argument in sect.~\ref{sec:XOR_and_heuristics} (where the theory was taken to be scalar) and in eq.~\eqref{lfmo}.

For gauge theories, the numerator may seem affected by performing Cutkosky cuts, since there is a distinction between physical photons, which are allowed to be asymptotic states, and unphysical photons, which are not. In the covariant gauge, the functional form of polarisation sums for spin-one bosons differ from that of propagators for spin-one bosons. This leads to apparent divergences at the local level associated with the non-physical degrees of freedom. Ward identities can be used to regularise such divergences, but this makes the proof of cancellation and the construction of interference diagrams more laborious.

Instead, we use the numerator of the usual Feynman propagators instead of polarisation sums for cut gluons, and allow ghosts to appear as final-state particles. This procedure relies on a known alternative to assigning a different Feynman rule to external bosons and internal bosons, and uses the fact that asymptotic states are defined up to states of the form $Q|\psi\rangle$, where $Q$ is the BRST operator. The vector $Q|\psi\rangle$ can then be written as a sum of vectors representing particles including non-physical bosons and ghosts. Rewriting transition amplitudes in this fashion, the polarisation sum rule for a boson can be substituted by the rule for a cut Feynman propagator, at the cost of including all the discontinuities of all supergraphs including those cutting through ghosts.
We explicitly tested for all supergraphs and each LO Cutkosky cut of complicated processes such as $\gamma^\star \rightarrow t \bar{t} g g g$ that the LU integrand is locally identical when computed with final-state ghosts instead of physical external gluons. We note that a symmetrisation over the final-state gluon momenta is necessary in order to carry out this exercise.

The sum over discontinuities of a fixed supergraph, although not gauge invariant, can be made locally IR finite through the procedure of sect.~\ref{sec:selfenergy_treatment_in_LU}. This sum includes thresholds corresponding to ghosts becoming on-shell. We now outline a procedure that determines the contributions necessary to reproduce gauge-invariant quantities: (1) generate all supergraphs with the chosen theory's particle content, including ghosts and Goldstone bosons if needed, (2) proceed to list all Cutkosky cuts of such supergraph, including those cutting through ghosts and Goldstone bosons and (3) express this sum locally, as done in eq.~\eqref{lfmo}, by aligning the loop momentum routing consistently for interference diagrams originating from the same supergraph. We stress that in this framework the action of a Cutkosky cut is equivalent to that of substituting the denominator of the propagator by two times its on-shell energy, and evaluating the numerator at the on-shell value. No polarisation sum rule is ever needed, and thus the interference diagram has the same numerator structure as the supergraph.

Finally, we observe that calculations in the Feynman gauge, i.e, with the gauge parameter set to one, are naturally the most convenient in our formalism, since the gluon propagators will not contain a term proportional to $p^\mu p^\nu/p^4$, which leads to raised propagators whose treatment is difficult to generalise and can lead to ad hoc manipulation of the local integrand. In the Feynman gauge, this means that we treat gluon propagators, both cut and uncut, as a scalar theory propagator multiplied with the tensor structure $g^{\mu\nu}$. 

The procedure outlined above is easily automated and shows that local IR cancellations become straightforward once physical and non-physical degrees of freedom, whether they belong to virtual or real contributions, are treated in a fully symmetrical manner.
We note that this section~\ref{sec:ghosts} is absent from the published version of this document, since we leave a thorougher investigation of this topic to future work.

}%

\newpage

\section{Generalisations}
\label{sec:generalisation}

In this section, we discuss future directions for extending the scope of Local Unitarity and we detail the path towards its fully automated implementation for the numerical computation of higher fixed-order corrections to generic differential cross sections.

\subsection{First steps towards Local Unitarity applicable to initial-state singularities}
\label{sec:ISR_outlook}

\vSWITCH{
The general route for extending the mechanism of FSR cancellations to ISR involves generalising eq.~(3.38) so that it includes Cutkosky cuts corresponding to initial-state thresholds. The guiding principle is that any extension of $\sigma_{\text{d}}$ should still be written as a weighted sum over the discontinuities of the supergraph that can be made locally IR-finite.

One possible attempt at the solution is to consider extending the sum of eq.~(3.38) to the initial-state thresholds. 
In doing so, we can keep our representation of supergraphs involving a fixed set of initial states, but then consider \emph{all} possible thresholds, including those in $\mathcal{E}_{\text{isr}}$. 
Although this may provide a valid cancellation mechanism, it imposes a fixed initial-state multiplicity which is at odds with the notion of backward evolution of initial-state partons.

One can instead tentatively consider a paradigm in which degenerate initial state configurations are accounted for by kinematic configurations featuring multiple incoming asymptotic states. This would corresponds to thresholds of vacuum graphs with altered $\mathrm{i}\epsilon$ prescriptions and identified with $N_{\text{in}}$ (resp. $N_{\text{out}}$) incoming (resp. outgoing) momenta. 

In both cases, the extra diagrams can be effectively considered as natural counter-terms. We leave a thorough investigation of this potential cancellation mechanism of initial state singularities to future publications.
}{ %
The general route for extending the mechanism of FSR cancellations to ISR involves generalising eq.~\eqref{lfmo} so that it includes Cutkosky cuts corresponding to initial-state thresholds. 
The guiding principle is that any extension of $\sigma_{\text{d}}$ should still be written as a weighted sum over the discontinuities of the supergraph that is locally IR-finite.
We note that the published version of this section is shortened and summarized, because it puts forth ideas that are of more speculative nature than the rest of our work.

A first attempt at the solution is to consider extending the sum of eq.~\eqref{lfmo} to the initial-state thresholds. 
In doing so, we can keep our representation of supergraphs involving a fixed set of initial states, but then consider \emph{all} possible thresholds, including those in $\mathcal{E}_{\text{isr}}$. 
Although this may provide a valid cancellation mechanism, it yields disconnected amplitudes and imposes a fixed initial-state multiplicity which is at odds with the notion of backward evolution of initial-state partons.
We instead tentatively consider a paradigm in which degenerate initial state configurations are accounted for by kinematic configurations featuring multiple incoming asymptotic states.
This corresponds to thresholds identified by $N_{\text{in}}$ (resp. $N_{\text{out}}$) incoming (resp. outgoing) momenta.
So far, our construction of Local Unitarity considered a \emph{constant} multiplicity $N_{\text{in}}=2$ or $N_{\text{in}}=1$ and we discuss here the first steps towards its generalisation for arbitrary (and varying) $N_{\text{in}}$.
The implicit equation of such generalised thresholds reads:
\begin{equation}\label{H-surf}
    \sum_{i\in N_{\text{in}}} E_i-\sum_{j\in N_{\text{out}}} E_j=0 \,,
\end{equation}
for the multiplicity $N_{\text{in}}$ and $N_{\text{out}}$ relevant to the perturbative order considered.
In order to give rise to such threshold, the $N_{\text{in}}$ external legs of our current supergraph construction must be closed onto themselves so as to form a vacuum diagram.

A first technical difficulty in addressing vacuum supergraphs is that their LTD representation only features internal E-surfaces, that is $\mathcal{E}_{\text{int}}=\mathcal{E}$, so that all thresholds of the form of eq.~\eqref{H-surf} are subject to dual cancellations.
The origin of dual cancellation stems from the fact that all propagators of the original representation of the vacuum supergraph in Minkowski space are assign the identical causal prescription $+\mathrm{i}\epsilon$ with a positive sign.
For this reason, we hypothesise that the generalisation of LU for initial-state singularities requires a modification of the sign of the causal prescription of some of the original loop propagators.
We therefore propose a candidate for an alternative definition of vacuum supergraphs that is locally finite for \emph{all} singularities and exhibits degeneracy in the multiplicity of \emph{both} initial and final states.
This alternative expression $M_G[v,\mathbf{s}]$ for the vacuum supergraph $G=(\mathbf{v},\mathbf{e})$ (with set of vertices $\mathbf{v}$ and oriented edges $\mathbf{e}$) depends on two new quantities: a set of signs $v$ and a set of vertices $\mathbf{s}$. It reads:
\begin{equation}
    M_G[v,\mathbf{s}]=\int \prod_i^L d k_i^0 \frac{N}{\prod_{j\in \mathbf{e}^+_{v,\mathbf{s}}}(q_j^2-m_j^2+i\epsilon)\prod_{j\in\mathbf{e}^-_{v,\mathbf{v}\setminus\mathbf{s}}}(q_j^2-m_j^2-i\epsilon) }, \hspace{0.2cm} v\in\{\pm 1\}^{|\delta(\mathbf{s})|},\label{eq:vacuum_supergraph}
\end{equation}
where $\mathbf{s}\subset \mathbf{v}$ and $\delta^{\circ}(\mathbf{s})$ is a connected subgraph. 
We stress that it is important to properly account for the symmetry factor of $G$ (part of the numerator $N$ in eq.~\eqref{eq:vacuum_supergraph}) in order to avoid double-counting.
Furthermore we defined:
\begin{equation}
    \mathbf{e}^\pm_{v,\mathbf{s}}=\delta^\circ(\mathbf{s})\cup\{e\in\delta(\mathbf{s})| v_e=\pm 1\} \,,
\end{equation}
which corresponds to the subset of edges within $\mathbf{s}$ and at its boundary when satisfying a particular choice for the sign of the energy flow across it.
The sign of the causal prescriptions for the propagators in $M_G[v,\mathbf{s}]$ are engineered such that after applying LTD, it contains (among other ones) a threshold corresponding to the particles in $\mathbf{e}^+_{v}$ being on-shell with positive energy and the particles in $\mathbf{e}^-_{v}$ being on-shell with negative energy.
These two sets are then naturally identified with the incoming and outgoing particles respectively, and participate in a threshold of the form of eq.~\eqref{H-surf}.
We note that the choice of $i\epsilon$-prescription for the edges not in $\delta(\mathbf{s})$ is set so as to reproduce complex conjugation (that is an opposite prescriptions for E-surfaces on either side of the Cutkosky cut).

We now construct a set containing all possible connected subgraphs and all physical thresholds associated to such subgraph:
\begin{equation}
\mathcal{H}=\left\{(\mathbf{s},v)\middle| \forall \mathbf{s}\subset \mathbf{v}, \ \delta^\circ(\mathbf{s}) \text{ is connected}, \ \forall v \in\{\pm 1\}^{|\mathbf{s}|}\right\} \,,
\end{equation}
so that each element $(\mathbf{s},v)\in\mathcal{H}$ can be associated to a threshold defined by the zeros of the equation
\begin{equation}
    \eta_{(\mathbf{s},v)}=\sum_{i\in \mathbf{e}^+_{\mathbf{s},v}} E_i-\sum_{j\in \mathbf{e}^-_{\mathbf{s},v}} E_j \,.
\end{equation}
We show in tab.~\ref{fig:vaccuum_isr_drawings} the list of all twenty thresholds in $\mathcal{H}$ contributing to the process $d \bar{d} \rightarrow Z$ for one vacuum supergraph $G$ relevant to that process.

\begin{table}[H]
\begin{center}
\resizebox{0.95\columnwidth}{!}{%
\begin{tabular}{ll@{\hskip 3em}|@{\hskip 3em} ll@{\hskip 3em}|@{\hskip 3em} ll}
\input{Figures/isr_diagrams/cut_1} &  \input{Figures/isr_diagrams/ext_cut_1} & \input{Figures/isr_diagrams/cut_5}& \input{Figures/isr_diagrams/ext_cut_5} &
\input{Figures/isr_diagrams/cut_2} & \input{Figures/isr_diagrams/ext_cut_2}\\ \input{Figures/isr_diagrams/cut_6}& \input{Figures/isr_diagrams/ext_cut_6}&
\input{Figures/isr_diagrams/cut_3} & \input{Figures/isr_diagrams/ext_cut_3}& \input{Figures/isr_diagrams/cut_7}& \input{Figures/isr_diagrams/ext_cut_7}\\
\input{Figures/isr_diagrams/cut_8} & \input{Figures/isr_diagrams/ext_cut_8}& \input{Figures/isr_diagrams/cut_4}& \input{Figures/isr_diagrams/ext_cut_4}&
\input{Figures/isr_diagrams/cut_9} & \input{Figures/isr_diagrams/ext_cut_10} \\ \input{Figures/isr_diagrams/cut_11} & \input{Figures/isr_diagrams/ext_cut_12} &
 \input{Figures/isr_diagrams/cut_10} & \input{Figures/isr_diagrams/ext_cut_9}& \input{Figures/isr_diagrams/cut_12} & \input{Figures/isr_diagrams/ext_cut_11} \\
 \input{Figures/isr_diagrams/cut_13} & \input{Figures/isr_diagrams/ext_cut_13} &  \input{Figures/isr_diagrams/cut_15} & \input{Figures/isr_diagrams/ext_cut_15}&
 \input{Figures/isr_diagrams/cut_14} & \input{Figures/isr_diagrams/ext_cut_14} \\ \input{Figures/isr_diagrams/cut_16} & \input{Figures/isr_diagrams/ext_cut_16}&
  \input{Figures/isr_diagrams/cut_17} & \input{Figures/isr_diagrams/ext_cut_17} &  \input{Figures/isr_diagrams/cut_19} & \input{Figures/isr_diagrams/ext_cut_19} \\
   \input{Figures/isr_diagrams/cut_18} & \input{Figures/isr_diagrams/ext_cut_18} &  \input{Figures/isr_diagrams/cut_20} & \input{Figures/isr_diagrams/ext_cut_20} & &
\end{tabular}
}
\end{center}
\caption{\label{fig:vaccuum_isr_drawings}
List of all twenty thresholds $(\mathbf{s},v)$ in $\mathcal{H}$ contributing to $p\bar{p}\rightarrow Z$ at NLO QCD accuracy, stemming from one of the two distinct supergraphs $G$ relevant to this process. The vertices in black are part of the set $\mathbf{s}$ while the grey ones are part of $\mathbf{v}\setminus\mathbf{s}$. The edges $e_i$ coloured in black are in $\mathbf{e}^+_{v,\mathbf{s}}$, that is either in $\delta^\circ(\mathbf{s})$ (non complex-conjugated internal propagators) or in $\delta(\mathbf{s})$ with $v_{e_i}=+1$ (final-states).
Similarly, the edges $e_j$ coloured in grey are in $\mathbf{e}^-_{v,\mathbf{v}\setminus\mathbf{s}}$, that is either in $\delta^\circ(\mathbf{v}\setminus\mathbf{s})$ (complex-conjugated internal propagators) or in $\delta(\mathbf{v}\setminus\mathbf{s})$ with $v_{e_j}=-1$ (initial-states).
We stress that we only retain contributions for which the final state includes a $Z$-boson, as per the process definition.
As for FSR supergraphs, the potential double-counting of interference diagrams is compensated for by the supergraph symmetry factor.
The embedding is chosen that the central vertex of the vacuum supergraph is always placed to the right of the Cutkosky cut, so that when it is black (i.e. part of $\mathbf{s}$) the resulting interference diagram needs to be flipped around the Cutkosky cut for complex conjugated propagators to sit on its right.
}
\end{table}

The only missing component in order to construct an analogue of $\sigma_\text{d}$ that also contains degenerate initial-state configurations is the corresponding causal flow. 
It should be constructed as the solution of an ODE whose defining vector field is causal on all surfaces in $\mathcal{H}$: 
\begin{equation}
\begin{cases}
    \partial_t \vec{\phi}=\kappa(\vec{\phi}) \\
    \vec{\phi}(0,\vec{k})=\vec{k}
    \end{cases} \,,
\end{equation}
where
\begin{equation}\label{cond-H-surf}
    \kappa\cdot \nabla \eta_{(\mathbf{s},v)}>0, \ \ \forall \vec{k} \ \text{  s.t. } \ \eta_{(\mathbf{s},v)}(\vec{k})=0, \ \forall (\mathbf{s},v)\in\mathcal{H} \,.
\end{equation}
When the energy flow vector $v$ does not have \emph{all} of its components set positive or negative, the set of requirements above is different than the case studied in ref.~\cite{Capatti:2019edf}.
We will however not discuss here whether eq.~\eqref{cond-H-surf} can be satisfied by a vector field $\kappa$; this investigation will be left as future work. Instead, we assume here that such a field exists and that it is thus possible to map any point of $\mathbb{R}^{3L}$ to points on distinct thresholds. We then construct the extension of eq.~\eqref{lfmo} by summing the discontinuity acquired by contour deforming around the threshold $\eta_{(\mathbf{s},v)}$ of $M_G[v,\mathbf{s}]$ for all possible elements of $(\mathbf{s},v)\in\mathcal{H}$. Along the causal flow one can expand the thresholds around the solutions $\eta_{(\mathbf{s},v)}(\vec{\phi}(t^\star_{\mathbf{s},v},\vec{k}))=0$ and carry out a similar study as the one performed in sect.~\ref{sec:general_IR_cancellation_proof} and thus arrive at the same conclusion regarding the pairwise cancellation of s-channel E-surfaces.
We observe that the residue of $M_G[v,\mathbf{s}]$ at $\eta_{(\mathbf{s},v)}$ factorises into two ordinary amplitude graphs $\delta^\circ(\mathbf{s})$ and $\delta^\circ(\mathbf{v}\setminus\mathbf{s})$ times a product of inverse energies for each element of $\delta(\mathbf{s})$. Thus, using insights from the factorisation property noted in sect.~\ref{sec:XOR_and_heuristics}, we see that the candidate expression for vacuum supergraphs of eq.~\eqref{eq:vacuum_supergraph} reproduces the known structure of cross sections for hadronic observables.

Finally, we can define a local representation of the differential cross section stemming from the vacuum supergraph $G$ that is locally finite at locations of both initial-state and final-state pinches (but excluding t-channel singularities~\cite{Melnikov_1997}):
\begin{equation}\label{isr-lfmo}
    \sigma_\text{d} = \sum_{(\mathbf{s},v)\in \mathcal{H}_{\vec{k},\vec{\phi}}} \lim_{t\rightarrow t^\star_{\mathbf{s},v}}(t-t^\star_{\mathbf{s},v}) M_G[v,\mathbf{s}](\vec{\phi}(t,\vec{k})) \mathcal{O}_{\mathbf{s},v} \,.
\end{equation}
where $\mathcal{H}_{\vec{k},\vec{\phi}}$ is the analogue of $\mathcal{E}_{\vec{k},\vec{\phi}}$ for $\mathcal{H}$ defined in eq.~\eqref{eq:intersected_e_surfaces}.
There is a crucial difference between the equation above and its final-state counterpart given in eq.~\eqref{lfmo} in which we include \emph{all} the thresholds that appear in the LTD representation of a supergraph specified with a \emph{fixed} $i\epsilon$ prescription. 
Instead, the expression of eq.~\eqref{isr-lfmo} considers the sum over a set of different prescriptions assigned to the same supergraph due to the nature of the thresholds singularities of vacuum diagrams. Moreover, for each choice $(\mathbf{s},v)$, \emph{only} the residue of the corresponding threshold is selected (i.e. Cutkosky cut $\mathbf{c}_{(\mathbf{s},v)}$).

The final expression of eq.~\eqref{isr-lfmo} also introduces the new symbol $\mathcal{O}_{\mathbf{s},v}$ which is the observable associated to the vacuum cut $\mathbf{s}$ for the chosen energy flow $v$. This observable can be set to zero for many of the ($\mathbf{s}$,$v$) cuts that does not match the process of interest (e.g. all $1\rightarrow N_{\text{out}}$ configurations in the case of hadronic scatterings).
Even in the context of our original construction of Local Unitarity, we can already find observable functions depending on the details of the Cutkosky cut. For example when computing the cross section for the process $\gamma^\star \rightarrow H t \bar{t}$, the observable function would select out all Cutkosky cuts involving only the top and anti-top quarks.
In the case of hadronic scatterings with initial-state singularities, the role of the observable function $\mathcal{O}_{\mathbf{s},v}$ is even more important. Indeed, it should also encode the details of the structure of the constituents of the colliding incoming bound states, that is parton distribution functions (PDFs). In particular, it is responsible to confine the contributions from (possibly degenerate) incoming momenta to lie within the close vicinity of an (anti-collinear) back-to-back configuration. In order to achieve this, the ``initial-state observable'' needs to contribute to the LU integrand with an actual weight (as opposed to just the common Heaviside) so as to be able to input the initial states density.
However, we note that in the framework proposed in this section, the treatment of parton distribution functions would likely be different than that of the traditional collinear factorisation paradigm~\cite{Collins:1984kg,Collins:1988ig,Collins:1989gx}.
Indeed, since $\sigma_{\text{d}}$ is free of \emph{all} s-channel infrared singularities, there is no direct analogue to the traditional PDF counterterms nor to the common factorisation scale $\mu_F$ driving the renormalisation group equations induced by the singularities of the PDF evolution kernels. 
Instead, we find an alternative perspective wherein initial states can vary in multiplicity, in the exact same way as final states do, and whose degeneracy scale is set dynamically from the requirement of an IR-safe definition of the initial-state observable $\mathcal{O}_{\mathbf{s},v}$.

This concludes the discussion of our preliminary investigation of the definition of
a differential cross section that is locally finite also on s-channel initial-state thresholds
We leave further developments of this construction to a future publication, including the regularisation of the remaining t-channel singularities and a quantitative investigation of its relation to traditional collinear factorisation and PDFs.
}%
\subsection{UV counterterms and renormalisation}

\label{sec:technical_UV}

The regularisation of the UV behaviour of the LU representation, together with the introduction of renormalisation contributions, lead to the following schematic modification of eq.~\eqref{eq:cross_section_master_formula} for the differential cross section:
\begin{eqnarray}
    \sigma &=& \sum_{\Gamma\in \mathcal{G}}\sum_{\mathbf{s}\in\mathcal{E}^\Gamma_{\text{s-ch}}}
    \left[ \mathcal{I}_{\Gamma,\mathbf{s}} 
    - \sum_{\mathbf{u}\in\mathcal{S}_{\Gamma,\mathbf{s}}^{(\text{left})}} 
        \left( \mathcal{I}^{(\text{UV})}_{\Gamma,\mathbf{s},\mathbf{u}} - \overline{\mathcal{I}}^{(\text{UV})}_{\Gamma,\mathbf{s},\mathbf{u}} \right)
    - \sum_{\mathbf{u}\in\mathcal{S}_{\Gamma,\mathbf{s}}^{(\text{right})}} 
        \left( \mathcal{I}^{(\text{UV})}_{\Gamma,\mathbf{s},\mathbf{u}} - \overline{\mathcal{I}}^{(\text{UV})}_{\Gamma,\mathbf{s},\mathbf{u}} \right)
    \right] \nonumber\\
    &+&
    \sum_{\Gamma^{\text{R}} \in \mathcal{R}(\mathcal{G})} 
    \sum_{\;\mathbf{s}\in \mathcal{E}^{\Gamma^{\text{R}}}_{\text{s-ch}}}
        \delta Z_{\Gamma^{\text{R}}} \mathcal{I}_{ \Gamma^{\text{R}},\mathbf{s} }\;, 
\label{eq:cross_section_master_formula_UV}
\end{eqnarray}
where we intentionally left implicit the precise definition of each integral $\mathcal{I}$, including its integration measure.
The quantity $\mathcal{S}_{\Gamma,\mathbf{s}}^{(\text{left})}$ (resp. $\mathcal{S}_{\Gamma,\mathbf{s}}^{(\text{right})}$) denotes a set of sets of loop edges in the subgraph to the left (resp. right) of the Cutkosky cut $\mathbf{c}_{\mathbf{s}}$ with a combined superficial degree of UV divergence equal to or greater than zero.
The terms $\mathcal{I}^{(\text{UV})}_{\Gamma,\mathbf{s},\mathbf{u}}$ are local counterterms and $\overline{\mathcal{I}}^{(\text{UV})}_{\Gamma,\mathbf{s},\mathbf{u}}$ are their counterparts integrated analytically using dimensional regularisation.
Finally, the operator $\mathcal{R}$ applies to the selected set of higher-order correction supergraphs $\mathcal{G}$ and generates the set of all corresponding renormalisation supergraphs. For each of them, we can identify a renormalisation counterterm $\delta Z_{\Gamma^{\text{R}}}$ factorising the LU representation of a lower-order supergraph.
In this section, we discuss the construction of each ingredient of eq.~\eqref{eq:cross_section_master_formula_UV} qualitatively.

By construction, interference diagrams in physical theories must have a UV singular structure that corresponds to the combined UV singularities of the two subgraph amplitudes to the left and right of the Cutkosky cut.
We choose to construct the local UV counterterms $\mathcal{I}^{(\text{UV})}$ in Minkowski space, and then convert them to the (c)LTD representation. This allows for the use of traditional analytic techniques and dimensional regularisation when computing their integrated counterpart $\overline{\mathcal{I}}^{(\text{UV})}$.

Beyond NLO, the local UV regularisation must be performed carefully so that the remaining UV divergences in one-loop UV counterterms locally cancel against the two-loop UV counterterms. In other words, the overlapping UV divergences in $\sum_{\mathbf{u}\in\mathcal{S}_{\Gamma,\mathbf{s}}^{(\text{x})}} \mathcal{I}^{(\text{UV})}_{\Gamma,\mathbf{s},\mathbf{u}}$
must cancel locally amongst the terms in this sum.
We achieve this by implementing $\mathcal{S}_{\Gamma,\mathbf{s}}^{(\text{x})}$ and $\mathcal{I}^{(\text{UV})}_{\Gamma,\mathbf{s},\mathbf{u}}$ according to the BPHZ forest formula treatment~\cite{Bogoliubov:1957gp,Hepp:1966eg, Zimmermann:1969jj}.
This procedure involves identifying all possible UV singular subgraphs and constructing an appropriate approximant of the integrand on UV limits. Proper subtraction of the original integrand and, specifically, the treatment of overlapping subdivergences is guaranteed by the forest formula. In particular, the study of overlapping subdivergences requires defining spinneys, which are collections of disjoint UV divergent subgraphs. 

The BPHZ forest formula is independent on the chosen renormalization scheme and only prescribes a Taylor expansion of the denominators of UV subgraphs in their external momenta. This operation is local, since it relies on a Taylor expansion of the four-dimensional integrand of the amplitude. It follows that the integrand can also be subtracted locally.

We identify the UV subgraphs as tensor graphs and assign them their superficial degree of divergence. After Taylor expanding in its external momenta, the scalar graph can be represented as a power series in scalar vacuum diagrams. We assign to each vacuum diagram propagator a UV mass $m_{\text{UV}}$. This UV counterterm is added to the integration and subtracts correctly the UV singularity represented by the subgraph. The analytically integrated counterpart $\overline{\mathcal{I}}^{(\text{UV})}_{\Gamma,\mathbf{s},\mathbf{u}}$ is computed as the integral of the four-dimensional Minkowsi representation of $\mathcal{I}^{(\text{UV})}_{\Gamma,\mathbf{s},\mathbf{u}}$, which always corresponds the massive vacuum diagram integral that can easily be computed using integration-by-parts identities and a finite set of known master integrals. 
In order to obtain a scalar expression, we tensor reduce the UV subgraph so that it completely factorises. Finally, we multiply the finite part of the integrated UV counterterm into the remaining diagram.

Ultimately, the integrated counterterm will be a Laurent series in $\epsilon=4-d$. These UV poles are cancelled by the renormalization supergraphs $\Gamma^{\text{R}}$ listed in $\mathcal{R}(\mathcal{G})$ and constructed by contracting the disjoint UV subgraphs in the spinneys to vertices. This forms separate renormalisation supergraphs with lower loop counts, and we assign to each vertex obtained in this fashion their respective renormalization constant. We stress that, beyond NLO, it is important that the poles in the dimensional regulator $\epsilon=(d-4)/2$ of the renormalization components $\delta Z_{\Gamma^{\text{R}}}$ multiply the $\mathcal{O}(\epsilon)$ parts of $\mathcal{I}_{ \Gamma^{\text{R}},\mathbf{s}}$ computed in $d$-dimensions.
This dynamic generation $\mathcal{R}(\mathcal{G})$ from the list of selected supergraphs considered $\mathcal{G}$ ensures the consistency of the renormalization procedure.
We can numerically test if the UV poles of the renormalisation graphs cancel with the ones from the integrated counterterm by numerically integrating the numerator dimensional regulator pole coefficients instead of its finite part.
We stress that this pole cancellation check can in principle also be made local by consistently rerouting the momenta of all supergraphs in $\{ \Gamma \in \mathcal{G} \;|\; \Gamma^{\text{R}} \in \mathcal{R}(\{\Gamma \})$ contributing to a particular renormalisation supergraph $\Gamma^{\text{R}}$.
In the example of the UV regularisation of the \DT{} supergraph that will be investigated in sect.~\ref{sec:DT_UV_renormalisation}, we find a one-to-one correspondence between the supergraph and its renormalisation contribution, leading to a direct cancellation of the UV poles (see eq.~\eqref{eq:DT_Final_UV_contribution}). Because this is in general not the case, we opt to account separately for higher order supergraphs and their renormalisation terms.

Often, an approach similar to the one by D. Soper~\cite{Soper:1999xk} is chosen in order to construct a modified local counterterm $\tilde{\mathcal{I}}^{(\text{UV})}_{\Gamma,\mathbf{s},\mathbf{u}}=\mathcal{I}^{(\text{UV})}_{\Gamma,\mathbf{s},\mathbf{u}}-\overline{\mathcal{I}}^{(\text{UV})}_{\Gamma,\mathbf{s},\mathbf{u}}$ which combines both local and integrated UV counterterms under the same integral measure.
This can be done by adjusting the higher-order terms of the original local UV counterterm of eq.~\eqref{eq:DT_UV_local_CT} obtained from the strict UV limit.
In this way, $\tilde{\mathcal{I}}^{(\text{UV})}_{\Gamma,\mathbf{s},\mathbf{u}}$ directly integrates into a finite part that is zero and automatically reproduces $\overline{\textrm{MS}}$ results, or alternatively allows one to add whichever finite part from the renormalisation constant of the preferred scheme without having to worry about the impact of the UV regularisation in Local Unitarity. 
Such an adjustment of the numerator is however typically not systematic and therefore not practical for automation.
Instead, we multiply the integrated counterterm $\overline{\mathcal{I}}^{(\text{UV})}_{\Gamma,\mathbf{s},\mathbf{u}}$ with a rational polynomial in the loop degrees of freedom that are analytically integrated over (e.g. a simple product of massive one-loop tadpoles) so that it can be systematically combined with $\mathcal{I}^{(\text{UV})}_{\Gamma,\mathbf{s},\mathbf{u}}$.
    
   Since we separately account for the contribution of the integrated UV counterterms and renormalisation constants, we must pay particular attention to the particular dimensional regularisation scheme considered when computing both terms.
    We identify four different parts in the expression of the integrated UV counterterms: the numerator $N^{\textrm{UV}}$ stemming from the edges part of the UV subdiagram, the external factorised ones $N^\textrm{E}$ and similarly for the denominators, $D^{\textrm{UV}}$ and $D^\textrm{E}$. As is well-known in the literature~\cite{tHooft:1972tcz,Kunszt:1993sd,Jack:1994bn,Kilgore:2011ta,Gnendiger:2017pys}, different choices regarding the dimensionality adopted for each of these pieces are possible and ultimately equivalent as long as they are performed consistently across terms whose poles cancel.
    We find it most convenient to build an automated procedure by treating the combination of $N^{\textrm{UV}}$, $N^\textrm{E}$ and $D^{\textrm{UV}}$ in $d$-dimension and $D^\textrm{E}$ in four dimensions. This choice also implies that no particular attention needs to be paid to the rational parts~\cite{Xiao:2006vr,Draggiotis:2009yb,Zoller:2019mla,Pozzorini:2020hkx,Lang:2020nnl} of the integral as they will then automatically be accounted for in this manner.

\subsection{Automation and numerical efficiency}

\label{sec:technical_implementation}

The relevance of the LU representation for collider phenomenology crucially depends on the performance of its implementation.\\
Our implementation leverages the framework of {\sc\small MadGraph5\_aMC@NLO}~\cite{Alwall:2014hca} ({\sc\small MG5aMC} henceforth) for the abstract representation of the user inputs such as the observable definition, process generation syntax and physics model considered.
In particular, the Feynman rules are provided through a model file following the Universal Feynrules Output (UFO) conventions~\cite{Degrande:2011ua}. 
This choice is motivated by our longer-term goal of offering the user an automated environment similar to {\sc\small MG5aMC} for steering its simulations.
However, the very different nature of LU requires an independent program.
We therefore decomposed the implementation of the LU representation into individual fundamental tasks that are efficiently carried out by the development of dedicated codes that we discuss in this section.

\subsubsection{Process generation}
\label{sec:technical_process_generation}

Given the specification of a scattering process and UFO model using {\sc\small MG5aMC} syntax, we steer {\sc\small QGRAF}~\cite{Nogueira:1991ex} to generate all contributing supergraphs and output them in a {\sc\small Python}-readable format.
Next, all isomorphic supergraphs are combined and for each resulting supergraph we enumerate  all Cutkosky cuts whose final state particles are compatible with the process definition.
We stress that it is important to also weight each supergraph with its symmetry factor provided by {\sc\small QGRAF} as it corrects for the over-counting due to identical particles (i.e. it plays a similar role as the customary normalisation of amplitudes by the symmetry factor stemming from identical particles in their final states).
We then identify external one-particle irreducible diagrams in the supergraphs (that is, all the 1PIs that are adjacent to a Cutkosky cut), and apply the propagator treatment of sect.~\ref{sec:selfenergy_treatment_in_LU}. 
We also detect all UV divergent subgraphs and construct local and integrated UV counterterms for them, together with renormalisation contributions, by following the procedure laid out in sect.~\ref{sec:technical_UV}. 
Finally, we collect all information on the supergraph that must be computed into input files ready to be processed and optimised by {\sc\small FORM} into {\sc\small C} compiled libraries. 

\subsubsection{Graph evaluation}
\label{sec:technical_numerator_treatment}
The treatment of the numerator is often the bottleneck for both generation and run-time efficiency when considering higher-loop Feynman diagrams. An efficient implementation in {\sc\small FORM}~\cite{Ruijl:2017dtg} is therefore crucial. 
Our {\sc\small FORM} code first substitutes the Feynman rules for each supergraph, contracts all indices, and applies polarisation sum rules.
After simplifying all Dirac traces and colour algebra, the output of this procedure is a sum of scalar products of loop momenta and external momenta. 

Next, a {\sc\small Python} code constructs the general Loop-Tree Duality expression derived in ref.~\cite{Capatti:2019ypt} and the Manifestly Causal LTD (cLTD) expression derived in ref.~\cite{Capatti:2020ytd} in a {\sc\small FORM}-readable format. This expression is added to the expression of the numerator. At this stage, the entire integrand per Cutskosky cut is represented in a {\sc\small FORM} expression. Next,
{\sc\small FORM} performs all derivatives necessary for the propagator treatment of sect.~\ref{sec:selfenergy_treatment_in_LU} as well as for the UV counterterms.
This integrand can be viewed as a polynomial in the scalar products and linear propagators.
We use {\sc\small FORM}'s optimisation~\cite{Kuipers:2013pba,Ruijl:2014spa} algorithms to generate highly efficient C code that can evaluate the entire integrand per Cutkosky cut by finding an optimal Horner scheme for the resulting polynomial and grouping common subexpressions. This optimisation step typically reduces the number of arithmetic operations by several orders of magnitude. 
Additionally, the C compiler further improves the implementation by vectorising (parts of) the resulting C code.
This finally results in evaluation times typically below 100~$\mu$s, even for complicated processes. 
For example, one particular supergraph with 11 Cutkosky cuts contributing to the NNLO correction to the process $H\rightarrow t\bar{t}gg$ can be evaluated for one sample point in only 25~$\mu$s (without multi-channeling). This means that the evaluation time of the integrand is not a limiting factor.

\subsubsection{Numerical stability}
\label{sec:technical_numerical_stability}

The LU integrand consists of terms that can feature large numerical cancellations, leading to instabilities when considering finite-size arithmetic.
Such unstable points can be large in magnitude (see, e.g. fig.~2 of ref.~\cite{Capatti:2020ytd} for numerical stabilities in the UV) and spoil the central value estimator of the Monte-Carlo integration.
It is therefore necessary to monitor the numerical stability of the integrand evaluation for each sample point and properly address cases that are insufficiently precise.
In order to estimate the numerical accuracy of the integrand for one particular sample point, we repeat its evaluation with all spatial directions subject to arbitrary rotations and compare the results obtained in this way, as they are analytically identical in virtue of the symmetries of the integrand.
The original sample point is then deemed unstable if its rotated evaluations differ by more than a certain threshold.
Since the LTD integrands may exhibit instabilities in the UV that are removed by cLTD (see ref.~\cite{Capatti:2020ytd}), we perform the same procedure using cLTD instead of LTD.
If the point is still considered unstable, we repeat this procedure using quadruple-precision arithmetic which is about 100 times slower but rescues the majority of unstable points. 
Exceptional sample points that are still deemed unstable at this stage are typically very close to infrared singularities or cancelling E-surfaces and we override the evaluation to zero. By varying the numerical stability requirements, we can verify that our procedure does not alter the result of the Monte-Carlo integration.

\subsubsection{Contour deformation}
\label{sec:contour_deformation}

When considering non-trivial observables or scattering process definitions that exclude certain Cutkosky cuts, the LU integrand may still be singular at the location of non-pinched thresholds. For these cases a complex contour deformation that follows the construction of ref.~\cite{Capatti:2019edf} is required. 
In the context of the LU representation, we deform the loop variables of the amplitude on the left and right side of the Cutkosky cut independently, to ensure that the observable function is evaluated with real-valued kinematics (see sect.~\ref{sec:pinched_E_surface_discussion}).
On pinched surfaces the deformation must be zero in order not to disrupt the cancellation of real and virtual Cutkosky cuts.
We further dampen the deformation magnitude on pinched E-surfaces by multiplying the deformation field with the following dampening factor $d_\text{IR}$:
\begin{equation}
    d_\text{IR}=T(\eta_\text{IR}, M),\qquad T(\eta_\text{IR},M) = \frac{\eta_\text{IR}^2}{\eta_\text{IR}^2 + M^2} \,,
\end{equation}
for each pinched E-surface with implicit equation $\eta_\text{IR}$, in addition to the dampening enforced by the complex pole constraint (see sect.~3.3.2 of ref.~\cite{Capatti:2019edf}).

The construction of ref.~\cite{Capatti:2019edf} requires a set of valid deformation sources to be constructed.
This problem is NP-hard and requires finding points in the internal region of overlapping E-surfaces. 
The difference between determining the overlap structure for an amplitude with fixed external momenta and determining it for the amplitudes participating in a supergraph is that in the latter case, the external momenta of the loop subgraphs change with every Monte Carlo sample point. 
Consequently, the overlap structure of the E-surfaces must be recomputed at run-time for each sample point and Cutkosky cut. 
Thus, it is important that the overlap structure can be determined efficiently.
Testing for overlap of E-surfaces in a set $\mathcal{E}$ can be achieved by finding a point in the interior of all E-surfaces in $\mathcal{E}$.
Such problems are known as a second-order cone program in the field of convex optimisation.
We use the ECOS solver~\cite{bib:Domahidi2013ecos} to determine such points.
ECOS takes about 0.5~ms for a test, and in the worst case $2^{|\mathcal{E}|}$ tests are required. 
In order to reduce the number of necessary overlap tests, we only test groups of E-surfaces if each pair in the group pairwise overlaps, which can be determined using fast heuristics. 
After these optimisations, even for complicated cases such as the two-loop amplitude \texttt{2L6P.e K1} from ref.~\cite{Capatti:2019edf}, which features 21 E-surfaces and 34 unique sources, the construction time is only about 18 milliseconds. 
We stress that the loop topologies most relevant to LU often involve many on-shell massless external momenta, for which the number of non-pinched E-surfaces is generally small since many E-surfaces are pinched in this case.
Additionally, we only have to contour deform around E-surfaces whose pair-wise cancellation is spoiled by either the observable or process definition. 
We thus find that for cases of phenomenological relevance the determination of the overlap structure, and therefore the complete deformation, typically requires about one millisecond per sample point.

\subsection{Phase space sampling and variance reduction}

Even once a local representation of a differential cross section free of both IR and UV singularities is established, it remains to be shown that its numerical integration is feasible in practice.
In this subsection, we discuss the main improvements that can help reduce the variance of the integrand.

\subsubsection{Multi-channelling}
\label{sec:technical_multi_channeling}
Our computation of the differential cross section in Local Unitarity is organised in three nested levels, as made clear by eq.~\eqref{eq:cross_section_master_formula} which we rewrite here with a notation appropriate to this section:
\begin{equation}
\sigma_\mathcal{O}=
\sum_{\Gamma \in \mathcal{G}}
\sum_{\mathbf{s}_{\Gamma}\in{\mathcal{E}^{\Gamma}_{\text{s-ch}}}} 
\int \left(\prod_{i=1}^{L} d^3\vec{k}_i \right) \; I_{\mathbf{s}_{\Gamma}}( \vec{k} ),
\label{eq:cross_section_formula_v2}
\end{equation}
which features:
\begin{itemize}
    \item A discrete sum over each supergraph $\Gamma\in \mathcal{G}$ of the scattering process of interest.
    \item A discrete sum over each Cutkosky cut $\mathbf{s}_{\Gamma} \in \mathcal{E}^{\Gamma}_{\text{s-ch}}$ appearing in supergraph $\Gamma \in \mathcal{G}$.
    \item An integral over the spatial loop degrees of freedom of the integrand $I_{\mathbf{s}_{\Gamma}}( \vec{k})$, which contains all elements of the LU representation, in particular the observable function $\mathcal{O}$ and the solution $t^\star_{\mathbf{s}_{\Gamma}}( \vec{k} )$ of the causal flow enforcing the conservation of on-shell energies involved in the Cutkosky cut $\mathbf{c}_{\mathbf{s}_{\Gamma}}$.
\end{itemize}
In order to improve the Monte-Carlo accuracy, one must reduce the variance of the estimator for $\sigma_\mathcal{O}$. 
This is achieved by using a discrete importance sampling technique for the sum of contributions and an educated choice for the parameterisation of the continuous degrees of freedom, whose Jacobian should approximate the (inverse of) integrand as closely as possible. We discuss both aspects in turn.

\begin{figure}[ht!]
\centering
\begin{subfigure}[b]{.49\linewidth}
\includegraphics[width=\linewidth]{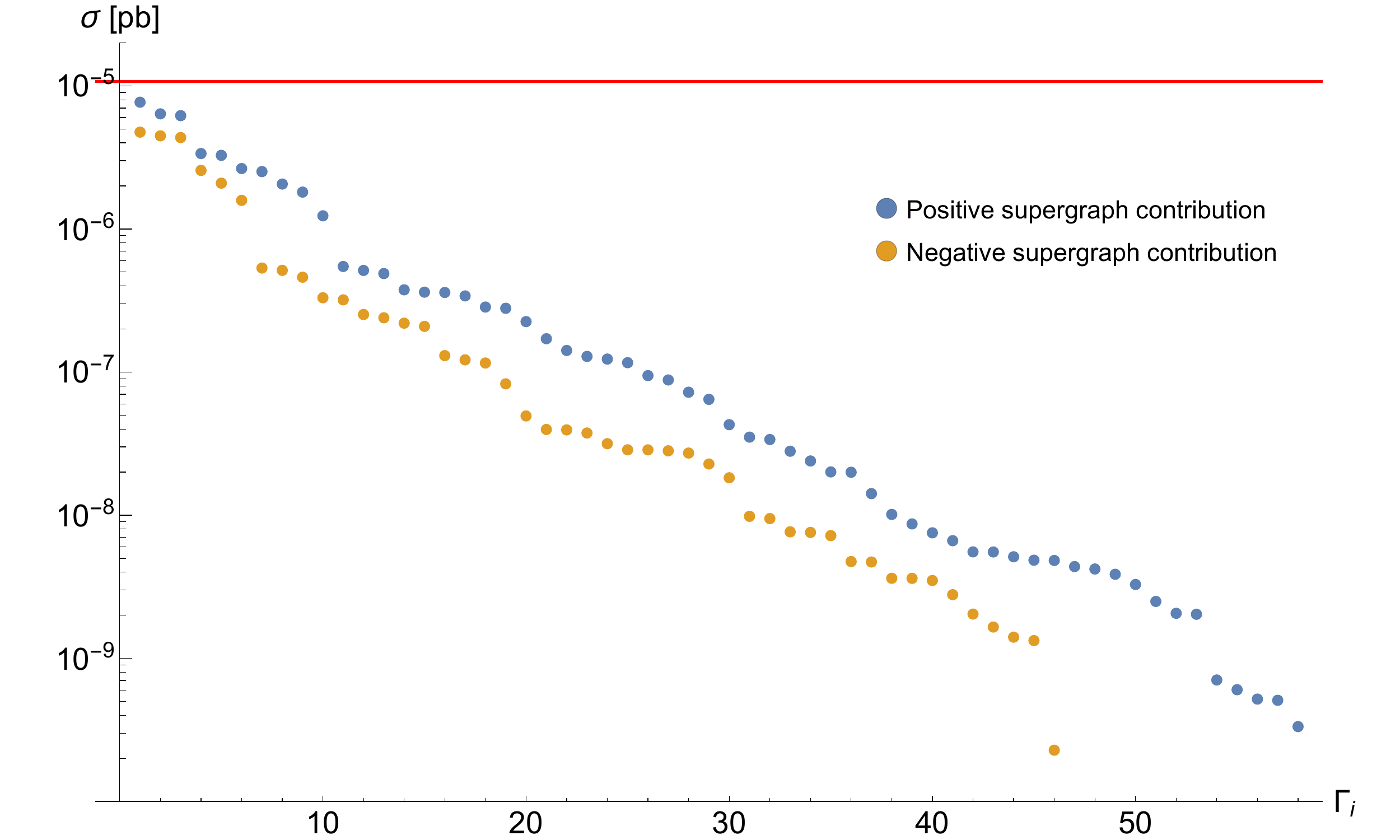}
\caption{Distribution of supergraph contributions}
\label{fig:supergraph_distribution}
\end{subfigure}
\begin{subfigure}[b]{.49\linewidth}
\includegraphics[width=\linewidth]{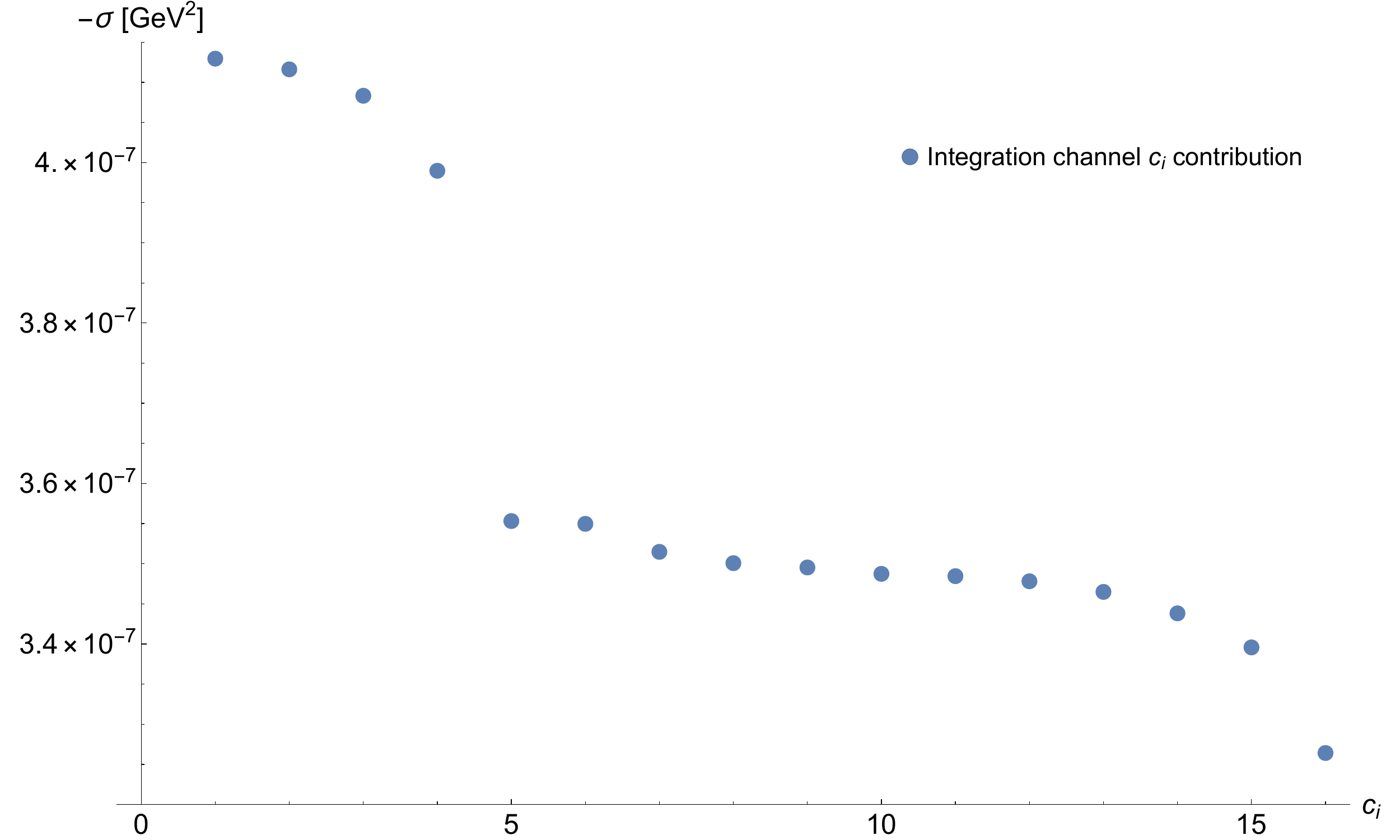}
\caption{Distribution of integration channel contributions}
\label{fig:integration_channel_distribution}
\end{subfigure}
\caption{
To the left, we show the distribution of the contributions to the LO cross section for the 104 distinct non-zero supergraphs for the processes $e^+ e^- \rightarrow \gamma^\star \rightarrow t \bar{t} g g g$ and $e^+ e^- \rightarrow \gamma^\star \rightarrow t \bar{t} g_h \bar{g_h} g$. Their sum, indicated by a red horizontal line, reproduces the total cross section of $1.071\cdot 10^{-5}$ [pb] obtained from {\sc\small MadGraph5\_aMC@NLO} with $\sqrt{s}=1 \textrm{TeV}$, $p_t(j_{i=1,2,3})>100 \textrm{GeV}$ and $\Delta R=0.4$.\\
To the right, we show the distribution of the negative of the contributions of each of the 16 integration channels for the 3-loop scalar supergraph a.3) of tab.~\ref{tab:scalar_supergraphs}, summing up to the {\sc\small forcer} result of $-5.754 \cdot 10^{-6}$ [$\text{GeV}^2$].
}
\end{figure}

Optimising the sampling of the sum over supergraphs is straightforward, since each contribution can in principle be integrated completely separately.
This aspect is unique to the LU representation, and so are the benefits discussed below stemming from a discrete importance sampling over supergraphs.
The optimal discrete probability distribution of sample points over supergraphs is obtained by following the inverse of their relative contribution to the total cross section. 
One typically starts with a uniform discrete grid that is progressively updated according to the new estimates of the relative cross sections for each supergraph.
While simple to implement, this sampling optimisation can be very powerful if the supergraph contributions are unevenly distributed. In fig.~\ref{fig:supergraph_distribution}, we show their distribution for the LO cross section from the scattering processes $e^+ e^- \rightarrow \gamma^\star \rightarrow t \bar{t} g g g$ and $e^+ e^- \rightarrow \gamma^\star \rightarrow t \bar{t} g_h \bar{g_h} g$ where $g_h$ denotes a QCD ghost.
We find that the relative contributions from supergraphs spans more than four orders of magnitude. While this observation depends on the scattering process, observable and gauge considered, we expect this uneven distribution to be quite generic and thus adaptive sampling can significantly help to counterbalance the increase in computation time resulting from the growth in the number of distinct supergraphs for more complicated processes.
Fig.~\ref{fig:supergraph_distribution} shows no apparent large (gauge) cancellations between supergraphs, showing that the supergraphs can be integrated independently. While there may still be benefits from integrating all supergraphs in a correlated manner, identifying a common parameterisation~\cite{Runkel:2019zbm} that leverages potential (gauge) cancellations is difficult and unlikely to outweigh the gain from discrete adaptive sampling over supergraphs. One exception being supergraphs involving QCD ghosts. These supergraphs always share the same topology as that of their gluonic counterpart, and can thus be integrated together with a common loop momentum basis.
However, in all LO cross sections we investigated thus far, we have never found large cancellations within such groups of supergraphs.

The large spread in the distribution of supergraph contributions is perhaps less surprising when placed in the context of more general computations of mixed QCD and EW corrections in the presence of resonant intermediate unstable particles.
In that case, one expects supergraphs encoding pure QCD corrections to dominate those corresponding to pure EW corrections.
Similarly, supergraphs involving Breit-Wigner enhancements are relevant for kinematic regions where the unstable particle is close to its mass shell but typically contribute less than non-resonant topologies to the inclusive cross section.
The separation of these various regimes within traditional methods is complicated, since only subsets of terms that are gauge invariant and that can be rendered finite within their computational paradigm can be used.
Instead, LU offers the perspective of systematically generating \emph{all} contributions and efficiently and dynamically home in onto the relevant ones for a particular observable of interest. 
This procedure is appealing both from a numerical and phenomenological perspective as it helps the identification of the underlying dynamics at play for any collider signature.

Once a particular supergraph $\Gamma$ is selected, we must establish a strategy for sampling its Cutkosky cut contributions $s_\Gamma$ and continuous degrees of freedom. The LU representation relies on pairwise cancellations between the terms $I_{\mathbf{s}_\Gamma}$ so that they must be integrated together with a common loop momentum basis and parameterisation.
At the same time, the LU representation still features integrable singularities on soft kinematic configurations and when internal propagators become on-shell, which we must render bounded with the Jacobian of an appropriate parameterisation so as to make the variance of the LU integrand finite.
Multi-channeling is a common technique (see, e.g. refs.~\cite{Soper:2001hu,Denner:2002cg,Maltoni:2002qb,Becker:2012aqa,Capatti:2019edf} for examples in a similar context) offering a solution to this problem by combining multiple parameterisations each designed to address the integrable singularities and enhancements of a single Cutkosky cut at a time.
In its most generic form, the multi-channeling approach applied to eq.~\eqref{eq:cross_section_formula_v2} for a $L$-loop supergraph yields:
\begin{equation}
\sigma_\mathcal{O}=
\sum_{\Gamma \in \mathcal{G}}
\sum_{a \in \Omega_\Gamma} \int_{[0,1]^{3L}} d \vec{x}\;\left| J^{a}(\vec{x}) \right|^{-1}
\frac{ \xi^{a}\left(\vec{\phi}^{a}_{\vec{x}}\right) }
{ \sum_{b \in \Omega_\Gamma} \xi^{a}\left(\vec{\phi}^{a}_{\vec{x}}\right) }
\sum_{\mathbf{s}_{\Gamma}\in{\mathcal{E}^{\Gamma}_{\text{s-ch}}}} 
I_{\mathbf{s}_{\Gamma}}\left( \vec{\phi}^{a}_{\vec{x}} \right),
\label{eq:cross_section_formula_v2_MC}
\end{equation}
where $\vec{x}$ is the vector of integration variables defined in the unit hypercube, $\Omega_\Gamma$ is the set of all integration channels chosen for the supergraph $\Gamma$, $\vec{\phi}^{a}_{\vec{x}}:=  \vec{k}^{a}(\vec{x}) $ is the parameterisation chosen for channel $a$, and $J^{a}:=\det\left[ \frac{ \partial \vec{x} }{ \partial \vec{\phi}^{a} } \right]$ is the Jacobian of said parameterisation. The scalar functions $\xi^{a}$ can be arbitrarily chosen\footnote{The functions $\xi_{a}$ must however be holomorphic if they are to be also analytically continued together with $I_{\mathbf{s}_{\Gamma}}$ when considering a contour deformation.}. 
We also define the inverse mapping $\vec{\Psi}^{a}_{\vec{k}} := \left[ \vec{\phi}^{a} \right]^{-1}(\vec{k})$. 
The rewriting in eq.~\eqref{eq:cross_section_formula_v2_MC} shows that if one chooses the parameterisations $\vec{\phi}^{i}$ and the functions $\xi^{i}$ such that:
\begin{eqnarray}
\left|{J^{a}}\right| &\sim& \xi^{a} \label{eq:jac_flattening}\\
\sum_{b \in \Omega_\Gamma} \xi^{b} &\sim& \sum_{\mathbf{s}_{\Gamma}\in{\mathcal{E}^{\Gamma}_{\text{s-ch}}}} I_{\mathbf{s}_{\Gamma}}, \label{eq:integrand_flattening}
\end{eqnarray}
all integrable singularities are removed and the variance of the overall resulting integrand is minimised.

In ref.~\cite{Capatti:2019edf}, the authors chose to have one channel per loop momentum basis of the graph $\Gamma$, so that $\Omega_\Gamma \equiv \mathcal{B}_{\Gamma}$ and the integration channels are then labelled with a basis $\mathbf{b}_i$ which map to the scalar index $i$.
Then, in order to accommodate eq.~\eqref{eq:integrand_flattening}, $\xi^{i}$ was built from the product of all (complex) on-shell energies of the edges 
in $\mathbf{b}_i$ whereas the spherical parameterisation adopted for the momenta in $\mathbf{b}_i$ yielded a Jacobian $\left|{J^{i}}\right|$ that dampens all integrable singularities in $\xi^{i}$.
In this work and for all the numerical results presented in sect.~\ref{sec:numerical_results}, we adopted a different strategy. We also consider spherical parameterisations for each of the integration channel associated to one loop momentum basis of the supergraph $\Gamma$, but
we choose $\xi^{i}\equiv J^{i}\left( \vec{\Psi}^{i}_{\vec{k}} \right)$ which dampens all integrable singularities as per eq.~\eqref{eq:integrand_flattening} but also \emph{exactly} fulfils eq.~\eqref{eq:jac_flattening}.
We note that we combine integration channels from loop momenta bases that are degenerate due to our particular choice of Lorentz frame, for which $\vec{Q}=\vec{0}$.

Our particular realisation of the generic multi-channeling strategy of eq.~\eqref{eq:cross_section_formula_v2} then reads:
\begin{equation}
\sigma_\mathcal{O}=
\sum_{\Gamma \in \mathcal{G}}
\sum_{\mathbf{b}_a \in \mathcal{B}_\Gamma} \int_{[0,1]^{3L}} d \vec{x}\; I_\Gamma^{a}(\vec{x}) ,\;\;\;\;\; I_\Gamma^{a}(\vec{x}):=
\frac{ 
\sum_{\mathbf{s}_{\Gamma}\in{\mathcal{E}^{\Gamma}_{\text{s-ch}}}}I_{\mathbf{s}_{\Gamma}}\left( \vec{\phi}^{a}_{\vec{x}} \right)
}
{
\sum_{\mathbf{b}_b \in \mathcal{B}_\Gamma} J^{b}\left( \vec{\Psi}^{b}_{ \vec{\phi}^{a}_{\vec{x}}} \right) 
}\;.
\label{eq:cross_section_formula_v2_us}
\end{equation}
The effect of this multi-channeling on the integrand $I_\Sigma^{\text{MC}}$ is visualised in sect.~\ref{sec:DT:integrand_visualisation}.

One can consider a discrete adaptive importance sampling similar to the one already discussed for the sum over supergraphs. 
However, we see in fig.~\ref{fig:integration_channel_distribution} that the distribution of the cross section stemming from each of the 16 multi-channel integrands $I_\Gamma^{a}$ of the 3-loop scalar supergraph a.3) of fig.~\ref{tab:scalar_supergraphs} is very even.
Even though the gain in variance reduction is more moderate in this case, sampling the integration channels instead of taking their sum yields a faster evaluation of each sample point and therefore a Monte-Carlo error reduction by a factor of up to $\sqrt{|\mathcal{B}_\Gamma|}$ for a fixed run-time. 
Furthermore, performing a \emph{separate} adaptive sampling for the kinematic dependency of \emph{each} integration channel allows the integrator to better adapt to their individual shape as opposed to when training directly on their sum.

The distribution of the cross section from each channel and the resulting variance of $I_\Gamma^{a}$ heavily depends on the particular choice of parameterisations $\vec{\phi}^{a}$. 
It is clear that the spherical parameterisations considered in this work are far from the optimal choice and prior work from D. Soper~\cite{Soper:2001hu} showcases the benefits of adopting more elaborate parameterisations. 
More specifically, our choice does not account for the details of the particular topology of the subgraphs on each side of a particular Cutkosky cut. 
It also ignores the sharp shape of the term $h\left(t^\star_{\mathbf{s}_{\Gamma}}(\vec{k})\right)$, part of $I_{\mathbf{s}_{\Gamma}}$, which strongly enhances kinematic configurations $\vec{k}$ close to the Cutkosky cut surface $\eta_{\mathbf{s}_{\Gamma}}$.
We now discuss a future enhancement that improves on these aspects. We start by factorising the dimensions sampled into three subsets of variables:
\begin{equation}
    \int_{[0,1]^{3L}} d \vec{x} \rightarrow \int_{[0,1]^{3L}}
    \left(\prod_{i=1,3L_\text{left}} dx_{\text{left},i}\right)
    \left( dx_t  \prod_{k=1,(3|\mathbf{c}_{\mathbf{s}_\Gamma}|-4)} dx_{\text{ext},k}\right)
    \left(\prod_{j=1,3L_\text{right}} dx_{\text{right},j}\right),
\end{equation}
where $L_\text{left}$ (resp. $L_\text{right}$) is the loop count of the subgraphs $\Gamma_\text{left}$ (resp. $\Gamma_\text{right}$) to the left (resp. right) of the Cutkosky cut $\mathbf{c}_{\mathbf{s}_\Gamma}$, whose number of edges is equal to $|\mathbf{c}_{\mathbf{s}_\Gamma}|=L-L_\text{left}-L_\text{right}+1$.
The complete parameterisation $\vec{\phi}^{a}$ can then be constructed in the following three \emph{successive} steps:
\begin{itemize}
    \item First, the external momenta $\vec{q}^{\;\mathbf{s}_\Gamma}:=\{ \vec{k}_e | e \in \mathbf{c}_{\mathbf{s}_\Gamma} \}$ with $\vec{k}\in \delta \eta_{\mathbf{s}_{\Gamma}}$ is generated so as to directly lie on the Cutkosky surface $\eta_{\mathbf{s}_{\Gamma}}$. To this end, we take advantage of the traditional phase-space factorisation already used by most event generators. This iterative construction of the phase-space parameterisation is organised such that the invariant mass of repeated propagators that appear\footnote{We must only consider propagators that are not part of any loop of $\Gamma_\text{left}$ or $\Gamma_\text{right}$. In case of competing Breit-Wigner resonances (e.g. hadroproduction of a Higgs decaying onto four leptons), one may consider assigning more than one integration channel to such topologies.} in both $\Gamma_\text{left}$ and $\Gamma_\text{right}$ is directly aligned to one variable $x_{\text{ext},k}$ which is sampled with a probability density flattening the corresponding resonant enhancement (see e.g. ref.~\cite{Denner:2002cg}). For ``interfering'' propagators, there is instead no good ansatz for the density with which they should be sampled (and those topologies are often suppressed), so that we can consider sampling them uniformly instead (see e.g. ref.~\cite{Platzer:2013esa}).
    
    \item Second, the configuration $\vec{q}^{\;\mathbf{s}_\Gamma}$ must be ``upscaled'' so as to cover the entire volume $\mathbb{R}^{3(|\mathbf{c}_{\mathbf{s}_\Gamma}|-1)}$. This aspect is crucial to the LU construction as the causal flow induces a mapping onto the Cutkosky surface of each contribution and is key to obtaining local IR cancellations between them. The most natural choice for achieving this upscaling is to use the causal flow of the supergraph $\Gamma$ itself, with boundary conditions given by $\vec{q}^{\;\mathbf{s}_\Gamma}$ and a value of $t^{\mathbf{s}_\Gamma}$ derived from the input variable $x_t$ and sampled with a density of $d t^{\mathbf{s}_\Gamma}/dx_t=1/h(1/t)=h(t)$. This ensures that the chosen parametrisation renders the integrand $I_{\mathbf{s}_{\Gamma}}( \vec{\phi}^{\mathbf{s}_\Gamma}_{\vec{x}} )$ independent of the choice of normalising function $h(t)$.
    The choice of the $\sigma$-tunable function $h(t)=\frac{2\sigma }{\pi}\cos\left(\frac{\pi}{2\sigma}\right)\frac{t^\sigma}{1+t^{2\sigma}}$ is particularly convenient so as to be able to analytically solve for the cumulative distribution function $H(t)=\int_0^{t} d t^\prime h(t^\prime)=\frac{2\sigma}{\pi(1+\sigma)}\cos\left(\frac{\pi}{2\sigma}\right) {}_2F_1\left(1,\frac{\sigma+1}{2\sigma},\frac{1}{2}\left(3+\frac{1}{\sigma}\right),-t^{2\sigma}\right)$, which can then easily be inverted numerically. However, in order to avoid integrable singularities in the UV, the power $\sigma$ would need to be increased depending on the process considered, so that retaining the exponential dampening featured in our original choice of normalised distribution (eq.~\eqref{eq:ht_definition}) may be preferable in then end. 

    The other Cutkosky cut integrands $I_{\mathbf{s}^\prime_{\Gamma}}( \vec{\phi}^{\mathbf{s}_\Gamma}_{\vec{x}} )$, with $\mathbf{s}^\prime_{\Gamma} \neq \mathbf{s}_{\Gamma}$, still retain their dependence on $h(t)$ and will be conveniently suppressed away from the intersection $\delta\eta_{\mathbf{s}_{\Gamma}}\cap \delta\eta_{\mathbf{s}^\prime_{\Gamma}}$ where pairwise E-surface cancellations occur. The spread of the function $h(t)$ must then be adjusted so as to find the optimal balance between making it narrow enough so as to efficiently focus each channel onto its defining Cutkosky surface and wide enough so as to maintain a good cancellation of threshold singularities in the neighbourhood of their intersection.
    
    \item Last, the loop spatial degrees of freedom must be assigned using the variables $\vec{x}_{\text{left}}$ and $\vec{x}_{\text{right}}$. 
    Note that the external kinematics of these loop subgraphs are now entirely determined from the quantities $\vec{q}^{\;\mathbf{s}_\Gamma}$ and $t^{\mathbf{s}_\Gamma}$ built in the previous two steps. 
    The situation is thus completely analogous to the sampling problem of normal loop integrals in their LTD representation already studied in ref.~\cite{Capatti:2019edf}. 
    A minimal solution is then to consider a separate channel for each loop momentum basis of the reduced subgraphs $\Gamma_\text{left}$ and $\Gamma_\text{right}$ which can be sampled with spherical parameterisations. This is sufficient to dampen all integrable singularities, though it may be beneficial to construct additional channels using elliptical parameterisations whose Jacobians can flatten possible enhancements in the vicinity of non-pinched threshold singularities (E-surfaces) which may not always cancel when considering differential observables.
\end{itemize}
The procedure described above yields complicated parameterisations $\vec{\phi}^{i}_{\vec{x}}$ that are nonetheless invertible and whose overall multi-channeling factor $\sum_{\mathbf{b}_b \in \mathcal{B}_\Gamma} J^{b}\left( \vec{\Psi}^{b}_{ \vec{\phi}^{a}_{\vec{x}}} \right)$ better approximates the shape of the overall integrand $\sum_{\mathbf{s}_{\Gamma}\in{\mathcal{E}^{\Gamma}_{\text{s-ch}}}}I_{\mathbf{s}_{\Gamma}}\left( \vec{\phi}^{a}_{\vec{x}} \right)$. This improvement is also a necessity when considering processes featuring narrow Breit-Wigner resonances. At LO, one could recover the same performance as that of traditional integration techniques since the $h(t)$ normalising function no longer impacts the variance of the integrands. Finally, notice that the number of integration channels are no longer $|\mathcal{B}_\Gamma|$, but instead $|{\mathcal{E}^{\Gamma}_{\text{s-ch}}}|\times|\mathcal{B}_{\Gamma_\text{left}}|\times|\mathcal{B}_{\Gamma_\text{right}}|$ (when ignoring the possibility of additional E-surface parameterisations for the sampling of $\Gamma_\text{left}$ and $\Gamma_\text{right}$), such that the sampling is automatically  simplified for observables selecting only a subset of the possible Cutkosky cuts. 

\subsubsection{Advanced adaptive sampling}
\label{sec:advanced_adaptive_sampling}

The techniques discussed in sect.~\ref{sec:technical_multi_channeling} aim at leveraging our \emph{prior} knowledge about the structure of the LU representation and the location of its enhancement so as to construct an integrand with the smallest possible variance. 
However, once we start sampling the integrand, we gain additional knowledge about its more detailed structure for the particular process and observable at hand. It is then possible to take advantage of this newly gained information to further reduce the integrand variance at run-time using machine-learning techniques implementing some form of iterative adaptive sampling which we discuss here.

The most straightforward way of improving on the parameterisation of the spatial loop degrees of freedom is to further refine it with a trainable integration grid assuming a factorised ansatz for the functional form of the integrand~\cite{Lepage:1980dq,Hahn:2005pf}, which can be complemented with the use of coarse hypercells~\cite{Lepage:2020tgj}.
Our use case of LU requires considering an individual continuous grid for \emph{each} possible choice of supergraph and integration channel, which are moreover subject to discrete importance sampling.
We developed a new integrator for this purpose, dubbed {\sc\small Havana}, that implements this hybrid adaptive sampling through a system of continuous grids nested within multiple layers of discrete ones. This will serve as a basis to accommodate further refinements and help to design custom parallelisation strategies in order to facilitate deployment on various computing architectures.

Discrete importance sampling over integration channels can also be used to improve upon the quality of the integrand approximation obtained from the multi-channeling factor of eq.~\eqref{eq:integrand_flattening}.
The a priori unknown relative magnitudes of the particular enhancements captured by the channel factors $\xi^{i}$ may significantly vary. 
One way to adjust for this is to consider weighted channels $\alpha^{i}\xi^{i}$, with weights $\alpha^{i}$ that can be trained~\cite{Kleiss:1994qy,Denner:2002cg} at run-time so as to minimise the resulting variance of the multi-channeled integrand $I_\Gamma$.

More recently, the development of advanced machine learning frameworks such as {\sc\small TensorFlow} and {\sc\small pyTorch} facilitated the exploration of sampling methods based on various neural networks architectures~\cite{Bendavid:2017zhk,Klimek:2018mza,Otten:2019hhl,Chen:2020nfb}. 
In particular, the use of \emph{Normalizing Flows} models, originally introduced in ref.~\cite{DBLP:journals/corr/abs-1808-03856} in the context of ray tracing and later refined in refs.~\cite{rezende2015variational,papamakarios2019normalizing}, offers a promising novel approach~\cite{Gao:2020zvv,Gao:2020vdv,Bothmann:2020ywa,ZunisToAppear:2020xxx} for improving adaptive sampling for Monte-Carlo integration beyond the factorised paradigm.
The investigation of the potential of these recent developments in the context of Local Unitarity can ideally be carried out in conjunction with the vectorisation of our implementation so as to render it suitable for accelerated hardware, such as graphics and tensor processing units (see e.g. ref.~\cite{Carrazza:2020rdn}).

\section{Numerical results}
\label{sec:numerical_results}

In these sections we showcase the performance of our implementation of the Local Unitarity representation in two different setups:
\begin{itemize}
    \item The computation of the NLO correction to the differential cross section of the scattering process $e^+ e^-\rightarrow \gamma^\star \rightarrow d \bar{d}$.
    \item The computation of subsets of IR finite interference diagrams (i.e. supergraphs) contributing to the N$^4$LO accurate cross-section of a $1\rightarrow2+X$ scalar scattering process.
\end{itemize}
All results reported in this section are obtained from our own implementation in a program called \alphaLoop{} for the computation of differential cross-sections within the Local Unitarity framework.

\subsection{NLO correction to $e^+ e^- \rightarrow d \bar{d}$}

In this section we present results for the NLO correction to $e^+ e^- \rightarrow d \bar{d}$. In sect.~\ref{sec:foundation_example} we have studied the two supergraphs of this process, namely the \DT{} supergraph and \SE{} supergraph, and have shown local IR cancellations for the \DT{} supergraph in sect.~\ref{sec:cancellation_example}.
Contrary to those sections, we label here the loop momenta of these supergraphs $(\vec{k},\vec{l})$ and not $(\vec{k}',\vec{l}')$.

In sect.~\ref{sec:DT_UV_renormalisation} we study the numerator and the renormalisation of the \DT{}, and visualise the integrand in sect.~\ref{sec:DT:integrand_visualisation}. In sect.~\ref{sec:SE_example} the self-energy treatment is applied to the \SE{} supergraph. Finally, we present results for the differential cross-section in sect.~\ref{sec:NLO_example_numerical_results}.

\subsubsection{The \DT{} supergraph}
\label{sec:DT_UV_renormalisation}

We write out the Feynman rules for the \DT{} supergraph and obtain the following numerator (common to all its Cutkosky cut contributions) in $d=4$:
\begin{eqnarray}
    N(k^\mu,l^\mu;p_1^\mu,p_2^\mu)&=&\frac{-64 (N_c^2-1) T_F}{2 Q^0}\frac{g_{\textrm{EW}}^4g_{\textrm{s}}^2Q_d^2Q_e^2 }{2\times2} \Big [ \nonumber\\
    && k^2 l^2 Q^2+l\cdot p_2 \left(k\cdot p_2 \left(l\cdot p_1-2 Q^2\right)-2 k^2 l\cdot p_1\right) \nonumber\\
    &+&k\cdot p_1 \left(k\cdot p_2 \left(-2 l^2+l\cdot p_1+l\cdot p_2\right)+l\cdot p_1 \left(l\cdot p_2-2 Q^2\right)-(l\cdot p_2)^2\right) \nonumber\\
    &+& k\cdot l \left(k\cdot p_2 \left(2 l\cdot p_1+Q^2\right)+k\cdot p_1 \left(2 l\cdot p_2+Q^2\right)+Q^2 (l\cdot p_1+l\cdot p_2)\right) \nonumber\\
    &-&k\cdot p_2\; l\cdot p_1 (k\cdot p_2+l\cdot p_1)- (k\cdot p_1)^2 l\cdot p_2-2 Q^2 (k\cdot l)^2
\Big],
\label{eq:N_DT}
\end{eqnarray}
where we included the averaging factor over incoming lepton helicity configurations as well as the flux factor. The additional factors of $\mathrm{i}$ and $-\mathrm{i}$ arising from the complex-conjugation of propagator numerators and vertices on the right of the Cutkosky cut always come in pairs so that we can ignore them for now\footnote{When considering a renormalisation scheme introducing complex-valued renormalised masses and couplings, such as the complex-mass scheme~\cite{Denner:1999gp,Denner:2005fg,Frederix:2018nkq}, the situation may be more involved. We however leave this aspect to a future investigation.}.
The expression of the numerator of supergraphs such the one above will need to be computed repeatedly during the Monte-Carlo integration and are optimised using the procedure described in sect.~\ref{sec:technical_numerator_treatment}. 

Given that the LU representation realises the local cancellations of all infrared divergences of the integrand, we only need to consider counterterms for the regularisation of its UV behaviour.
As is laid out in sect.~\ref{sec:technical_UV}, this is achieved by implementing a local UV subtraction procedure. The local UV counterterms are built from the original (unexpanded) numerator in (4-2$\epsilon$)-dimensions and from denominators expanded around the UV point. This expansion is made systematic by scaling all external momenta involved by a parameter $\lambda$ and expanding around $\lambda=0$.
This procedure is well suited for automation, but
to demonstrate the 
interplay between renormalisation and integrated UV counterterms, we factorise the triangle loop correction and apply the UV regularisation procedure to it only.
We therefore rewrite the integrand stemming from the numerator of eq.~\eqref{eq:N_DT} for the Cutkosky cut $\mathbf{c}_{\mathbf{s}_1^{\text{v}}}$ as follows (the case of $\mathbf{c}_{\mathbf{s}_2^{\text{v}}}$ is fully analogous):
\begin{equation}
    I_{\mathbf{s}_1^{\text{v}}} = K^{\left(\mathbf{s}_1^{\text{v}},\textrm{left}\right)}_\mu \ \Gamma^\mu_{i j} \  K^{\left(\mathbf{s}_1^{\text{v}},\textrm{right}\right)}_{i j} \,,\label{eq:DT_complete_Itriangle_loop}
\end{equation}
where $K^{\left(\mathbf{s}_1^{\text{v}},\textrm{left}\right)}$ and $K^{\left(\mathbf{s}_1^{\text{v}},\textrm{right}\right)}$ are the factorised pieces that are irrelevant for this discussion, while the one-loop corrected $\gamma d \bar{d}$-vertex $\Gamma^\mu_{ij}$ reads:
\begin{eqnarray}
\Gamma^\mu_{ij} &=& g_\textrm{s}^2 C_F \left [ (-\textrm{i})  g_{\gamma d\bar{d}} \right] I^{\mu}_{ij}(k,l,Q) \label{eq:DT_vertex_correction}\\
I^{\mu}(k,l,Q) &=& \frac{ \gamma^{\rho} (\slashed{k}-\slashed{Q})\gamma^\mu \slashed{k} \gamma^{\rho}  }{k^2 (k-l)^2 (k-Q)^2} \nonumber\\
&=& \frac{(d-2) \left( k^2 \gamma^\mu -2 k^\mu \slashed{k} \right) + 2 \slashed{k}\gamma^\mu \slashed{Q} + (d-4) \slashed{Q}\gamma^\mu\slashed{k}  }{k^2 (k-l)^2 (k-Q)^2} \label{eq:DT_triangle_loop_expression}.
\end{eqnarray}
Given that the superficial degree of UV divergence of the one-loop vertex correction is zero, the local UV counterterm denoted by $\overline{I}^{\mu}$ can be constructed by setting $d=4$ in eq.~\eqref{eq:DT_triangle_loop_expression} and expanding the denominators around their UV limit to first order. We must also assign a mass $M^2_{\textrm{UV}}$ to the resulting denominators in order to prevent introducing new IR divergences in the UV counterterm: 
\begin{equation}
\overline{I}^{\mu}(k,Q) = \frac{2 \left( k^2 \gamma^\mu -2 k^\mu \slashed{k} \right) + 2 \slashed{k}\gamma^\mu \slashed{Q} }{(k^2-M^2_\textrm{UV})^3} \,. \label{eq:DT_UV_local_CT}
\end{equation}
The denominator structure $(k^2-M^2_\textrm{UV})$ is now guaranteed to never yield a pole in its (c)LTD representation since the absences of any shift in the denominator implies that the existence conditions of the UV counterterm E-surfaces can never be fulfilled.
This is different from similar UV local counterterms in four-dimensional Minkowski space where the squared mass term in $(k^2-M^2_\textrm{UV})$ needs to be set negative or imaginary (see, e.g.~\cite{Becker:2012bi}) in order to avoid complications.

The integrated counterpart of eq.~\eqref{eq:DT_UV_local_CT}, denoted by $\Delta \overline{I}_{\mathbf{s}_1^{\text{v}}}$ is straightforward and is derived in appendix.~\ref{app:UV_DT} for completeness. We obtain:
\begin{eqnarray}
\Delta \overline{I}_{\mathbf{s}_1^{\text{v}}} &=& K^{\left(\mathbf{s}_1^{\text{v}},\textrm{left}\right)}_\mu \ 
g_\textrm{s}^2 C_F \left [ (-\textrm{i})  g_{\gamma d\bar{d}} \right] \ 
K^{\left(\mathbf{s}_1^{\text{v}},\textrm{right}\right)}_{i j} \
\int \frac{\tilde{d} k}{(2\pi)^4} \overline{I}^{\mu}(k,Q)
\nonumber \\
&=& \frac{\alpha_s}{\pi} I^{(\textrm{LO})} \left[ \frac{1}{4} C_F \left( \frac{1}{\epsilon} - 2 + \log_{M_{\textrm{UV}}} \right) \right] \label{eq:DT_DeltaUV_main},
\end{eqnarray}
where we defined $\log_{M_{\textrm{UV}}}:= \log{\frac{\mu^2}{-M_\textrm{UV}^2}}$ and the $d$-dimensional measure $\int \tilde{d}k\; := \left(\frac{\mu^2}{4\pi e^{-\gamma}}\right)^{\epsilon} \int \frac{d^{4-2\epsilon}k}{(2\pi)^{-2\epsilon}} $, with the additional factor $ \left(\frac{1}{4\pi e^{-\gamma}}\right)^{\epsilon} $ designed so as to automatically reabsorb all the terms normally removed by the finite part of the renormalisation counterterms in the $\overline{\textrm{MS}}$ scheme into a redefinition of the renormalisation scale.
This integrated UV counterterm then needs to be combined with the $\overline{\textrm{MS}}$ vertex renormalisation constant $\delta Z^{(\overline{\textrm{MS}})}_{\Gamma_{\gamma d \bar{d}}}$:
\begin{equation}
 \delta Z_{\Gamma_{\gamma d \bar{d}}} 
 = Z_{\alpha_\textrm{QED}} \left(Z^{(\overline{\textrm{MS}})}_q\right)^{\frac{1}{2}} \left(Z^{(\overline{\textrm{MS}})}_{\bar{q}}\right)^{\frac{1}{2}}-1
 = \delta \alpha_\textrm{QED} +\frac{1}{2} \delta Z^{(\overline{\textrm{MS}})}_q + \frac{1}{2} \delta Z^{(\overline{\textrm{MS}})}_{\bar{q}} + o(\alpha_s^2) \,.
\end{equation}
Traditionally, the wavefunction counterterms are provided in the on-shell scheme (see, e.g. eq.~(6.8) of ref.~\cite{Gnendiger:2017pys}) and the renormalisation of the strong coupling in the mixed scheme where massless quarks are renormalised in $\overline{\textrm{MS}}$ and massive ones at zero momentum (see, e.g. ref.~\cite{Beenakker:2002nc} or eq.~B.5 of~\cite{Hirschi:2011pa}).
The renormalisation constant $\delta \alpha_\textrm{QED}$ of the QED charge from QCD correction is of course zero.
In order to convert the massless quark renormalisation constant from the on-shell scheme to the $\overline{\textrm{MS}}$ one, it is important to separately keep track of the dimensional regularisation pole of UV and IR origin in order obtain:
\begin{eqnarray}
\delta Z_q^{(OS)}&=&\left( \frac{\alpha_s}{4\pi}\right)C_F \left[\frac{1}{\epsilon_{\textrm{IR}}} - \frac{1}{\epsilon_{\textrm{UV}}}\right] 
\ \rightarrow\ \delta Z_q^{(\overline{\textrm{MS}})}= - \frac{1}{\epsilon_{\textrm{UV}}} \left( \frac{\alpha_s}{4\pi}\right)C_F \label{eq:delta_Zdown_MSbar}\\
\delta \alpha_\textrm{QED} &=& 0 \label{eq:delta_alpha_QED}\\
\rightarrow\ \delta Z_{\Gamma_{\gamma d \bar{d}}} &=& - \frac{1}{\epsilon_{\textrm{UV}}} \left( \frac{\alpha_s}{4\pi}\right)C_F \,. \label{eq:delta_Zdda}
\end{eqnarray}
We then find that the combination of the integrated local UV counterterm $\Delta \overline{I}_{\mathbf{s}_1^{\text{v}}}$ in eq.~\eqref{eq:DT_DeltaUV_main} and of the renormalisation contribution $\delta I_{(\gamma d \bar{d})^{(\textrm{left})}}$ is finite, as expected:
\begin{equation}
    \Delta \overline{I}_{\mathbf{s}_1^{\text{v}}}+ \delta I_{(\gamma d \bar{d})^{(\textrm{left})}} = \frac{\alpha_s}{\pi} I^{(\textrm{LO})} \frac{1}{4} C_F \left( \log_{M_{\textrm{UV}}} - 2 \right) \,. \label{eq:DT_Final_UV_contribution}
\end{equation}

\subsubsection{Double-triangle supergraph integrand visualisation}
\label{sec:DT:integrand_visualisation}

\begin{landscape}
\begin{figure}[ht]
\centering
\begin{subfigure}[t]{.32\linewidth}
\includegraphics[width=\linewidth]{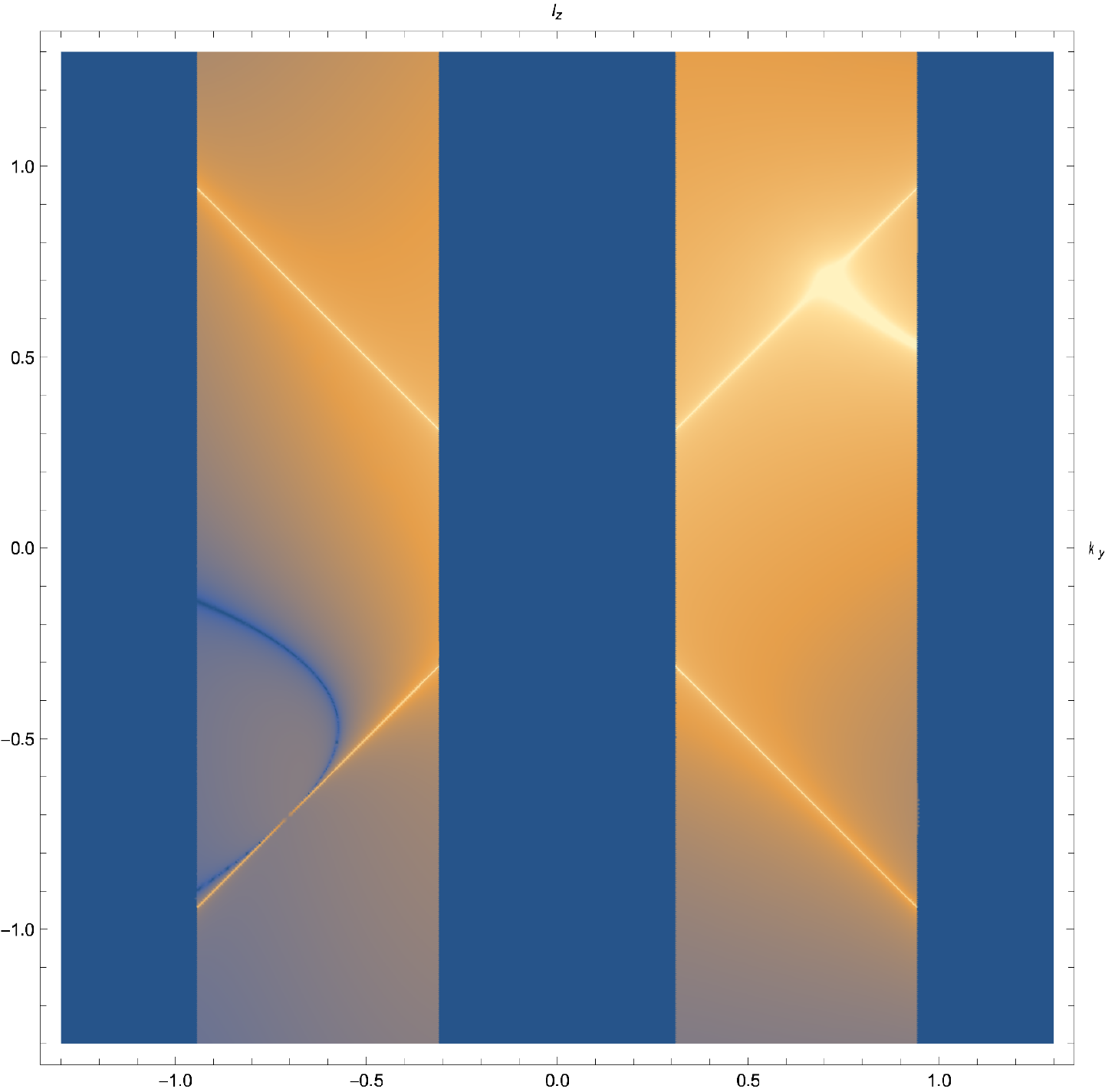}
\caption{$I_{\mathbf{s}_{2}^{\text{v}}}$}\label{DT:fig:DT_no_deformation_density:LxB}
\end{subfigure}
\begin{subfigure}[t]{.32\linewidth}
\includegraphics[width=\linewidth]{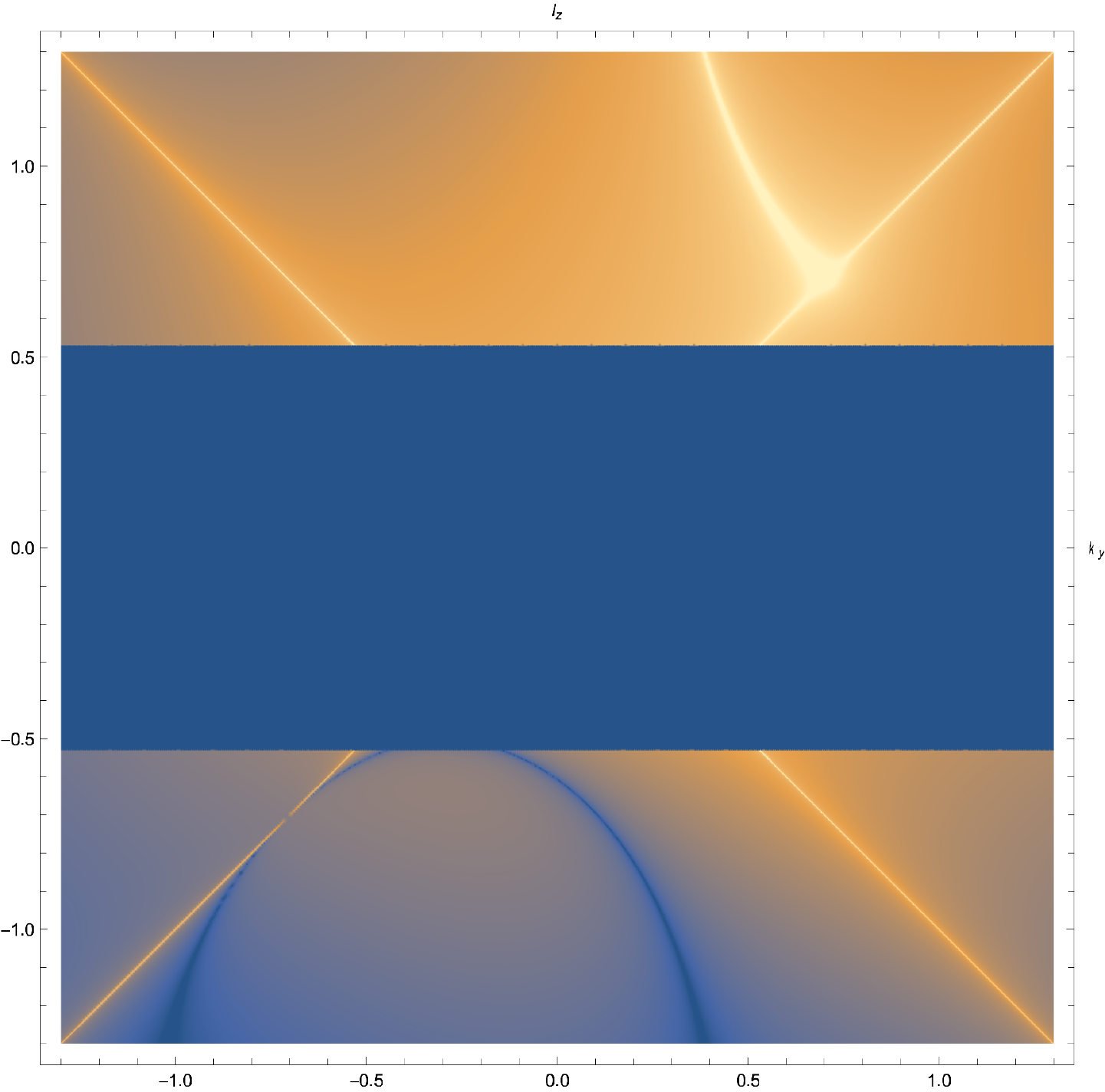}
\caption{$I_{\mathbf{s}_{1}^{\text{v}}}$}\label{DT:fig:DT_no_deformation_density:BxL}
\end{subfigure}
\begin{subfigure}[t]{.32\linewidth}
\includegraphics[width=\linewidth]{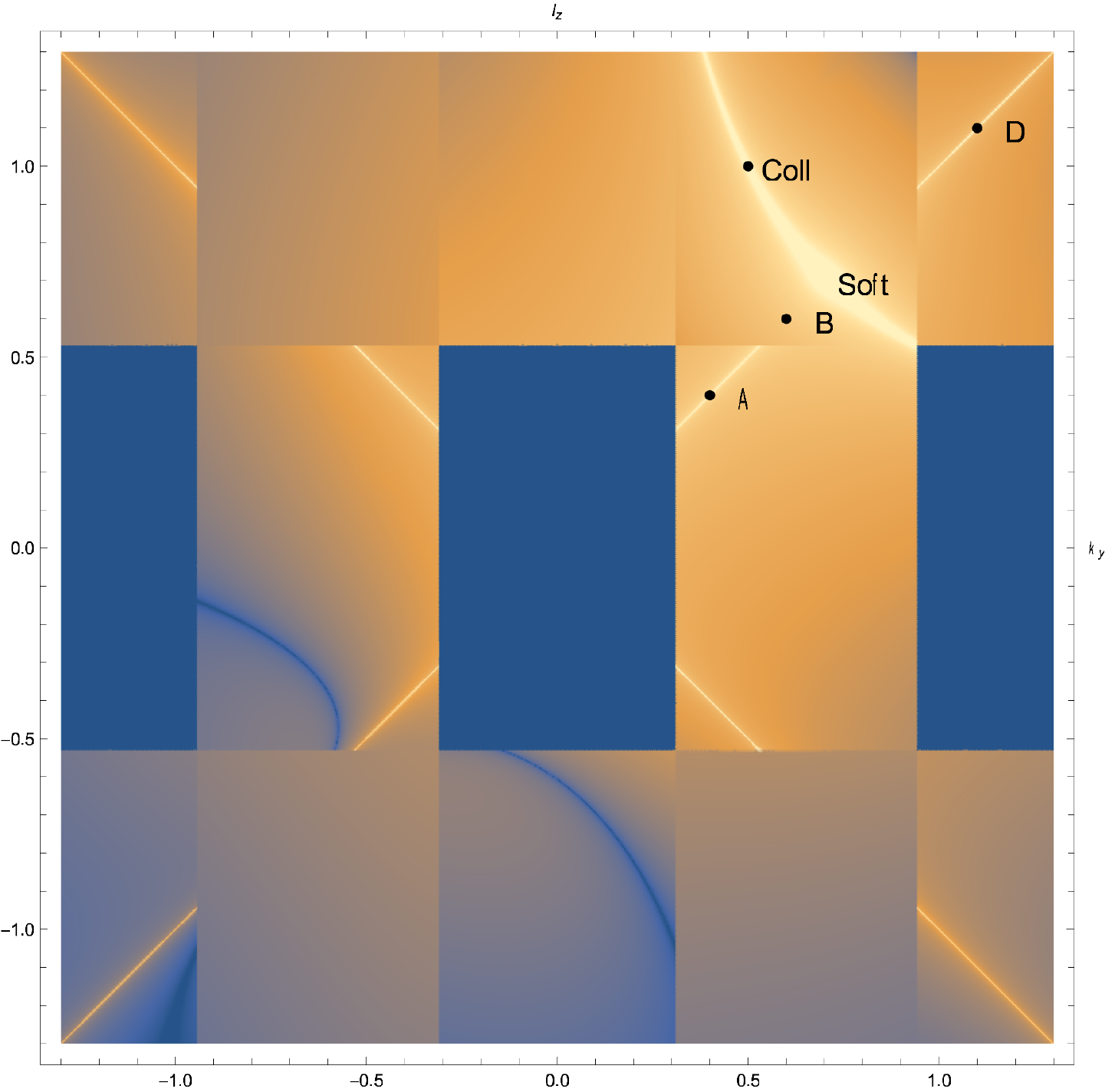}
\caption{$I_{\mathbf{s}_{1}^{\text{v}}}+I_{\mathbf{s}_{2}^{\text{v}}}$}\label{DT:fig:DT_no_deformation_density:LxB_plus_BxL}
\end{subfigure}
\begin{subfigure}[t]{.02\linewidth}
\includegraphics[width=\linewidth]{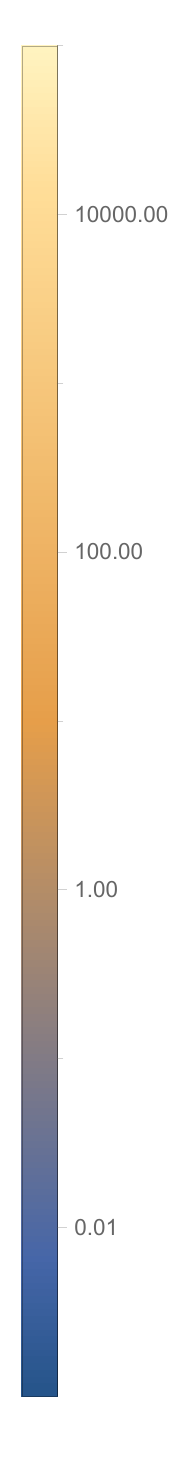}
\end{subfigure}
\\
\begin{subfigure}[t]{.32\linewidth}
\includegraphics[width=\linewidth]{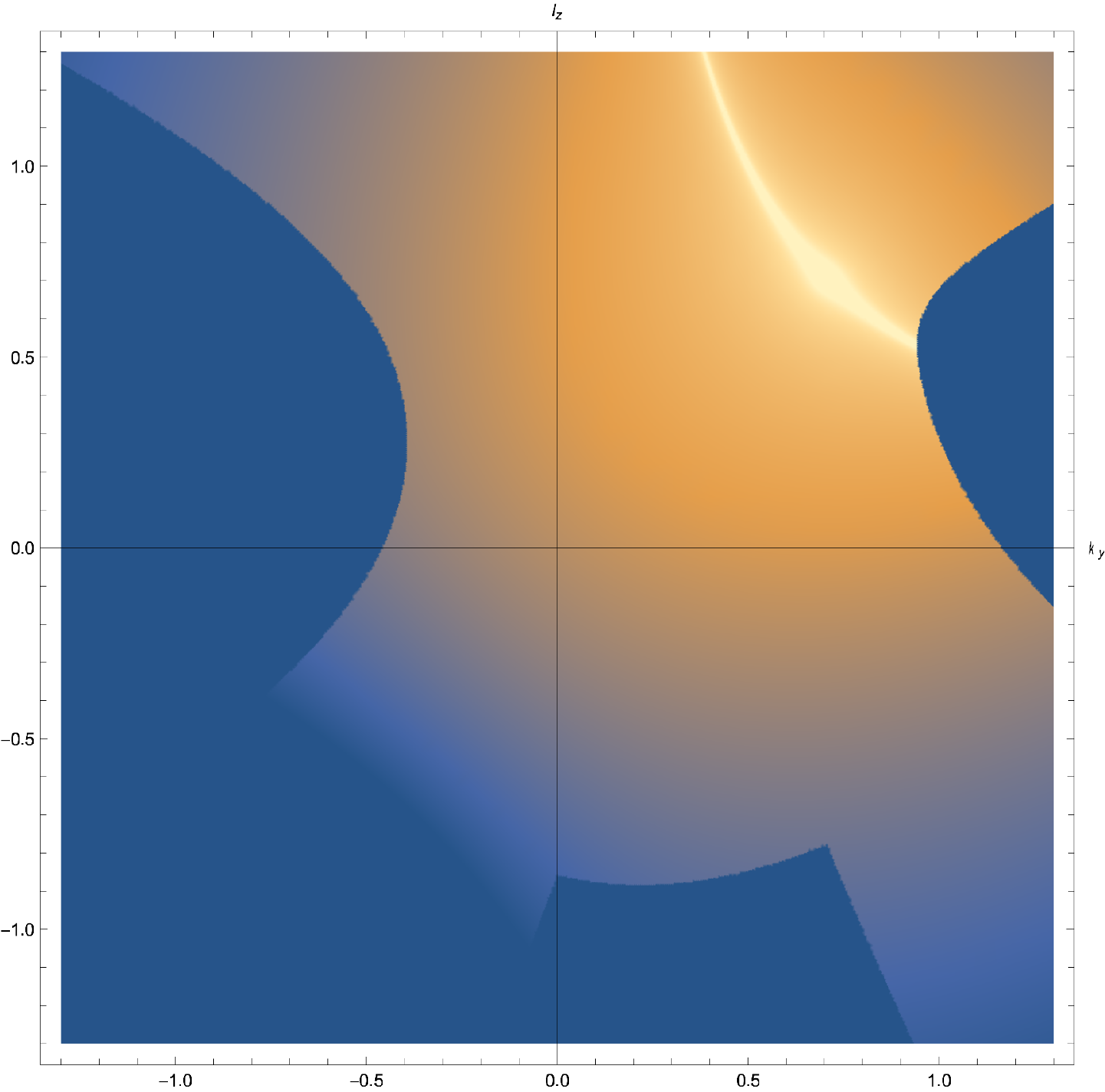}
\caption{$2I_{\mathbf{s}_{1}^{\text{r}}}=I_{\mathbf{s}_{1}^{\text{r}}}+I_{\mathbf{s}_{2}^{\text{r}}}$ }\label{DT:fig:DT_no_deformation_density:RA_plus_RB}
\end{subfigure}
\begin{subfigure}[t]{.32\linewidth}
\includegraphics[width=\linewidth]{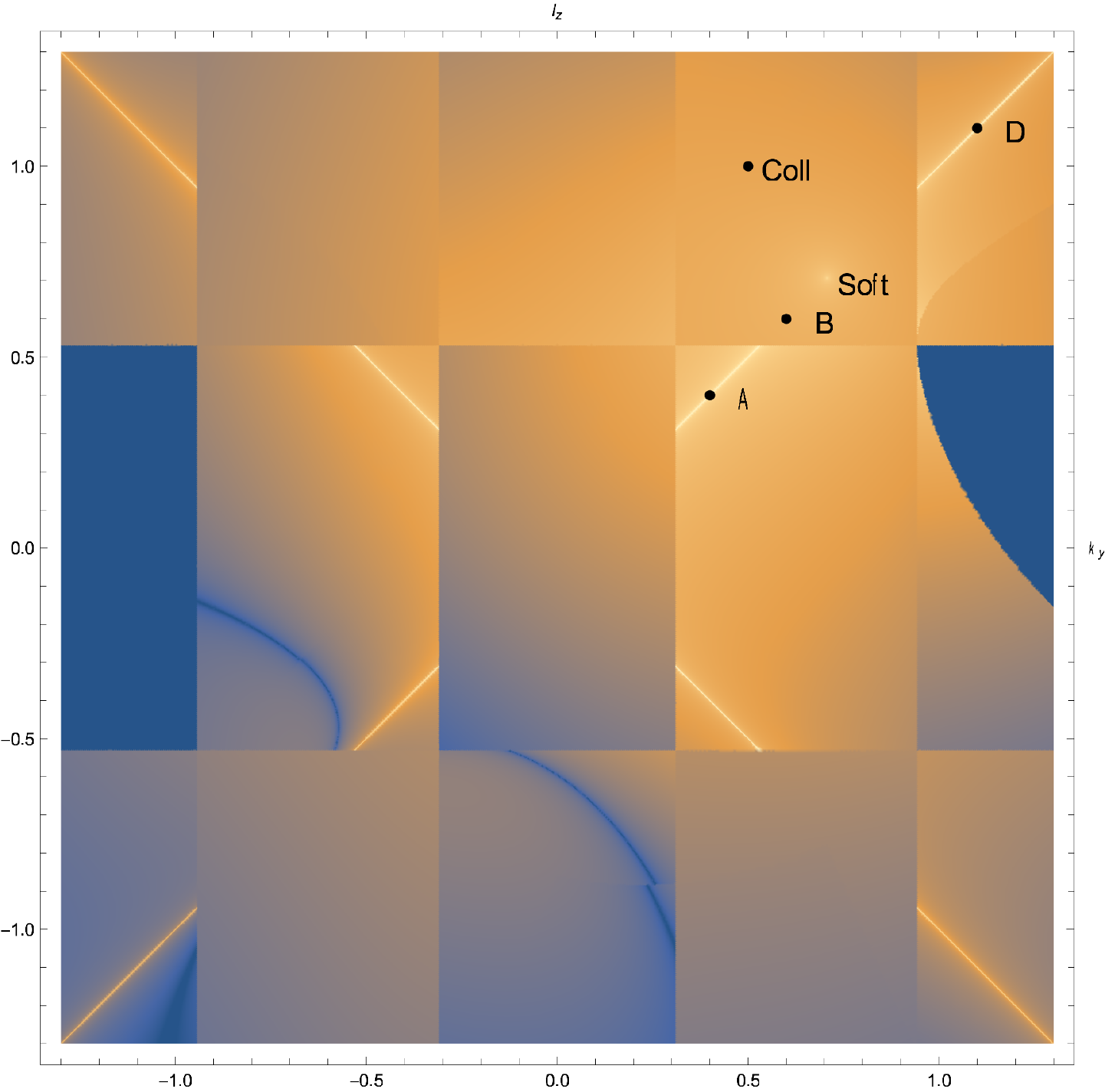}
\caption{$I_\Sigma$ (sll Cutkosky cut contributions) }\label{DT:fig:DT_no_deformation_density:RA_plus_RB_plus_LxB_plus_BxL}
\end{subfigure}
\begin{subfigure}[t]{.32\linewidth}
\includegraphics[width=\linewidth]{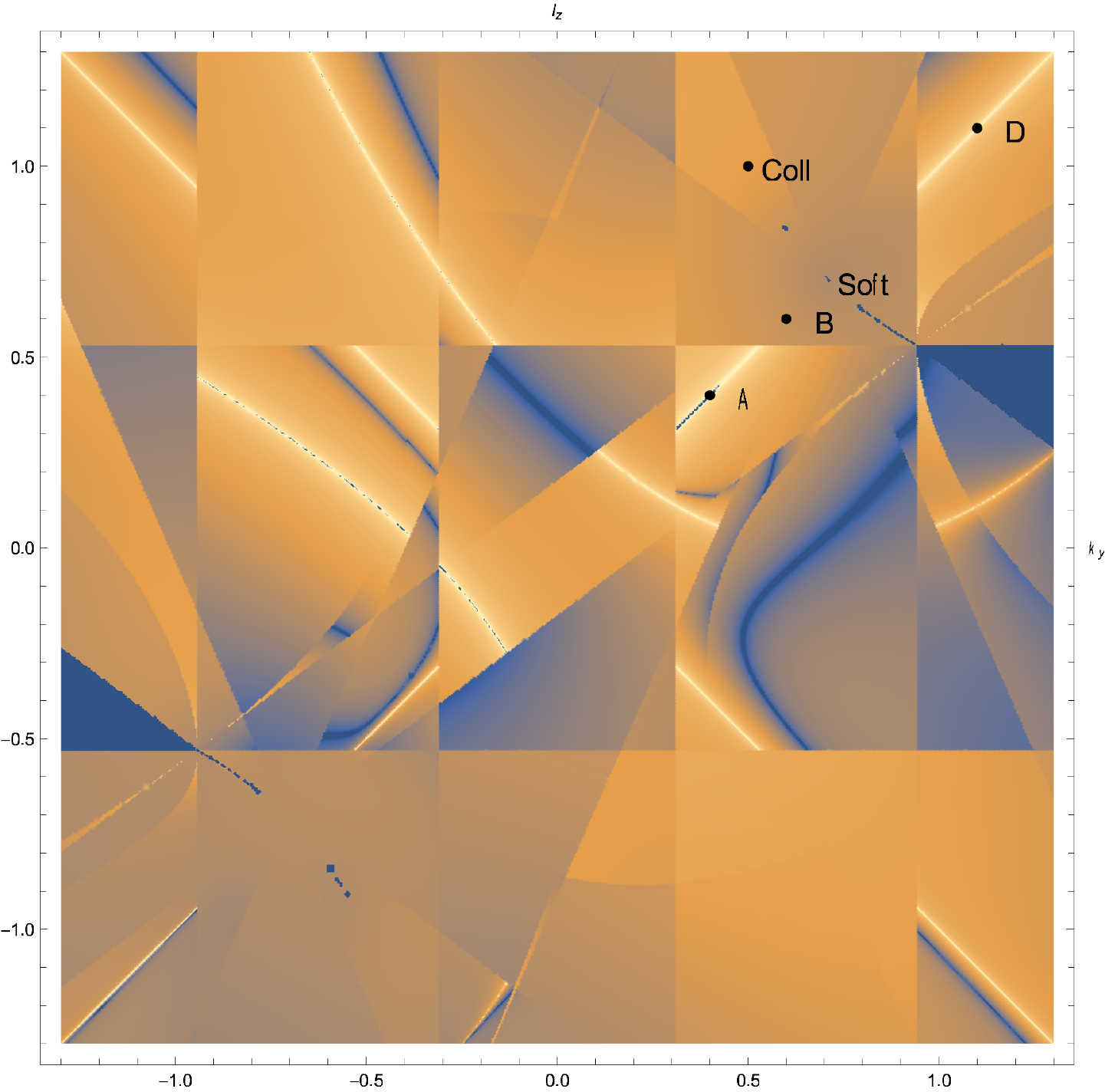}
\caption{$I_\Sigma^{(\textrm{MC})}$ (multi-channeling and incl. param. jac.)  }\label{DT:fig:DT_no_deformation_density:RA_plus_RB_plus_LxB_plus_BxL_MC}
\end{subfigure}
\begin{subfigure}[t]{.02\linewidth}
\includegraphics[width=\linewidth]{Figures/MathematicaPlots/NoDeformation/NoDeformation_legend_bar.pdf}
\end{subfigure}
\caption{\label{DT:fig:DT_no_deformation_density}
Absolute value of the Local Unitarity integrand from \alphaLoop\ evaluated at $(\vec{k},\vec{l})=((0,k_y,\frac{1}{\sqrt{2}}),(0,\frac{1}{\sqrt{2}},l_z))$ for the semi-inclusive cross-section of the DT supergraph and various combination of Cutkoksy cuts and with no contour deformation. The log scale has been capped within the range $[10^{-3},10^5]$.
}
\end{figure}
\end{landscape}

In order to visualise the important features of the \DT{} supergraph integrands, we must choose a different projection than that of sect.~\ref{sec:DT_causal_flow} so as to allow the approach of the soft and collinear configurations of the gluon. We choose here again $Q^\mu=p_1^\mu+p_2^\mu=(1,0,0,1)+(1,0,0,-1)$ with the momentum routing of fig.~\ref{fig:DT_LTD} and set
\begin{equation}
    (\vec{k},\vec{l})=((0,k_y,\frac{1}{\sqrt{2}}),(0,\frac{1}{\sqrt{2}},l_z))\label{eq:DT_approach_projection} \,.
\end{equation}
Moreover, we consider the integrand for the semi-inclusive cross-section defined by the following phase-space cut:
\begin{equation}
    0.4 < p_{t,j_1} < 0.8 \label{eq:DT_PS_cut}
\end{equation}
applied to the $p_t$-leading jet reconstructed according to the anti-kt algorithm~\cite{Ellis:1993tq,Cacciari:2006sm} with cone radius parameter $\Delta R$ set to 0.4.
It is clear that this observable is not phenomenologically relevant for lepton collider experiments but it serves as case-study of a non-trivial observables.
We start by showing in figs.~\ref{DT:fig:DT_no_deformation_density} various combinations of the integrands of the interference diagrams produced by \alphaLoop{} when turning off the complex contour deformation. Except when showing results for multi-channelling, the Jacobians from the loop momentum parameterisations is not included.

The first fig.~\ref{DT:fig:DT_no_deformation_density:LxB} shows the contribution stemming solely from the Cutkosky cut $\mathbf{c}_{\mathbf{s}_{1}^{\text{v}}}$. A first observation is the linear bands that correspond to the selection cuts of eq.~\eqref{eq:DT_PS_cut}. The observable function for this cut reads $\mathcal{O}(\vec{l},\vec{l}-\vec{Q})$ so that the jet $p_t$ is dependent on $l_z$ only for our projection and the selection bands appear vertical. 
It may appear counter-intuitive at first that the transverse momentum depends on the z-component of the $\vec{l}$ loop momentum. This is however expected because the space presented in the figures corresponds to the dependency of the Local Unitarity integrand \emph{before} the change of variables of eq.~\eqref{eq:Soper_change_of_variables} (i.e. the sampling space probed by the Monte-Carlo integration). This change of variables results in a $l_z$-dependent rescaling of the fixed component $l_y=\frac{1}{\sqrt{2}}$ which in turn induces an indirect dependence of $p_{t,j_1}$ on $l_z$.
The situation is opposite for the contribution of the cut $\mathbf{c}_{\mathbf{s}_{2}^{\text{v}}}$, shown in fig.~\ref{DT:fig:DT_no_deformation_density:BxL}, since in that case the observable reads $\mathcal{O}(\vec{k},\vec{k}-\vec{Q})$ so that the jet $p_t$ depends only $k_y$, thus yielding an horizontal band.
The boundaries of the phase-space cuts are more complicated for the real-emission contribution, as expected from its complicated three-body observable dependence $\mathcal{O}(\vec{l},\vec{l}-\vec{k},\vec{k}-\vec{Q})$. 
It is interesting to note however that the central region is only populated by the real-emission kinematics, making it effectively LO accurate. This feature bears resemblance with that of other more common differential quantities in lepton collisions, such as for example the C jet-shape parameter~\cite{Ellis:1980nc,Gardi:2003iv} which only receives contributions at $C=0$ from $e^+ e^- \rightarrow j j$ and for $C<\frac{3}{4}$ from $e^+ e^- \rightarrow j j j$.

Another important feature from the set of figs.~\ref{DT:fig:DT_no_deformation_density} is the behaviour of the integrands around the non-pinched E-surface located at $k_y=l_z$, since Soper's rescaling  will simultaneously bring both norms $|\vec{k}|$ and $|\vec{l}|$ to $Q^0/2=1$, thus placing the configuration on both the E-surface of the triangle loop and the Cutkosky cut $\mathbf{c}_{\mathbf{s}_{1}^{\text{v}}}$ or $\mathbf{c}_{\mathbf{s}_{2}^{\text{v}}}$ (see fig.~\ref{DT:fig:threshold_singularities}).
In the absence of a deformation, the singularity along the E-surface is left unregulated and this is reflected by the bright diagonal lines in figs.~\ref{DT:fig:DT_no_deformation_density:LxB} and~\ref{DT:fig:DT_no_deformation_density:BxL}. The \emph{local} cancellation of these \emph{non-pinched} E-surfaces is then made apparent in fig.~\ref{DT:fig:DT_no_deformation_density:LxB_plus_BxL} where the areas accessible to both Cutkosky cuts $\mathbf{c}_{\mathbf{s}_{1}^{\text{v}}}$ and $\mathbf{c}_{\mathbf{s}_{2}^{\text{v}}}$ is devoid of any particular features along the diagonal (point labelled ``B''). However, the E-surface cancellation breaks down on points labelled ``A'' and ``D'' where the observable function (i.e. phase-space cut in our present case) retains only one of the two cancelling terms.
This shows that \emph{fully}\footnote{``Fully'' in this context implies that \emph{all} possible Cutkosky cuts of each contributing supergraph is considered in the entirety of the phase-space available to it. By definition, this therefore excludes any $1\rightarrow N,\;N>2$ process in the Standard Model} inclusive cross-sections can be computed without a contour deformation.

We now consider the soft and collinear gluon kinematic configuration. Due to Soper's rescaling, the collinear subspace does not only lie at the soft end-point $k_y=l_z$ but instead lies along an arc denoted by ``Coll'' in the upper-left portion of figs.~\ref{DT:fig:DT_no_deformation_density}. Each of the two virtual Cuktkosky cuts covers only ``half'' of the collinear space since our particular choice of signs for the fixed spatial components of $\vec{k}$ and $\vec{l}$ only allows one of the two pinched E-surfaces of the loop triangle to be reached (either the $qg$ or the $\bar{q}g$ collinear one). Together however, the virtual contributions of fig.~\ref{DT:fig:DT_no_deformation_density:LxB_plus_BxL} reproduce the complete collinear singularity to be locally cancelled by the sum of real-emission contributions shown in fig.~\ref{DT:fig:DT_no_deformation_density:RA_plus_RB}. The overall sum in fig.~\ref{DT:fig:DT_no_deformation_density:RA_plus_RB_plus_LxB_plus_BxL} presents no particular feature close to collinear regions any longer.

The soft singularity in fig.~\ref{DT:fig:DT_no_deformation_density:RA_plus_RB_plus_LxB_plus_BxL} appears dampened but is still unbounded, as expected from the power counting arguments of sect.~\ref{sec:power_counting_soft_integrable_singularities}. 
This integrable singularity is particularly harmful for a numerical implementation since for the choice of momentum routing of fig.~\ref{fig:Two_SG_epemddx_NLO}, the soft singularity lies at $\vec{k}-\vec{l}=\vec{0}$ which spans a full three-dimensional volume of the phase-space parameterised with $(\vec{k},\vec{l})$. In our particular example, this can be remedied simply by adopting a different loop momentum basis containing the gluon propagator as a defining edge. 
This solution is not generic, since for a triple gluon vertex it is not possible to consider all three connected gluons to have an independent momentum. %
The general solution then involves combining multiple parameterisations using a \emph{multi-channeling} technique discussed in sect.~\ref{sec:technical_multi_channeling} and analogous to the one we already presented in sect.~5.2 of ref.~\cite{Capatti:2019edf}.
Fig.~\ref{DT:fig:DT_no_deformation_density:RA_plus_RB_plus_LxB_plus_BxL_MC} showcases the suppression of the soft-singularity from multi-channeling. 
Notice that in all figures not involving multi-channeling, the integrand weight considered did not include the parameterisation Jacobian, since the objective was to focus on the behaviour of the integrand only. For our default choice of momentum routing, the Jacobian is anyway relatively shallow within our chosen sampling space projection.
For multi-channeling, the combination of the three non-degenerate loop momentum bases of the supergraph results in a parameterisation Jacobian which is precisely the quantity responsible for achieving the dampening of the integrable singularities and it is therefore crucial that it be included in the weights displayed by the multi-channeling density plot.
Furthermore, the multi-channelling figures are generated using the exact \alphaLoop\ integrand function exposed to the integrator, which implies that it includes the numerical stability estimate and rescue system described in sect.~\ref{sec:technical_numerical_stability} which exclaims the sparse spots in the density plot where the integrand is sent to zero as it failed the numerical stability requirement threshold.
We find the expected suppression of the soft integrable singularity in fig.~\ref{DT:fig:DT_no_deformation_density:RA_plus_RB_plus_LxB_plus_BxL_MC}, but also see a complicated mangling resulting from the combined contributions of our input $\vec{k}$ and $\vec{l}$ re-interpreted for all three relevant loop momentum bases of the \DT{} supergraph. 
We stress that the visual complexity of the integrand can be deceptive since in practice highly discontinuous bounded functions with a lot of structures often converge significantly faster than smoother integrands that have integrable singularities, especially when considering adaptive Monte-Carlo sampling.
We note however that, as shown in fig.~\ref{DT:fig:DT_deformation_3D_zoomin_MC}, the multi-channeling strategy renders the integrand bounded only in the absence of a deformation. When a deformation is enabled, there remains an integrable singularity along the non-pinched E-surface cancelation diagonal line, which is however much less severe.
The local UV counterterm discussed in sect.~\ref{sec:DT_UV_renormalisation} is active as part of the integrands shown, but has no visible impact for our chosen value of $M^2_{\textrm{UV}}=2Q^2$ and kinematic range $(k_y,l_z)$ considered.

\begin{figure}[ht!]
\centering
\begin{subfigure}[t]{.32\linewidth}
\includegraphics[width=\linewidth]{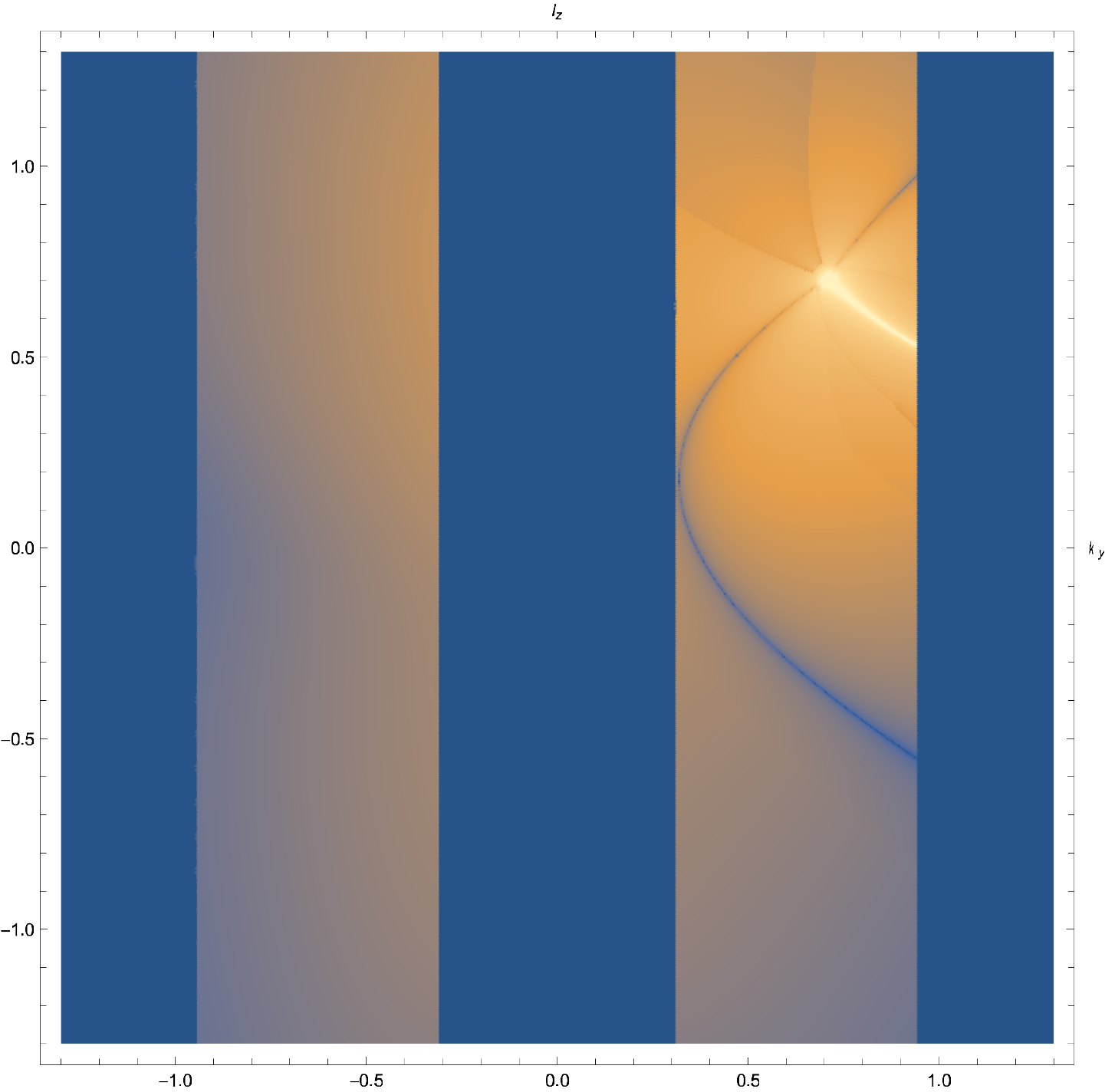}
\caption{$\Re[ I_{\mathbf{s}_{1}^{\text{v}}} ] $}\label{DT:fig:DT_deformation_density:LxB_RE}
\end{subfigure}
\begin{subfigure}[t]{.32\linewidth}
\includegraphics[width=\linewidth]{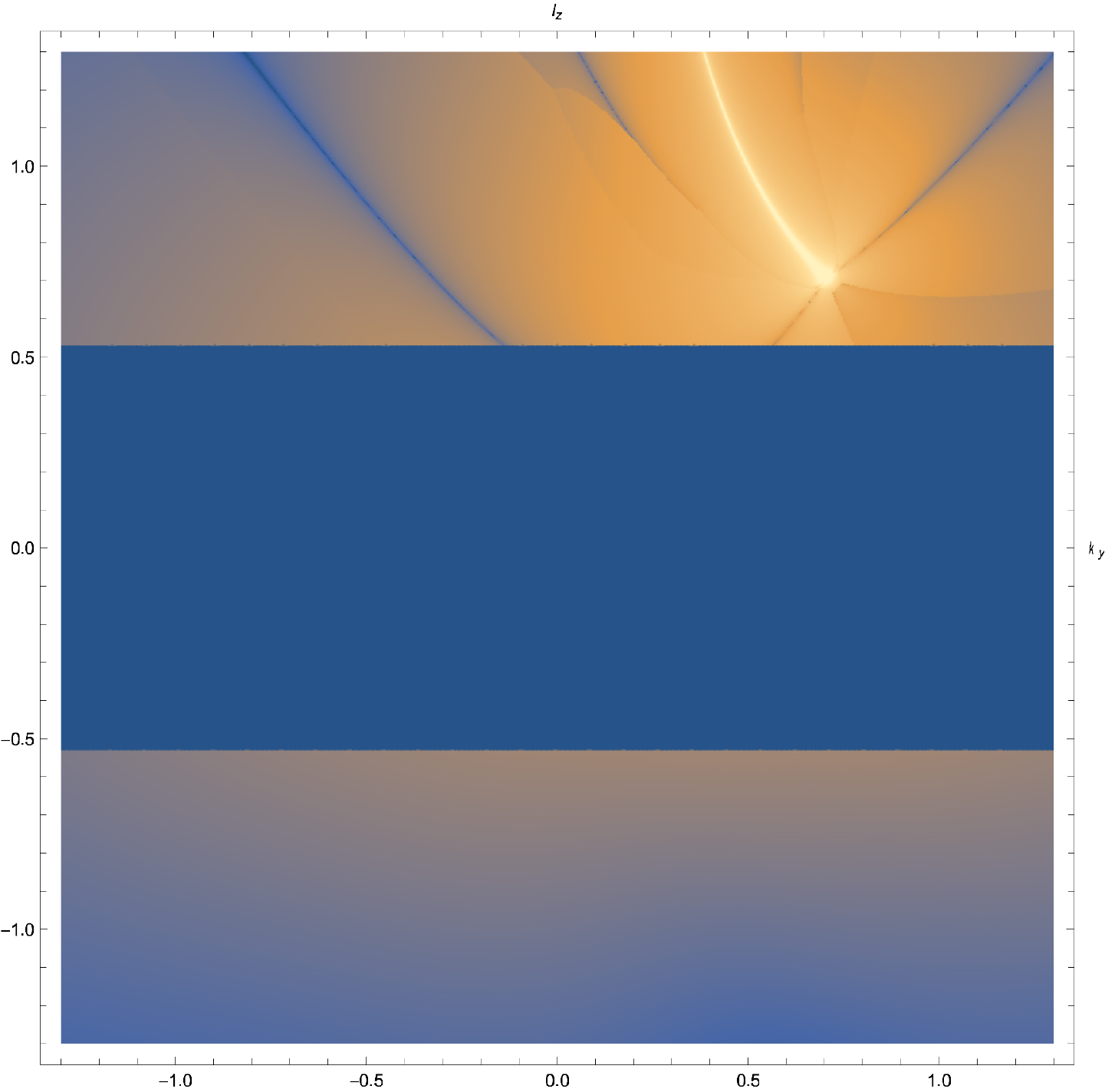}
\caption{$\Re[I_{\mathbf{s}_{2}^{\text{v}}} ]$}\label{DT:fig:DT_deformation_density:BxL_RE}
\end{subfigure}
\begin{subfigure}[t]{.32\linewidth}
\includegraphics[width=\linewidth]{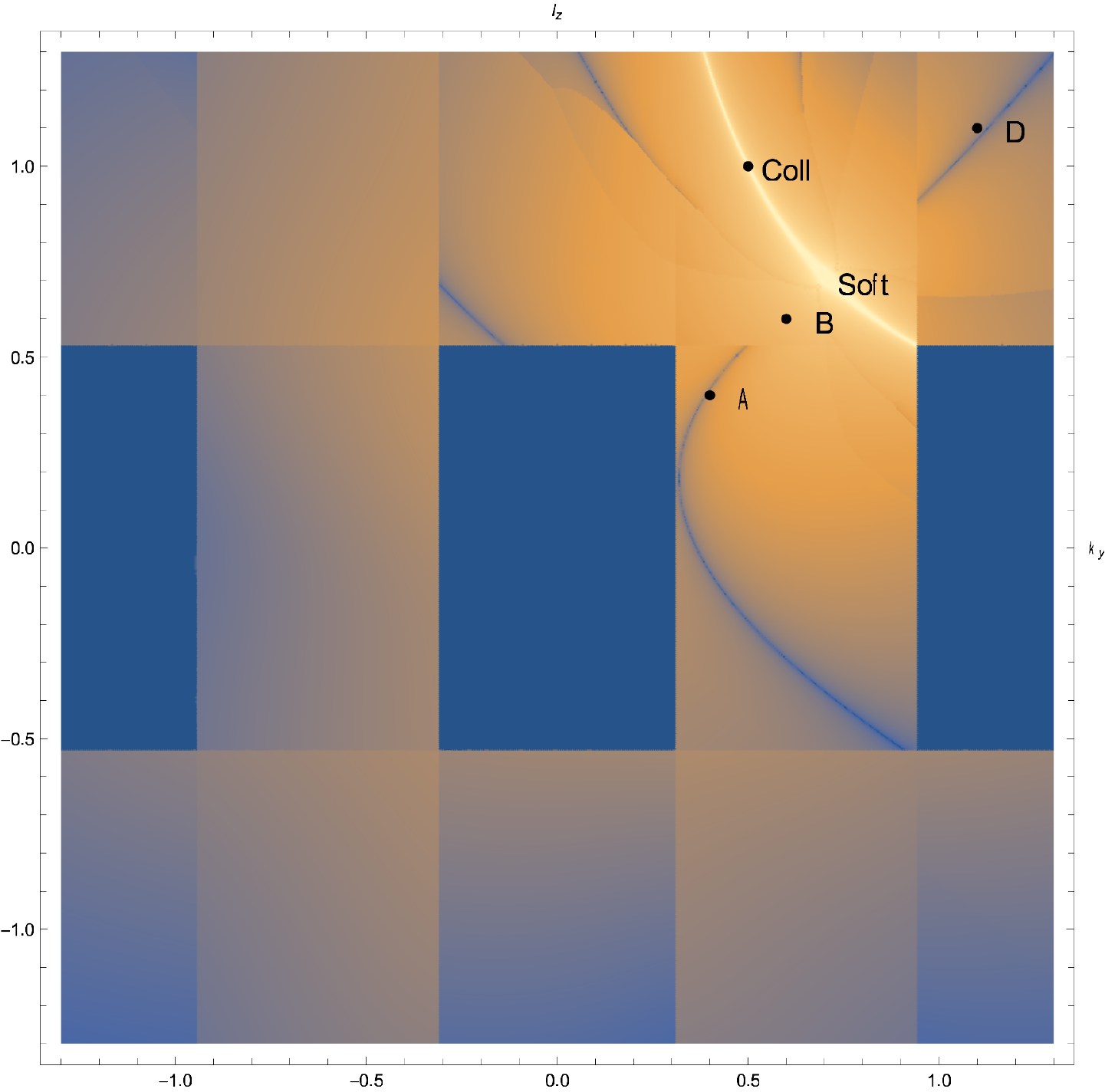}
\caption{$\Re[I_{\mathbf{s}_{1}^{\text{v}}}+I_{\mathbf{s}_{2}^{\text{v}}}]$}\label{DT:fig:DT_deformation_density:LxB_plus_BxL_RE}
\end{subfigure}
\\
\begin{subfigure}[t]{.32\linewidth}
\includegraphics[width=\linewidth]{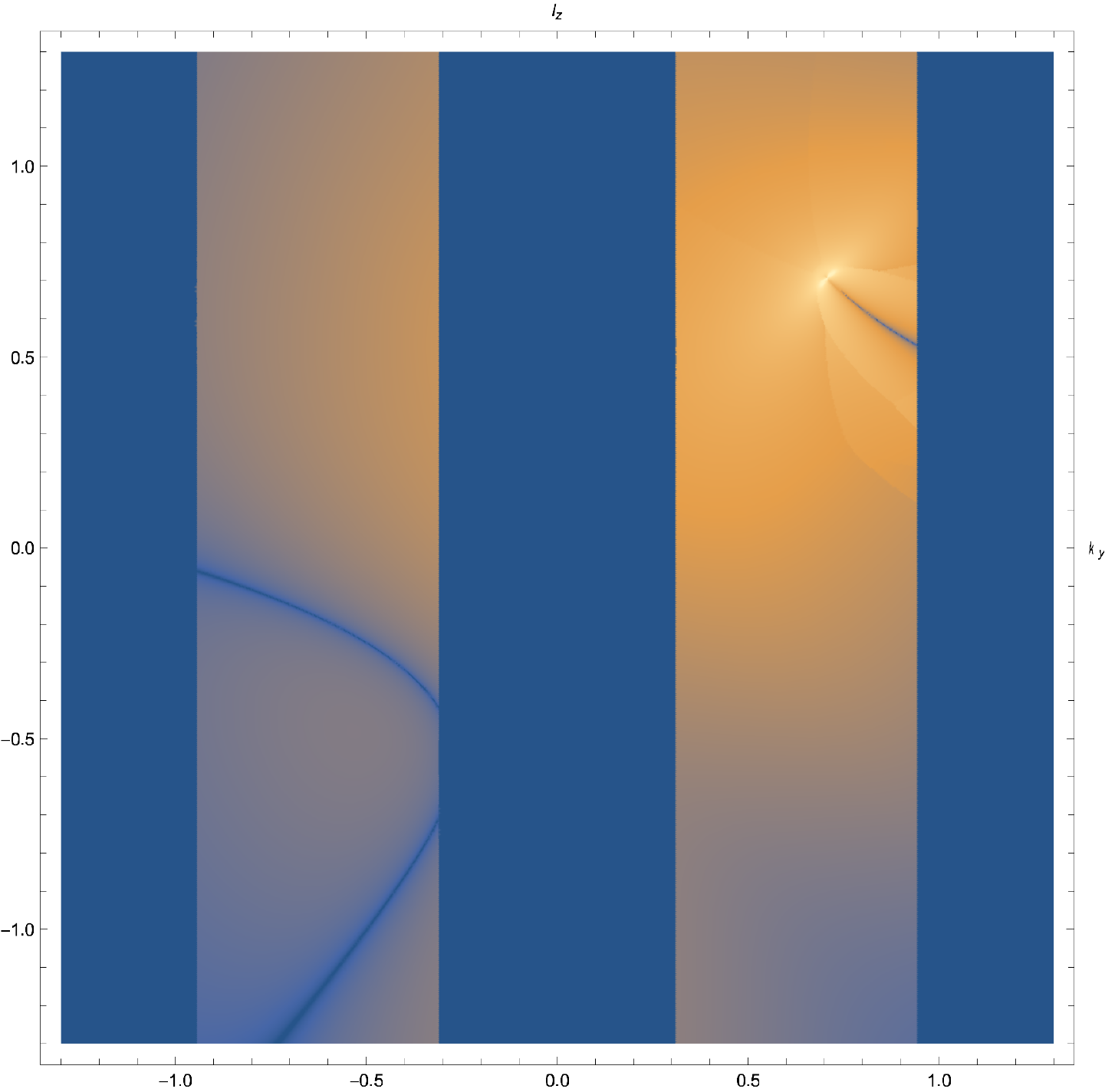}
\caption{$\Im[ I_{\mathbf{s}_{1}^{\text{v}}} ] $}\label{DT:fig:DT_deformation_density:LxB_IM}
\end{subfigure}
\begin{subfigure}[t]{.32\linewidth}
\includegraphics[width=\linewidth]{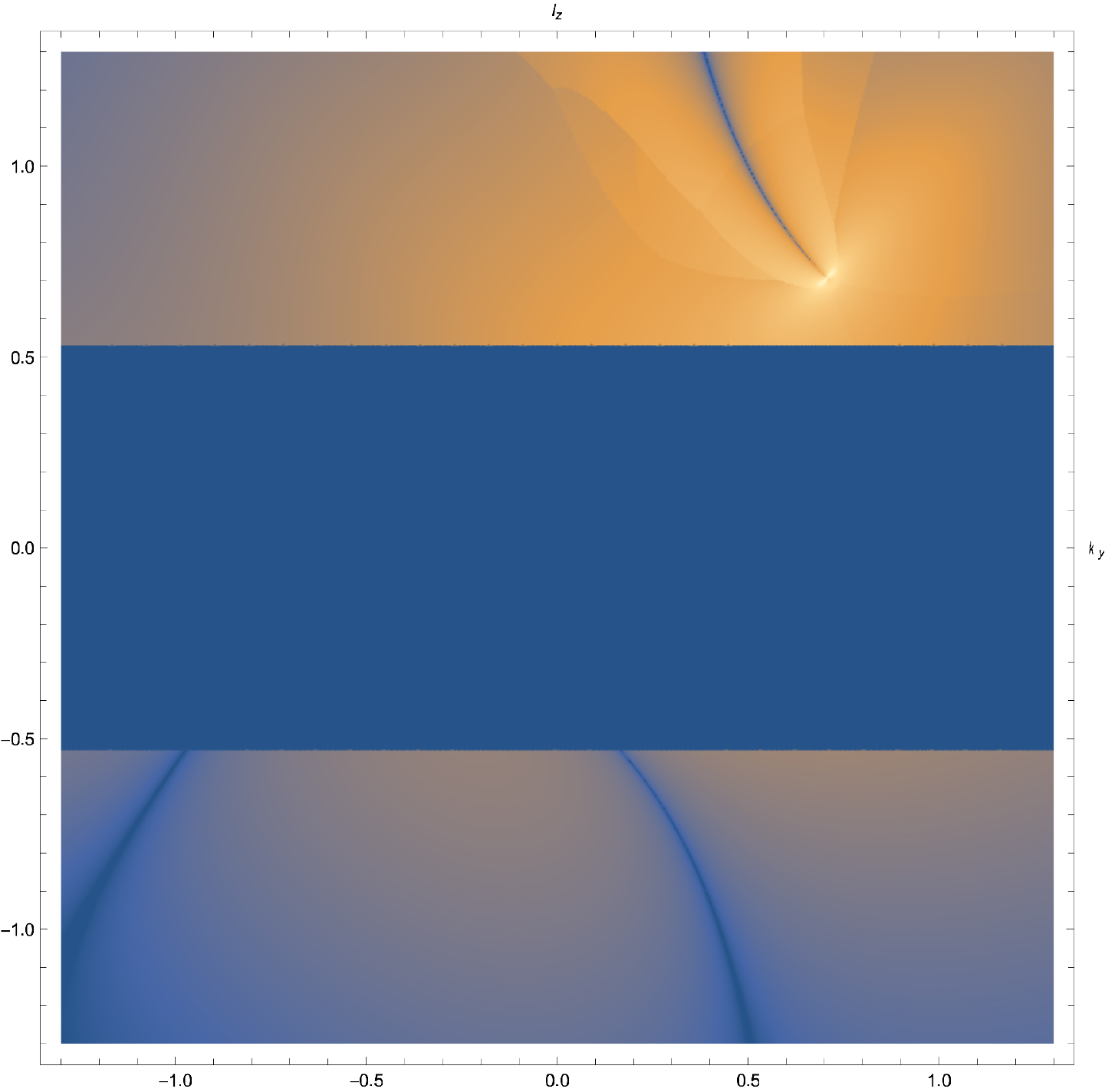}
\caption{$\Im[I_{\mathbf{s}_{2}^{\text{v}}} ]$}\label{DT:fig:DT_deformation_density:BxL_IM}
\end{subfigure}
\begin{subfigure}[t]{.32\linewidth}
\includegraphics[width=\linewidth]{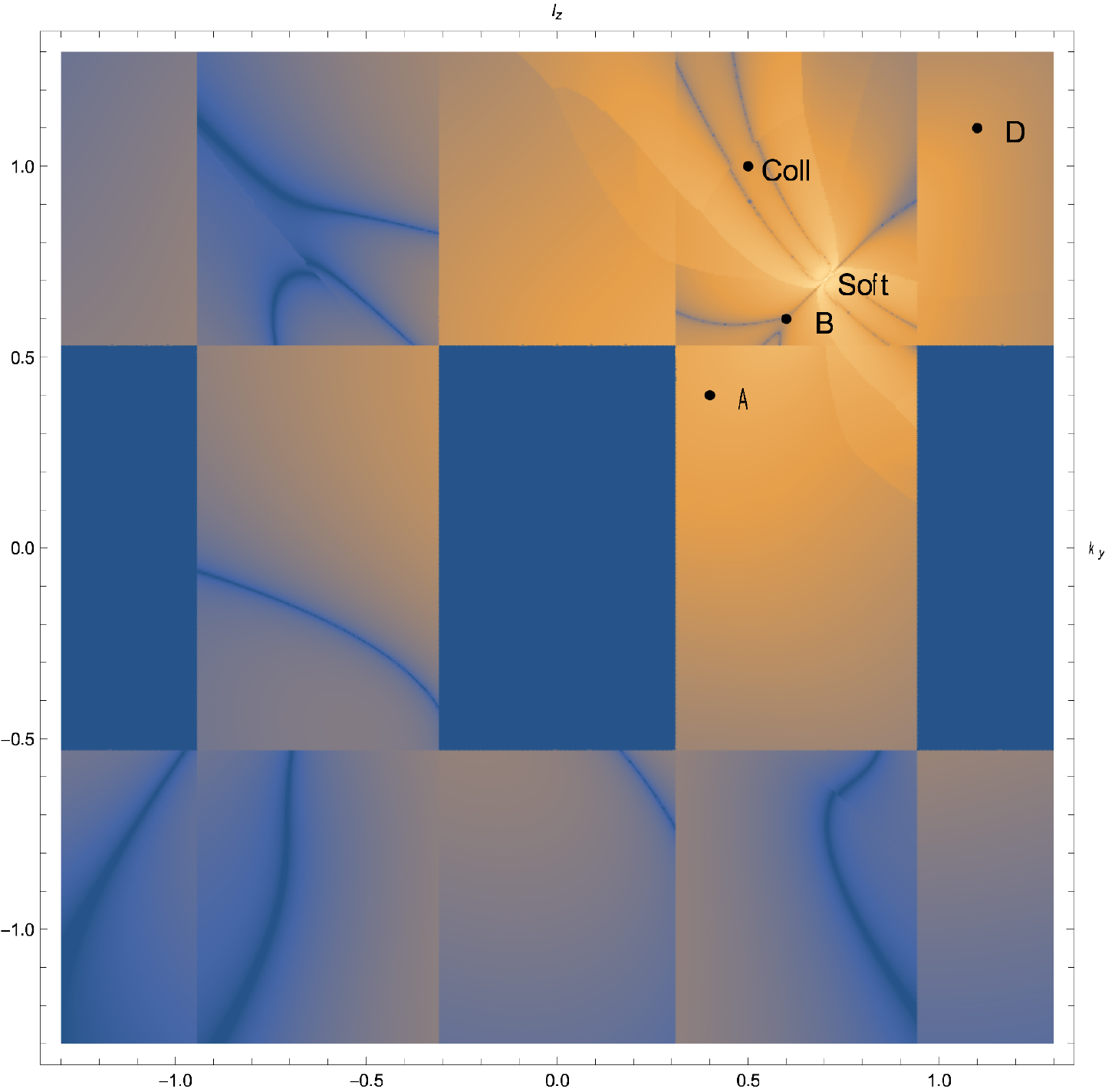}
\caption{$\Im[I_{\mathbf{s}_{1}^{\text{v}}}+I_{\mathbf{s}_{2}^{\text{v}}}]$}\label{DT:fig:DT_deformation_density:LxB_plus_BxL_IM}
\end{subfigure}
\\
\begin{subfigure}[t]{.32\linewidth}
\includegraphics[width=\linewidth]{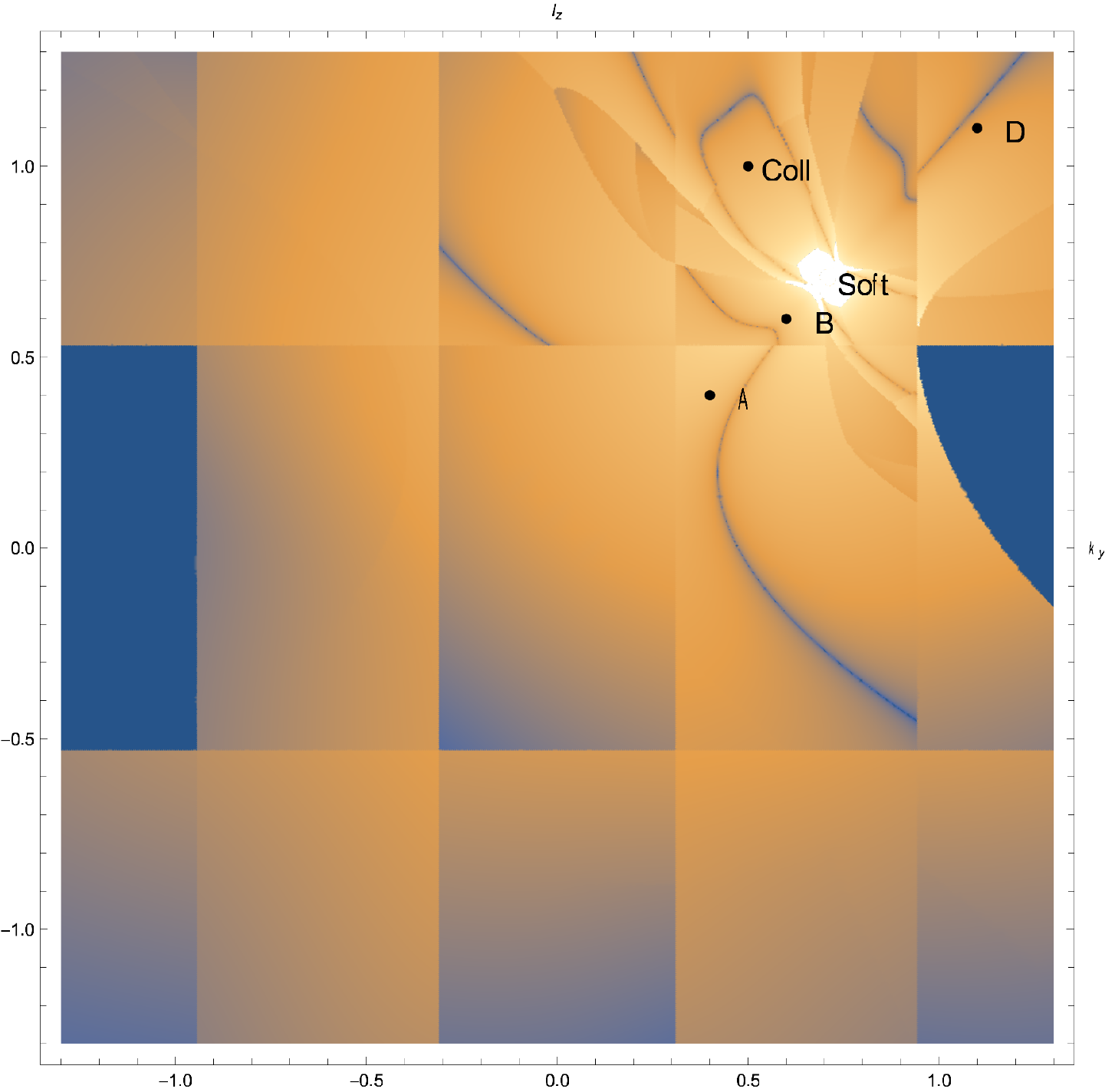}
\caption{$\Re[ I_\Sigma ]$ (all integrands) }\label{DT:fig:DT_deformation_density:RA_plus_RB_plus_LxB_plus_BxL_RE}
\end{subfigure}
\begin{subfigure}[t]{.32\linewidth}
\includegraphics[width=\linewidth]{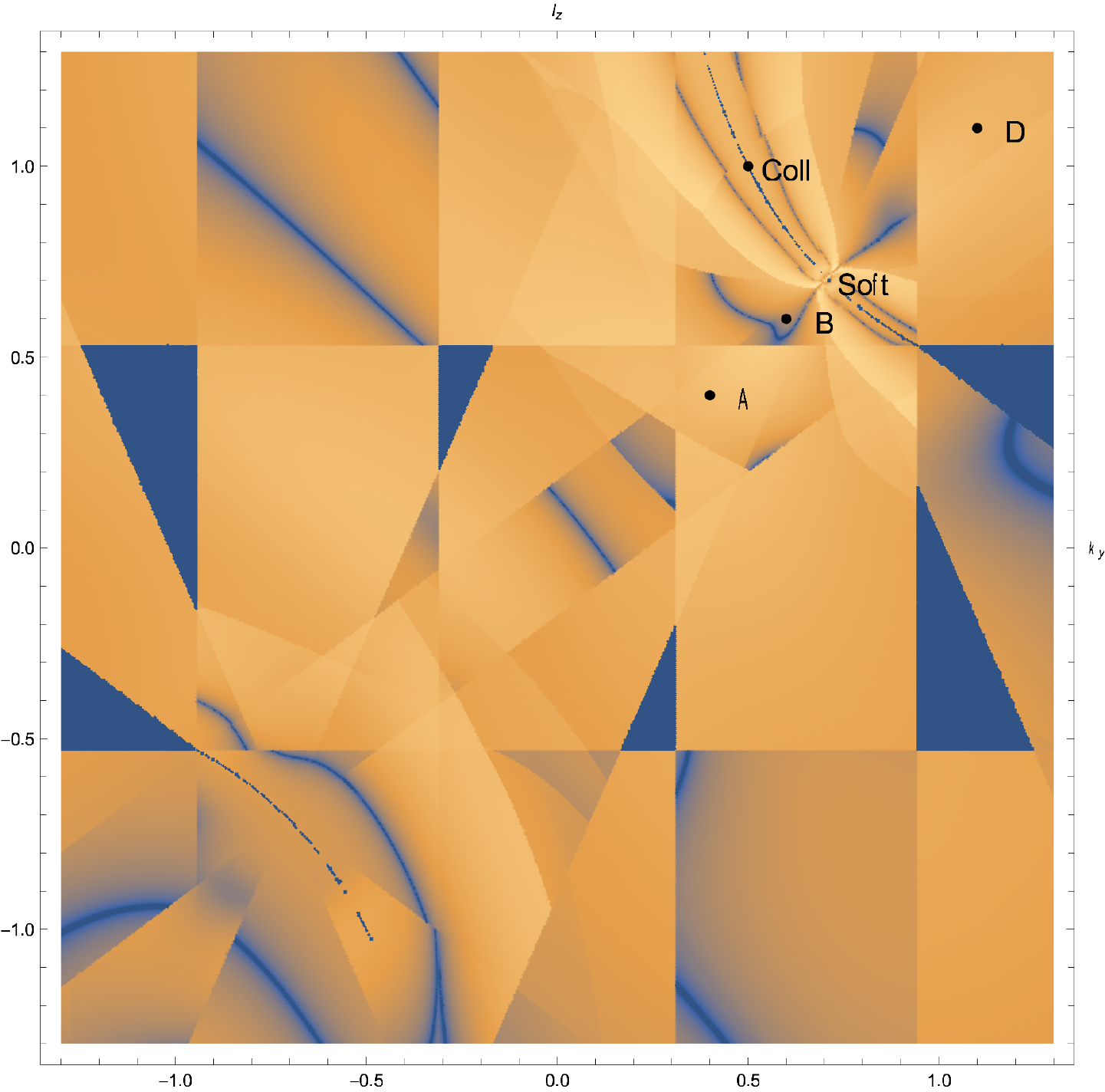}
\caption{$\Im[ I_\Sigma^{(\textrm{MC})} ] $ (multi-channeling)}\label{DT:fig:DT_deformation_density:LxB_plus_BxL_IM_MC}
\end{subfigure}
\begin{subfigure}[t]{.32\linewidth}
\includegraphics[width=\linewidth]{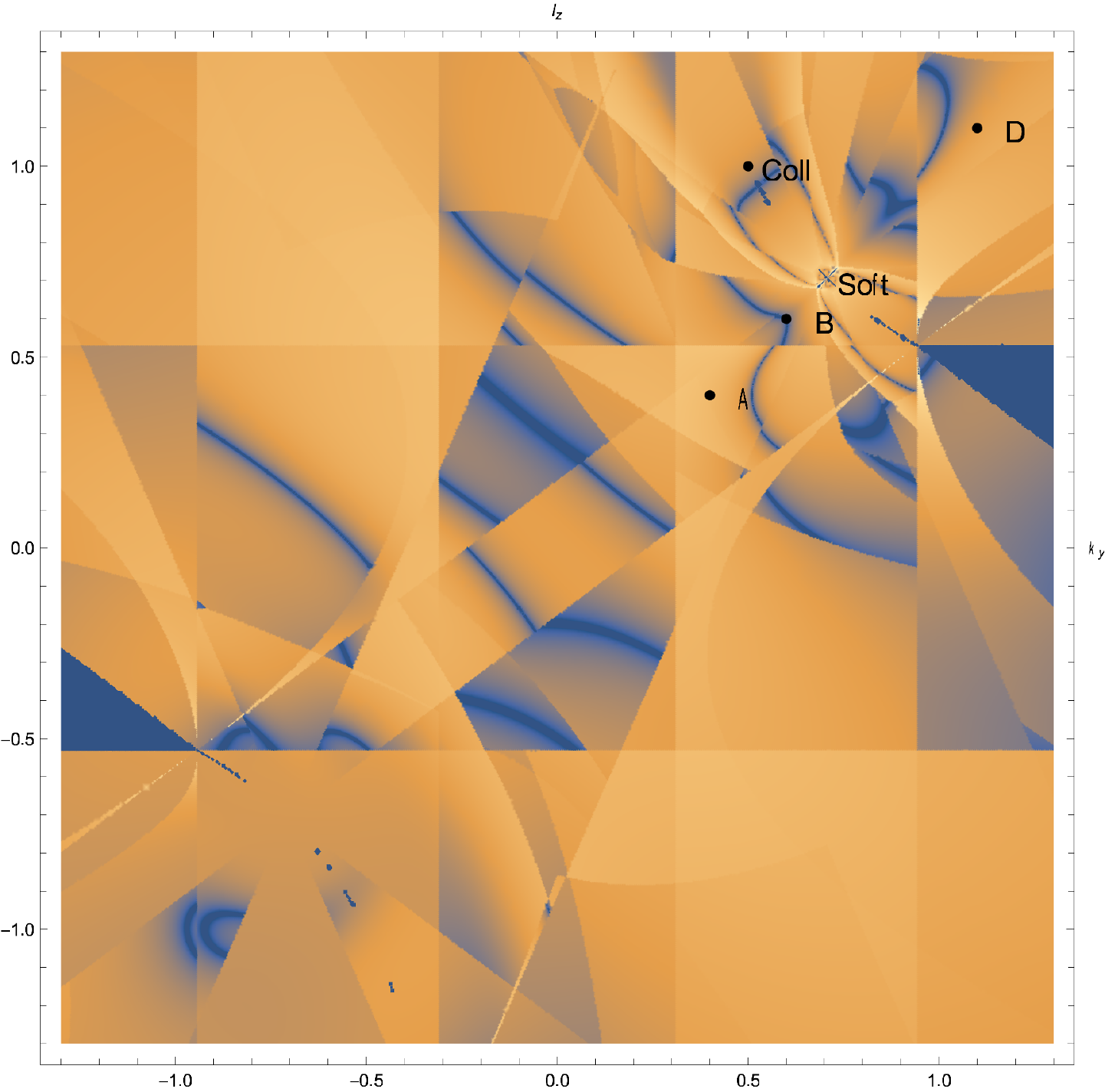}
\caption{{$\Re[ I_\Sigma^{(\textrm{MC})} ]$ (multi-channeling) }}\label{DT:fig:DT_deformation_density:RA_plus_RB_plus_LxB_plus_BxL_RE_MC}
\end{subfigure}
\caption{\label{DT:fig:DT_deformation_density} 
Similar visualisations as in figs.~\ref{DT:fig:DT_no_deformation_density} but for the absolute value of the real and imaginary parts of the Local Unitarity integrands of the \DT{} supergraph when enabling the contour deformation discussed in sect.~\ref{sec:contour_deformation}. The contribution from the real-emission $2I_{\mathbf{s}_{1}^{\text{r}}}$ is not shown since it is not subject to a contour deformation and it is therefore identical to that of fig.~\ref{DT:fig:DT_no_deformation_density:RA_plus_RB}. For the same reason, we have that $\Im[ I_\Sigma ]$ is not shown since it is equal to $\Im[I_{\mathbf{s}_{1}^{\text{v}}}+I_{\mathbf{s}_{2}^{\text{v}}}]$ of fig.~\ref{DT:fig:DT_deformation_density:LxB_plus_BxL_IM}. 
}
\end{figure}

In figs.~\ref{DT:fig:DT_deformation_density} we visualise the \DT{} integrands with the \emph{dynamic deformation} of sect.~\ref{sec:contour_deformation} enabled.
The integrand visualised corresponds to the integrand $I_\Gamma (\vec{k})$ introduced in eq.~\eqref{eq:cross_section_formula_v2}. When multi-channeling is considered, the resulting integrand $I_\Gamma^{(\text{MC})}(\vec{k})$ is defined to be $\sum_{\mathbf{b}_a \in \mathcal{B}_\Gamma} I_\Gamma^{(a)}( \vec{\Psi}^{(b)}_{\vec{k}} ) $ (see eq.~\eqref{eq:cross_section_formula_v2_us}), with $\mathbf{b}_b$ being the loop momentum basis shown in fig.~\ref{fig:Two_SG_epemddx_NLO:DT_RxRA}.
We remind the reader that the contour deformation is not the one constructed for the entire \DT{} supergraph (see vector field of fig.~\ref{fig:DT_causal_flow}),  but it is constructed \emph{independently} for each loop integral remaining in each Cutkosky cut contribution, so as to have real-valued momenta in the observable function and allow to separately accommodate the complex-conjugated causal prescription applying to loops appearing on the right-hand side of the Cutkosky cut.
This implies that in the case of the cuts $\mathbf{s}_{1}^{\text{v}}$ and $\mathbf{s}_{2}^{\text{v}}$ the remaining loops are triangles and their contour deformation must be constructed dynamically for each three-point external kinematics induced by the particular sampling point considered.
The general solution for this problem discussed in sect.~\ref{sec:contour_deformation} is greatly simplified in the case of the \DT{} supergraph as one can show that the maximal overlap structure (see ref.~\cite{Capatti:2019edf}) contains only a single E-surface (the only one present) which moreover always admits the origin in its interior. Thus a radial deformation field with the origin as a source is a valid deformation.
The integrands of figs.~\ref{DT:fig:DT_deformation_density} share most of the features already seen in figs.~\ref{DT:fig:DT_no_deformation_density} with the main difference being the expected regularisation of the E-surfaces, for example at the points labelled ``A'' and ``D''.

When following the collinear arc, we find blue regions (i.e. values approaching zero) in the density plots of $\Im[ I_{\mathbf{s}_{1}^{\text{v}}} ]$ (fig.~\ref{DT:fig:DT_deformation_density:LxB_IM}) and $\Im[I_{\mathbf{s}_{2}^{\text{v}}} ]$ (fig.~\ref{DT:fig:DT_deformation_density:BxL_IM}) resulting from our dampening of the deformation on infrared singularities so as to retain the cancellation with their real-emission counterpart. This results in a non-trivial structure of the complex phase of the complete integrand (fig.~\ref{DT:fig:DT_deformation_density:RA_plus_RB_plus_LxB_plus_BxL_RE}) close to the integrable soft singularity. The multi-channeling procedure successfully dampens that soft singularity and yields an integrand whose real part (fig.~\ref{DT:fig:DT_deformation_density:RA_plus_RB_plus_LxB_plus_BxL_RE_MC}) is now bounded and suitable for numerical integration.
Note that the many discontinuities of the integrand are now not only related to the phase-space cuts but also stem from the Jacobian of our contour deformation, which is discontinuous due to the presence of a {\texttt min} function in the functional form of the dynamic normalisation of our deformation (see eq. 3.33 of ref.~\cite{Capatti:2019edf}).

Despite the introduction of a contour deformation, the differential cross-section is guaranteed to be real-valued as it eventually corresponds to the norm of an overall complex-valued amplitude. In the particular case of LU, this reality condition is realised through the existence of a complex-conjugated partner of each loop present in each Cutkosky cut contribution of each supergraph. The cancellation of the imaginary component of the cross-section only holds at the integrated level since in general it even involves terms belonging to different supergraphs (contrary to the accidentally symmetric \DT{} case where the two complex-conjugated triangle-loop partners happen to belong to the same supergraph).
We use this fact to our advantage as a strong cross-check of the correctness of the LU implementation in \alphaLoop{} as well as an independent assessment of the reliability of the numerical accuracy reported by the Monte-Carlo integrator.

\begin{landscape}
\begin{figure}[ht!]
\centering
\begin{subfigure}[t]{.32\linewidth}
\includegraphics[width=\linewidth]{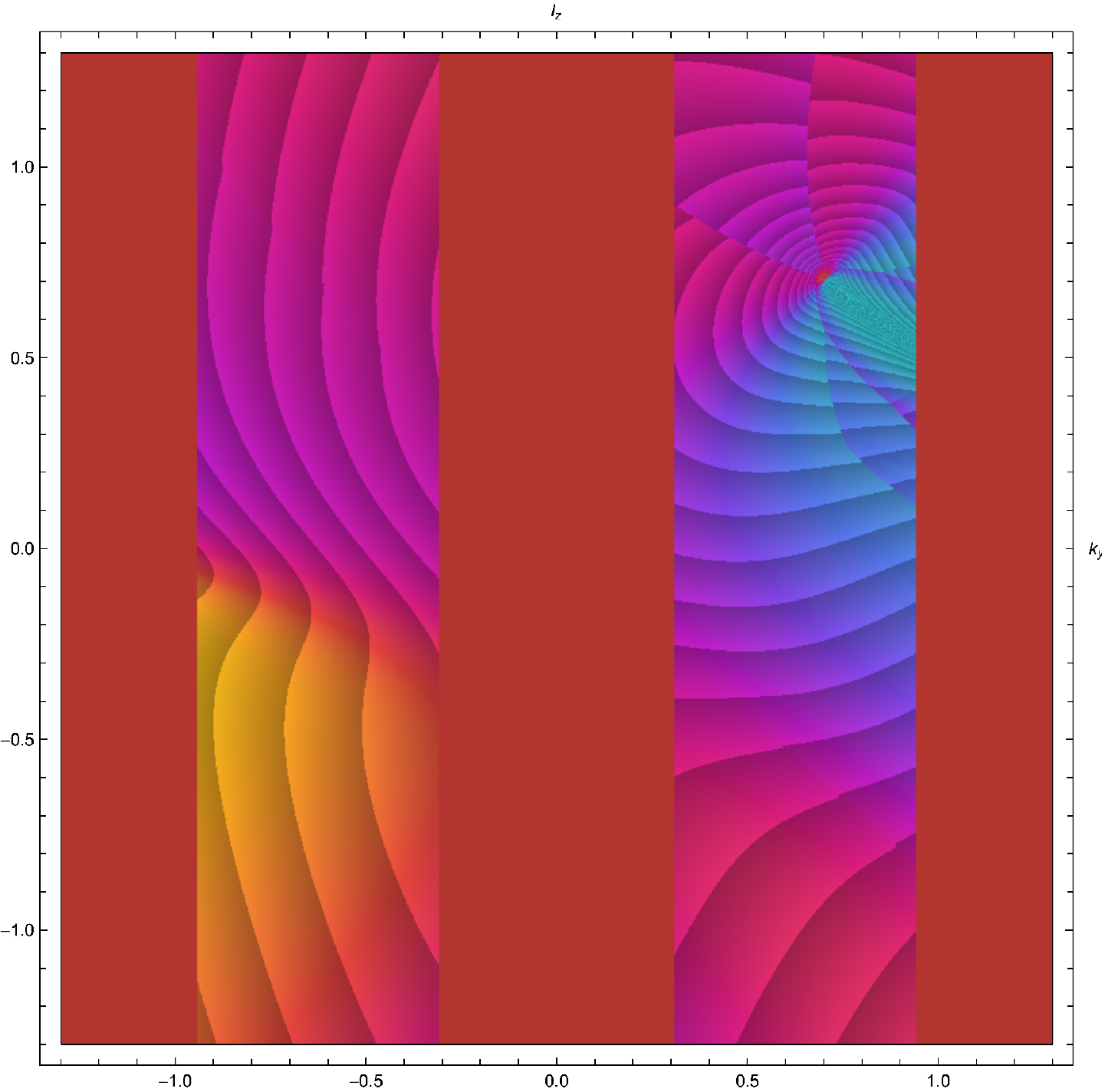}
\caption{$I_{\mathbf{s}_1^{\text{v}}}$}\label{DT:fig:DT_deformation_phase:LxB}
\end{subfigure}
\begin{subfigure}[t]{.32\linewidth}
\includegraphics[width=\linewidth]{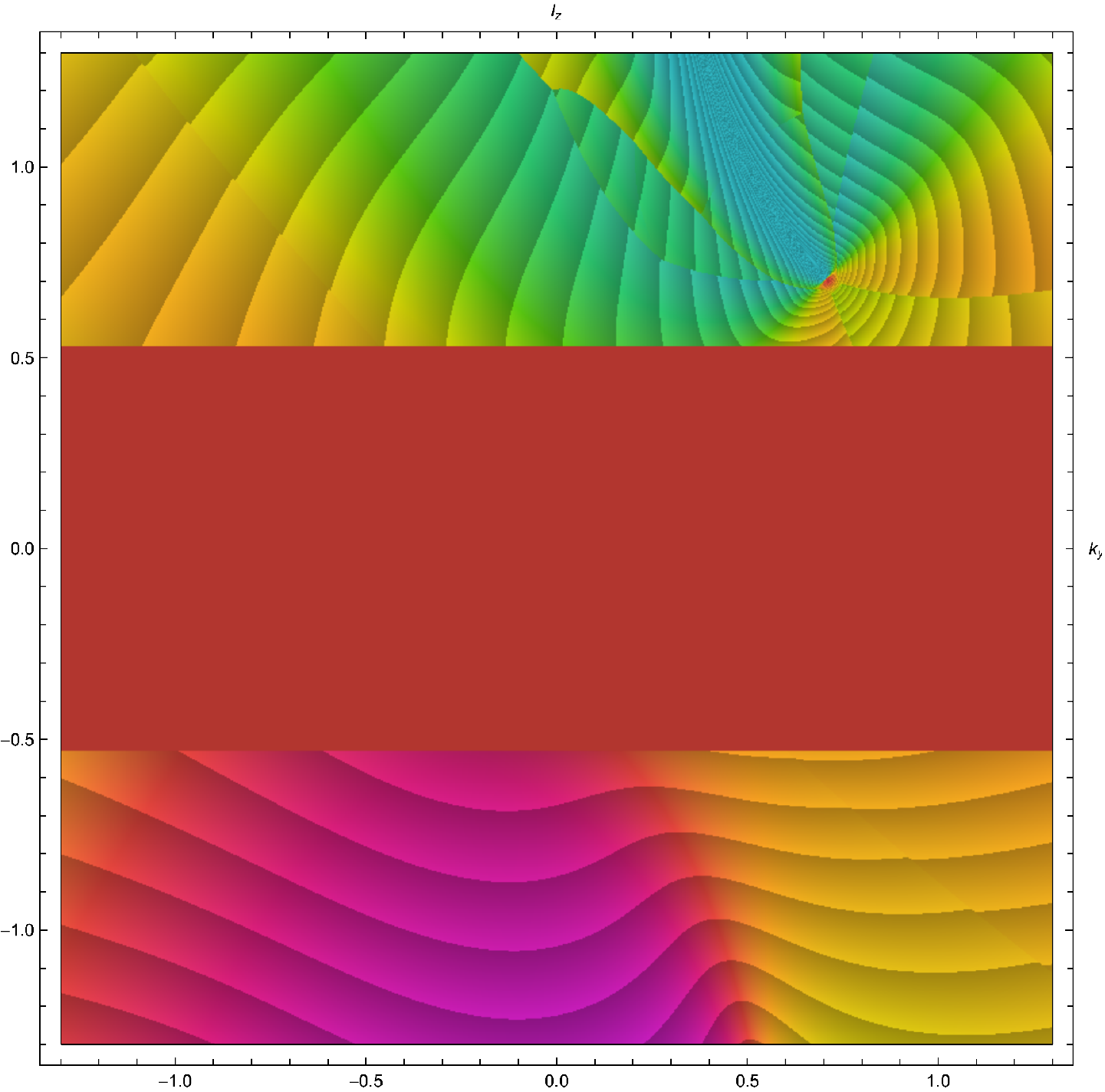}
\caption{$I_{\mathbf{s}_2^{\text{v}}}$}\label{DT:fig:DT_no_deformation_phase:BxL}
\end{subfigure}
\begin{subfigure}[t]{.32\linewidth}
\includegraphics[width=\linewidth]{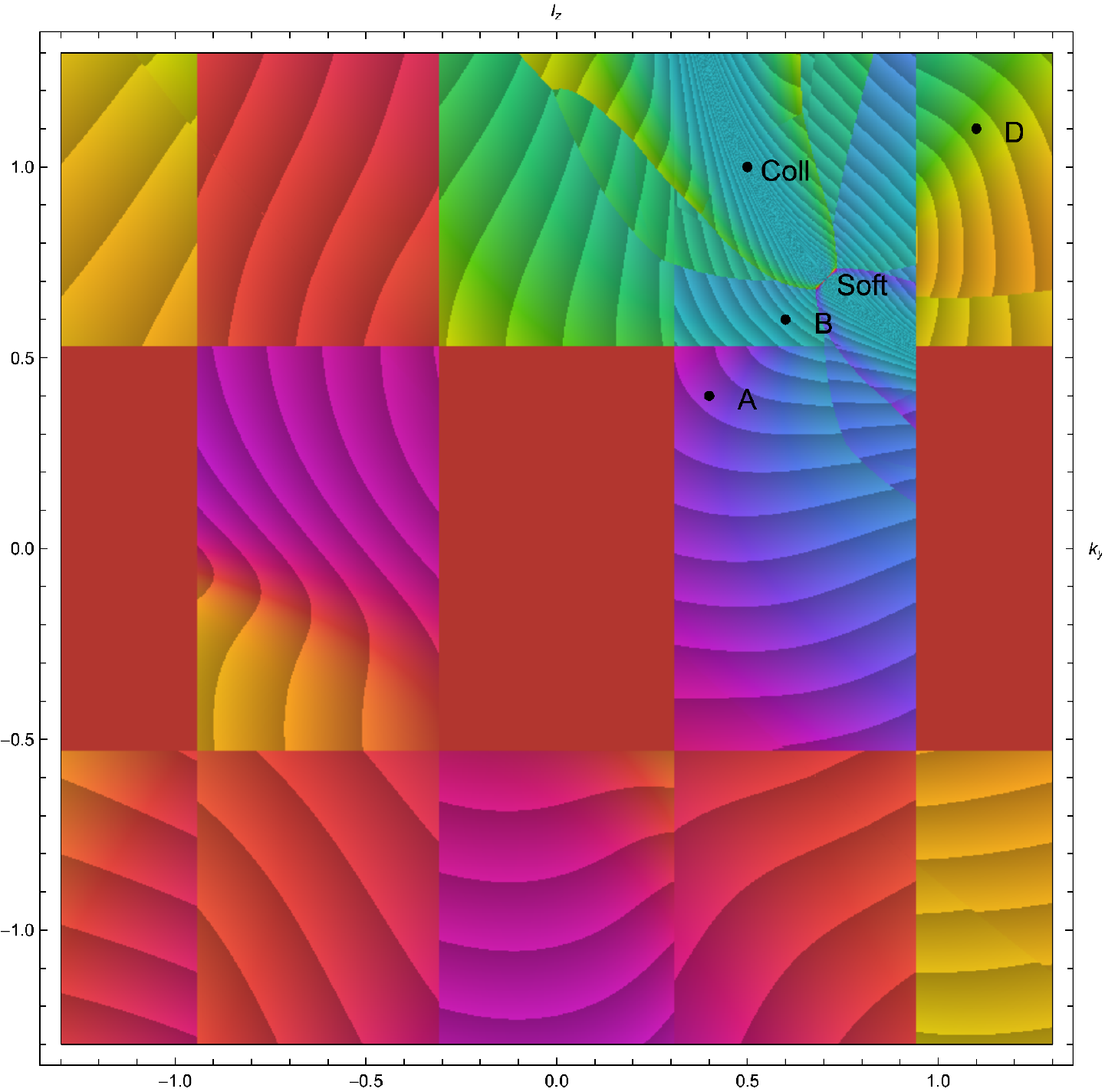}
\caption{$I_{\mathbf{s}_1^{\text{v}}}+I_{\mathbf{s}_2^{\text{v}}}$}\label{DT:fig:DT_deformation_phase:LxB_plus_BxL}
\end{subfigure}
\begin{subfigure}[t]{.02\linewidth}
\includegraphics[width=\linewidth]{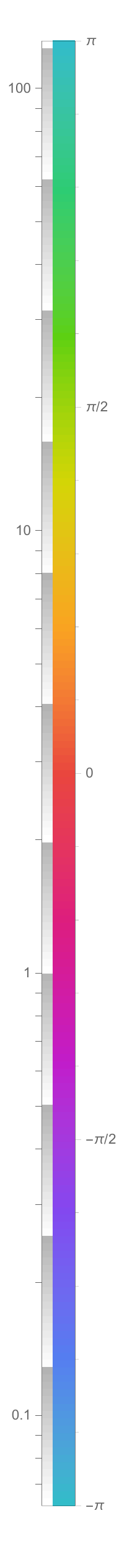}
\end{subfigure}
\\
\begin{subfigure}[t]{.32\linewidth}
\includegraphics[width=\linewidth]{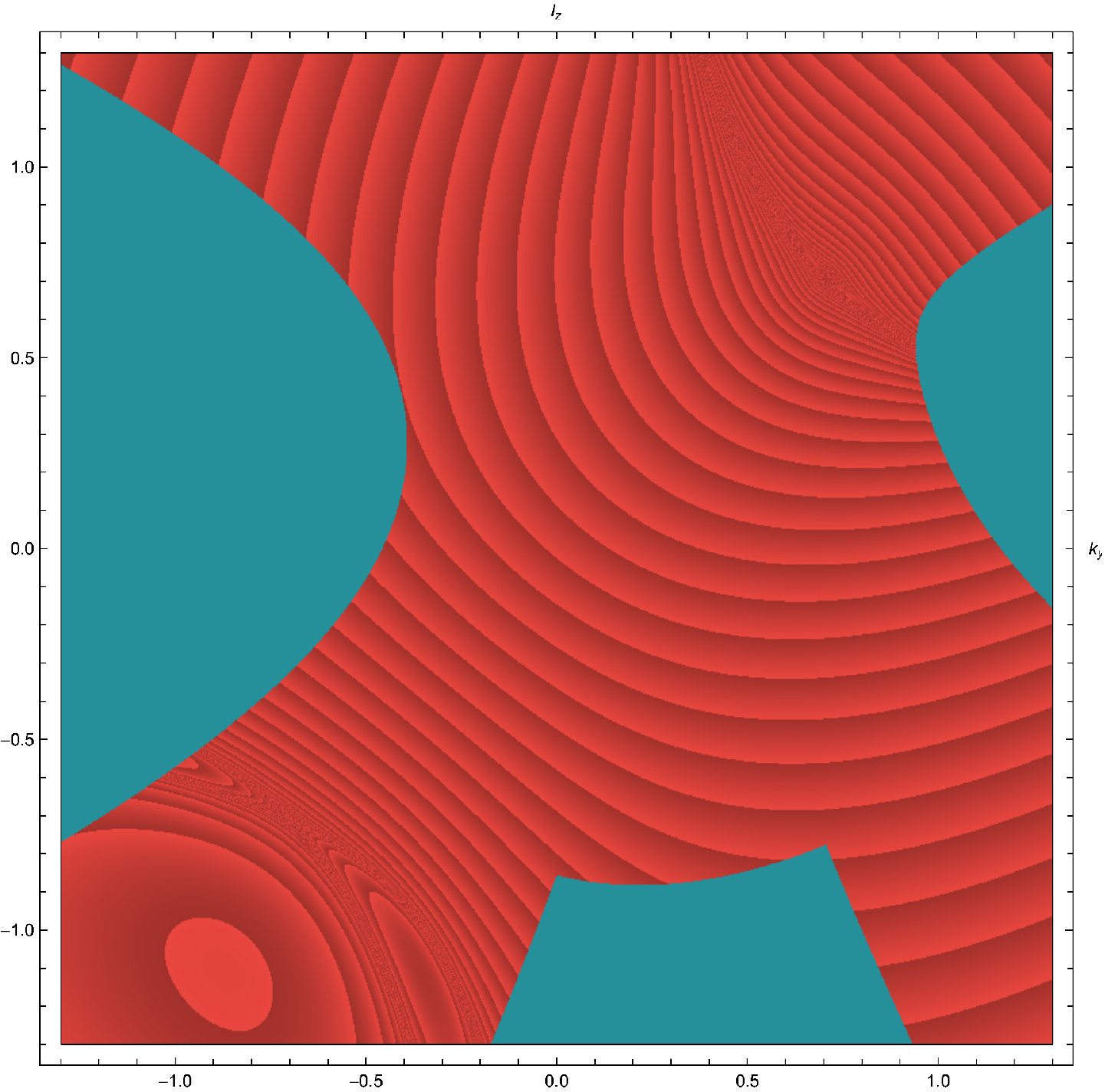}
\caption{$2I_{\mathbf{s}_1^{\text{r}}}$
(cyan regions are PS cuts) }\label{DT:fig:DT_deformation_phase:RA_plus_RB}
\end{subfigure}
\begin{subfigure}[t]{.32\linewidth}
\includegraphics[width=\linewidth]{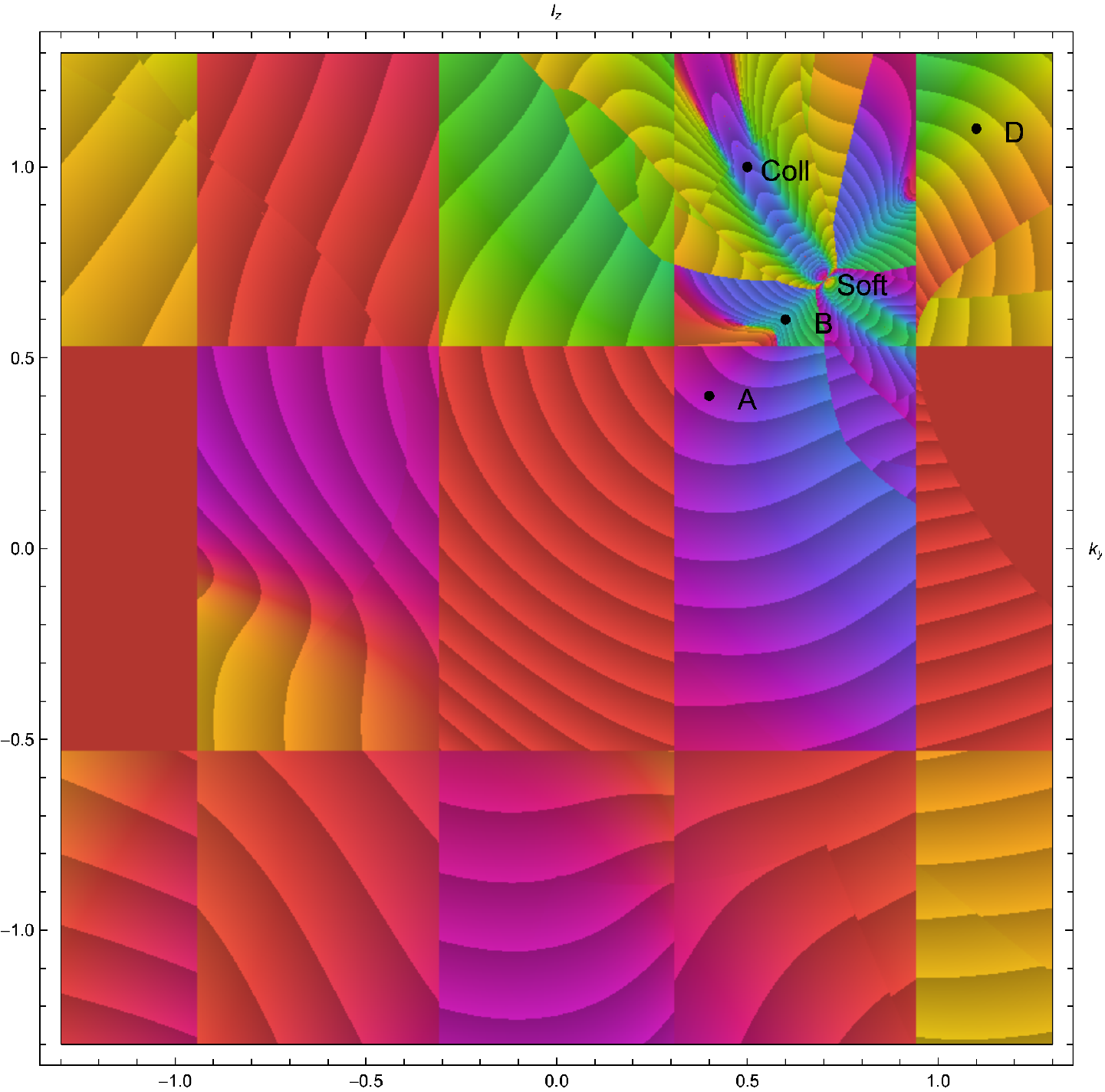}
\caption{$I_\Sigma$ (all Cutkosky cut contributions) }\label{DT:fig:DT_deformation_phase:RA_plus_RB_plus_LxB_plus_BxL}
\end{subfigure}
\begin{subfigure}[t]{.32\linewidth}
\includegraphics[width=\linewidth]{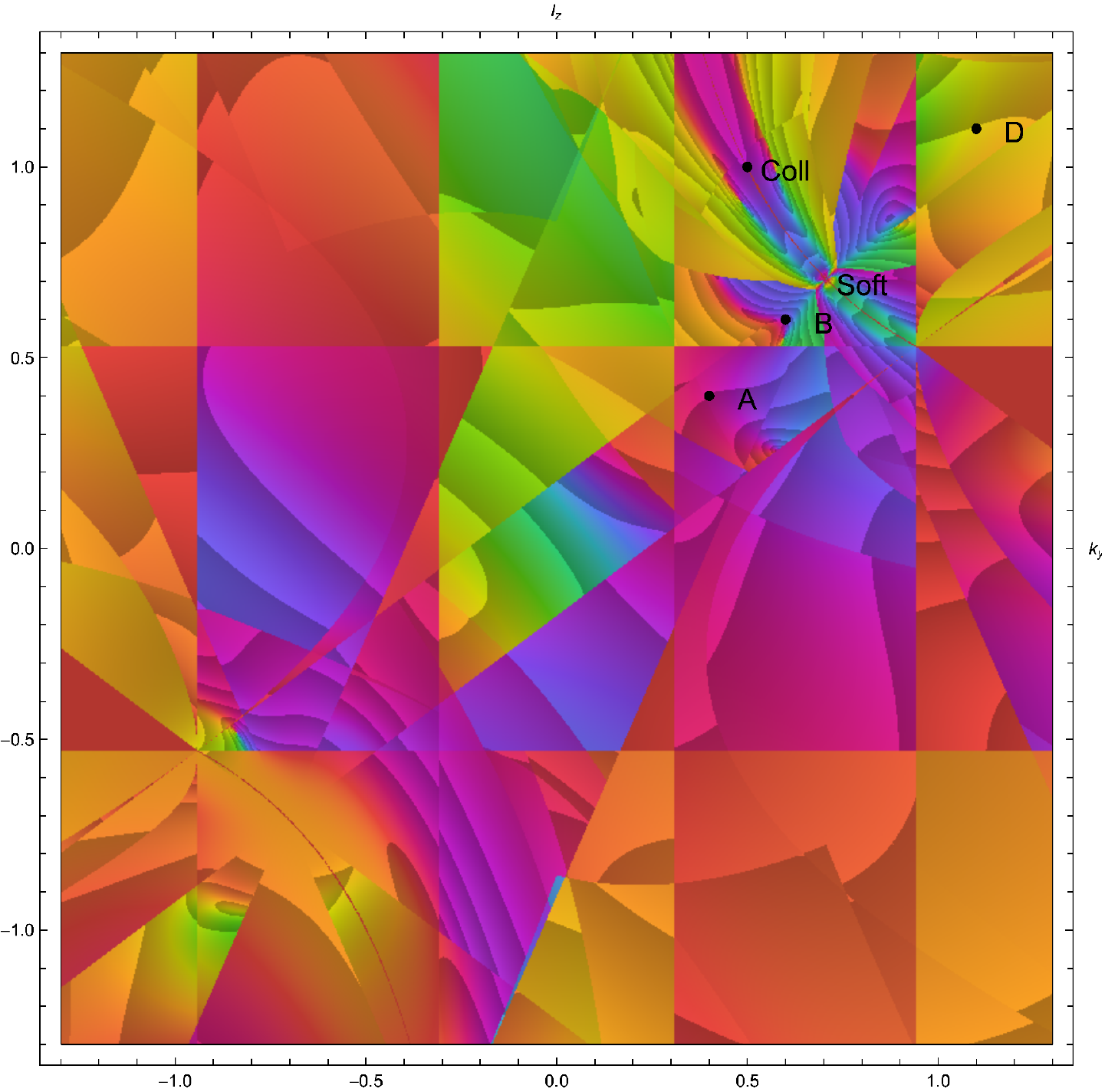}
\caption{$I_\Sigma^{(\textrm{MC})}$ (multi-channeling and incl. param. jac.)  }\label{DT:fig:DT_deformation_phase:RA_plus_RB_plus_LxB_plus_BxL_MC}
\end{subfigure}
\begin{subfigure}[t]{.02\linewidth}
\includegraphics[width=\linewidth]{Figures/MathematicaPlots/DeformationPhase/DeformationPhase_legend_bar.pdf}
\end{subfigure}
\caption{\label{DT:fig:phase}
Complex 2D plots of the Local Unitarity double-triangle supergraph integrand from \alphaLoop{} evaluated at $(\vec{k},\vec{l})=((0,k_y,\frac{1}{\sqrt{2}}),(0,\frac{1}{\sqrt{2}},l_z))$ with a contour deformation enabled. The colour shading denotes logarithmic isolines of the norm of the integrand while the colour indicate the value of its complex phase.
}
\end{figure}
\end{landscape}

\begin{landscape}
\begin{figure}[ht!]
\centering
\begin{subfigure}[t]{.32\linewidth}
\includegraphics[width=\linewidth]{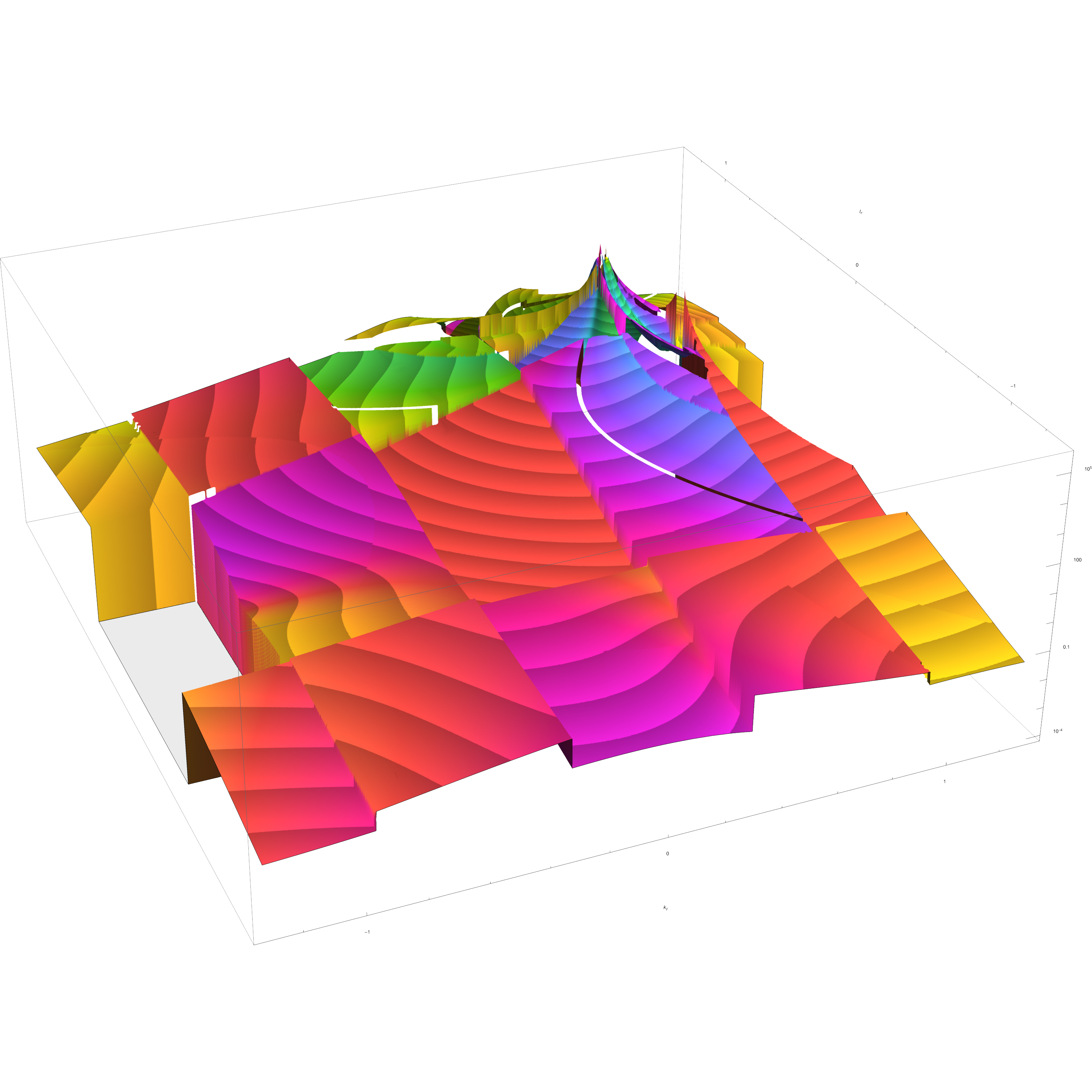}
\caption{3D overview of $I_\Sigma$ within $[-1.3,1.3]\times[-1.3,1.3]$}\label{DT:fig:DT_deformation_3D_overview}
\end{subfigure}
\begin{subfigure}[t]{.32\linewidth}
\includegraphics[width=\linewidth]{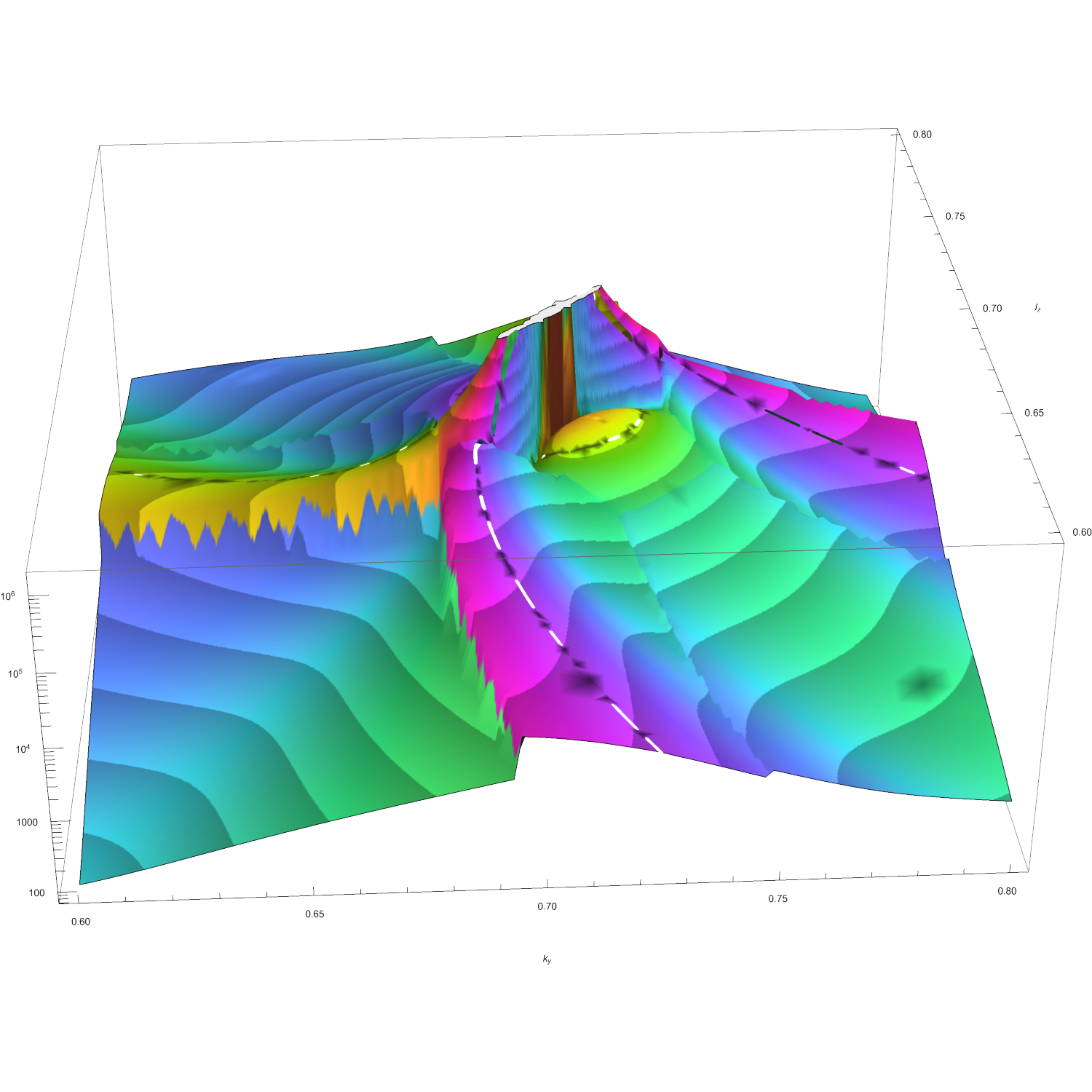}
\caption{3D close-up of $I_\Sigma$ within $[0.6,0.8]\times[0.6,0.8]$}\label{DT:fig:DT_deformation_3D_zoomin}
\end{subfigure}
\begin{subfigure}[t]{.32\linewidth}
\includegraphics[width=\linewidth]{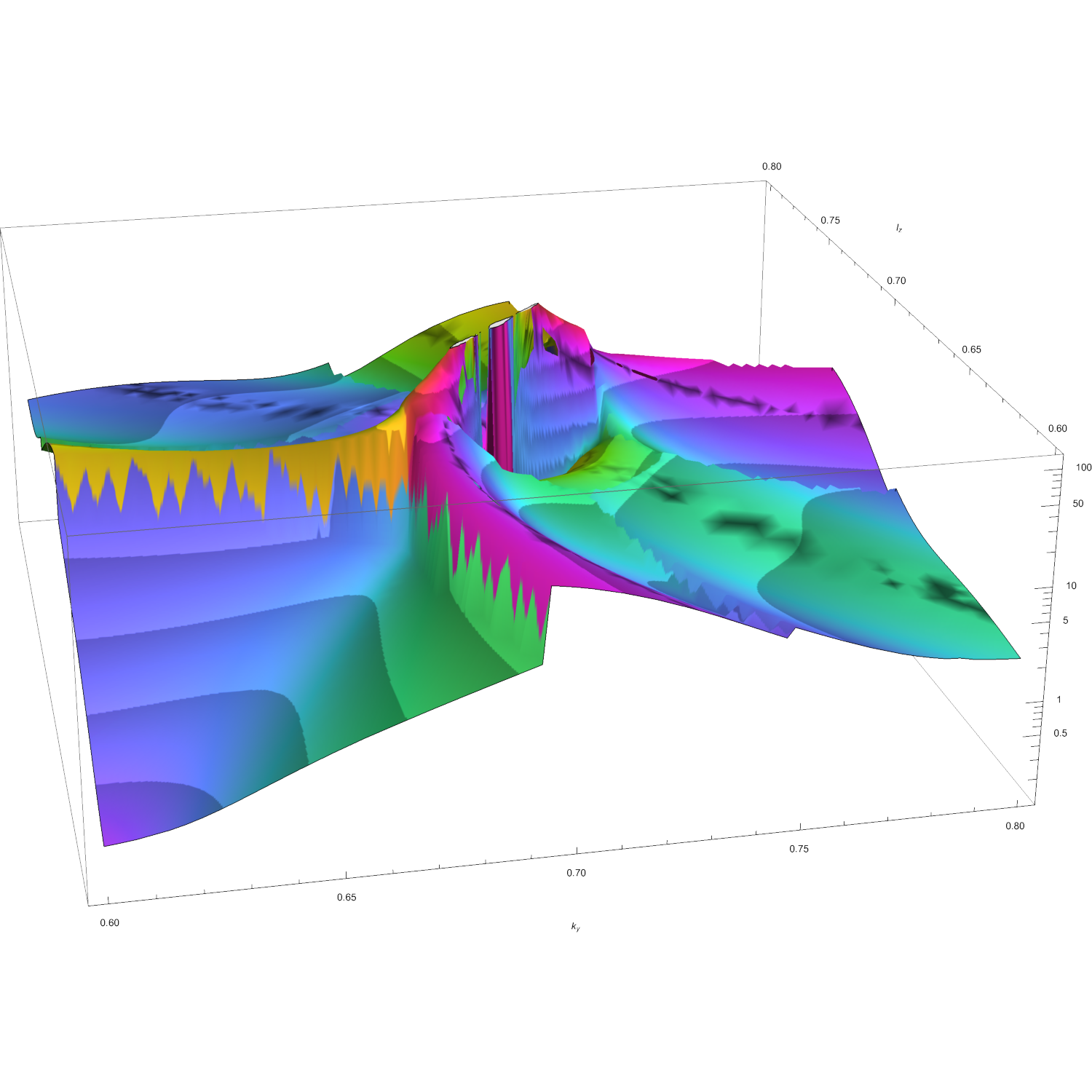}
\caption{3D close-up of $I_\Sigma^{(\textrm{MC})}$ within $[0.6,0.8]\times[0.6,0.8]$}\label{DT:fig:DT_deformation_3D_zoomin_MC}
\end{subfigure}
\begin{subfigure}[t]{.02\linewidth}
\includegraphics[width=\linewidth]{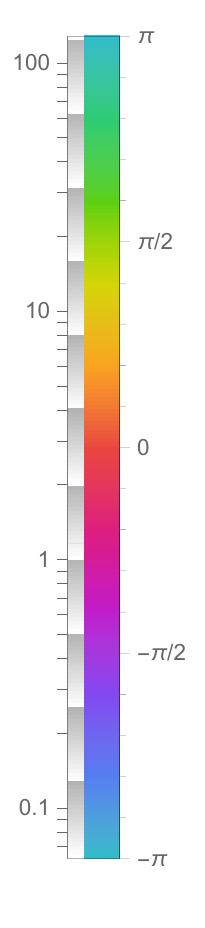}
\end{subfigure}
\\
\begin{subfigure}[t]{.32\linewidth}
\includegraphics[width=\linewidth]{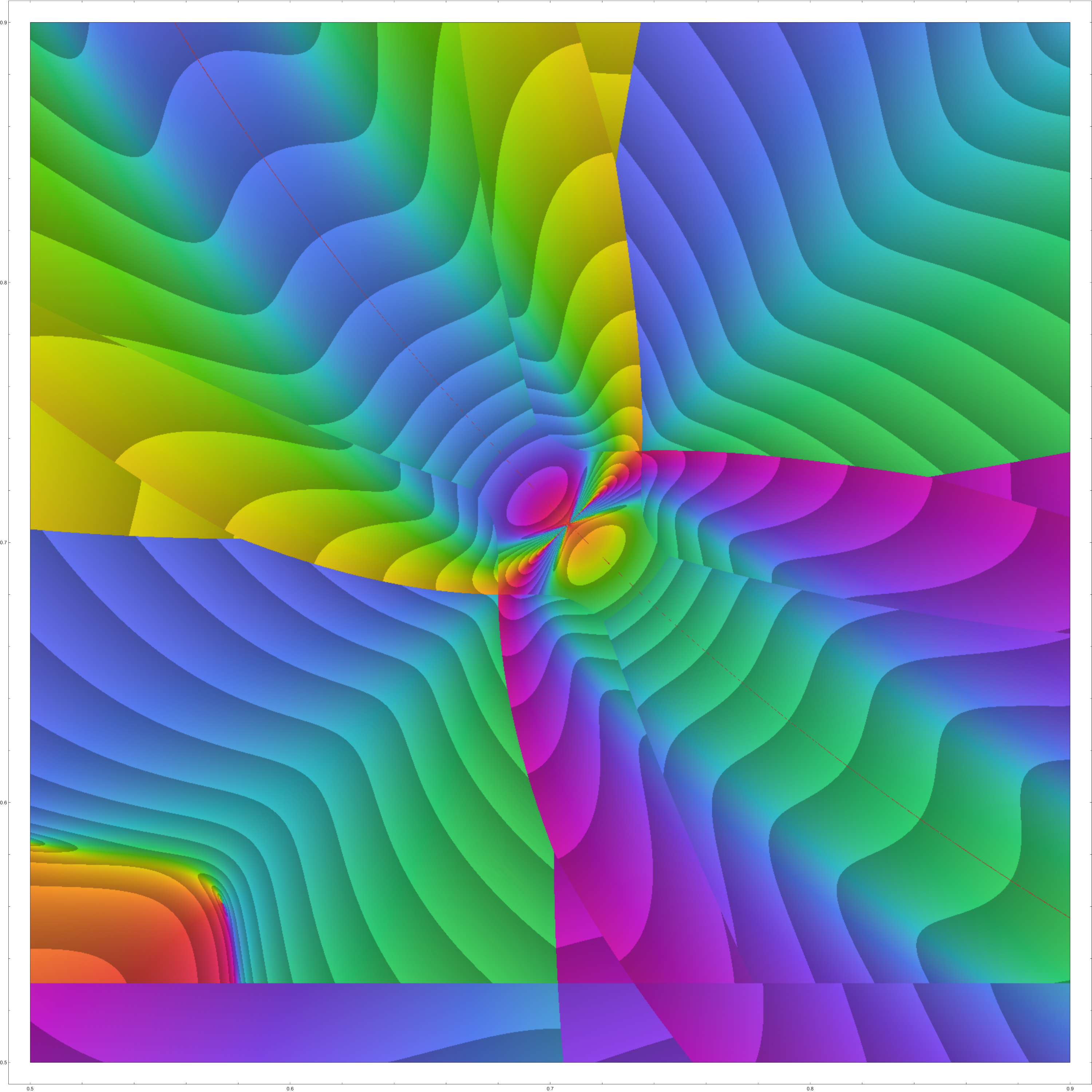}
\caption{Zoom-in of $I_\Sigma$ within $[0.5,0.9]\times[0.5,0.9]$ }\label{DT:fig:DT_deformation_2D_zoomin}
\end{subfigure}
\begin{subfigure}[t]{.32\linewidth}
\includegraphics[width=\linewidth]{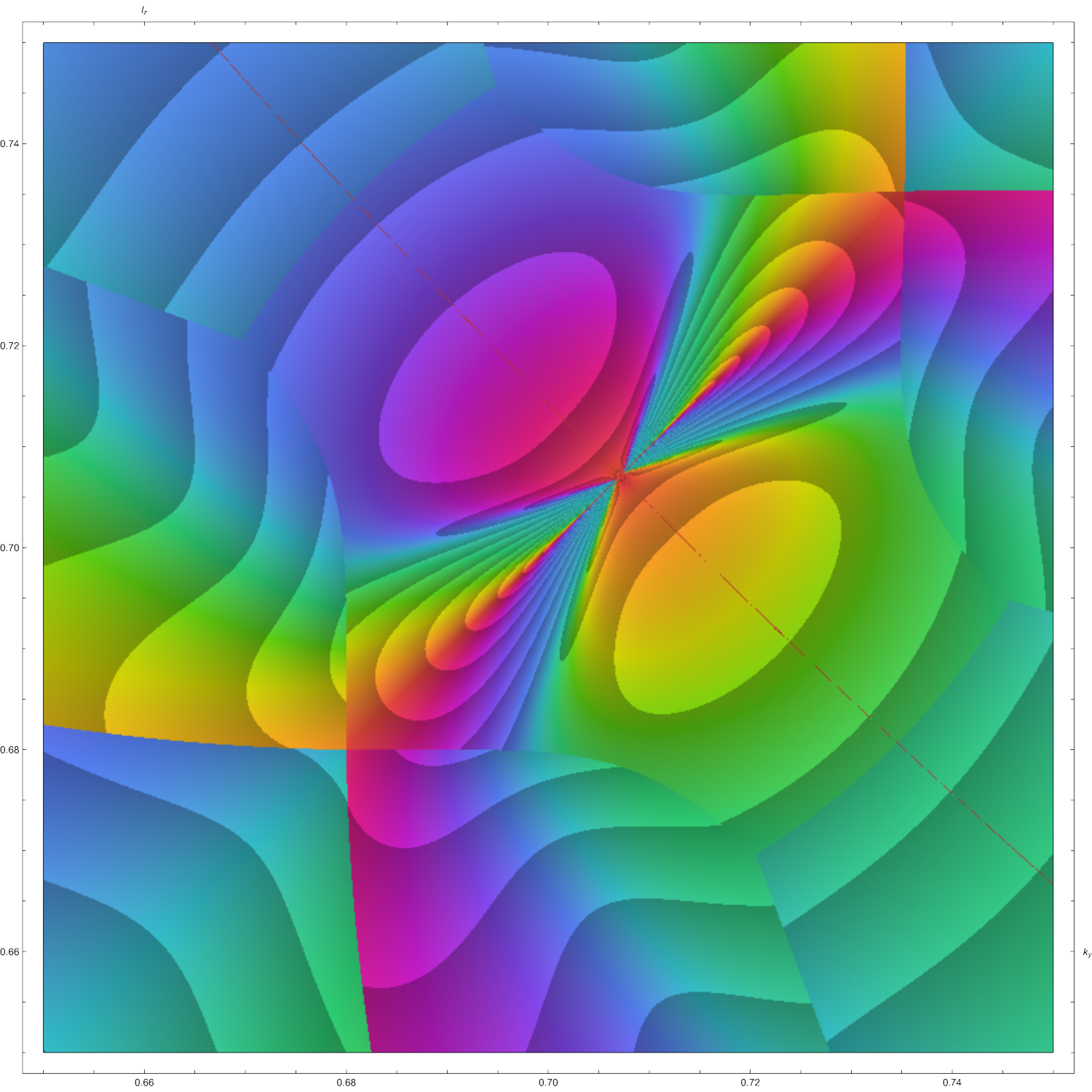}
\caption{Close-up of $I_\Sigma$ within $[0.65,0.75]\times[0.65,0.75]$ }\label{DT:fig:DT_deformation_2D_closeup}
\end{subfigure}
\begin{subfigure}[t]{.32\linewidth}
\includegraphics[width=\linewidth]{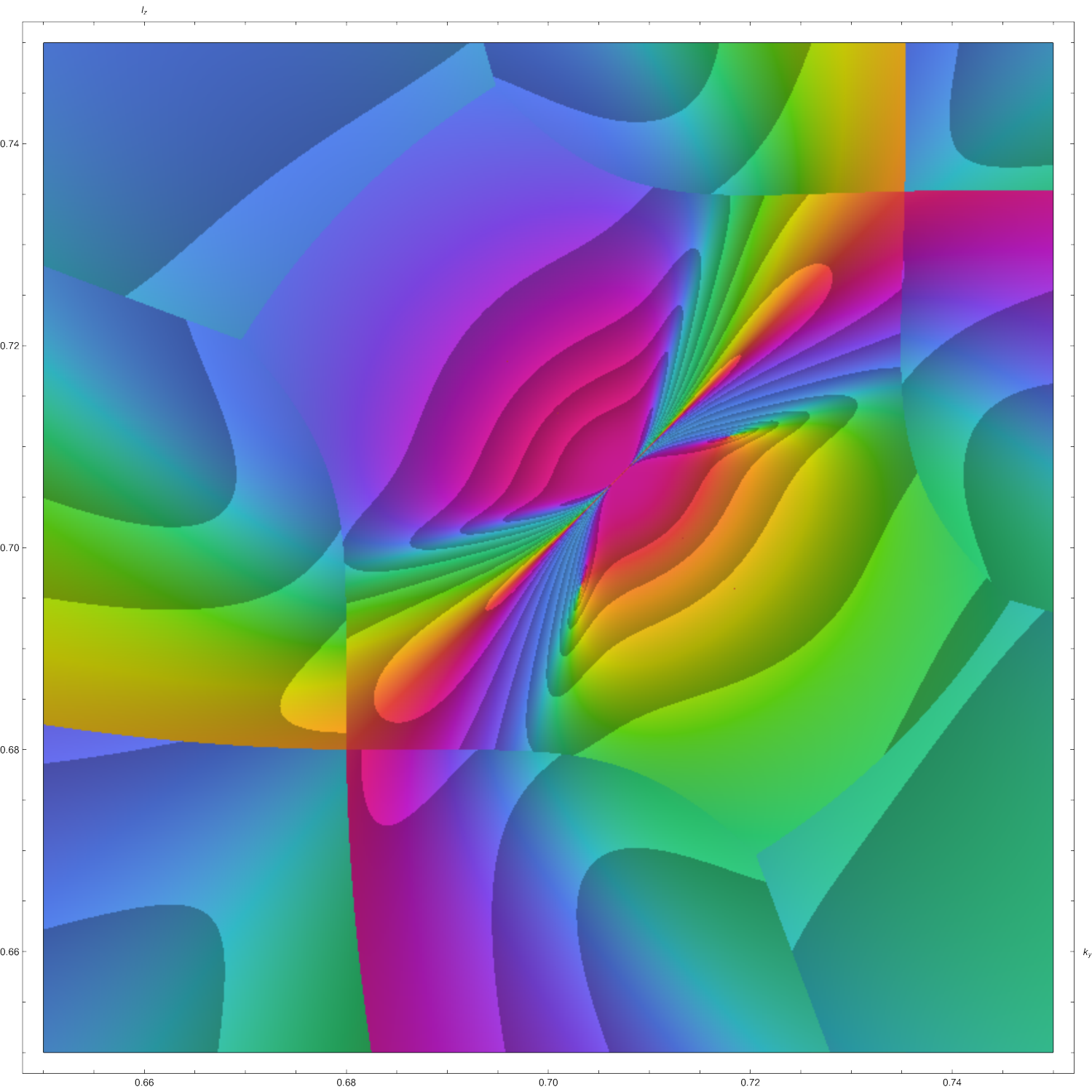}
\caption{Close-up of $I_\Sigma^{(\textrm{MC})}$ within $[0.65,0.75]\times[0.65,0.75]$  }\label{DT:fig:DT_deformation_2D_closeup_MC}
\end{subfigure}
\begin{subfigure}[t]{.02\linewidth}
\includegraphics[width=\linewidth]{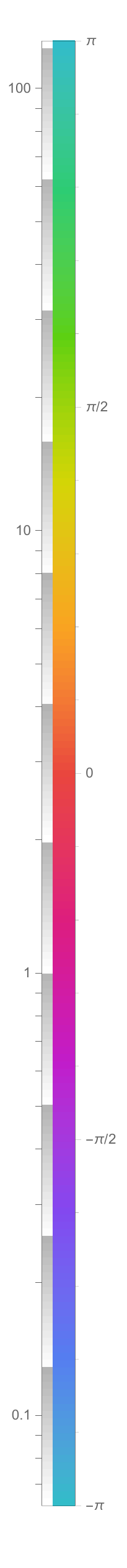}
\end{subfigure}
\caption{\label{DT:fig:soft_zoom_ins}
Various 2D and 3D visualisations of the Local Unitarity double-triangle supergraph contour-deformed integrand from \alphaLoop\ for various ranges of the input values $(k_y,l_z)$.
The colour shading denotes logarithmic isolines of the norm of the integrand while the colour indicates the value of its complex phase.
}
\end{figure}
\end{landscape}

The series of figs.~\ref{DT:fig:DT_deformation_density} capture the core features of the LU integrand that we want to highlight, but it does not reveal more detailed aspects such as the complex phase of the argument since we always considered absolute value of its real and imaginary component. 
We therefore complement our visualisation with figs.~\ref{DT:fig:phase}, where the colouring reveals the phase of the integrand and its shading marks isolines reveals the logarithmic progression of its magnitude across the plane. We find the expected phase-shift of $\pi/2$ of the virtual integrands when crossing the soft singularity along the collinear line as the deformation infrared dampening kicks in.
When comparing the phase of $I_{\mathbf{s}_{2}^{\text{v}}}$ in fig.~\ref{DT:fig:DT_deformation_phase:LxB} and $I_{\mathbf{s}_{1}^{\text{v}}}$ in fig.~\ref{DT:fig:DT_no_deformation_density:BxL} on their respective side of the collinear singularity where their deformation is active, we find a phase-shift of $\pi$ stemming from the complex-conjugation of the causal prescription of the triangle loop placed on the left-hand side of the Cutkosky cut, which flips the sign of the deformation vector $\vec{\kappa}$.
The combination of the virtual contribution shown in fig.~\ref{DT:fig:DT_deformation_phase:LxB_plus_BxL} has a constant phase along the collinear singularity since the real part is divergent and negative. Once the positive divergent real-emission contribution of fig.~\ref{DT:fig:DT_deformation_phase:RA_plus_RB} is added, we find in fig.~\ref{DT:fig:DT_deformation_phase:RA_plus_RB_plus_LxB_plus_BxL} that the cancellation of this large negative real part leaves off a remnant with a complicated phase structure.
We observe in fig.~\ref{DT:fig:DT_deformation_phase:RA_plus_RB_plus_LxB_plus_BxL_MC} that the multi-channeling procedure dampens the integrable singularity without disrupting much of the phase-structure of the overall integrand $I_\Sigma$ around that region, as one expects since the multi-channeling Jacobian responsible for the dampening is purely real.

Finally, the interesting detailed structure around the soft configuration is hard to resolve within the selected range. For this reason, we also present alternative visualisations homing in on the region of interest in figs~\ref{DT:fig:soft_zoom_ins}. This reveals many discontinuity lines that result from the interplay of the various competing constraints on the deformation normalisation realised through the different functional forms of the scaling factors $\lambda_i$ of eq. 3.33 of ref.~\cite{Capatti:2019edf} which eventually send the deformation to zero on the soft point.
The clipped regions of the 3D visualisation stem from discontinuities of $I_\Sigma$ while the few ``spots'' on the multi-channeling plots come from evaluations deemed unstable by our strict precision requirements and consequently sent to zero.
Comparing the normalisation of the z-axis scale between figs.~\ref{DT:fig:DT_deformation_3D_zoomin} and~\ref{DT:fig:DT_deformation_3D_zoomin_MC} highlights the impact of the multi-channeling on the integrable soft singularity. We note however that when a deformation is enabled (and only then), the multi-channeling treatment still leaves the integrand unbounded when approaching the soft singularity from the diagonal E-surface cancelling direction. 
This remaining integrable singularity does not severely increase the variance of the integrand for the \DT{} supergraph.

\subsubsection{The \SE{} supergraph}
\label{sec:SE_example}
We now consider the \SE{} supergraph. %
The main novelty for the \SE{} supergraph compared to the \DT{} supergraph is the presence of a repeated propagator, which requires special treatment since both of them are put on-shell by the Cutkosky cut.
We describe our general treatment of the self-energy renormalisation in sect.~\ref{sec:selfenergy_treatment_in_LU} which addresses this problem by considering a combination of the on-shell scheme renormalisation conditions together with a Taylor expansion of the self-energy around its on-shell point. 
We observe that the self-energy insertion does not require a complex contour deformation, since the pairwise cancellation of \emph{all} E-surfaces of the self-energy corrections present on either side of the Cutkosky cut in $\mathbf{c}_{\mathbf{s}_{1}
^{\text{v}}}$ (fig.~\ref{fig:Two_SG_epemddx_NLO:SE_LxBstar}) and $\mathbf{c}_{\mathbf{s}_{2}
^{\text{v}}}$ (fig.~\ref{fig:Two_SG_epemddx_NLO:SE_BxLstar}) can never be spoiled by the observable function since it has the same kinematic dependence for both cuts. That is:
\begin{equation}
\mathcal{O}_{\mathbf{s}_{1}
^{\text{v}}}\big(\{q_e\}_{e\in \mathbf{c}_{\mathbf{s}_{1}
^{\text{v}}}}\big) = 
\mathcal{O}_{\mathbf{s}_{2}
^{\text{v}}}\big(\{q_e\}_{e\in \mathbf{c}_{\mathbf{s}_{2}
^{\text{v}}}}\big)  \ \forall (\vec{k},\vec{l}).
\end{equation}
This property holds for all self-energy corrections at any order whose external propagators are cut by a Cutskosky cut.

In this section, we explicitly derive the local and integrated UV counterterm of the self-energy routed as in fig.~\ref{fig:Two_SG_epemddx_NLO:SE_LxBstar} together with the corresponding renormalisation contribution.
Similarly to our treatment of the triangle loop of the \DT{} supergraph in~\ref{eq:DT_complete_Itriangle_loop}, we start by writing the factorising of the \emph{bare} self-energy contribution from the rest of the \SE{} supergraph as follows:
\begin{eqnarray}
    I^{(\textrm{bare})}_{\mathbf{s}_{1}
^{\text{v}}} &=& 
    K^{\left(\mathbf{s}_{1}
^{\text{v}},\textrm{left}\right)}_i \ 
    \frac{\mathrm{i}\slashed{k}_{ix}}{k^2}\Sigma_{x j} \ 
    K^{\left(\mathbf{s}_{1}
^{\text{v}},\textrm{right}\right)}_j
    \label{eq:SE_complete_bubble_loop}\\
    \Sigma(k,l)  &=& (-\mathrm{i} g_s) (\mathrm{i}) (-\mathrm{i} g_s )(-\mathrm{i} ) C_F \gamma^\mu 
    \left[ \frac{\slashed{l}}{l^2 (l-k)^2} \right] \gamma^\mu.
\end{eqnarray}
The propagator $\frac{\slashed{k}_{ix}}{k^2}$ is put exactly on-shell by the Cutkosky cut $\mathbf{c}_{\mathbf{s}_{1}
^{\text{v}}}$ and we show in sect.~\ref{sec:selfenergy_treatment_in_LU} that as a result the treatment of the bare self-energy $I^{(\textrm{bare})}_{\mathbf{s}_{1}^{\text{v}}}$ and its corresponding mass renormalisation counterterm $I^{(\delta m)}_{\mathbf{s}_{1}
^{\text{v}}}$, we obtain the following integrand:
\begin{eqnarray}
I_{\mathbf{s}_{1}
^{\text{v}}} &:=& 
I^{(\textrm{bare})}_{\mathbf{s}_{1}^{\text{v}}}
+I^{(\delta m)}_{\mathbf{s}_{1}
^{\text{v}}} \nonumber\\
I_{\mathbf{s}_{1}^{\text{v}}} &=&
K^{\left(\mathbf{s}_{1}
^{\text{v}},\textrm{left}\right)}
\frac{\mathrm{i}\slashed{k}}{k^0+\lVert\vec{k}\rVert} 
\left[ \frac{ \partial }{\partial k^0} \Sigma(k,l) \right]
K^{\left(\mathbf{s}_{1}^{\text{v}},\textrm{right}\right)} 
\label{eq:I_SE_explicit_expression}\\
\frac{ \partial }{\partial k^0} \Sigma(k,l) &=& (-\mathrm{i})g_s^2 C_F \gamma^\mu I^{\text{se}}\gamma^\mu \label{eq:SE_partial_Sigma_def}\nonumber\\
I^{(\text{se})}&=& \frac{\partial}{\partial k^0} \frac{\slashed{l}}{l^2 (l-k)^2}
= \frac{ 2 \slashed{l}(l^0-k^0) }{l^2 \left[(l-k)^2\right]^2},
\label{eq:SE_partial_derivative_complete_bubble_loop}
\end{eqnarray}
which completes the description of the integrand $I_{\mathbf{s}_{1}
^{\text{v}}}$ of the Local Unitarity representation for the cut $\mathbf{s}_{1}
^{\text{v}}$ of the SE supergraph.
A key aspect of our construction is to perform the on-shell expansion around $k^0-\lVert\vec{k}\rVert$ in eq.~\eqref{eq:I_SE_explicit_expression} of \emph{only} the factorised self-energy energy $\Sigma_{i j}(k,l)$ and \emph{not} the complete integrand, as it would then include the observable function which depends on the external momentum of the self-energy.

The expression derived here is different from the one considered in ref.~\cite{Soper:1999xk} where the dispersive representation of the self-energy was used in order to \emph{locally} extract its $k^2$ dependence necessary for regulating the on-shell propagator. 
We find the Taylor expansion approach preferable as it is better suited to manifestly accommodate on-shell renormalisation conditions while at the same time being systematic and suitable for automation.

The local UV counterterm is easily obtained from eq.~\eqref{eq:SE_partial_derivative_complete_bubble_loop}, since its superficial degree of divergence of zero implies that only the leading term of the UV expansion needs to be retained while keeping the original numerator unexpanded:
\begin{equation}
    \overline{I}^{(\textrm{se})}=\frac{ 2 \slashed{l}(l^0-k^0) }{(l^2-M_{\textrm{UV}}^2)^3} \,. \label{eq:SE_local_UVCT}
\end{equation}
It is important to build the UV counterterm only after the on-shell Taylor expansion has been constructed, since it does not commute with the UV expansion.

In appendix~\ref{app:UV_SE} we provide the details of the computation of the integrated counterpart of the local UV counterterm of eq.~\eqref{eq:SE_local_UVCT}. In order to write the result with the LO contribution factorised, we sum the two integrated UV counterterms contributions arising from \emph{both} self-energy loops  to the left ($\mathbf{c}_{\mathbf{s}_{1}}$) and right ($\mathbf{c}_{\mathbf{s}_{2}}$) of the Cutkosky cuts:
\begin{equation}
\Delta \overline{I}^{(\textrm{se})}_{\mathbf{s}_{1}
^{\text{v}}}+\Delta \overline{I}^{(\textrm{se})}_{\mathbf{s}_{2}
^{\text{v}}} =
\frac{\alpha_s }{\pi} \left(
\frac{C_F}{4} \Big[ -\frac{1}{\epsilon} + 1 - \log_{M_{\textrm{UV}}} +\mathrm{o}(\epsilon) \Big]
\right)I^{(\textrm{LO})} \,. \label{eq:SE_complete_DeltaUV_main}
\end{equation}
We then add the renormalisation contribution $\delta \overline{I}^{(\textrm{se})}_{q}$ stemming from the same wavefunction renormalisation constant $Z_q^{(\overline{\textrm{MS}})}$ already given in~\eqref{eq:delta_Zdown_MSbar}:
\begin{eqnarray}
&&\Delta \overline{I}^{(\textrm{se})}_{\mathbf{s}_{1}
^{\text{v}}}+\Delta \overline{I}^{(\textrm{se})}_{\mathbf{s}_{2}
^{\text{v}}}
+\delta \overline{I}^{(\textrm{se})}_{q}
=
\frac{\alpha_s }{\pi} \left(
\frac{C_F}{4} \Big[ -\frac{1}{\epsilon} + 1 - \log_{M_{\textrm{UV}}} \Big]
- \delta Z_q^{(\overline{\textrm{MS}})}
\right)I^{(\textrm{LO})} \nonumber\\
&=&
I^{(\textrm{LO})} \frac{\alpha_s }{\pi} C_F \left(
\frac{1}{4} \Big[ -\frac{1}{\epsilon} + 1 - \log_{M_{\textrm{UV}}} \Big]
+\frac{1}{4}\left[ \frac{1}{\epsilon} \right]
\right)
=
\frac{\alpha_s}{\pi} I^{(\textrm{LO})} \frac{1}{4} C_F
\left( - \log_{M_{\textrm{UV}}} + 1 \right) \,.\label{eq:SE_Final_UV_contribution}
\end{eqnarray}
We stress that we must consider the wavefunction renormalisation in the $\overline{\textrm{MS}}$ scheme and \emph{not} in the on-shell scheme. Our procedure still realises renormalisation in the on-shell scheme, in virtue of our particular choice for the self-energy on-shell expansion given in eq.~\eqref{eq:general_UV_regularised_SE}.

When combining the integrated UV counterterms and renormalisation of eq.~\eqref{eq:DT_Final_UV_contribution} for the triangle loop appearing in each of the \emph{two} virtual Cutkosky cuts of the \DT{} supergraph as well as eq.~\eqref{eq:SE_Final_UV_contribution} for \emph{both} isomorphic \SE{} supergraphs, we find the overall contribution from the UV regularisation to be $\left( -\frac{1}{2} \right) C_F \frac{\alpha_s }{\pi} I^{(\textrm{LO})}$.

The cancellation of the logarithmic contributions $\log_{M_{\textrm{UV}}}$ results directly from the gauge cancellation between the vertex and quark wavefunction corrections.
It also implies that the corresponding NLO QCD correction to $e^+ e^- \rightarrow d \bar{d}$ has no residual dependency in the renormalisation scale as expected, since its Born contribution does not depend on $\alpha_s(\mu_r^2)$.

\subsubsection{Results for the differential cross-section}
\label{sec:NLO_example_numerical_results}
For all numerical results and visualisations presented in this section, we choose $Q^\mu = p_1^\mu+p_2^\mu=(1,0,0,1)+(1,0,0,-1)$, $Q_d=-\frac{1}{3}$, $Q_e=-1$, $g_{\textrm{EW}}=0.307954$ and $g_{\textrm{s}}=1.21772$. The choice of the renormalisation scale is irrelevant for the process $e^- e^+ \rightarrow \gamma^\star \rightarrow d \bar{d}$ studied in this section, but matters for the contribution of the individual \SE{} and \DT{} supergraph which we report here as well. We therefore specify that we set $\mu_r$ equal to our choice of $M_{\textrm{UV}}=\frac{Q^0}{2}=1$, such that the integrated UV counterterms in eq.~\eqref{eq:DT_Final_UV_contribution} and eq.~\eqref{eq:SE_Final_UV_contribution} feature no logarithmic terms.

We present our results in table~\ref{tab:DT_NLO_example_results} for the inclusive and semi-inclusive cross-section within the acceptance cuts of eq.~\eqref{eq:DT_PS_cut}. When a deformation is enabled, we keep the defaults for the deformation of ref.~\cite{Capatti:2019edf} with an overall maximal deformation magnitude scaling $\lambda_\textrm{max}=1$, together with its dampening on infrared singularities described in sect.~\ref{sec:contour_deformation}.
\begin{table}[ht!]
\centering
\texttt{
\begin{tabular}{lccrclllcc}
\hline
\multirow{2}{*}{Contrib.} & 
\multirow{2}{*}{C.D.} & 
\multirow{2}{*}{$\mathtt{N_{\text{p}} \ [10^6]}$} & 
\multicolumn{2}{c}{$\mathtt{\sfrac{t}{p} \ [\mu \text{s}]}$} & 
\multirow{2}{*}{{\sc\small MG5aMC} $[\text{pb}]$} &
\multirow{2}{*}{\alphaLoop\ $[\text{pb}]$} &
\multirow{2}{*}{$\mathtt{\Delta \ [\sigma]}$} &
\multirow{2}{*}{$\mathtt{\Delta \ [\%]}$} \\
& & & min & avg \\
\hline
\multicolumn{9}{c}{Inclusive cross-section}\\
\hline
LO & No & 1035 & 0.06 & 6 & 7741.102(5) & 7741.133(32) & 1.0 & 0.04 \\
SE & No & 2485 & 34 & 240 & N/A  & -91.7966(95) & N/A & 0.01 \\
DT & Yes & 780 & 72 & 658 & N/A & 474.34(11) & N/A & 0.02 \\
DT & No & 780 & 37 & 202 & N/A & 474.303(30) & N/A & 0.006 \\
2$\times$SE$+$DT & No & 1035 & 108 & 752  & 290.768(6) & 290.732(28) & 1.3 & 0.01 \\
2$\times$SE$+$DT & Yes & 1035 & 124 & 1290 & 290.768(6) & 290.78(11) & 0.1 & 0.04 \\
\hline
\multicolumn{9}{c}{$\mathtt{0.4<p_t(j_1)<0.8}$}\\
\hline
LO & No & 1035 & 0.06 & 20 & 2909.53(14) & 2909.520(30) & 0.07 & 0.005 \\
SE & No & 780 & 25 & 304 & N/A & 3.624(11) & N/A & 0.3 \\
DT & Yes & 300 & 19 & 632 & N/A & 552.25(18) & N/A & 0.03 \\
2$\times$SE$+$DT & Yes & 1035 & 48 & 830 & 559.41(72) & 559.38(21) & 0.04 & 0.13 \\
\hline
\end{tabular}
}
\caption{\label{tab:DT_NLO_example_results} Numerical results for the NLOQCD accurate (semi-)inclusive cross-section for the process $e^+ e^- \rightarrow \gamma^\star \rightarrow d \bar{d}$. The second column entitled ``C.D.'' stands for ``Contour Deformation''.}%
\end{table}

The timing indicated by {$\mathtt{\sfrac{t}{p}}$ \texttt{min} corresponds to the single-core time spent in the {\texttt{C}} routine for the double-precision evaluation of each numerator involved per sampling point. In this sense, this is the minimal time that one evaluation could possibly take. The timing indicated by {\texttt{avg}} corresponds to the effective actual timing per sampling point, including the overhead from the integrator, observable functions, the deformation construction, and most importantly additional processing in \alphaLoop{} such as the stability tests that necessitate additional evaluations in double and sometimes quadruple precision arithmetic (see sect~\ref{sec:technical_numerical_stability}).
At LO, the timing is dominated by the overhead in \alphaLoop{}, which explains the significant increase when enabling the phase-space cut selection code.
Both timings reported are for one sampling point \emph{including the multi-channeling treatment}, which in the case of the process $e^+ e^- \rightarrow \gamma^\star \rightarrow d \bar{d} g$ implies three independent evaluation of the integrands for each of its three channels integrated together.
The imaginary part of the integrated cross-section is not reported but was found to always be compatible with zero given the Monte-Carlo uncertainty.
The comparison of the result with and without a contour deformation enabled (only when the latter is possible) shows a moderate impact of a factor of about three in the variance of the integrand.
The increase in the runtime per sample when enabling the deformation is mostly due to the increase in fraction of unstable point that require a quadruple precision rescue. Note that since the heuristics described in sect.~\ref{sec:contour_deformation} for the computation of the deformation centres always apply for this simple process, there is no need for numerically solving a second-order cone program at run-time and therefore not a lot of time is spent in constructing the deformation.
As can be seen in tab.~\ref{tab:DT_NLO_example_results}, the infamous NLO K-factor of $\alpha_s/\pi$ (for entry 2$\times$SE$+$DT) is very well reproduced by our implementation of numerical Local Unitarity.

We complement the (semi-)inclusive results of tab.~\ref{tab:DT_NLO_example_results} with the binned distribution of the transverse momentum of the leading reconstructed jet in fig.~\ref{fig:NLO_differential_pt_validation}. This observable is of little interest in the case of lepton collisions but was chosen to demonstrate the applicability of Local Unitarity to arbitrary observables.
To the best of our knowledge, this result of humble appearance stands as the first NLOQCD accurate differential binned distribution ever computed without relying on IR counterterm and/or dimensional regulator.

\begin{figure}[t]
\centering
\includegraphics[width=1.0\linewidth]{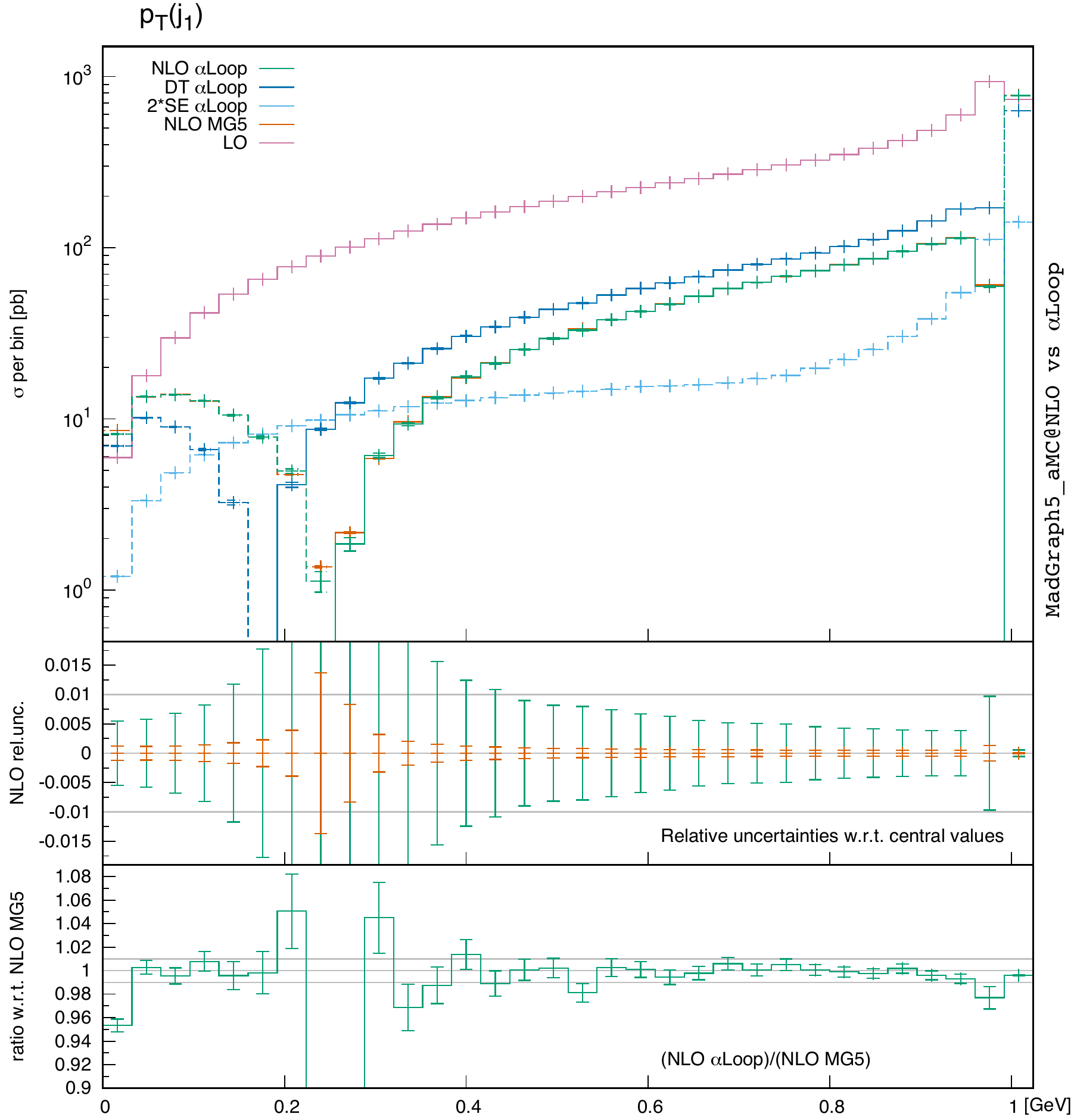}
\caption{\label{fig:NLO_differential_pt_validation}
Comparison of the differential prediction for NLOQCD correction to the observable $p_t(j_1)$ for the process $e^+ e^- \rightarrow \gamma^\star \rightarrow d \bar{d}$ as obtained from {\sc\small MadGraph5\_aMC@NLO} and \alphaLoop\ . The Monte-Carlo sampling for this run with \alphaLoop\ is $5 \cdot 10^9$ points. The dashed pattern indicates when a contribution is negative. 
}
\end{figure}

Overall, we find excellent agreement between \alphaLoop{} and {\sc\small MadGraph5\_aMC@NLO} in both tab.~\ref{tab:DT_NLO_example_results} and fig.~\ref{fig:NLO_differential_pt_validation}.
In particular, discrepancies are always within a couple of Monte Carlo standard deviations, with a \emph{relative} accuracy of the NLO correction always below the percent mark, except for the region $0.2<p_t(j_1)<0.4$ where the distribution changes sign and therefore neighbours zero.
These results and especially the reliability of the Monte Carlo accuracy reported, are indicative of the predictive power of our numerical implementation of Local Unitarity.

\subsection{Individual scalar supergraphs}

An important verification of the local properties of the LU representation is by testing whether the IR cancellation pattern is realised in practice in a Monte Carlo integration.%
The most direct test consists of integrating the overall contributions obtained by summing together all the interference diagrams corresponding to a fixed, higher-loop scalar supergraph $\Gamma$ in the LU representation, $\sigma_\Gamma=d\sigma_\Gamma/d\mathcal{O}$ with constant observable.
Scalar graphs do not have a numerator that could improve the IR behaviour and are not helped by gauge invariance or any property specific to physical theories in order to realise IR cancellations.

In table~\ref{tab:scalar_supergraphs}, we enumerate the massless scalar supergraphs whose LU representation can be numerically integrated using a Monte Carlo procedure.
We divide these scalar supergraphs in three series: the diagrams labelled by $\mathtt{a}$ correspond to all of the three-loop supergraphs whose $\sigma_\Gamma$ is non-zero and has no self-energy insertions; the diagrams labelled by $\mathtt{b}$ correspond to all of the four-loop supergraphs whose $\sigma_\Gamma$ is finite, non-zero and has no self-energy insertions; and we selected three five-loop supergraphs, labelled by $\mathtt{c}$.

The integration results, obtained via our implementation named \alphaLoop{}, are exhibited in table~\ref{tab:scalar_results}, along with the analytic result obtained through {\sc forcer}~\cite{Ruijl:2017cxj}. We also report the total number of sample points used for the integration, $\mathtt{N_{\text{p}}}$, the minimum evaluation time per sample and per core (that is, the time for one single evaluation of the integrand in double precision on one Intel(R) Xeon(R) Gold 6136 CPU @ 3.00GH core) and the average evaluation time per sample, which includes the sum over all integration channels, the numerical stability checks and the dynamic stability rescue mechanisms.  We also report the number of integration channels $\mathtt{N_{\text{ch}}}$ and the ratio $\mathtt{\Delta \ [\sigma]}$ of the discrepancy w.r.t the exact result with the Monte Carlo error, and the relative size $\mathtt{\Delta \ [\%]}$ of this discrepancy.

We find that all numerical integrations yield results that are within three sigma of the analytic result with Monte Carlo errors at the percent level or better. Overall, the results undoubtedly confirm that the integrands are stable and well-behaved. We observe that the $\mathtt{b.16}$ supergraph requires double the sampling points to achieve a relative error below 2\%.

The differences in the evaluation time across the topologies considered, both minimal and average, is related to a variety of factors, including the number of interference diagrams arising from the supergraph and the number of integration channels. Another determining factor is the overall numerical stability of the integrand: indeed, a worse stability implies that a larger fraction of sampling points trigger the numerical stability rescue mechanism, which significantly slows does the evaluation of the integrand which is then performed using quadruple precision arithmetics.

\clearpage
\begin{table}[ht!]
\vspace{-1.0cm}
\begin{center}
\begin{tabular}{lclclclclc}
a.1) & \input{Figures/scalar_diagrams/STF_3L_0} &
a.2) & \input{Figures/scalar_diagrams/STF_3L_1} &
a.3) & \input{Figures/scalar_diagrams/STF_3L_2} &
b.1) & \input{Figures/scalar_diagrams/STF_4L_0} &
b.2) & \input{Figures/scalar_diagrams/STF_4L_1} \\
b.3) & \input{Figures/scalar_diagrams/STF_4L_2} &
b.4) & \input{Figures/scalar_diagrams/STF_4L_3} &
b.5) & \input{Figures/scalar_diagrams/STF_4L_4} &
b.6) & \input{Figures/scalar_diagrams/STF_4L_5} &
b.7) & \input{Figures/scalar_diagrams/STF_4L_7} \\
b.8) & \input{Figures/scalar_diagrams/STF_4L_8} &
b.9) &\input{Figures/scalar_diagrams/STF_4L_9} &
b.10) & \input{Figures/scalar_diagrams/STF_4L_10} &
b.11) & \input{Figures/scalar_diagrams/STF_4L_11} &
b.12) & \input{Figures/scalar_diagrams/STF_4L_12} \\
b.13) & \input{Figures/scalar_diagrams/STF_4L_14} &
b.14) & \input{Figures/scalar_diagrams/STF_4L_15} &
b.15) & \input{Figures/scalar_diagrams/STF_4L_16} &
b.16) & \input{Figures/scalar_diagrams/STF_4L_17} &
c.1) & \input{Figures/scalar_diagrams/STF_5L_0} \\
c.2) & \input{Figures/scalar_diagrams/STF_5L_1} &
c.3) & \input{Figures/scalar_diagrams/STF_5L_3} &
 &  \\
\end{tabular}
\caption{\label{tab:scalar_supergraphs}List of scalar supergraphs whose corresponding cross-section contribution is computed in~\ref{tab:scalar_results}. They are divided in three series: a) 3-loop supergraphs, b) 4-loop supergraphs, c) 5-loop supergraphs. Grey edges and black edges do not intersect each other, but overlap due to the planar embedding.}%
\end{center}

\centering
\texttt{
\begin{tabular}{lcrrrlllll}
\hline
\multirow{2}{*}{$\Gamma$} & 
\multirow{2}{*}{$\mathtt{N_{\text{p}} \ [10^6]}$} & 
\multicolumn{2}{c}{$\mathtt{\sfrac{t}{p} \ [\mu \text{s}]}$} &
\multirow{2}{*}{$\mathtt{N_{\text{ch}}}$} &
\multirow{2}{*}{{\sc\small forcer} $[\text{GeV}^2]$} &
\multirow{2}{*}{\alphaLoop\ $[\text{GeV}^2]$} &
\multirow{2}{*}{exp.} &
\multirow{2}{*}{$\mathtt{\Delta \ [\sigma]}$} &
\multirow{2}{*}{$\mathtt{\Delta \ [\%]}$} \\
& & min & avg  \\
\hline
\multicolumn{9}{c}{Inclusive cross-section per supergraph}\\
\hline
a.1 & 1 & 5 & 450 & 16 & \phantom{-}5.75396 & \phantom{-}5.7530(46) & -6 & 0.21 & 0.00017 \\
a.2  & 1 & 10 & 690 & 16 & -5.75396 & -5.763(11) & -6 & 0.82 & 0.0016 \\
a.3  & 1 & 25 & 1400 & 16 & -5.75396 & -5.771(23)  & -6 & 0.74 & 0.0039 \\
\hline
b.1  & 1 & 150  & 6600  & 45 &  -1.04773 & -1.0459(23)  & -7 & 0.79 & 0.0017 \\
b.2  & 1 & 270  & 39000  & 45 &  -1.04773 & -1.0457(21)  & -7 & 0.97 &0.0029 \\
b.3  & 1 & 320  & 52000  & 81 &  -1.04773 & -1.0448(21)  & -7 & 1.4 & 0.0028 \\
b.4  & 1 & 740  & 96000  & 75 &  -1.04773 & -1.0455(22)  & -7 & 1.0 & 0.0021 \\
b.5  & 1 & 340  & 20000  & 45 &  -1.04773 & -1.0441(23)  & -7 & 1.6 & 0.0035 \\
b.6  & 1 & 350  & 12000  & 45 &  -1.04773 & -1.0434(26)  & -7 & 1.7 & 0.0042 \\
b.7  & 1 & 1800  & 180000  & 81 &  -1.04773 & -1.0563(51)  & -7 & 1.7 & 0.0081 \\
b.8  & 1 & 1400 & 120000  & 75 &  -1.04773 & -1.0526(42)  & -7 & 1.2 & 0.0046 \\
b.9  & 1 & 1200  & 36000  & 45 &  -1.04773 & -1.0439(27)  & -7 & 1.4 & 0.0037 \\
b.10  & 1 & 1100  & 32000  & 45 &  -1.04773 & -1.0488(29)  & -7 & 0.37 & 0.0010 \\
b.11  & 1 & 1100  & 54000  & 45 &  -1.04773 & -1.0516(35)  & -7 & 1.1 &0.0037 \\
b.12  & 1 & 1100  & 30000  & 45 &  -1.04773 & -1.0473(30)  & -7 & 0.14 & 0.00041 \\
b.13  & 1 & 2700  & 83000  & 45 &  -1.04773 & -1.040(15)  & -7 & 0.51 & 0.0074 \\
b.14  & 1 & 3100  & 110000  & 75 &  -2.09546 & -2.123(12)  & -7 & 2.3 & 0.0130 \\
b.15  & 1 & 3100  & 210000  & 81 &  -2.09546 & -2.1045(67)  & -7 & 1.3 & 0.0043 \\
b.16  & 2 & 1800  & 120000  & 75 &  -5.23865 & -5.312(65)  & -8 & 1.1 & 0.014 \\
\hline
c.1  & 1 & 1100   & 49000   & 128 & \phantom{-}1.66419 & \phantom{-}1.6691(79)   & -9 & 0.62  & 0.0029 \\
c.2  & 1 & 900  & 46000  & 130 & \phantom{-}1.77832 & \phantom{-}1.7752(71)  & -9 & 0.44  & 0.0018 \\
c.3  & 1 & 1600  & 69000  & 130 & \phantom{-}1.77832  & \phantom{-}1.7797(33)  & -9 & 0.42 & 0.00077 \\
\hline
\end{tabular}
}
\caption{\label{tab:scalar_results} Values of $\sigma_\Gamma$ (that is, the integrated sum of all interference diagrams arising from $\Gamma$, or inclusive cross-section per supergraph) for the supergraphs listed in~\ref{tab:scalar_supergraphs} computed with \alphaLoop{} for $Q^\mu=(1,\vec{0})$ and compared with benchmark results from {\sc\small forcer}~\cite{Ruijl:2017cxj}. The table includes the total number of samples used for each integration, $\mathtt{N_p}$, the number of integration channels $\mathtt{N_{ch}}$ (see sect.~\ref{sec:technical_multi_channeling}), the minimum and average per sample evaluation time of the integrand $\mathtt{t/p}$, and the relative discrepancy with respect to the standard deviation $\mathtt{\Delta \ [\sigma]}$ and numerical result $\mathtt{\Delta \ [\%]}$.}%
\end{table}
\clearpage

\newpage
\section{Conclusion}
\label{sec:conclusion}
In this paper, we have presented a general and systematic way to construct a representation of differential cross-sections that is locally free of IR divergences. Given an observable and scattering process, the procedure yields an integrand free of (non-integrable) singularities and whose integral reproduces physical differential cross-sections. We call the new expression of the differential cross-section its \emph{Local Unitarity} (LU) representation. Our framework ultimately achieves two objectives: first, it shows that the appearance of infrared singularities is a feature of traditional calculation methods and not a fundamental attribute of physical theories (a statement already implicitly contained in the KLN theorem) and second, it provides a concrete method well-suited for an automated and fully numerical computation of physical observables.

Achieving both of these objectives inherently requires an in-depth study of the singular structure of Feynman integrals, of phase space integrals and their relationship. We use the Loop-Tree Duality (LTD framework) to express loop integration on the same footing as phase space integration.
Once amplitudes are re-expressed using their LTD representation, a clear picture emerges: all the interference diagrams that correspond to a particular Cutkosky cut of the same forward scattering topology (i.e. supergraph) can be collected under the same integral sign, and their sum is finite. 
This principle effectively establishes how to collect and group interference diagrams into classes that are infrared finite. 
We even identify that cancellations between IR singularities happens pairwise. More specifically, if an interference diagram has an IR singularity, there exists a unique distinct other interference diagram which exhibits the exact same IR singularity so that they sum up to an integrable quantity on the location of said singularity.

One of the core issues in constructing an integrand that realises IR cancellation locally while simultaneously yielding differential cross-sections once integrated, is that of aligning the phase space measures. Part of this task is performed through the LTD formalism, along with the choice of a consistent loop momentum across all interference diagrams corresponding to the same supergraph. 
The other fundamental component allows one to solve simultaneously all of the Dirac delta functions imposing the energy conservation of on-shell external particles: the causal flow. The simplest form of this flow amounts to a trivial rescaling of all loop variables, but its most generic solution naturally follows from the general solution to a contour deformation encoding the Feynman prescription associated to each propagator. 
The causal flow is an essential ingredient to the construction of the function whose integral yields the physical cross-section, and is used for the all-order proof of local cancellations of final-state IR divergences in the LU representation.
\vSWITCH{
}{%
A corollary of our investigation reveals how gauge invariance, and more specifically the IR behaviour of diagrams featuring ghosts, together with the LSZ formalism accommodate IR cancellations in external self-energy corrections.
}%

Aligning the integration measures also provides a solution to the problem of automating the calculation of differential cross-sections, as it allows for the systematic construction of an integrand especially well-suited for Monte Carlo integration. In order to corroborate and demonstrate the workings of the LU representation in practical cases, we provided the results of the numerical integration of the LU representation for a variety of example scalar supergraphs. Furthermore, we explicitly unfolded our theoretical construction for the NLO correction to the differential cross-section of the physical process $e^+ e^-\rightarrow d \bar{d}$.

This paper opens up new horizons both in the understanding of the analytical properties of observable quantities calculated perturbatively in QFTs and in the ambitious goal of achieving the complete automation of the calculation of physical cross-sections for collider experiments beyond NLO accuracy. It goes without saying that, ultimately, these two goals are inextricable, and can potentially lead to a radically different understanding of QFTs as we know them.

We expect the foundational work laid out in this paper to serve as a stepping stone for its many implications, both regarding the theoretical understanding of initial and final state infrared singularities and its concrete applications to the computation of higher-order perturbative corrections for collider phenomenology.

\section{Acknowledgements}

We thank Armin Schweitzer for useful input and exchanges regarding the UV treatment in Local Unitarity.
We also thank Charalampos Anastasiou, Stefano Frixione and Dario Kermanschah for fruitful discussions. 
We are grateful to Davison Soper for his attentive reading of our work and insightful comments. 
This project has received funding from the European Research Council (ERC) under grant agreement No 694712 (PertQCD) and SNSF grant No 179016.

\clearpage
\appendix

\section{Integrated UV counterterms for the NLO correction to $e^+ e^- \rightarrow d \bar{d}$}

\subsection{Double-triangle supergraph}
\label{app:UV_DT}

In this section, we explicitly derive the result given in~\eqref{eq:DT_DeltaUV_main}. We first drop the term linear in $k$ that integrates to zero and arrive at:
\begin{eqnarray}
\Delta\bar{I}^{\mu} &=& \int \frac{\tilde{d} k}{(2\pi)^4} \frac{(d-2) \left( k^2 \gamma^\mu -2 k^\mu \slashed{k}\right) }{\left( k^2-M_{\textrm{UV}}^2\right)^3} \nonumber \\
&=& \left [ 2 \gamma^\mu g^{\rho_1 \rho_2} -4 \gamma^{\rho_2} g^{\mu \rho_1} \right ] (d-2) \int \frac{\tilde{d} k}{(2\pi)^4} \frac{k_{\rho_1}k_{\rho_2}}{\left( k^2-M_{\textrm{UV}}^2\right)^3} \label{eq:UVDT_tensor_decomp}.
\end{eqnarray}
The tadpole tensor integral in~\eqref{eq:UVDT_tensor_decomp} can be evaluated using the usual decomposition in the only manifestly Lorentz invariant structure $g^{\rho_1\rho_2}$:
\begin{equation}
\int \frac{\tilde{d} k}{(2\pi)^4} \frac{k^{\rho_1}k^{\rho_2}}{\left( k^2-M_{\textrm{UV}}^2\right)^3} = A g^{\rho_1\rho_2}\ ,\ A = \frac{1}{d} \int \frac{\tilde{d} k}{(2\pi)^4} \frac{k^2}{\left( k^2-M_{\textrm{UV}}^2\right)^3}, \label{eq:DT_tensor_integral_result}
\end{equation}
followed by an elementary reduction step:
\begin{eqnarray}
A&=&\frac{1}{d} \int \frac{\tilde{d} k}{(2\pi)^4} \frac{ k^2 }{\left[ k^2-M_\textrm{UV}^2\right]^3} \nonumber\\
&=&\frac{1}{d} \left [ 
\int\frac{\tilde{d} k}{(2\pi)^4} \frac{ 1 }{\left[ k^2-M_\textrm{UV}^2\right]^2} 
-(-M_\textrm{UV}^2) \int \frac{\tilde{d} k}{(2\pi)^4} \frac{ 1 }{\left[ k^2-M_\textrm{UV}^2\right]^3}
\right ],
\end{eqnarray}
together with the general expression for a scalar tadpole integral (see for instance eq.~(3) of ref.~\cite{Steinhauser:2000ry}):
\begin{equation}
V^{(d)}_{a} := \int \frac{d^d k}{(2\pi)^d} \frac{ 1 }{\left[ k^2+M^2\right]^a} =\frac{(M^2)^{\frac{d}{2}-a}}{(4\pi)^\frac{d}{2}}\frac{\Gamma(a-\frac{d}{2})}{\Gamma(a)} \label{eq:tadpole_volume_integral},
\end{equation}
eventually yielding:
\begin{eqnarray}
A&=&\frac{1}{4-2\epsilon}\left(\frac{\mu^2}{4\pi e^{-\gamma}}\right)^{\epsilon} \frac{(-M_\textrm{UV}^2)^{\frac{4-2\epsilon}{2}-2}}{(4\pi)^\frac{d}{2}} \left [ 
\frac{\Gamma(2-\frac{4-2\epsilon}{2})}{\Gamma(2)} - \frac{\Gamma(3-\frac{4-2\epsilon}{2})}{\Gamma(3)}
\right] \nonumber \\
&=&\frac{1}{4-2\epsilon} \frac{1}{(4\pi)^2}\left(\frac{\mu^2}{-M_\textrm{UV}^2 e^{-\gamma}}\right)^{\epsilon}\left [ 
\frac{\Gamma(\epsilon)}{\Gamma(2)} - \frac{\Gamma(1+\epsilon)}{\Gamma(3)}
\right] 
\end{eqnarray}
which we can expand to order $\mathrm{o}(\epsilon)$, arriving at:
\begin{equation}
A=\frac{1}{64\pi^2}\Big[ \frac{1}{\epsilon} + \underbrace{\log{\frac{\mu^2}{-M_\textrm{UV}^2}}}_{:=\log_{M_{\textrm{UV}}}} +\mathrm{o}(\epsilon) \Big],\label{eq:NLO_example_A_factor}
\end{equation}
where we introduced the shorthand notation $\log_{M_{\textrm{UV}}}$ for the omnipresent logarithm involving the UV mass regulator $M^2_{\textrm{UV}}$ since the tadpole nature of all integrated counterterms is such that it will always appear together with $\mu_r^2$ and never any other scale.

The presence of the negative sign in $-M_\textrm{UV}^2$ calls for a regulator in order to set the sign of the imaginary part of the logarithm. The local UV counterterm of~\eqref{eq:DT_UV_local_CT} will be cast into its (c)LTD representation while assuming a causal Feynman prescription $+\mathrm{i}\delta$ in the denominator of the UV propagator. This choice dictates what is the correct branch to select for the evaluation of $\log_{M_{\textrm{UV}}}$. For this process at NLO we note that the particular sign of this imaginary part is irrelevant as it would always cancel between the two integrated counterterms arising from the loop integrals sitting to the left and right of the Cutkosky cut, since in the latter case the complex conjugation flips the sign of the causal prescription.

Substituting~\eqref{eq:NLO_example_A_factor} in~\eqref{eq:DT_tensor_integral_result} gives:
\begin{eqnarray}
\Delta\bar{I}^{\mu}(l,Q) &=& \left [ \gamma^\mu g^{\rho_1 \rho_2} -2 \gamma^{\rho_2} g^{\mu \rho_1} \right ] g_{\rho_1\rho_2} (d-2) \left[ \frac{1}{64\pi^2} \left(\frac{1}{\epsilon} + \log_{M_{\textrm{UV}}} + \mathcal{O}\left(\epsilon\right) \right) \right] \nonumber \\
&=& \gamma^\mu  \frac{1}{64 \pi^2} \left(\frac{1}{\epsilon} + \log_{M_{\textrm{UV}}} \right) \left( d - 2 \right)^2+\mathcal{O}\left(\epsilon\right)= \gamma^\mu \frac{1}{16 \pi^2} \left( \frac{1}{\epsilon} - 2  + \log_{M_{\textrm{UV}}} + \mathcal{O}\left(\epsilon\right) \right).\nonumber\\ 
\label{eq:DT_DeltaI}
\end{eqnarray}
We can then insert~\eqref{eq:DT_DeltaI} in~\eqref{eq:DT_vertex_correction} and~\eqref{eq:DT_complete_Itriangle_loop} in order to obtain the final expression of the integrated UV counterterm $\Delta \overline{I}_{\mathbf{s}_{1}^{\text{v}}}$ of the complete DT supergraph:

\begin{eqnarray}
\Delta \overline{I}_{\mathbf{s}_1^{\text{v}}}  &=& K^{\left(\mathbf{s}_1^{\text{v}},\textrm{left}\right)}_\mu 
\left( 
g_\textrm{s}^2 C_F \left [ (-\textrm{i}) g_{\gamma d\bar{d}} \right] \gamma^\mu_{ij}\frac{1}{16 \pi^2} \left( \frac{1}{\epsilon} - 2 + \log_{M_{\textrm{UV}}} \right) 
\right) 
K^{\left(\mathbf{s}_1^{\text{v}},\textrm{right}\right)}_{i j}
\nonumber\\
&=& \underbrace{\frac{\alpha_s}{\pi}}_{\textrm{K-factor}}
\left[
K^{\left(\mathbf{s}_1^{\text{v}},\textrm{left}\right)}_\mu 
\underbrace{ (-\mathrm{i})  g_{\gamma d\bar{d}} \gamma^\mu }_{\textrm{LO form-factor}} 
K^{\left(\mathbf{s}_1^{\text{v}},\textrm{right}\right)}
\right]
\frac{1}{4} C_F \left( \frac{1}{\epsilon} - 2 + \log_{M_{\textrm{UV}}}  \right) \nonumber\\
&=& \frac{\alpha_s}{\pi} I^{(\textrm{LO})} \left[ \frac{1}{4} C_F \left( \frac{1}{\epsilon} - 2 + \log_{M_{\textrm{UV}}} \right) \right]. \label{eq:DT_DeltaUV}
\end{eqnarray}

\newpage
\subsection{Self-energy supergraph}
\label{app:UV_SE}

We provide here a detailed computation of the integrated counterpart of the local UV counterterms given in~\eqref{eq:SE_local_UVCT} for the self-energy supergraph appearing in the NLO QCD correction of the process $e^+ e^- \rightarrow d \bar{d}$.
To this end, it is useful to write it in a manifesly Lorentz invariant fashion by introducing the reference vector $\eta^\mu=(1,0,0,0,\dots)$, where the dots denote the ($d-4$)-dimensional orthogonal subspace. We find
\begin{eqnarray}
    \Delta \overline{I}^{(\textrm{se})} &=& \int \frac{\tilde{d} k}{(2\pi)^4} \frac{ 2 \slashed{l} l\cdot \eta }{(l^2-M_{\textrm{UV}}^2)^3} 
    = \gamma^\mu \eta^\nu \int \frac{\tilde{d} k}{(2\pi)^4} \frac{ 2 l^\mu l^\nu }{(l^2-M_{\textrm{UV}}^2)^3}
    =  \slashed{\eta} \frac{1}{32\pi^2}\Big[ \frac{1}{\epsilon} + \log_{M_{\textrm{UV}}} +\mathrm{o}(\epsilon) \Big], \nonumber\\
    \label{eq:SE_integrated_UV_factorised_piece}
\end{eqnarray}
where we have immediately removed the linear contribution from the numerator since it integrates to zero and recycled the expression for the tensor integral found in~\ref{eq:DT_tensor_integral_result}.
In order to obtain the complete integrated UV counterterm, we must re-insert $\Delta \overline{I}^{(\textrm{se})}$ into~\eqref{eq:SE_partial_Sigma_def} and~\eqref{eq:I_SE_explicit_expression}:
\begin{eqnarray}
    \Delta \overline{I}^{(\textrm{se})}_{\mathbf{s}_{1}
^{\text{v}}} &=& 
K^{\left(\mathbf{s}_{1}^{\text{v}},\textrm{left}\right)}
\frac{\mathrm{i}\slashed{k}}{k^0+\lVert\vec{k}\rVert} 
\Bigg[
(-\mathrm{i})g_s^2 C_F \gamma^\mu 
\left(
\slashed{\eta} \frac{1}{32\pi^2}\Big[ \frac{1}{\epsilon} + \log_{M_{\textrm{UV}}} +\mathrm{o}(\epsilon) \Bigg]
\right)
\gamma^\mu
\Big]
K^{\left(\mathbf{s}_{1}^{\text{v}},\textrm{right}\right)} \nonumber\\
&=&
K^{\left(\mathbf{s}_{1}^{\text{v}},\textrm{left}\right)}
\frac{\slashed{k}\slashed{\eta}}{k^0+\lVert\vec{k}\rVert} 
\left(
\frac{(2-d) g_s^2 C_F }{32\pi^2}\Big[ \frac{1}{\epsilon} + \log_{M_{\textrm{UV}}} +\mathrm{o}(\epsilon) \Big]
\right)
K^{\left(\mathbf{s}_{1}^{\text{v}},\textrm{right}\right)} \nonumber\\
&=&
K^{\left(\mathbf{s}_{1}^{\text{v}},\textrm{left}\right)}_i
\frac{\slashed{k}\slashed{\eta}}{k^0+\lVert\vec{k}\rVert} 
\frac{\alpha_s }{\pi} \left(
\frac{C_F}{4} \Big[ -\frac{1}{\epsilon} + 1 - \log_{M_{\textrm{UV}}} +\mathrm{o}(\epsilon) \Big]
\right)
K^{\left(\mathbf{s}_{1}^{\text{v}},\textrm{right}\right)}
\label{eq:SE_integreated_UV_LxB}
\end{eqnarray}
where we used the identity $\gamma^\mu \gamma^\nu \gamma^\mu=(2-d) \gamma^\nu$. In order to demonstrate the factorisation of the LO supergraph, we must consider the sum of the contributions from both Cutkosky cuts $\mathbf{c}_{\mathbf{s}_{1}^{\text{v}}}$ and $\mathbf{c}_{\mathbf{s}_{2}^{\text{v}}}$. The cut $\mathbf{c}_{\mathbf{s}_{1}^{\text{v}}}$ yields a derivation fully analogous to the one carried out in this section, with the exception that the on-shell propagator being regulated is sitting to the right of self-energy, so that one naturally arrives at the same expression as~\eqref{eq:SE_integreated_UV_LxB} but with the Lorentz structure $\slashed{\eta}\slashed{k}$ in place of $\slashed{k}\slashed{\eta}$.  We then observe that:
\begin{equation}
\frac{
\slashed{k}\slashed{\eta} + \slashed{\eta}\slashed{k}
}{
k^0+\lVert\vec{k}\rVert
} = \frac{2 k\cdot \eta}{k^0+\lVert\vec{k}\rVert} = \frac{2k^0}{k^0+\lVert\vec{k}\rVert}  = 1,
\end{equation}
where we used the fact that both Cutkosky cuts of the SE supergraph force the on-shell condition  $k^0=\lVert\vec{k}\rVert$. Using the factt hat that $K^{\left(\mathbf{s}_{1}^{\text{v}},\textrm{left}\right)}_i
K^{\left(\mathbf{s}_{1}^{\text{v}},\textrm{right}\right)}_i=
K^{\left(\mathbf{s}_{2}^{\text{v}},\textrm{left}\right)}_i
K^{\left(\mathbf{s}_{2}^{\text{v}},\textrm{right}\right)}_i
=
I^{(\textrm{LO})}$, we can finally rewrite $\Delta \overline{I}^{(\textrm{se})}_{\mathbf{s}_{1}
^{\text{v}}}+\Delta \overline{I}^{(\textrm{se})}_{\mathbf{s}_{2}
^{\text{v}}}$ as follows:
\begin{equation}
\Delta \overline{I}^{(\textrm{se})}_{\mathbf{s}_{1}
^{\text{v}}}+\Delta \overline{I}^{(\textrm{se})}_{\mathbf{s}_{2}
^{\text{v}}} =
\frac{\alpha_s }{\pi} \left(
\frac{C_F}{4} \Big[ -\frac{1}{\epsilon} + 1 - \log_{M_{\textrm{UV}}} +\mathrm{o}(\epsilon) \Big]
\right)I^{(\textrm{LO})}. \label{eq:SE_complete_DeltaUV}
\end{equation}

\clearpage
\bibliographystyle{JHEP}

\bibliography{biblio}

\end{document}